%% file: ms.tex
\documentclass[twocolumn,useAMS,usenatbib]{mn2e}
\usepackage{natbib}
\usepackage[T1]{fontenc}
\usepackage{amssymb,amsmath}
\usepackage{graphicx,color}
\usepackage{hyperref}
\usepackage{enumerate}
\usepackage{comment}
\usepackage[usenames,dvipsnames]{xcolor}
\usepackage[OT2,T1]{fontenc}
\usepackage[draft]{todonotes}
\usepackage{rotating}
\newcommand{\sersic}{S\'{e}rsic}

\newcommand{\referee}[1]{#1}
\newcommand{\newtext}[1]{#1}

\newcommand{\code}[1]{{\sc #1}}
\newcommand{\psfex}{\code{PSFEx}}
\newcommand{\sextractor}{\code{SExtractor}}
\newcommand{\galsim}{\code{GalSim}}
\newcommand{\gfit}{\code{gfit}}

\title[GREAT3 Results I]{GREAT3 results I: systematic errors in shear
  estimation and the impact of real galaxy morphology}
\author[Mandelbaum, Rowe, et al.]{
Rachel Mandelbaum$^1$\thanks{E-mail: {\tt rmandelb@andrew.cmu.edu}}, 
Barnaby Rowe$^2$\thanks{E-mail:
{\tt browe@star.ucl.ac.uk; barnabytprowe@gmail.com}}, 
Robert Armstrong$^3$,
Deborah Bard$^{4,5}$,
\newauthor
Emmanuel Bertin$^6$,
James Bosch$^3$,
Dominique Boutigny$^{5,7}$,
Frederic Courbin$^8$,
\newauthor
William A. Dawson$^9$,
Annamaria Donnarumma$^6$,
Ian Fenech Conti$^{10}$,
\newauthor
Rapha\"el Gavazzi$^6$,
Marc Gentile$^8$,
Mandeep S.~S. Gill$^{4,5}$,
David W. Hogg$^{11}$,
\newauthor
Eric M. Huff$^{12}$,
M. James Jee$^{13}$,
Tomasz Kacprzak$^{2,14}$,
Martin Kilbinger$^{15}$,
\newauthor
Thibault Kuntzer$^{8}$,
Dustin Lang$^1$,
Wentao Luo$^{16}$,
Marisa C. March$^{17}$,
Philip J. Marshall$^4$,
\newauthor
Joshua E. Meyers$^{4}$,
Lance Miller$^{18}$,
Hironao Miyatake$^{3,19}$,
Reiko Nakajima$^{20}$,
\newauthor
Fred Maurice Ngol\'{e} Mboula$^{15}$,
Guldariya Nurbaeva$^{8}$,
Yuki Okura$^{21}$,
\newauthor
St\'ephane Paulin-Henriksson$^{15}$,
Jason Rhodes$^{22,23}$,
Michael D. Schneider$^9$,
\newauthor
Huanyuan Shan$^{8}$,
Erin S. Sheldon$^{24}$,
Melanie Simet$^1$,
Jean-Luc Starck$^{15}$,
\newauthor
Florent Sureau$^{15}$,
Malte Tewes$^{20}$,
Kristian Zarb Adami$^{10,18}$,
Jun Zhang$^{25}$,
Joe Zuntz$^{26}$
\\$^1$McWilliams Center for Cosmology, Department of
  Physics, Carnegie Mellon University, 5000 Forbes Ave., Pittsburgh,
  PA 15213, USA
\\$^2$Department of Physics \& Astronomy, University College London,
Gower Street, London, WC1E 6BT, UK
\\$^3$Department of Astrophysical Sciences, Princeton University,
Peyton Hall, Princeton, NJ 08544, USA
\\$^4$Kavli Institute for Particle Astrophysics and Cosmology,
Department of Physics, Stanford University, Stanford, CA 94305, USA
\\$^5$SLAC National Accelerator Laboratory, 2575 Sand Hill Road, Menlo
Park, CA 94025, USA
\\$^6$Institut d'Astrophysique de Paris, UMR 7095 CNRS -- Universit\'e
Pierre et Marie Curie, 98bis Bd Arago, 75014 Paris, France
\\$^7$Centre de Calcul de l'IN2P3, USR 6402 du CNRS-IN2P3, 43 Bd. du
11 Novembre 1918, 69622 Villeurbanne Cedex, France
\\$^8$Laboratoire d'astrophysique, Ecole Polytechnique F\'ed\'erale de
Lausanne (EPFL), Observatoire de Sauverny, CH-1290 Versoix, Switzerland 
\\$^9$Lawrence Livermore National Laboratory, P.O. Box 808 L-210,
Livermore, CA, 94551, USA
\\$^{10}$Institute of Space Sciences \& Astronomy (ISSA), University of
Malta, Msida, MSD 2080, Malta
\\$^{11}$Center for Cosmology and Particle Physics, Department of
Physics, New York University, 4 Washington Pl \#424, NY, NY 10003, USA
\\$^{12}$Center for Cosmology and AstroParticle Physics (CCAPP) and
Department of Physics, The Ohio State University, 191 W. Woodruff
Ave., \\Columbus, OH 43210, USA
\\$^{13}$Department of Physics, University of California, Davis, One Shields
Avenue, Davis, CA 95616, USA
\\$^{14}$Institut f\"{u}r Astronomie, ETH Z\"{u}rich, Wolfgang-Pauli-Str. 27, 8093 Z\"{u}rich, Switzerland
\\$^{15}$Laboratory AIM, UMR CEA-CNRS-Paris 7, Irfu, SAp SEDI, Service d'Astrophysique, CEA Saclay, F-91191 Gif-sur-Yvette Cedex, France
\\$^{16}$Key Laboratory for Research in Galaxies and Cosmology, Shanghai
Astronomical Observatory; Nandan Road 80, Shanghai 200030, China
\\$^{17}$David Rittenhouse Laboratory, University of Pennsylvania, 209 South 33rd Street, Philadelphia, PA 19104 USA
\\$^{18}$Department of Physics, University of Oxford, Denys Wilkinson Building, Keble Road, Oxford OX1 3RH, UK
\\$^{19}$Kavli Institute for the Physics and Mathematics of the Universe (Kavli IPMU, WPI), The University of Tokyo, Kashiwa, \\Chiba 277-8582, Japan
\\$^{20}$Argelander-Institut f\"ur Astronomie, Auf dem H\"ugel 71,
D-53121 Bonn, Germany
\\$^{21}$National Astronomical Observatory of Japan, Tokyo 181-8588, Japan
\\$^{22}$Jet Propulsion Laboratory, California Institute of Technology, Pasadena, CA 91109,
\\$^{23}$California Institute of Technology, Pasadena, CA 91125, USA
\\$^{24}$Brookhaven National Laboratory, Bldg 510, Upton, New York 11973
\\$^{25}$Center for Astronomy and Astrophysics, Department of Physics
and Astronomy, Shanghai Jiao Tong University, 955 Jianchuan Road,\\
Shanghai, 200240, China
\\$^{26}$Jodrell Bank Centre for Astrophysics, University of Manchester, Manchester M13 9PL, UK
}

\date{\today}

\begin{document}
\maketitle
\clearpage
\begin{abstract}
  We present first results from the third GRavitational lEnsing
  Accuracy Testing (GREAT3) challenge, the third in a sequence of
  challenges for testing methods of inferring weak
  gravitational lensing shear distortions from simulated galaxy images.
  GREAT3 was divided into experiments to test three specific
  questions, and included simulated space-
  and ground-based data with constant or cosmologically-varying
  shear fields.
  The simplest (control) experiment included parametric galaxies with a
  realistic distribution of signal-to-noise, size, and ellipticity, and a 
  complex 
  point spread function (PSF).  The other experiments tested the
  additional impact of realistic galaxy morphology,
  multiple exposure imaging, and the 
  uncertainty about a spatially-varying PSF; \newtext{the last two questions will be explored in Paper II.} 
 The 24 participating teams 
  competed to estimate lensing shears to within
  systematic error tolerances for upcoming Stage-IV dark energy surveys, making 1525 submissions overall.
\newtext{ GREAT3 saw considerable variety and innovation in the types of methods applied.} 
  Several teams now meet or exceed the targets in many of the tests
  conducted (to within the statistical errors). 
  \newtext{We conclude} that the presence of
  realistic galaxy morphology in simulations 
  changes shear calibration biases 
  by $\sim 1$ per cent for a wide range of methods.  Other effects
  such as truncation biases due to finite galaxy postage stamps, and the
  impact of galaxy type as measured by the \sersic\ index, are
  quantified for the first time.  Our results generalize previous
  studies regarding 
  sensitivities to galaxy size and signal-to-noise,
  and to PSF properties such as seeing and defocus.  Almost all
  methods' results support the simple model in which additive shear biases depend
  linearly on PSF ellipticity.
\end{abstract}


\section{Introduction}

  Weak gravitational lensing, the small but coherent deflections of
  light from distant objects due to the gravitational field of
  more nearby matter \citep[for a review,
  see][]{2001PhR...340..291B,2003ARA&A..41..645R,schneider06,2008ARNPS..58...99H,2010RPPh...73h6901M},
  has emerged in the past two decades as a promising way to constrain
  cosmological models, to study the relationship between visible and
  dark matter, and even to constrain the theory of gravity on cosmological
  scales \citep[e.g.,][]{2002PhRvD..65b3003H,
    2002PhRvD..65f3001H,2003PhRvL..91d1301A,2007PhRvL..99n1302Z}.
  Because of this promise, gravitational lensing has already been
  measured in many datasets, and there are several large
  surveys planned for the next few decades to measure weak lensing
  even more precisely, including
  Euclid\footnote{\url{http://sci.esa.int/euclid/},
  \url{http://www.euclid-ec.org}}
  \citep{2011arXiv1110.3193L},
  LSST\footnote{\url{http://www.lsst.org/lsst/}}
  \citep{2009arXiv0912.0201L}, and
  WFIRST-AFTA\footnote{\url{http://wfirst.gsfc.nasa.gov}}
  \citep{2013arXiv1305.5422S}, all of which are Stage IV
  dark energy experiments according to the Dark Energy Task Force
  \citep{2006astro.ph..9591A} definitions.

  The most common type of weak lensing measurement involves measuring
  coherent distortions (``shear'') in the shapes of galaxies.  In
  order for the aforementioned surveys to make the most of their
  ability to measure these distortions with sub-per cent 
  statistical errors, they must ensure adequate control of systematic
  errors.  While a full systematic error budget for weak lensing 
  includes both astrophysical and instrumental systematic errors, a problem 
  that has occupied much attention in the community for over a 
  decade is ensuring accurate measurements of the shear
  distortions of galaxies given that they have been convolved with a
  point spread function (PSF) and rendered into noisy images.

  With the rapid proliferation of shear estimation methods, the
  weak lensing community began a series of blind community challenges,
  with simulations that included a lensing shear (known only
  to the organizers) that participants must measure.  This
  served as a way to benchmark different shear estimation methods.
  The earliest of these challenges were the first Shear TEsting Programme
  \citep[STEP1:][]{2006MNRAS.368.1323H} and its successor \citep[STEP2:][]{2007MNRAS.376...13M}.
  Then it became apparent that many complex aspects of the
  process of shear estimation would benefit from simpler
  and more controlled simulations, which led to the GRavitational
  lEnsing Accuracy Testing (GREAT08) challenge
  \citep{2009AnApS...3....6B,2010MNRAS.405.2044B}, followed by the
  GREAT10 challenge
  \citep{2010arXiv1009.0779K,2012MNRAS.423.3163K,2013ApJS..205...12K}.

  Each of these challenges has been informative in its own way,
  illuminating important issues in shear estimation while also
  generating significant improvement in the accuracy of weak lensing
  shear estimation.  For example, both the GREAT08 and GREAT10 challenges highlighted
  the role played by pixel noise in biasing shear estimates.  While
  this $S/N$- and resolution-dependent ``noise bias'' was studied
  in specific contexts before 
  GREAT08 and GREAT10 \citep[e.g.,][]{2002AJ....123..583B,2004MNRAS.353..529H}, \newtext{the landscape changed
  after GREAT08, with several more general studies
  \citep{2012MNRAS.427.2711K,2012MNRAS.424.2757M,2012MNRAS.425.1951R}, some of which used the GREAT10 simulations as a test for calibration schemes}.
  However, despite the progress encouraged by these 
  challenges, there remained a number of outstanding issues in
  shear estimation that needed to be addressed for the community to ensure
  its ability to measure weak lensing in near-term and future surveys.
  These issues include the impact of realistic galaxy morphology: a
  number of studies have convincingly demonstrated that when
  estimating shears in a way that assumes a particular galaxy model,
  the shears can be biased if the galaxy light profiles are not
  correctly described by that model \citep[termed ``model
  bias'':][]{2010MNRAS.404..458V, 2010A&A...510A..75M}.  More
  generally, any method based on the use of second
  moments to estimate shears cannot be completely independent of the
  details of the galaxy light profiles, such as the overall galaxy
  morphology and presence of detailed substructure
  \citep{2007MNRAS.380..229M,2010MNRAS.406.2793B,2011MNRAS.414.1047Z}.
  Thus, the question of the impact of realistic galaxy morphology (and
  the way that galaxies deviate from simple parametric
  models) on shear estimation is important to address in a
  community-wide challenge.  This is one of the key questions of the
  GREAT3 challenge.

  The GREAT3 challenge was also designed to address two additional
  questions.  One of these is the combination of multiple exposures,
  which is necessary to analyze the data from nearly any current or
  upcoming weak lensing survey.  For Nyquist-sampled data this is
  relatively straightforward, but for data that are not
  Nyquist-sampled (such as some images from space telescopes), the
  problem is more challenging
  \citep[e.g.][]{1999PASP..111..227L,2011ApJ...741...46R,2011PASP..123..497F}.
  The final problem addressed in GREAT3 is the impact of PSF estimation
  from stars and interpolation to the positions of the galaxies.
  However, this paper will focus predominantly on the question of
  shear estimation in general and realistic galaxy morphology in
  particular, leaving the other questions for Paper II.

  In
  Sec.~\ref{sec:challenge}, we describe how the challenge was designed
  and run, how submissions were evaluated, and a basic summary of the
  submissions that were made.  We discuss the methods used by
  participants to analyze the simulated data in
  Sec.~\ref{sec:methods}.  For certain methods for which the teams
  made many submissions, we derive lessons related to those methods in
  Sec.~\ref{sec:specific}.  We then present the overall results for
  all teams in Sec.~\ref{sec:overall}.  Sec.~\ref{sec:lessons}
  describes some lessons learned about shear estimation from GREAT3,
  and we conclude in Sec.~\ref{sec:conclusions}.  Finally, there are
  appendices with some further technical details related to the
  challenge simulations, and lengthier descriptions of the methods
  used by each team.

\section{The challenge}\label{sec:challenge}

\subsection{Theoretical background}\label{subsec:theory}

Gravitational lensing distorts the images of distant galaxies.
When this distortion can be described as a locally linear
transformation, then the lensing effect is described as ``weak''.  In
this case, it relates unlensed coordinates ($x_u$, $y_u$; with the
origin at the center of the distant light source) and the observed,
lensed coordinates ($x_l$, $y_l$; with the origin at the center of the
observed image), via
\begin{equation}\label{eq:lensingshear}
  \left( \begin{array}{c} x_u \\ y_u \end{array} \right)
  = \left( \begin{array}{cc} 1 - \gamma_1 - \kappa & - \gamma_2 \\ -\gamma_2 & 1+\gamma_1 - \kappa \end{array} \right)
  \left( \begin{array}{c} x_l \\ y_l \end{array} \right).
\end{equation}
The two components of the lensing shear $(\gamma_1, \gamma_2)$
describe the stretching of galaxy images due to lensing, whereas the
convergence $\kappa$ describes a change in apparent size and
brightness for lensed objects. 
This transformation is often recast as
\begin{equation}
\left( \begin{array}{c} x_u \\ y_u \end{array} \right)
= (1 - \kappa) \left( \begin{array}{cc} 1 - g_1 & - g_2 \\ -g_2 & 1+g_1\end{array} \right)
\left( \begin{array}{c} x_l \\ y_l \end{array} \right),
\end{equation}
in terms of the reduced shear, $g_i = \gamma_i / (1 - \kappa) \simeq
\gamma_i$ in most cosmological applications.  Typically it is the
stretching described by the reduced shear that is actually observed.
We often encode the two components of shear (reduced shear) as a
single complex number, $\gamma = \gamma_1 + {\rm i}\gamma_2$ ($g
= g_1 + {\rm i} g_2$).

The lensing shear causes a change in estimates of the
\emph{ellipticity} of distant galaxies.  In practice, the effect is
estimated statistically by measuring galaxy properties that transform
in simple ways under a shear.  One method is to model the galaxy image
using a profile with a well-defined ellipticity, written as
$\varepsilon = \varepsilon_1 + {\rm i} \varepsilon_2$, with magnitude
\begin{equation}\label{eq:shear}
|\varepsilon| = \frac{1 - b/a}{1 + b/a}
\end{equation}
for semi-minor and semi-major axis lengths $b$ and $a$,
and orientation angle determined
by the major axis direction.  For a population of randomly-oriented source intrinsic
ellipticities, the ensemble average ellipticity after lensing gives an
unbiased estimate of the shear: $\langle \varepsilon \rangle \simeq g$.

Another common choice of shape parametrization is based on second brightness moments of the
galaxy image,
\begin{equation}\label{eq:qij}
Q_{ij} = \frac{\int {\rm d}^2 x I({\bf x}) W({\bf x}) x_i x_j }
{\int {\rm d}^2 x I({\bf x}) W({\bf x}) },
\end{equation}
where ($x_1$, $x_2$) correspond to the ($x$, $y$)
directions, $I({\bf x})$ denotes the galaxy image light
profile, $W({\bf x})$ is an optional\footnote{Optional for the purpose
  of this definition; but in practice, for images with noise, some
  weight function that reduces the contribution from the wings of the
  galaxy is necessary to avoid moments being dominated by noise.}
weight function (see, e.g., \citealp{schneider06}), and the coordinate
origin is placed at the galaxy image center. A second
ellipticity definition (sometimes called the
\emph{distortion} to distinguish it from the ellipticity that
satisfies Eq.~\ref{eq:shear}) can be written as
\begin{equation}\label{eq:ellipticity}
e = e_1 + {\rm i} e_2 = \frac{Q_{11} - Q_{22} + 2 {\rm i} Q_{12}}{Q_{11} + Q_{22}}.
\end{equation}
The ellipticity $\varepsilon$ can also be related to the moments by
replacing the denominator in Eq.~\eqref{eq:ellipticity} with $Q_{11} +
Q_{22} + 2 (Q_{11}Q_{22}-Q_{12}^2)^{1/2}$.

If the weight function $W$ is constant or brightness-dependent, an image with
elliptical isophotes has
\begin{equation}\label{eq:ellipticity-ba}
|e| = \frac{1 - b^2 / a^2}{1 + b^2 / a^2}.
\end{equation}
For a randomly-oriented population of source distortions,
the ensemble average $e$ after lensing
gives an unbiased estimate of shear that depends on the
population root mean square (RMS) distortion $\langle (e^{(s)})^2\rangle$ as $\langle e \rangle
\simeq 2[1-\langle (e^{(s)})^2\rangle] g $.

See e.g. \cite{2002AJ....123..583B} for further details on
commonly-used shear
and ellipticity definitions.

\subsection{Summary of challenge structure}\label{subsec:structure}

Here we describe how the GREAT3 challenge was
structured; more details are
given in the handbook, \cite{2014ApJS..212....5M}.

The GREAT3 challenge was designed to address how three issues affect
shear estimation: (a) the impact of realistic galaxy morphology, (b)
the impact of the image combination process, and (c) the effect of
errors due to estimation and interpolation of the PSF.  To this end,
the challenge consisted of five experiments:
\begin{enumerate}[(1)]
\item Control: Parametric (single or double \sersic) galaxy models
  based on fits \citep{2012MNRAS.421.2277L} to {\em HST}
  data from the COSMOS
  \citep{2007ApJS..172..196K,2007ApJS..172....1S,2007ApJS..172...38S}
  survey, meant to represent the galaxy population in a typical weak
  lensing survey, including appropriate size vs.\ galaxy flux
  signal-to-noise ($S/N$) relations, morphology distributions, and so
  on.  In each image, the non-trivially complex PSF was provided for
  the participants \newtext{as a set of nine images with different centroid offsets.}
\item Real galaxy: Differed from the control experiment 
  only in the use of the actual images from the {\em HST} COSMOS
  dataset instead of the best-fitting parametric models.
\item Multiepoch: Differed from the control experiment only in that
  each field contained six images (representing 
  observations that must be combined) instead of one. \newtext{ For the space
  branches, the six images were not Nyquist sampled.}
\item Variable PSF: Differed from the control experiment only in that
  the PSF varied across the image in a realistic way, and had to be estimated from star images.
\item Full: Included the complications of the real galaxy, multiepoch,
  and variable PSF experiments all together.
\end{enumerate}

In all cases, the goal was to estimate the lensing shear\footnote{This
  is not the same as testing the ability to measure a
  per-galaxy {\em shape}.  Two different methods 
  can recover a different per-galaxy shape, while still
  estimating the overall shear accurately.}.  For each experiment,
there were four branches, which came from the combination of two types
of simulated data (ground, space) and two types of shear fields
(constant, variable).  For 
convenience, we will refer to branches by their
combinations of \{experiment\}-\{observation type\}-\{shear type\}, e.g.,
control-ground-constant, and will use the unique abbreviations CGC, CGV, and so
on.  Of the 20 branches (five experiments $\times$
two data types $\times$ two shear types), participants could 
submit results for as many or few as they chose
\citep[see][figure 5]{2014ApJS..212....5M}.  A given branch included 200 subfields, each
with $10^4$ galaxies on grids.  To reduce statistical errors on the shear biases, galaxies were arranged such that the intrinsic noise due to non-circular galaxy shapes (`shape
noise') was nearly cancelled out.

Submissions to the challenge were
evaluated according to metrics described in
Sec.~\ref{subsec:diagnostics}.  Within a branch, teams were ranked
based on their best submission in that branch.  Per-branch rankings
were used to award teams points, which were then added up across
multiple branches to give an overall leaderboard ranking.  While the
leaderboard ranking was necessary for the purpose of carrying out a
challenge, the goal of this work is to study how teams performed and
derive lessons for the future based on analysis 
that goes far beyond a simple ranking scheme.  

There are a number of online resources related to the challenge and
the simulations.  The main challenge
website\footnote{\url{http://www.great3challenge.info}} contains
overall information.  The leaderboard
website, linked from the main challenge website, 
contains the archived challenge leaderboards, and additional
post-challenge boards to which submissions were made after the end of
the challenge.  It also links to download the GREAT3
simulations and truth tables.  The GitHub
site\footnote{\url{https://github.com/barnabytprowe/great3-public}}
contains software to reproduce the simulated data and to analyze it
using simple methods, and a wiki with information for the
participants.  Finally, \galsim\footnote{\url{https://github.com/GalSim-developers/GalSim}}
is the simulation software that was used to make the GREAT3
simulations, and its algorithms, design, and functionality are
described in \cite{2014arXiv1407.7676R}.

Some physical effects that are not tested in the
challenge include object detection, selection, and deblending,
because the galaxies are located on grids; wavelength-dependent
effects; instrumental and detector defects or non-linearities; star/galaxy separation;
background estimation; complex pixel noise models; cosmic rays and
other image artifacts; redshift-dependent
shear calibration; \newtext{shear estimation for galaxies with sizes comparable
to the PSF;} non-weak shear signals (e.g.\ cluster lensing); and flexion.

Appendix~\ref{app:challenge} contains more detailed information
about some aspects of the challenge that were not in the
handbook.  These include Appendix~\ref{app:intrinsic-pe}, on the
intrinsic ellipticity distribution ($p(\varepsilon)$) of the galaxies;
Appendix~\ref{app:shear}, which describes the distributions from which
the lensing shears were drawn; Appendix~\ref{app:psf}, which 
presents distributions of optical and atmospheric PSF properties; and
Appendix~\ref{app:sn}, which shows the actual $S/N$ distributions
for galaxies in GREAT3.  The last point is particularly relevant for how
pixel noise should affect shear estimates in the challenge.

Finally, the GREAT3 Executive Committee\footnote{The Executive
  Committee created the simulations,
  ranking scheme, and other aspects of the challenge, and had access 
  to privileged information about the simulations.  Because of
  this access, teams to which they
  made significant contributions did not receive points in the challenge, and were not ranked.  Those teams appear on
  the leaderboard with an asterisk for their score.}
(EC) distributed example
scripts to automatically process the challenge data, including shear
estimation, coaddition of multi-epoch data, and variable PSF
estimation.  While the latter two will be discussed in Paper II, we
describe the algorithms in the shear estimation example script in
Appendix~\ref{app:scripts}.

\subsection{Diagnostics}\label{subsec:diagnostics}

Here we describe the diagnostics used to quantify the performance of
each submission to the challenge.
The metrics for constant- and variable-shear branches, 
discussed in detail in \cite{2014ApJS..212....5M}, were used to rank
submissions.  Here we briefly define the 
equations used.

\subsubsection{Constant shear}
For constant-shear simulations, each field has a particular
value of shear applied to all galaxies (App.~\ref{app:shear}).  Participants
submitted estimated (``observed'') shears for each constant shear value in the branch.  We relate biases in observed shears
$g^{\rm obs}$ to the true shear $g^{\rm true}$ using a 
linear model in each component:
\begin{equation}\label{eq:linbias}
g^{\rm obs}_i - g^{\rm true}_i = m_i g^{\rm true}_i + c_i
\end{equation}
where $i$ denotes the shear component, and $m_i$ and $c_i$ are
the multiplicative and additive biases, respectively.
From user-submitted estimates of all $g^{\rm obs}_i$ 
in a branch, the metric calculation begins with an
unweighted least-squares linear regression to provide estimates of $m_i$, $c_i$ given
the true shears (in Sec.~\ref{subsec:outliers} we discuss
the role of outliers in affecting the $m_i$ and $c_i$
estimates).  The regression is done in a coordinate frame rotated to
be aligned with the mean PSF ellipticity in each field, so that $c$
values will properly reflect the contamination of galaxy shapes by the PSF
anisotropy. 

Having estimated $m_i$ and $c_i$, 
we constructed the metric,
$Q_{\rm c}$, by comparison with `target' values $m_{\rm target}$, $c_{\rm target}$.
These come from requirements for upcoming weak lensing experiments; 
we use $m_{\rm target} = 2 \times 10^{-3}$ and $c_{\rm
  target} = 2 \times 10^{-4}$, motivated by a recent estimate
of requirements \citep{2013MNRAS.431.3103C,2013MNRAS.429..661M} for the 
Euclid space mission.  The constant-shear
metric is then defined as
\begin{equation}\label{eq:q_c}
Q_{\rm c} = \frac{2000 \times \eta_{\rm c}}{\sqrt{
 \sigma^2_{\text{min}, {\rm c}} + \displaystyle\sum_{i=+,\times} \left[\left(
      \frac{m_i}{m_{\rm target}} \right)^2 + \left( \frac{c_i}{c_{\rm target}} \right)^2\right] }}.
\end{equation}
The indices $+$, $\times$ refer to the two shear 
  components in the rotated reference frame described above.
  We adopt $\sigma_{\text{min}, {\rm c}}^2 = 1~(4)$ for
  space~(ground) branches, corresponding to the typical dispersion
  in the quadrature sum of $m_i/m_{\rm target}$ and $c_i/c_{\rm
    target}$ due to pixel noise. This metric is normalized by
  $\eta_{\rm c}$ such that 
methods that meet our chosen targets on
$m_i$ and $c_i$ in space-based data should achieve $Q_{\rm c} \simeq 1000$. 
In the ground branches $Q_{\rm c}$ is 
  slightly lower for submissions reaching target bias levels,
  reflecting their larger $\sigma^2_{\text{min}, {\rm c}}$ due to
  greater uncertainty in individual shear estimates for ground data.
  However, $Q_{\rm c}$ scores are consistent between space
  and ground branches where biases are significant.

\newtext{Given the nature of this metric definition, the uncertainty in $Q_{\rm
  c}$ is larger at high $Q_{\rm c}$ than at small $Q_{\rm c}$.  For
the level of pixel noise in the simulations from ground (space), the effective
uncertainty on $Q_{\rm c}$ for $Q_{\rm c}$ values of $[100, 300, 500,
1000]$ is $[3, 28, 80, 328]$ ($[2, 19, 55, 229]$).}

\subsubsection{Variable shear}

For variable-shear simulations, the key test is the
reconstruction of the shear correlation function. 
Submission of results for these branches begins with
calculation of correlation functions 
by the participant\footnote{Software for this purpose was distributed
  publicly at \url{https://github.com/barnabytprowe/great3-public}.}. The
submission consists of estimates of the
aperture mass dispersion 
\citep[e.g.,][]{schneider06,map_schneider}, which are constructed 
from two-point correlation function estimates, and 
allows a separation into contributions from $E$ and $B$
modes\footnote{\newtext{For more discussion of the limitations on $E$- and
  $B$-mode separation in GREAT3, please see \cite{2014ApJS..212....5M}.}}.  We label these $E$ and
$B$ mode aperture mass dispersions $M_E$ and $M_B$.

The submissions were estimates of $M_{E,j}$ for
each of ten fields labelled by index $j$; this estimate is 
constructed using  twenty subfields in a given field.  This
choice provides a large dynamic range of spatial scales in the
correlation function, and thereby probes a greater range of
shear signals.  The $M_{E,j}$ are estimated 
in $N_{\rm bins}$ logarithmically spaced annular bins of galaxy pair
separation $\theta_k$, from the smallest available angular scales in
the field to the largest.

The metric $Q_{\rm v}$ for the variable-shear branches was 
constructed by comparison to the known, true value of the aperture
mass dispersion for the realization of $E$-mode shears in each field. 
These we label $M_{E, {\rm true}, j}(\theta_k)$.  The
variable-shear branch metric is then calculated as
\begin{equation}\label{eq:q_v}
Q_{\rm v} = \frac{1000 \times \eta_{\rm v}}{\sigma_{\text{min}, {\rm
      v}}^2 + \frac{1}{N_{\rm norm}}
  \displaystyle\sum_{k=1}^{N_{\rm bins}} \left|
    \displaystyle\sum_{j=1}^{N_{\rm fields}} \left[M_{E, j} (\theta_k)
    - M_{E, {\rm true}, j}(\theta_k)\right] \right| }
\end{equation}
where $N_{\rm norm} = N_{\rm fields} N_{\rm bins}$,
\newtext{$\sigma_{\text{min}, {\rm v}}^2 = 4~(9) \times 10^{-8}$ for space
(ground) branches,} and 
  $\eta_{\rm v}$ is a normalization factor designed to yield $Q_{\rm
    v} \simeq 1000$ for a method achieving $m_1=m_2=m_{\rm target}$
  and $c_1=c_2=c_{\rm target}$.

  The primary source of noise in the $M_{E, j} (\theta_k)$ is
  pixel noise, with some residual shape noise playing a role despite
  the shape noise cancellation scheme.  After the end of the
  challenge we found that a small additional source of noise comes
  from the interplay between the $\theta_k$ bin size, the galaxy
  grid configuration, and approximations used in the calculation of
  the correlation function and aperture mass dispersion in \code{corr2}\footnote{\url{https://code.google.com/p/mjarvis/}}.  While this is a
  subdominant source of noise \newtext{($\sim 1/4$ of that 
  due to measurement error)}, it does mean that participants will find that
  their $Q_{\rm v}$ results depend slightly on the ordering of galaxies in their catalog.

\newtext{ For
the level of pixel noise in the simulations from ground (space), the effective
uncertainty on $Q_{\rm v}$ for $Q_{\rm v}$ values of $[100, 300, 500,
1000]$ is $[6, 47, 118, 418]$ ($[5, 36, 91, 326]$).}

\subsubsection{Other diagnostics}

For the constant-shear branches, we have a clean way to directly study
additive and multiplicative biases in the form of $m_i$ and $c_i$,
where $i=+,\times$ (defined in the frame aligned with the PSF
ellipticity, and at 45 degree angles with respect to that direction).
However, also of interest are the $m_i$ and $c_i$ defined in the frame
defined by the pixel coordinates, for $i=1,2$.
In the STEP2 challenge \citep{2007MNRAS.376...13M},
many methods exhibited 
coherent differences in shear systematics along the pixel
axes and at 45 degrees with respect to them, presumably due to the
different effective sampling of the galaxy and PSF profiles.  Since
the PSF ellipticity direction has a random orientation with respect to
the pixel axes, differences between $m_1$ and $m_2$ will 
average out, giving $m_+\approx m_\times$.  Since differences between
$m_1$ and $m_2$ may be interesting in understanding the performance of
a method, we will use $m_1$ and $m_2$ for some of our plots.

In addition, $c_1$ and $c_2$ may be of interest.  While $c_+$ 
shows the influence of PSF anisotropy, additive
systematics due to PSF anisotropy will have a random sign and
direction for each subfield in the pixel coordinate frame, so $c_1$ and $c_2$
have an expectation value of zero.   Nonzero values  
may indicate selection biases with respect to the
pixel direction, or asymmetric numerical artifacts.

Given the more fundamental nature of $m_1$ and $m_2$, and the
need to use $c_+$ to identify additive PSF systematics, we also
consider what we will call a ``mixed metric'', $Q_{\rm mix}$, defined
in analogy to $Q_{\rm c}$ (Eq.~\ref{eq:q_c}) as
\begin{equation}\label{eq:q_mix}
Q_{\rm mix} = \frac{2000 \times \eta_{\rm c}}{\sqrt{
 \sigma^2_{\text{min}, {\rm c}} +  \displaystyle\sum_{i=1,2}\left(
      \frac{m_i}{m_{\rm target}} \right)^2 + \displaystyle\sum_{i=+,\times} \left( \frac{c_i}{c_{\rm target}} \right)^2}}.
\end{equation}

\subsection{Challenge process}

During the challenge period, there were 1525
submissions\footnote{The leaderboard website shows
  1532 submissions, but seven had an incorrect submission
  format, giving $Q=0$.} with nonzero score, from 24
distinct teams.  Of these, two teams were actually members of the
GREAT3 EC making submissions based on simple test
scripts to validate the simulations or submission process; sixteen were
teams of participants; and six were teams that included at
least one member of the GREAT3 EC, and were thus 
excluded from winning any points or the challenge itself.

\begin{figure}
\begin{center}
\includegraphics[width=0.9\columnwidth,angle=0]{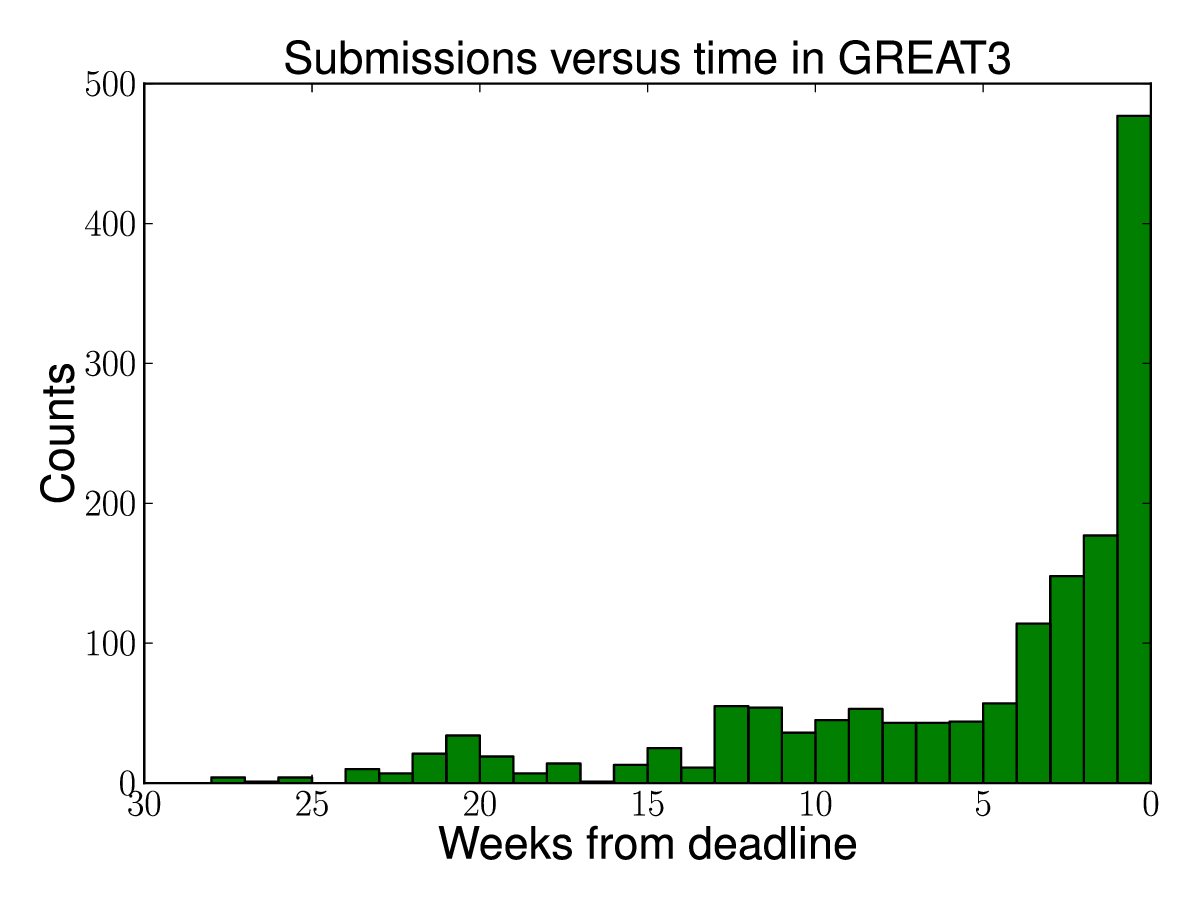}
\caption{\label{fig:subs_vs_time} Number of submissions to the GREAT3
  challenge as a function of time, expressed in terms of weeks until
  the deadline.  \newtext{The rules for the number of submissions per team per day were relaxed in the final week of the challenge.}}
\end{center}
\end{figure}

Fig.~\ref{fig:subs_vs_time} shows the number of submissions to the
challenge as a function of time, expressed in terms of weeks
until the deadline.  The first entries were submitted
near the beginning of the challenge period, which ran from
mid-October 2013 until April 30 2014.  The submission rate was an
increasing function of time particularly in the last month; the spike
in entries in the last week was partly due to a 
relaxation of the rules on the number of entries per team per day.

Two teams
entered all twenty branches, and 7/24 (30\%) of the teams entered more
than half the branches.  Not surprisingly, many teams chose to focus
on the control and realistic galaxy branches, which required the least
amount of software infrastructure to participate.

Table~\ref{T:overall} shows the results for each branch, including the
winning team, the winning score (defined in 
Sec.~\ref{subsec:diagnostics}), the
number of participating teams, and the number of entries.  As shown, a
variety of teams with different methods won individual 
branches, rather than one team dominating everything.  For all but
two branches, VGV and FGV, the winning scores were $\gtrsim 800$,
meaning that within the ability of the simulations to determine shear
systematic errors, the winning submissions were effectively unbiased.  Not only
the winning team but also typically several other teams had scores in
this range, representing an unprecedented quality of submissions
in a weak lensing community challenge.  We will discuss why the
combination of variable PSF and variable shear was more
difficult in Paper II.

\begin{table}
\begin{tabular}{lllll}
\hline\hline
Branch & Winning & Winning & \# of & \# of \\
       & team    & score   & teams & entries \\
\hline
CGC & CEA-EPFL & 1211 & 22 \newtext{(4)} & 250 \\
CGV & CEA-EPFL & 1068 & 16 \newtext{(5)} & 160 \\
CSC & Amalgam@IAP & 1516 & 16 \newtext{(3)}& 110 \\
CSV & Amalgam@IAP & 1199 & 11 \newtext{(4)}& 96 \\
RGC & Amalgam@IAP & 1121 & 20 \newtext{(4)}& 195 \\
RGV & CEA-EPFL & 791 & 14 \newtext{(4)}& 93 \\
RSC & Fourier\_Quad & 1919 & 12 \newtext{(3)}& 92 \\ 
RSV & MegaLUT & 1667 & 9 \newtext{(4)}&  83 \\
MGC & sFIT & 1017 & 9 \newtext{(3)}& 71 \\
MGV & MegaLUT & 1131 & 7 \newtext{(2)}&  53 \\
MSC & sFIT & 841 &  6 \newtext{(1)}& 48 \\
MSV & CEA-EPFL & 1605 & 6 \newtext{(5)}& 45 \\
VGC & sFIT & 884 & 7 \newtext{(1)}& 60 \\
VGV & Amalgam@IAP & 230 & 6 \newtext{(0)}& 60 \\
VSC & Amalgam@IAP & 1183 & 4 \newtext{(1)}& 25 \\
VSV & sFIT & 1276 &  4 \newtext{(2)}& 17 \\
FGC & sFIT & 800 & 2 \newtext{(1)}& 11 \\
FGV & sFIT & 379 & 2 \newtext{(0)}& 17 \\
FSC & sFIT & 1184 & 2 \newtext{(2)}& 17 \\
FSV & sFIT & 856 & 2 \newtext{(2)}& 25 \\
\hline
\end{tabular}
\caption{For each branch, this table shows the winning team and its score, the number of teams that 
  submitted to that branch \newtext{(with the number having scores above 500
  for the submissions analyzed in Sec.~\ref{sec:overall} 
  shown in parenthesis)}, and the total number of entries in the
  branch.\label{T:overall}}
\end{table}

\begin{figure}
\begin{center}
\includegraphics[width=\columnwidth,angle=0]{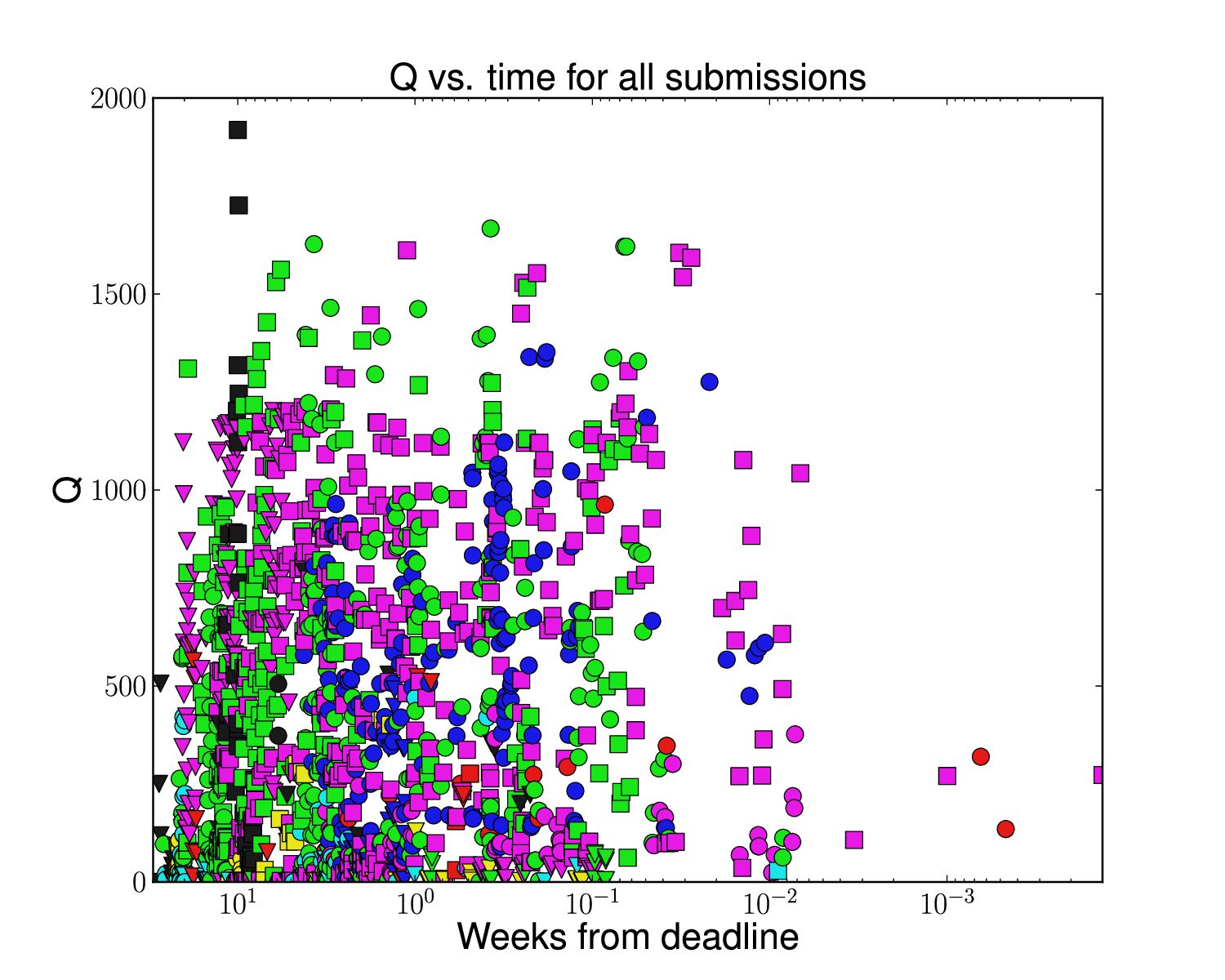}
\caption{\label{fig:bigmess} \newtext{$Q$ for all submissions as a function of
  time, expressed in terms of weeks until 
  the deadline.  Later submissions by the same team that appear to perform worse than earlier submissions typically went to more challenging branches.}}
\end{center}
\end{figure}

\newtext{To motivate the approach we take for the analysis,
Fig.~\ref{fig:bigmess} shows a scatter plot of metric $Q$ (either
$Q_{\rm c}$ or $Q_{\rm v}$ as appropriate) as a function of time, for
all submissions across all branches.  Point styles
indicate the team; the legend has been suppressed because our purpose
is only to show that (1) there are a huge number of submissions
with a wide range of performance, and (2) sometimes even within a
given team, the results varied a great deal.  We thus approach the
analysis in two stages.  Our first step, in Sec.~\ref{sec:specific},
is to analyze the results for specific teams that made many 
submissions, to understand the trends for that method and
identify a fair subset of their submissions (one per branch) to
compare with those from other teams.  Then, in Sec.~\ref{sec:overall},
we use this fair subsample of submissions, one per team per branch, to
learn lessons from the overall challenge results.}

\section{Shear estimation methods}\label{sec:methods}

\begin{table*}
\begin{tabular}{lllllllllll}
\hline\hline
Team & Class & Weighting & Calibration & Limitations & $N_{\rm branch}\!\!\!\!$
& Rank & Exact & New & Time per$\!\!\!\!$ \\
 & & scheme & philosophy & & & & PSF? & software & galaxy \\
\hline
Amalgam@IAP$\!\!$ & Maximum  & Inverse & Ellipticity & None & 16 &
2 & Yes & 
Some & 0.1--1~s \\
           & likelihood & variance & penalty & & & & & \\
\hline %
BAMPenn & Bayesian & Implicit & $p(\varepsilon)$ from & Variable     &
2 & - & Yes & Yes & $<1$~s \\
        & Fourier  &          & deep data   & shear & &
        & & \\
\hline %
EPFL\_gfit & Maximum & Constant $+$ & None & None & 8 & 6 & Yes & Yes & 1--3~s \\
        & likelihood & rejection &  & & & & & \\
\hline %
CEA-EPFL & Maximum    & Various & None & None & 20 & 3 & Yes & Yes & 1--3~s \\
         & likelihood & & &  & & & & \\
\hline %
CEA\_denoise & Moments & Constant & None & None & 8 & - & Yes &
No & 0.03~s \\
\hline %
CMU & Stacking & Constant & External & Variable & 2 &
N/A & Yes & Some & 0.03~s \\
experimenters & & & simulations & shear & & & & \\
\hline %
COGS & Maximum & Constant & External & None & 12 & N/A & Yes & 
Yes & 1~s \\
(\code{im3shape}) & likelihood & & simulations & & & & & & & \\
\hline %
E-HOLICS & Moments & Constant $+$ & External & None & 12 & 8 & Yes
& No & 1--3~s \\
 &  & rejection & simulations &  & & & & \\
\hline %
EPFL\_HNN & Neural & Constant & None & None & 7 & - & Yes & Yes & 2-3~s \\
 &  network & & & & & & & \\
\hline %
EPFL\_KSB & Moments & Inverse & None & None & 4 & - & Yes & No & 0.001-0.002~s \\
 &  & variance & & & & & & \\
\hline %
EPFL\_MLP / & Neural & Constant & None & None & 5 & - & Yes & Yes & 2-3~s \\
EPFL\_MLP\_FIT & network & & & & & & & \\
\hline %
FDNT & Fourier & Inverse & External & None & 12 & N/A & Yes & Some & $\sim 1$~s \\
 & moments & variance & simulations & & & & & \\
\hline %
Fourier\_Quad & Fourier & Various & None & None & 6 & 5 & Yes & No & 0.001-0.002~s \\
 & moments & & &  & & & & \\
\hline %
HSC/LSST-HSM & Moments & Inverse & External & None & 4 & N/A & Yes &
Some & 0.05~s \\
& & variance & simulations & & & & & \\
\hline %
MBI & Bayesian & Implicit & Inferred & Variable & 4 & 9 & No & Some & 10~s \\
 & hierarchical & & $p(\varepsilon)$ & shear, PSF & & & & \\
\hline %
MaltaOx & Partially & Inverse & Self- & None & 3 & 7 & Yes & Some
& 0.05~s \\
(\code{LensFit}) & Bayesian & variance & calibration & & & & & \\
\hline %
MegaLUT & Supervised & Constant $+$ & External & None & 16 & 4 & Yes & Some & 0.02~s \\
 & ML & rejection & simulations &  & & & & \\
\hline %
MetaCalibration & Moments $+$ & Inverse & Self- & Variable & 1 & N/A &
Yes & Yes &
 0.3~s \\
 & self-calibration & variance & calibration & shear & & & & \\
\hline %
Wentao\_Luo & Moments & Inverse & None & None & 4 & - & Yes & Yes & 1-2~s \\
 & & variance & & & & & & \\
\hline %
ess & Bayesian & Implicit & $p(\varepsilon)$ from & Variable & 2 & - & No & Yes & 1~s \\
 & model-fitting & & deep data & shear & & & & \\
\hline %
sFIT & Maximum & Inverse & External & None & 20 & 1 & Yes & Yes &
0.8~s \\
 & likelihood & variance & simulations & & & & \\
 & & & (iterative) & & & & \\
\hline
\end{tabular}
\caption{\newtext{Table summarizing the methods used by teams that participated in the
  challenge, including basic information such as team name; class 
  (overall type of method); weighting scheme; calibration philosophy
  (discussed in the text); and number of
  branches entered in the challenge ($N_{\rm branch}$).  ``Limitations'' refers to 
  types of data to which the implementation used here
  is not applicable without significant further development.  ``Rank''
  is the leaderboard ranking for those that
  received points (``-'' for those that did not, and ``N/A'' for those
  that were ineligible due to participation of a GREAT3 EC 
  member). ``exact PSF?'' indicates whether they used the
  exact PSF or an approximation to it (e.g., sums of Gaussians).  ``New
  software'' indicates whether the software used to analyze the GREAT3
  simulations was newly developed (``yes''), included some existing
  infrastructure with new software of non-trivial complexity
  (``some''), or was entirely pre-existing (``no'').  Finally, we show
  the approximate processing 
  time per galaxy per exposure (on a single core) for science-quality shear estimates.
  Several fields are discussed in detail in Sec.~\ref{sec:methods}.} \label{T:methods}}
\end{table*}

In this section, we broadly categorize and describe the methods used
to analyze the GREAT3 data.  
Appendix~\ref{app:methods} contains a more detailed description of all methods.  The main aspects of the methods used by the teams in GREAT3
are summarized in Table~\ref{T:methods}, which forms the basis for
the discussion in this section\footnote{A few teams listed on the
  GREAT3 challenge website are not in this table, either because
  they did not make any submissions, because the team 
  solely existed to demonstrate the use of the example scripts
  (Appendix~\ref{app:scripts}) distributed by the GREAT3 EC 
  (team ``GREAT3\_EC''), or because the team was created by a
  GREAT3 EC member only to check 
  the GREAT3 simulations as part of the validation process (team
  ``miyatake-test'').}.

We have assigned each of the 21 teams to a ``class'' (listed in
Table~\ref{T:methods}) that describes how the method essentially
works.  There are several options for the class:
\begin{enumerate}[(1)]
\item Maximum likelihood: maximum-likelihood model-fitting methods, of
  which there are five.
\item Bayesian methods: there are four of these, each with 
  different labels (e.g., ``Bayesian hierarchical'', ``Bayesian Fourier'',
  etc.) indicating differences in how they work.  \newtext{The ``Partially
  Bayesian'' label for MaltaOx is meant to indicate a Bayesian
  marginalization over nuisance parameters combined with mean
  likelihood estimation, rather than a fully Bayesian approach.}
\item Moments: there are eight methods that work by combining estimates of
  galaxy and PSF moments in some way.  Of these, six are real-space
  moments methods (called ``Moments'') and two are Fourier-space moments
  methods (``Fourier moments'').  Of the six real-space moments methods,
  one involves as a key aspect of the method a self-calibration scheme
  (``Moments $+$ self-calibration''), and that self-calibration
  could be extended to non-moments-based methods.
\item Stacking: a single team used image stacking.
\item Neural network and supervised machine learning (ML): three 
  methods rely heavily on machine learning.
\end{enumerate}

The table also lists the weighting scheme that was used.  Here there
are a few options.  Several teams used constant (equal) weighting, 
in some cases allowing optional rejection using certain
selection criteria (``Constant $+$ rejection'').  Many teams used
inverse variance weighting, where the variance is a combination of
shape noise and measurement error due to pixel noise.  In the Bayesian
methods, the weights are often implicit rather than explicitly
assigned.  Some teams experimented with multiple weighting schemes,
in which case their entry in the table is ``Various'', and details 
are in the Appendix.

Another important entry in Table~\ref{T:methods} is ``Calibration
philosophy'', which relates to how or whether a team tries to
calibrate out systematic errors, versus attempting to be unbiased {\em
  a priori}.  Here there are a few options:
\begin{enumerate}[(a)]
\item None: These teams apply no calibration
  corrections.
\item External simulations: These teams generate their own
  simulations in order to calibrate their shears. In one case (sFIT),
  these are produced iteratively until they are found to sufficiently
  match the data that are being analyzed (``External simulations
  (iterative)'').
\item \newtext{Ellipticity penalty term: One team, rather than applying
  calibrations after the fact, uses a penalty term on high ellipticity to reduce 
  certain calibration biases.  This penalty term must be calibrated in
  some way, making it somewhat different in nature from the next option.}
\item $p(\varepsilon)$ from deep data: Some methods require an input intrinsic
  ellipticity distribution from deep data \newtext{(or more precisely, for
  BAMPenn, the full distribution of unnormalized moments)}.  This is qualitatively
  different from requiring external simulations, since many 
  surveys will have a deeper subset of the data that could be used to
  derive this prior.
\item \newtext{Inferred $p(\varepsilon)$: One team tried to hierarchically
  infer the $p(\varepsilon)$ and the shear from the data itself.}
\item Self-calibration: Finally, two teams (MetaCalibration and
  MaltaOx) implemented a self-calibration scheme to derive
  calibration corrections from the data itself.
\end{enumerate}

Table~\ref{T:methods} also lists other useful pieces of information about these methods, as described in the caption.

\section{Informative results for specific methods}\label{sec:specific}

Before exploring the overall results of the challenge, we
first consider several methods in detail.  For methods with many
submissions, it is important to understand overall behavior of the
method before comparing with others.  For this reason, we carry out two
types of tests:

\begin{enumerate}[(1)]
\item Controlled tests of the performance of the method as a function
  of the various initial settings and parameter values that determine its
  performance, for multiple submissions {\em in a
  given branch}.
\item A comparison of submissions for that method {\em across multiple
    branches}, while holding its initial settings and parameters fixed
  (instead of using those that happened to give the best metric score in each branch).
\end{enumerate}

These results then serve as a basis for the fair comparison between
methods and across branches, which will be performed later in the paper.
For all the methods discussed, see Appendix~\ref{app:methods}
for a more detailed description.

\subsection{\code{gfit}}\label{subsec:focus:gfit}

\subsubsection{Controlled tests of variation in \code{gfit} parameters}\label{subsubsec:gfitvariation}

In this section, we show results of a more detailed exploration of the
\code{gfit} software used by the EPFL\_gfit and CEA-EPFL
teams (see method descriptions in Appendices~\ref{subsec:gfit}
and~\ref{subsec:cea-epfl}).  In particular, we investigate the
dependence of the results on choices made in the course of estimating
the per-object shears, or the 
weighting used to estimate an average shear for the entire field).
Our comparison focuses on the constant-shear branches, where we have 
additional diagnostics such as the multiplicative and
additive biases (see
Sec.~\ref{subsec:diagnostics} for definitions).

This comparison uses the submissions from EPFL\_gfit, but the results
are also applicable to CEA-EPFL submissions.  The
factors that were considered in the comparison are the galaxy model,
the postage stamp size, precision on the total flux and
centroid, maximum half-light radii of the bulge and disk, filtering of the galaxy catalog, constraints on positivity of bulge and disk
  flux, and occasional other experiments, such as stacking the 9 PSFs in the
  starfield images, or running a denoising scheme. 

\begin{figure}
\begin{center}
\includegraphics[width=0.93\columnwidth,angle=0]{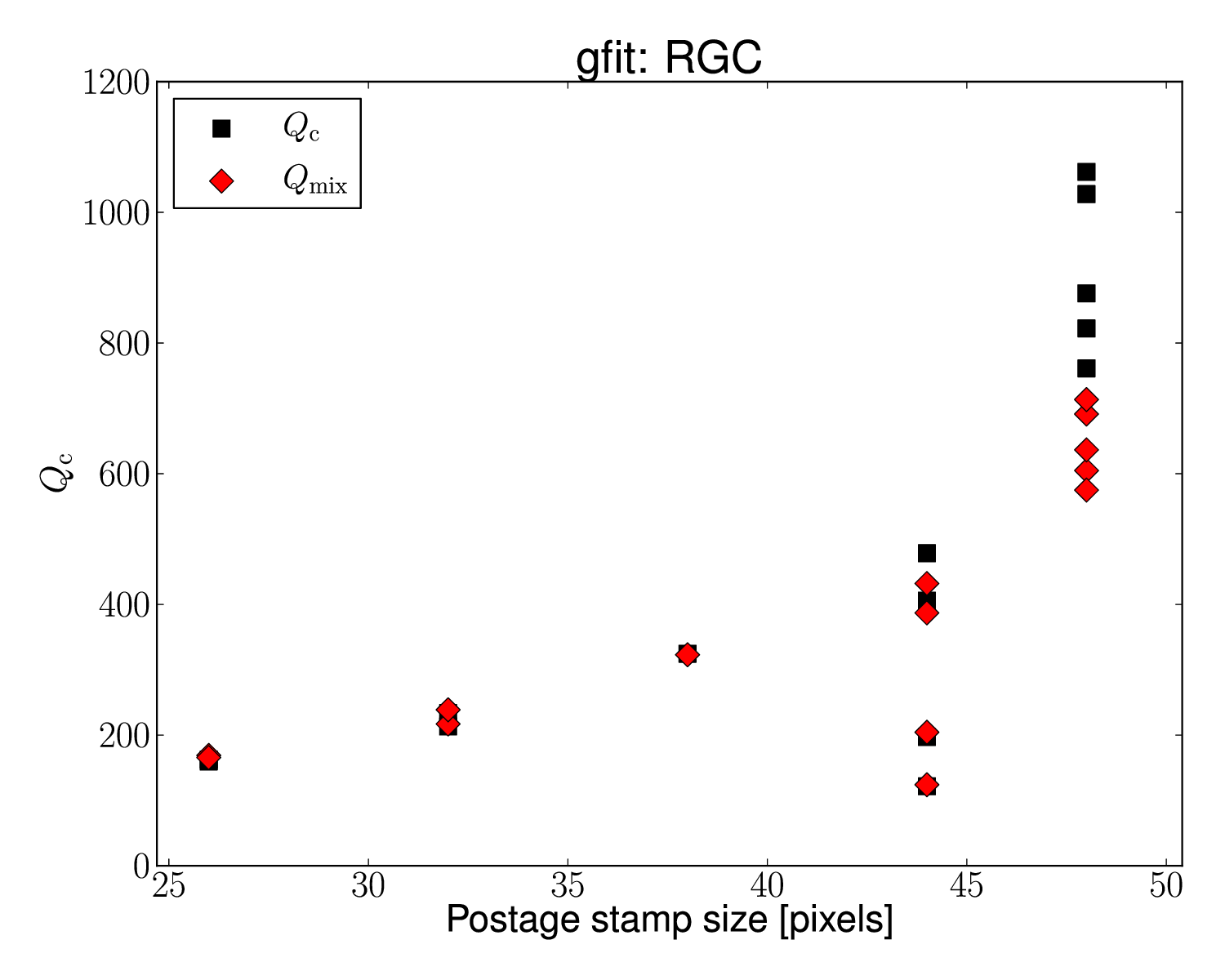}
\includegraphics[width=0.93\columnwidth,angle=0]{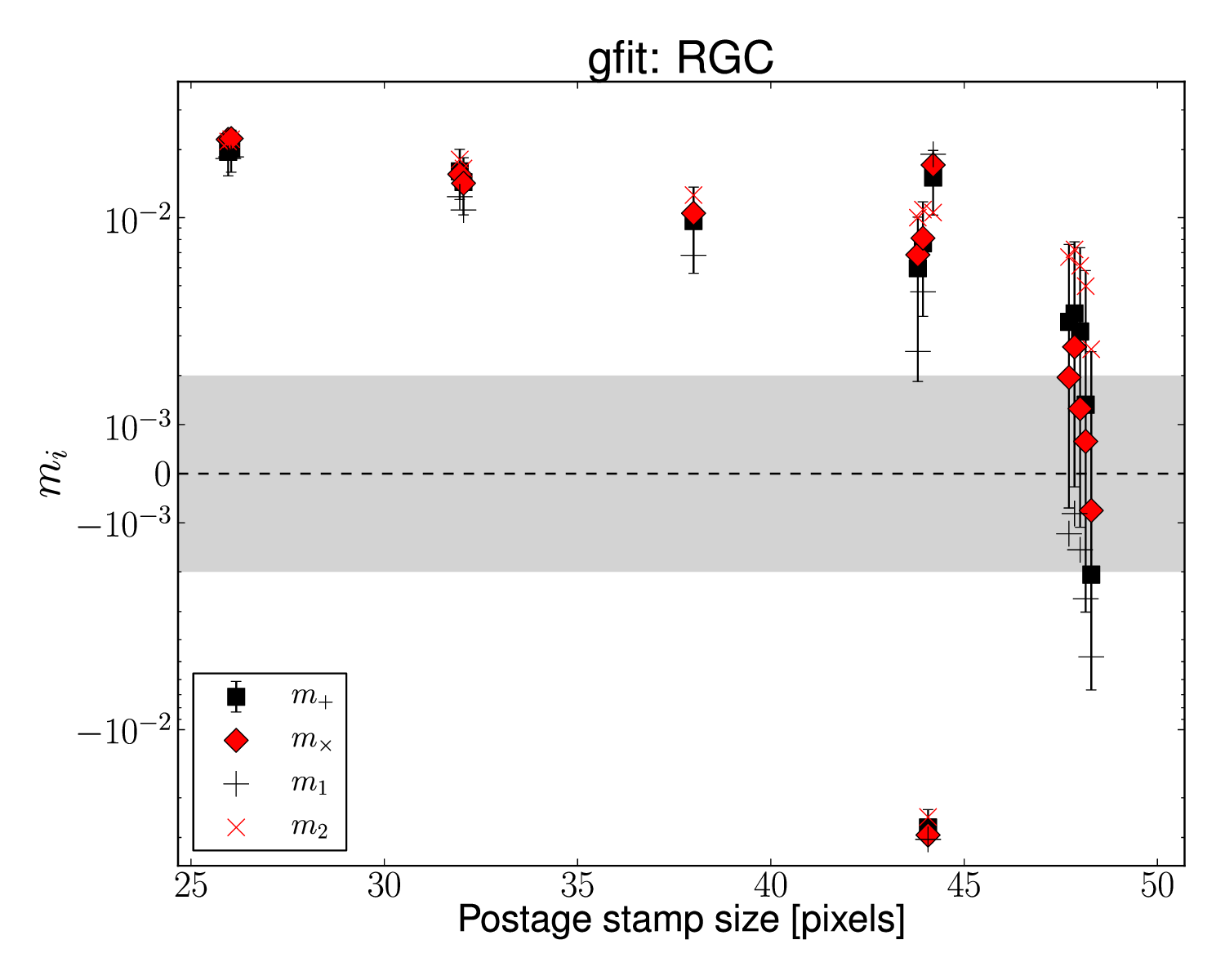}
\includegraphics[width=0.93\columnwidth,angle=0]{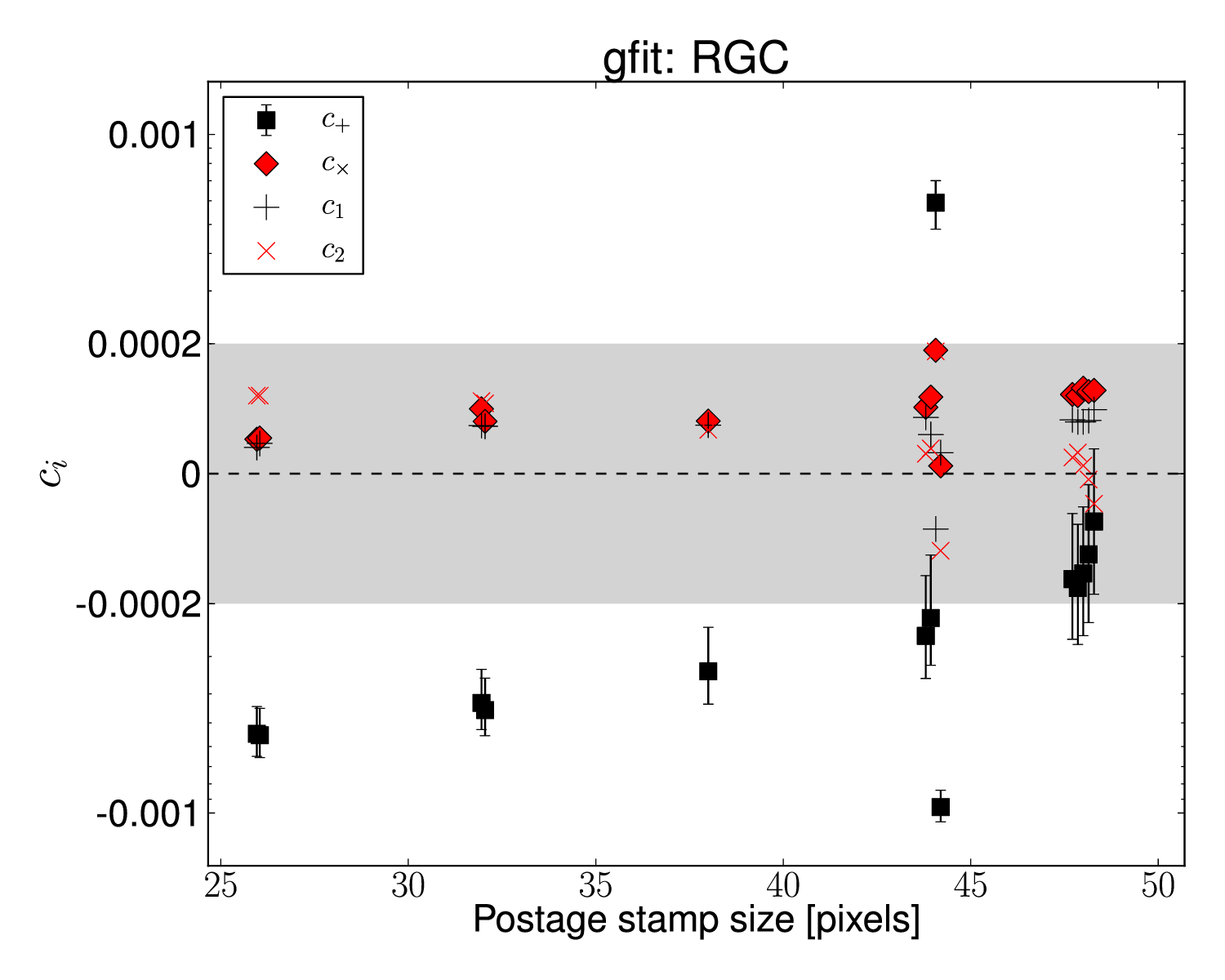}
\caption{\label{fig:gfit_focus_rgc}
  $Q_{\rm c}$ and $Q_{\rm mix}$ (top), and the bias components $m_i$ (middle) and $c_i$ (bottom), for the
  \code{gfit} method as a function of the postage stamp size used for
  modeling the galaxy images in the RGC branch.  The target
  regions are shown as a grey shaded region, within which the vertical
  axis has a linear scaling; outside of the shaded region, the scaling
  is logarithmic.  Multiple submissions with the same stamp size have
  slight horizontal offsets for clarity. 
  The errorbars are correlated between the submissions, so the figure
  cannot be used to assess statistical significance of differences
  between them.  See the discussion in the text for quantitative
  calculations of statistical significance.  The $m_i$ and $c_i$ panels
  only show errors on a single quantity ($i=+$), for clarity.
}
\end{center}
\end{figure}

We begin by analyzing the fourteen submissions in RGC.  Correlating the $Q_{\rm c}$ values
with the settings that vary for these submissions, 
we find that the parameter that most directly predicts
$Q_{\rm c}$ is the postage stamp size used for the model
fitting (see top panel of Fig.~\ref{fig:gfit_focus_rgc}).  As shown, 
using the full $48\times 48$ postage stamp maximizes the $Q_{\rm c}$ score.
\begin{figure}
\begin{center}
\includegraphics[width=0.93\columnwidth,angle=0]{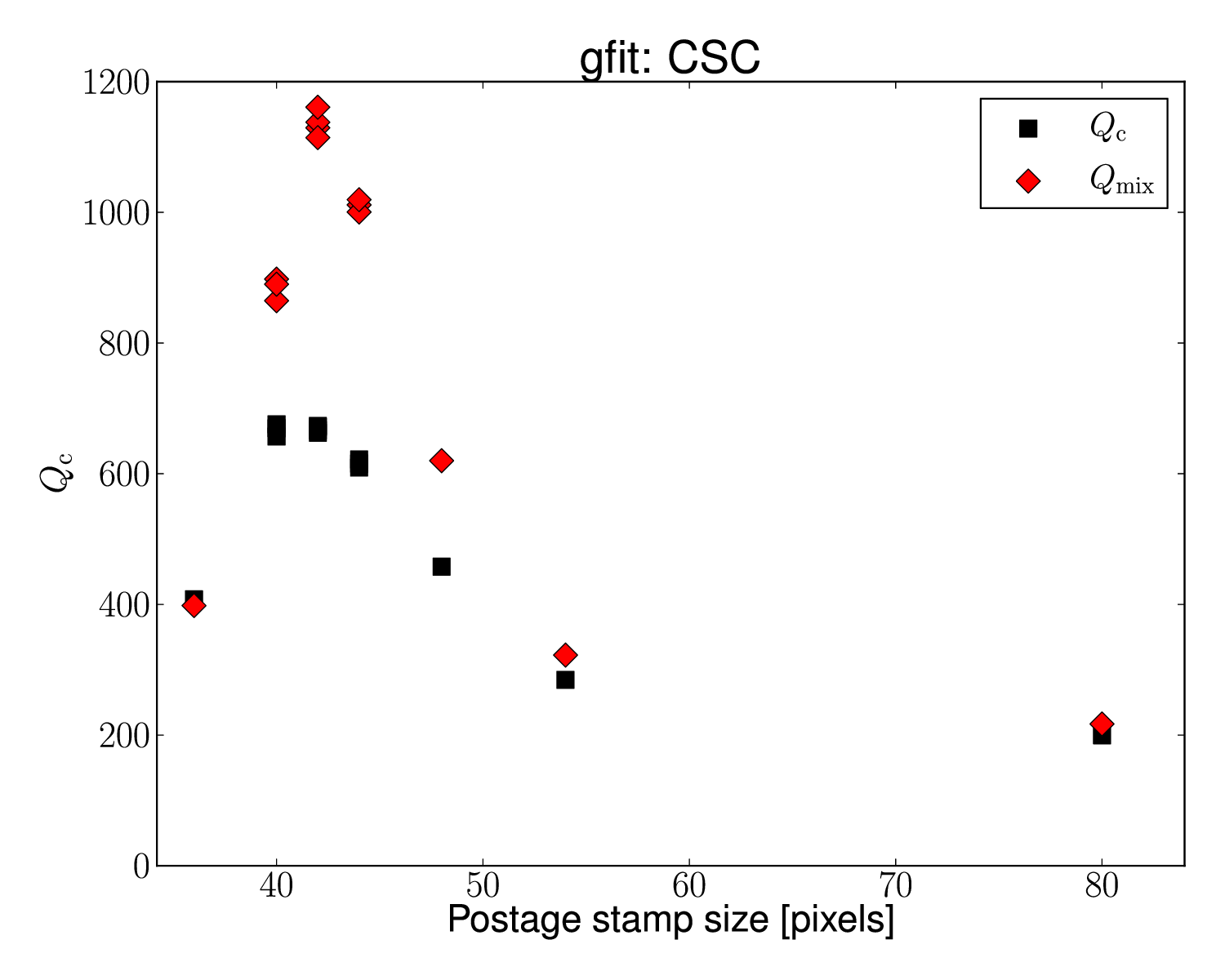}
\caption{\label{fig:gfit_focus_csc}$Q_{\rm c}$ and $Q_{\rm mix}$ for \code{gfit}
  as a function of the postage stamp size used for modeling the
  galaxy images in the CSC branch.}
\end{center}
\end{figure}

To understand this correlation, we consider the multiplicative bias as
a function of postage stamp size (middle panel of
Fig.~\ref{fig:gfit_focus_rgc}).  As shown, except for a few outliers,
the multiplicative biases $m_+$ and $m_\times$ that contribute to
$Q_{\rm c}$ increase from being consistent with zero to $2.0\pm 0.4$
and $2.2\pm 0.5$ per cent, respectively, as 
the postage stamp is reduced to half of its (linear) size.  
The statistical significance of the difference between the results
with the maximum and minimum stamp size is more than the $3\sigma$
that it appears to be in Fig.~\ref{fig:gfit_focus_rgc}; given the high
($\sim 0.75$) correlation coefficient between the submissions, the
change in $m$ is detected at approximately $8\sigma$
significance.

For maximum-likelihood fitting methods, we expect  a
calibration bias due to the effects of noise (``noise bias''). 
One interpretation of the RGC results at the maximal postage stamp
size is therefore a (cancelling) combination of noise bias with other
potential biases, such as those expected due to an imperfect galaxy
model.

As the postage stamp size is reduced, the likelihood surface for the
shear estimate changes due to reduced information about the light
profile, and this change will generally depend on the galaxy size and
shape, postage stamp size, and the noise level.  This change in the
likelihood surface will in general change the location of the maximum
likelihood, causing a potential bias for such methods.  We refer to the
resulting bias on ensemble shear estimates\footnote{\newtext{Note that with
  perfect models and in the absence of noise, truncation 
should not in general cause a bias.  Truncation bias could therefore
be seen as a modulation of the model and/or noise biases as the
weighting of the pixels changes.}} as ``truncation bias''. 
For this method, the sign of the effect is apparently increasingly
positive as stamp sizes decrease, though that does not necessarily
have to be the case for all methods.

We can also see signs that $m_1$ and $m_2$,
the calibration biases defined in the  pixel coordinate system, \newtext{may be related as  $m_2 \approx m_1+0.007$ ($1.5\sigma$ significance).} 
A difference between the calibration bias along the pixel directions
($m_1$) and along the diagonals ($m_2$) would be consistent with the results
of previous work \citep{2007MNRAS.376...13M,2007PASP..119.1295H}, and
could plausibly be explained either by the 
different effective sampling of the galaxy and PSF profile along those
directions, or by the fact that postage stamp itself extends further
in the diagonal directions. 
For the maximal postage stamp size, $m_1$ and $m_2$ have
opposite signs, which yields $m_+$ and $m_\times$ near zero.
For this reason, $Q_{\rm c} > Q_{\rm mix}$ for the maximal
postage stamp; in this case, $Q_{\rm mix}$ is a better estimator of the
level of systematics in \code{gfit}.

We also investigated the additive bias and its variation with postage
stamp size in the bottom panel of Fig.~\ref{fig:gfit_focus_rgc}.
Results consistent with zero, $c_+ = (-1\pm 1)\times 10^{-4}$, are
achieved at the maximal postage stamp size, but additive bias becomes
steadily more negative until it exceeds our target value for the
smallest postage stamp sizes, where $c_+= (-5\pm 1)\times 10^{-4}$.
This result suggests that additive systematics also exhibit truncation
bias (with $7\sigma$ significance after accounting for the correlation
between submissions).  However, the best-fitting values of $c_1$, $c_2$,
and $c_\times$ are within the target region and statistically
consistent with zero.

Fig.~\ref{fig:gfit_focus_rgc} also shows that a few submissions with large
postage stamp sizes had worse than typical results.  For the largest postage stamp size, these
variations in $Q_{\rm c}$ are due to variations in the amount of
filtering imposed on the output catalog before averaging to get a mean
shear for the field.  The filtering typically involves the value of
the best-fit radii, the sum of the fit residuals (related to fit
quality), and the $S/N$, and usually involves removing several per
cent of the galaxies in each field.  For the next-largest stamp size
(44), the submissions with worse results involved experimenting with
fit settings (e.g., allowing components with negative flux), with use
of denoised images, and with stacking the nine provided PSF images
instead of using just one.

Among the space branches, CSC has many \code{gfit} entries with
different postage stamp sizes, though the maximum is $80\times 80$
(out of a possible $96\times 96$).  As for the ground, postage stamp
size is the most important factor, with $Q_{\rm c}$ as a function of
this parameter in Fig.~\ref{fig:gfit_focus_csc}.  In this case, the
best postage stamp size of $40\times 40$ does non-negligibly truncate
the light profiles of a fair fraction of the galaxies, whereas the
largest postage stamp size used ($80\times 80$) has a substantially
lower $Q_{\rm c}$ due to its multiplicative calibration bias of $m_+=
-2.0\pm 0.3$ per cent and $m_\times=-1.3\pm 0.3$ per cent.  These
biases are reduced to $m_+ = -0.3\pm 0.3$ per cent and $m_\times =
+0.5\pm 0.3$ per cent for the best stamp size, an $>11\sigma$ change
when accounting for the strong correlation between the submissions.

The natural interpretation is that the various sources of bias in
the space simulations for the largest stamp size result in a negative
multiplicative bias of $\langle m\rangle \simeq -1.7\pm 0.3$\% (where
$\langle m\rangle = [m_+ + m_\times]/2$), but a \emph{positive}
truncation bias cancels this out for smaller postage stamp sizes. 
\newtext{The fact that the bias becomes more positive for smaller stamp sizes is consistent across ground and space simulations.}

The potential sources of bias in the $80\times 80$ case include noise
bias, some truncation bias compared to the full $96\times 96$ case,
and model bias due to an inexact match between the parametric model in
the simulations versus those used by \code{gfit}.  In all cases, there
is a detection of additive systematics, with $c_+$ ranging from $(7\pm
1)\times 10^{-4}$for the $80\times 80$ stamp size, to $(3\pm 1)\times
10^{-4}$ for stamps smaller than $60\times 60$.  The decrease in $c_+$
due to truncation bias is significant at the $9\sigma$ level.

\subsubsection{Fair cross-branch comparison}\label{subsubsec:gfit-cb}

The best
results from the \code{gfit} team used quite different postage stamp
sizes for each branch.  Since the galaxy populations are, in a
statistical sense, consistent when comparing across all ground
branches and all space branches, a fair cross-branch comparison would
use consistent settings for all ground branches and for all
space branches.  Here we present the results of this 
comparison.  

\begin{figure*}
\begin{center}
\includegraphics[width=0.99\columnwidth,angle=0]{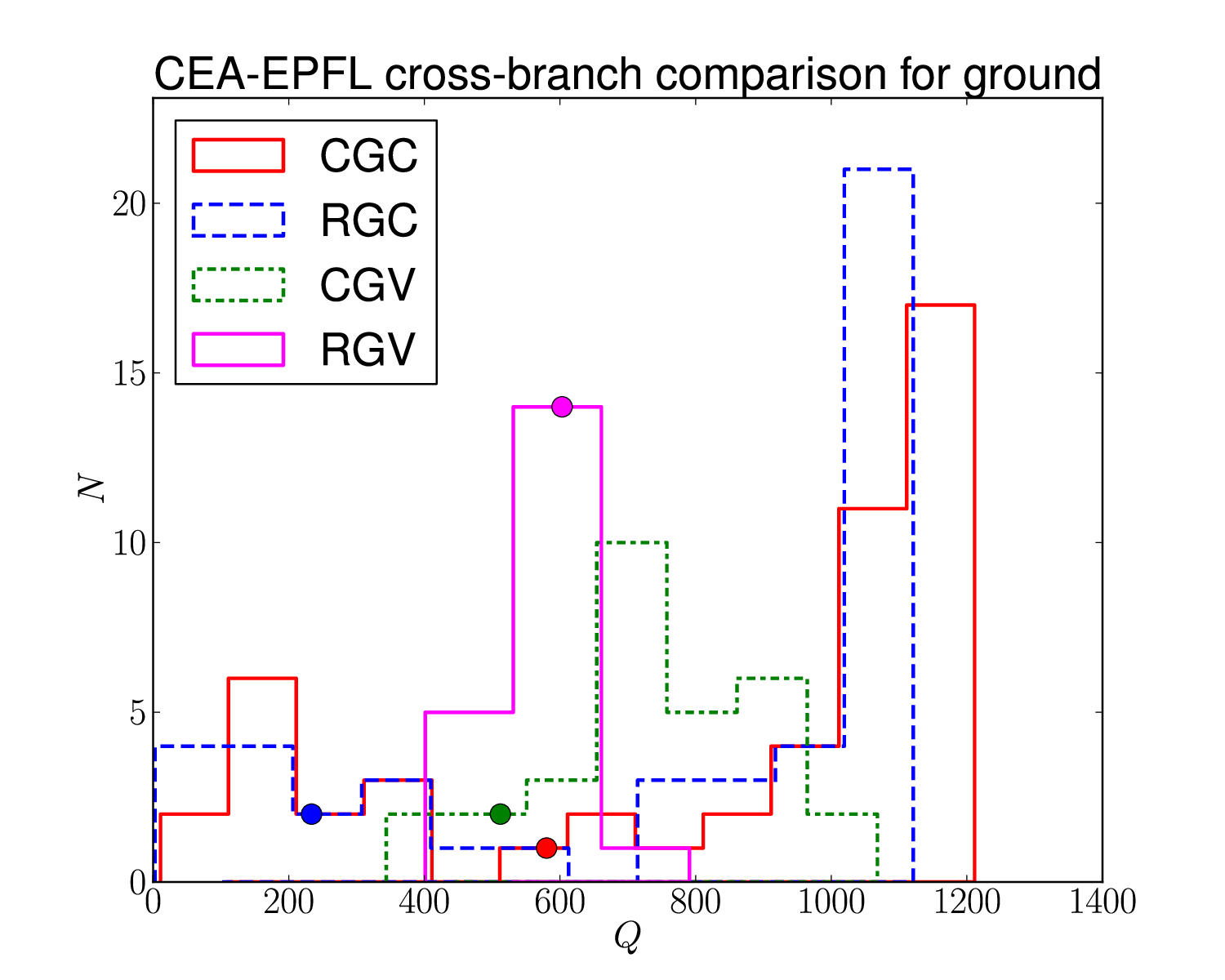}
\includegraphics[width=0.99\columnwidth,angle=0]{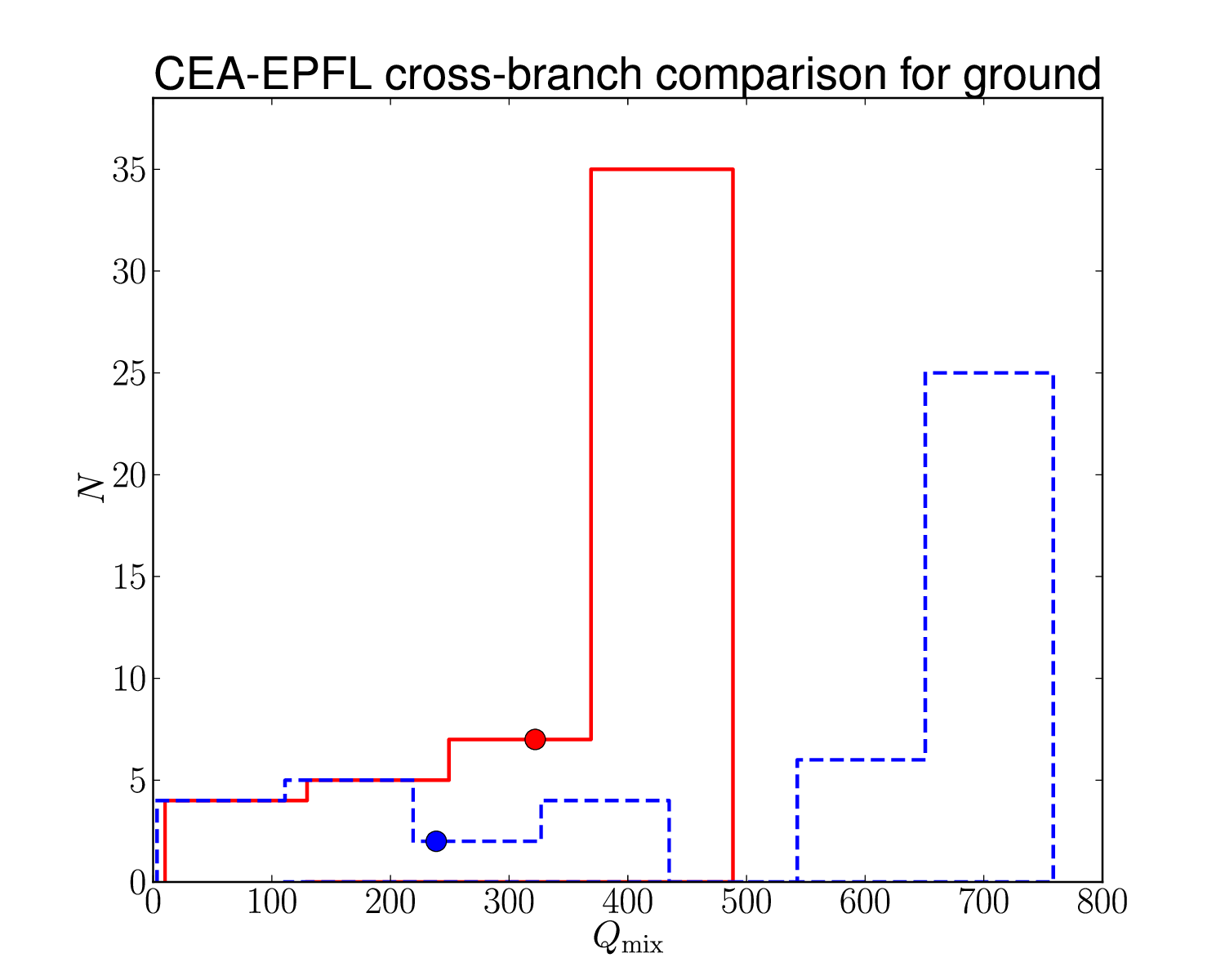}
\includegraphics[width=0.99\columnwidth,angle=0]{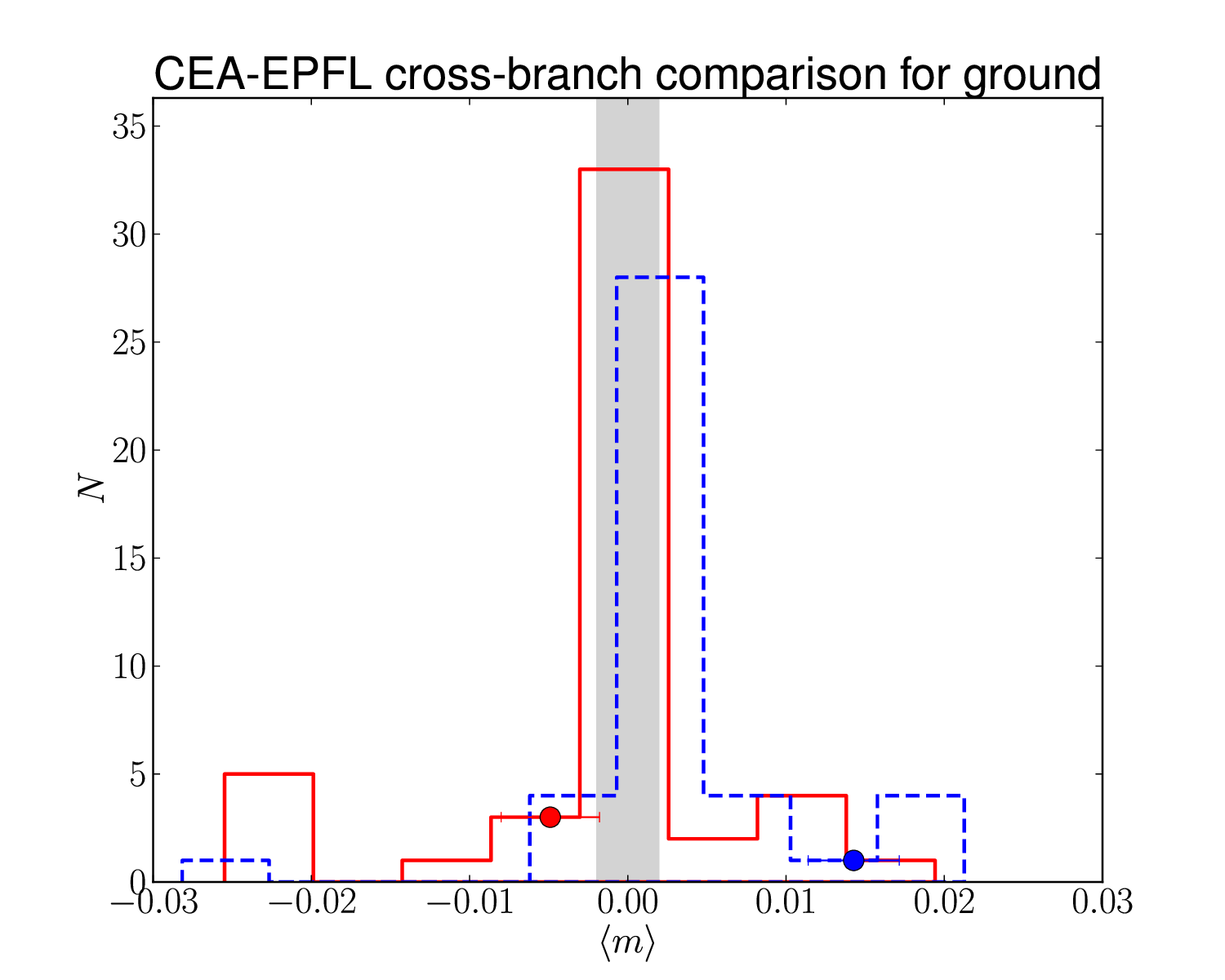}
\includegraphics[width=0.99\columnwidth,angle=0]{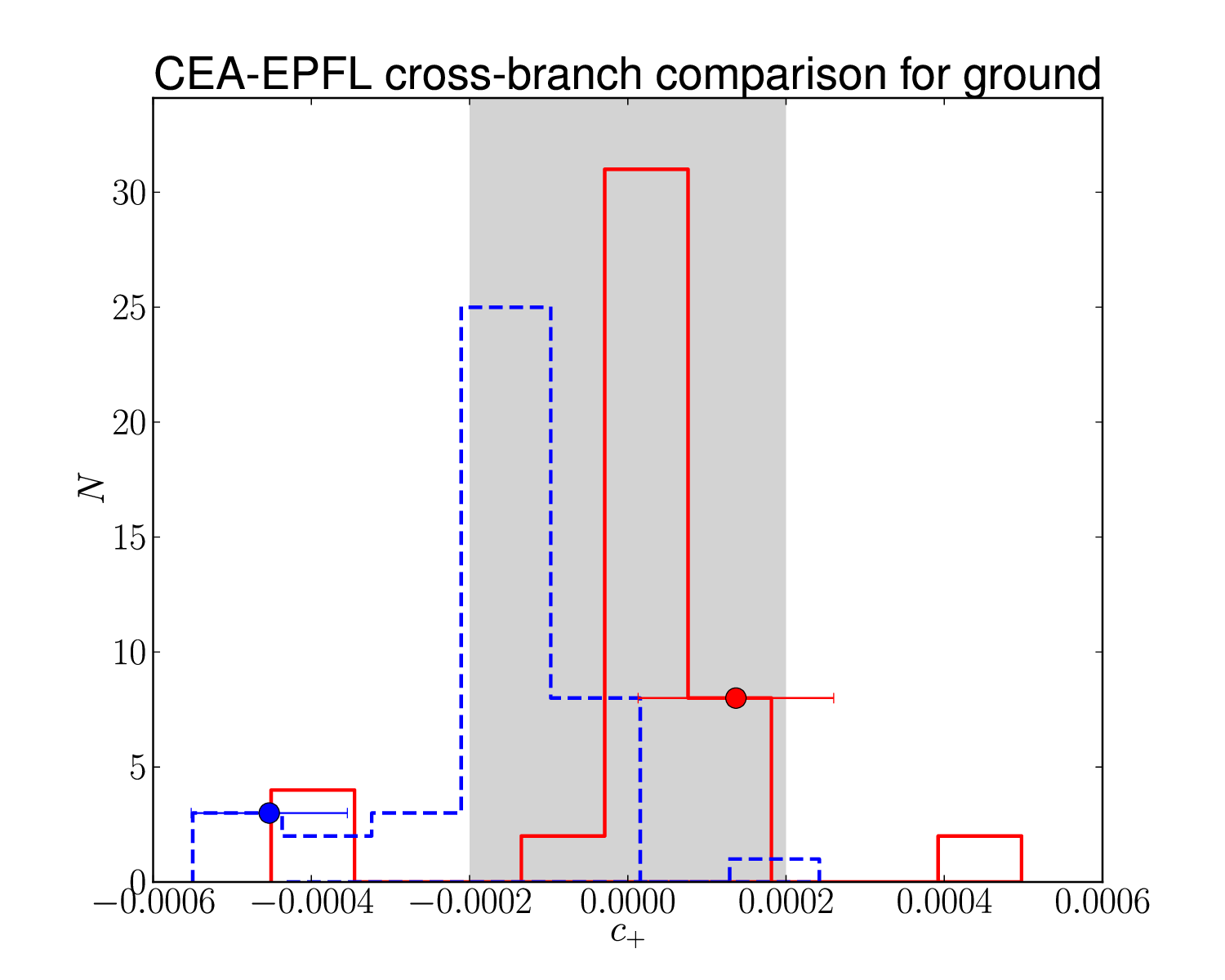}
\caption{\label{fig:gfit_cross_ground}\referee{ {\em Top left:} Histogram of $Q$ value
  (either $Q_{\rm c}$ or $Q_{\rm v}$ depending on the branch) for the
  \code{gfit} method for all submissions in ground branches from
  CEA-EPFL and EPFL\_gfit teams.  The large dots located on the histograms indicate the
  submissions that are part of the fair cross-branch comparison, with the same choice of postage stamp size. {\em
    Top right, bottom left, bottom right:} The same, but for $Q_{\rm
    mix}$, $\langle m\rangle$, and $c_+$ (respectively), for 
  constant-shear branches.  In the bottom plots, the points have horizontal errorbars indicating
  their statistical uncertainty, and the shaded regions indicate the target values of $\langle
  m\rangle$ and $c_+$.  Outliers have been removed from the bottom two
  panels so that the main part of the distribution can be clearly seen.}
}
\end{center}
\end{figure*}

For ground branches, all branches except for CGC had a
submission with stamp size of $32\times32$, and CGC has one with
$30\times30$, which is close enough for this
comparison.  Fig.~\ref{fig:gfit_cross_ground} shows the $Q$ values
for all \code{gfit} submissions in all ground branches, particularly
indicating those submissions that are part of the fair cross-branch
comparison.  Note that the
$Q_{\rm c}$ and 
$Q_{\rm v}$ values do not relate to shear systematics in quite the
same way, so we cannot directly compare across constant and variable
shear branches.  However, it is clear in general that the submissions
in this fair comparison \referee{sample perform} respectably (\referee{$200\lesssim Q\lesssim 600$})
but do not typically include the best submission in each branch.  The
results for the mixed metric $Q_{\rm mix}$  in that figure (top right)
for constant-shear branches actually shows 
consistency across branches for the selected submissions, with $250\lesssim Q_{\rm mix}\lesssim 350$.

The bottom row of Fig.~\ref{fig:gfit_cross_ground} shows the distribution of 
multiplicative biases averaged over both components, $\langle m\rangle =
(m_+ + m_\times)/2$, and additive biases aligned with the PSF ($c_+$;
no significant $c_\times$ was detected for this or any method) \referee{for all submissions in CGC and RGC}.  For
$\langle m\rangle$, given the fixed \code{gfit} analysis settings, the
differences between the red points in CGC and RGC indicate additional
multiplicative model bias due to real galaxy morphology \newtext{of $\langle m\rangle_{\rm RGC} - \langle m\rangle_{\rm
  CGC} = 1.9\pm 0.4$ per cent. There}
  may also be model bias in CGC due to the parametric models used by
  \code{gfit} not precisely matching the ones in the GREAT3
  simulations.  The CGC vs.\ RGC comparison therefore reflects only
  additional model bias due to real galaxy morphology, rather than all sources of
  model bias.

\referee{When considering the points that indicate the submissions in the fair comparison sample, t}he additive biases are consistent with zero for CGC but a significant
detection for RGC is seen, suggestive that model bias due to realistic
morphology can result in additive errors from imperfect PSF
deconvolution. However, it is worth bearing in mind that the postage
stamps used in this cross-branch comparison are significantly
truncated.  In all these ground branch submissions there will thus be
some truncation bias that might interact with other biases such as
model biases.  The individual effects cannot be wholly isolated, but
the compound effects are clear.

For space branches, the ``fair comparison'' submissions had postage
stamp sizes of $44\times44$, representing significant truncation compared to the full
size of $96\times96$.  The fair comparison results do not exhibit the
very high $Q$ values of the best submissions ($>1000$) but are,
however, in the range $500<Q<800$.  Comparing CSC and RSC suggests a
multiplicative model bias due to realistic galaxy morphology of
$\langle m\rangle_{\rm RSC} - \langle m\rangle_{\rm
  CSC} = 0.7\pm 0.2$ per cent, but no additive model bias.

\subsubsection{Summary}

In summary, \code{gfit} results are significantly affected by the
postage stamp size used for modeling, with small stamp sizes resulting
in what we call truncation bias.  This (generally positive) truncation
bias can offset the negative noise bias that is a natural consequence
of using a maximum-likelihood fitting method.  The next most
interesting factor is the filtering of the catalog to exclude galaxies
on the basis of fit quality or fit parameters, with typically a few
per cent of galaxies being excluded.

Results for a consistent choice of stamp size suggest differences in
$\langle m \rangle$ between the control and realistic galaxy
experiments of order $\Delta \langle m \rangle \simeq
1$--$2$\% (greater for ground than for space) due to model bias from
realistic galaxy morphology.  \newtext{This conclusion is based on the fact 
that the galaxy and data properties in these 
branches are the same, except for the way of representing the light
profiles (parametric models vs.\ \emph{HST} images).  Thus truncation, noise, and other biases should be}
consistent between the two sets of results.  Differences in $c_+$ for
the control and realistic galaxy experiments depend on
whether the simulated data represents a space survey or a ground
survey.

We note the general point that, using this dataset,
we cannot cleanly separate model bias in true isolation,
as compounding interplays may exist between model bias, truncation bias, noise
bias and other biases. This would be an interesting subject
for future study.
For the purpose of controlling for the effects found in this analysis
of \code{gfit} results, in the general analysis in Sec.~\ref{sec:overall}, we will use a set
of \code{gfit} submissions with consistent postage stamp sizes (one
set for ground, and another for space).  These will be the same submissions used in
Sec.~\ref{subsubsec:gfit-cb}.

\subsection{Amalgam@IAP}

\subsubsection{Controlled tests of variation in Amalgam@IAP options}

The Amalgam@IAP analysis pipeline (see Appendix~\ref{app:amalgam}) has a significant number of
parameters that can change.  These include the postage stamp size,
subpixel resolution, and order of interpolation used to combine star
images for PSF estimation; the type of filtering of the galaxy
catalogs; the modeling window (the maximum allowed region to use for
modeling, which was either fixed to the postage stamp size or was
permitted to vary with a maximum value equal to the postage stamp
size); the use of regular vs.\ modified $\chi^2$ \newtext{to mitigate the
effects of galaxy blends} (see
Appendix~\ref{app:amalgam}); the use of an additional penalty term on \sersic\
index and/or aspect ratio, see Eq.~\eqref{eq:amalgam-cost-function};
and the choice of effective shape noise $\sigma_s$ in the weighting
used to combine individual galaxy shape estimates (see
Appendix~\ref{app:amalgam-weight}).

Early in the challenge it was found that increasing the sampling
density of both the PSF and the galaxy models ($\approx 2.5\times$ on
each axis compared to the values that would automatically be set by
the regular versions of PSFEx and SExtractor) significantly improved
the scores, at the price of increasing computing
time by an order of magnitude.

In RGC, we carried out multi-factor ANOVA to
understand the most important factors determining the performance of
the Amalgam@IAP team.  Unfortunately, even
with nearly 40 submissions, the 8-dimensional parameter space was not
sampled well enough to get a clear answer.  The results 
suggest that $\sigma_s$ was the most important factor
determining performance, with
choice of form for the $\chi^2$ (regular preferred over modified) and
use of penalty term (penalty on aspect ratio preferred over not) being
important with marginal significance.

\begin{figure}
\begin{center}
\includegraphics[width=0.975\columnwidth,angle=0]{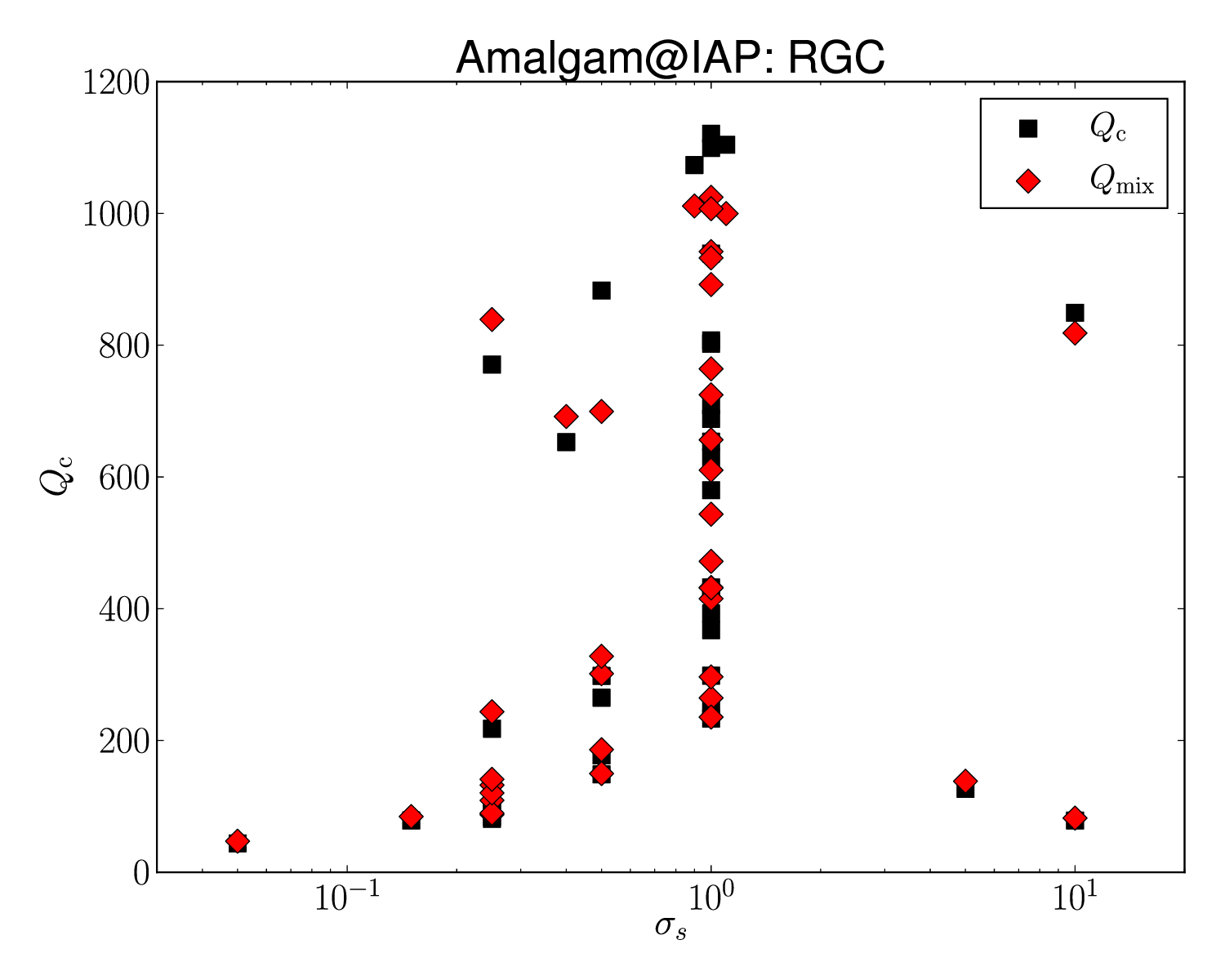}
\includegraphics[width=0.975\columnwidth,angle=0]{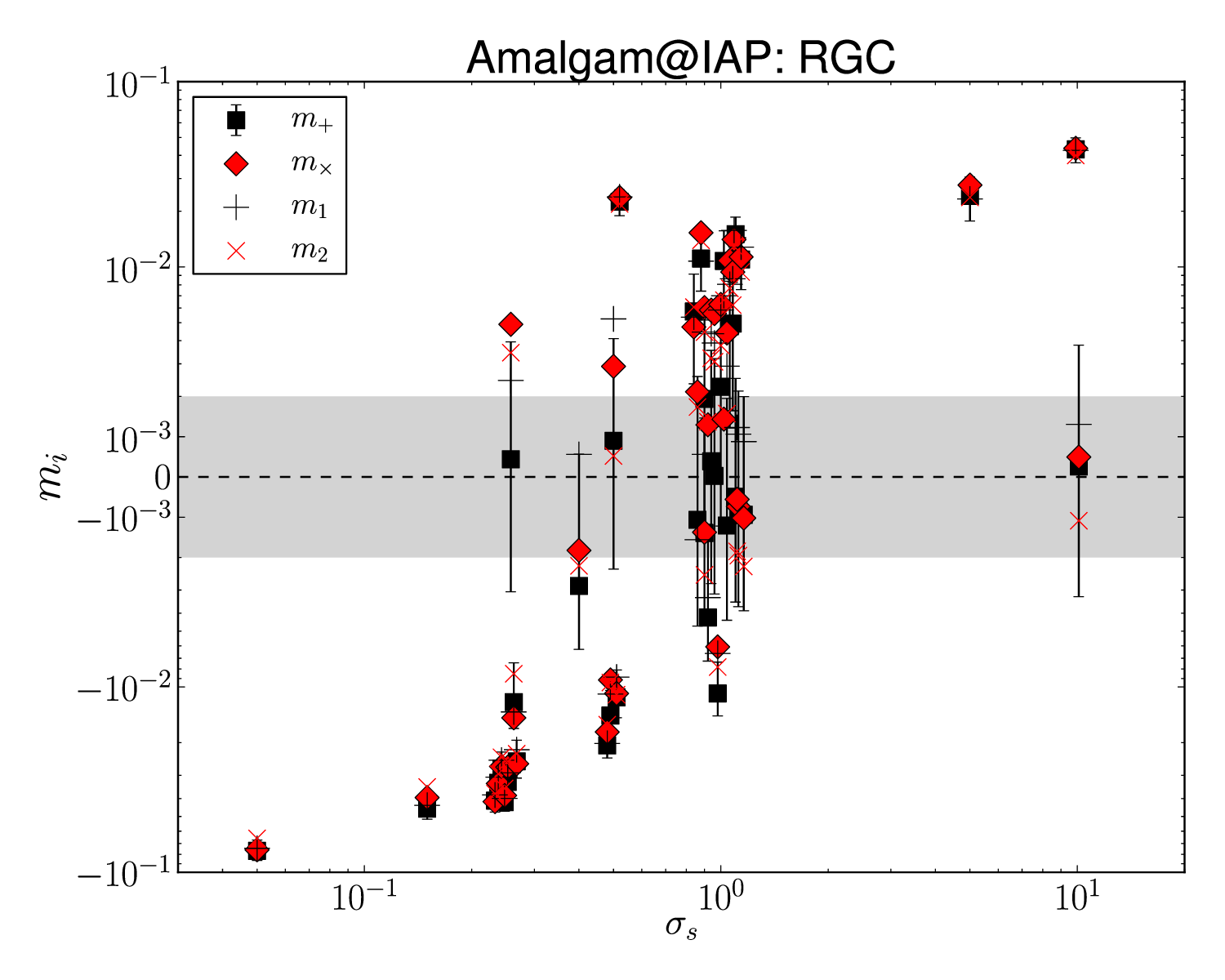}
\includegraphics[width=0.975\columnwidth,angle=0]{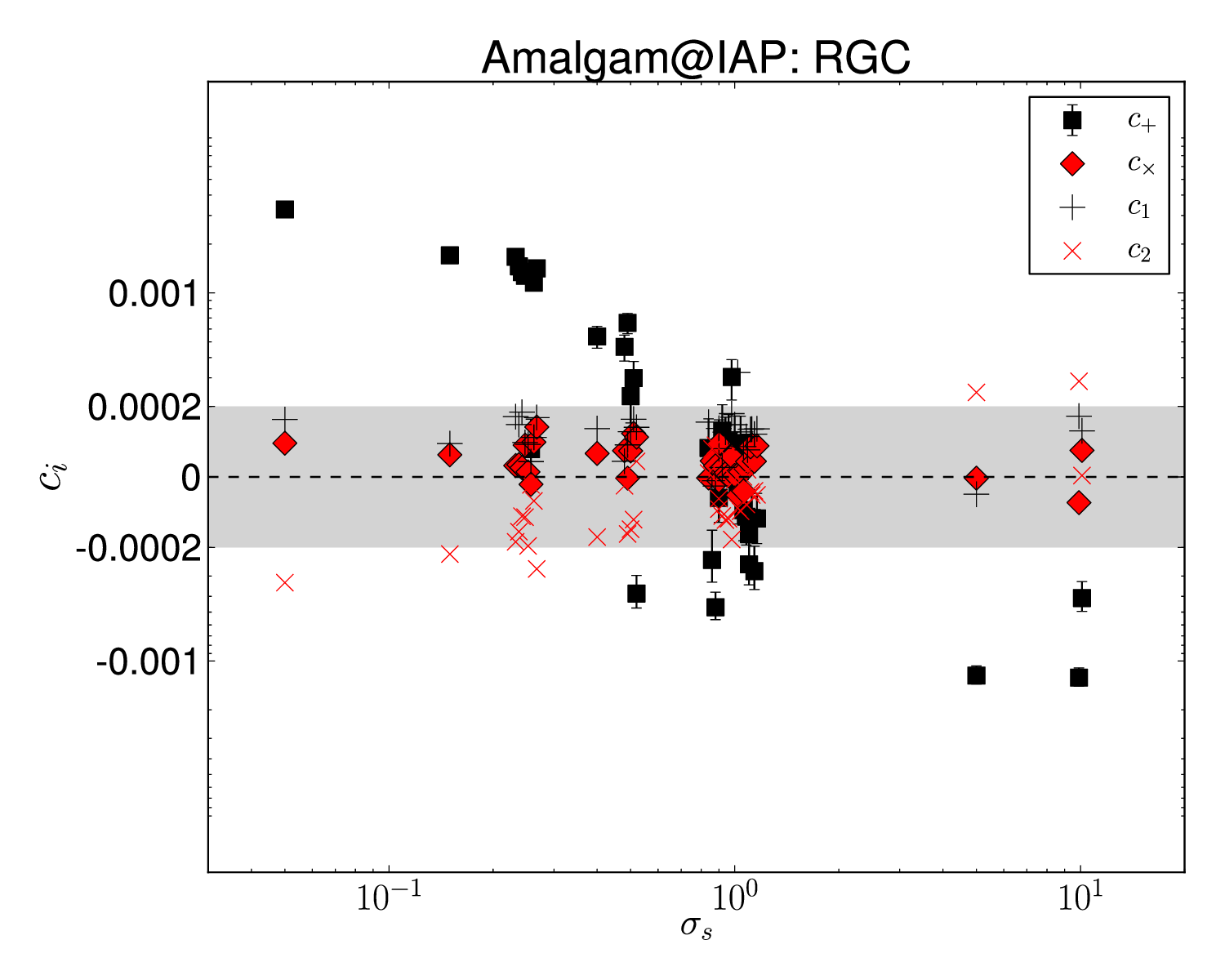}
\caption{\label{fig:amalgam_focus_rgc}From top to bottom, we show 
  $Q_{\rm c}$, $m_i$, and $c_i$ for the Amalgam@IAP team submissions 
  as a function of the $\sigma_s$ used in the
  weighting scheme, for all submissions in RGC.  The target regions are shown
  as a grey shaded region, within which the vertical axis has a linear
scaling; outside of the shaded region, the scaling is logarithmic.
Note that the entries shown at $\sigma_s=10$ actually had
$\sigma_s=\infty$, i.e., completely equal weighting for all galaxies.
Multiple submissions with the same $\sigma_s$ have slight horizontal
offsets for clarity. The $m_i$ and $c_i$ panels
  only show errors on a single quantity, for clarity.
}
\end{center}
\end{figure}

Given the importance of $\sigma_s$,
Fig.~\ref{fig:amalgam_focus_rgc} shows the variation of our metrics with this parameter.
As shown in the top panel, $Q_{\rm c}$ sharply decreases for
very small $\sigma_s$, and reaches a maximum for $\sigma_s\approx 1$.
For infinite $\sigma_s$ 
(constant weighting), there are two submissions with quite different
$Q_{\rm c}$ values, $849$ and $78$, which we discuss in more detail below.

The decrease in $Q_{\rm c}$ for very low $\sigma_s$ is quite
interesting.  As $\sigma_s$ approaches zero, the weighting
scheme gives a strong preference to very high $S/N$
galaxies. In real data, there is no advantage to giving such a
preference because of shape noise.  However, in GREAT3, we have
canceled out the shape noise by including 90$^\circ$ rotated pairs, so
in principle, a perfect shear estimate for just the two highest $S/N$
galaxies would perfectly determine the shear for the
whole field.  The
low $Q_{\rm c}$ in this case implies that either the covariance matrix
used for the weighting is poorly determined or
has some correlation with shear direction, {\em or} that the
shear estimates for high-$S/N$ galaxies are poor. 
The high-$S/N$ galaxies should have little noise bias, but may have
model 
bias due to a mismatch between the input parametric models and the ones
fitted by the Amalgam@IAP team. Another possible explanation relates to the
adaptive selection of modeling window size (up to but not beyond the
size of the input postage stamps).  If the algorithm chooses
too-small postage stamps for the highest-$S/N$ galaxies, it could
introduce truncation biases as seen in \code{gfit} results (see
Sec.~\ref{subsec:focus:gfit}).
Since a similar trend in $Q_{\rm c}$ was seen in CGC, the problem
is not plausibly due solely to realistic galaxy morphology.
Unfortunately given the data that we have, we are unable to tease
apart these effects.

The other panels in Fig.~\ref{fig:amalgam_focus_rgc} show the $m_i$
and $c_i$ values as a function of $\sigma_s$, to explain the trends in
the $Q_{\rm c}$ plot.  For very low $\sigma_s$ (upweighting the high
$S/N$ galaxies), the multiplicative biases can be as bad as $-7.6\pm
0.5$ per cent, with a very high detection significance for the trends in $m_i$.
For constant weighting, the submission with near-zero $m_i$ and $c_i$
includes a penalty term on the aspect ratio, whereas the
poorly-performing submission does not (giving a $10\sigma$ change in
$m_i$).  In the bottom panel,
as $\sigma_s$ goes from 0.05 up to 1 and
finally to $\infty$ (corresponding to strong $S/N$ upweighting,
weighting with a substantial shape noise term, and constant
weighting, respectively),
$c_+$ goes from $(3.2\pm 0.2) \times 10^{-3}$, to consistent with zero, to negative
values, $(-4\pm 1)\times 10^{-4}$.  The statistical significance
of these changes is $>10\sigma$.  This suggests that $c_+$ for this
method is positive (negative) for the high- (low-)
$S/N$ galaxies.

\begin{figure}
\begin{center}
\includegraphics[width=\columnwidth,angle=0]{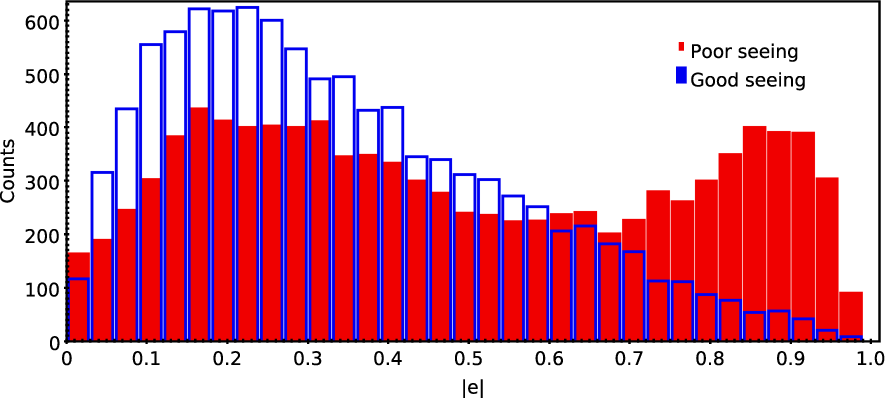}
\includegraphics[width=\columnwidth,angle=0]{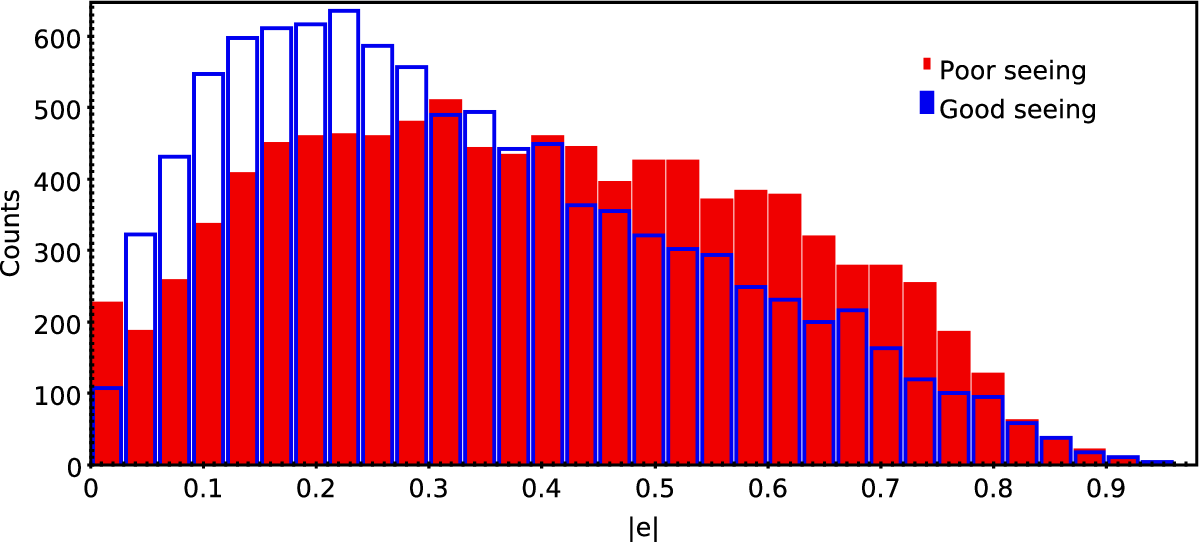}
\caption{\label{fig:amalgam_focus_penalty_cgc}Fitted ellipticity
  distributions for the Amalgam@IAP team for a good-seeing (blue) and
  poor-seeing (red) subfield in GREAT3, in the CGC branch.  The top
  (bottom) panel shows the results without  (with) a penalty term
  on aspect ratio.
}
\end{center}
\end{figure}

We now address the issue of the penalty term on aspect ratio, another
parameter of interest that causes highly significant changes in
multiplicative and additive biases as discussed above.  The idea of the penalty term is that for
galaxies that have low $S/N$ and poor resolution, the ellipticity is
so poorly determined that there is a very large tail to high
ellipticity (which is a manifestation of noise bias).  Hence the idea
is to penalize high ellipticity values by adding a term to the
$\chi^2$, which will have little effect on high-ellipticity objects
with high $S/N$. 
This was important particularly for fields with
poor seeing and/or substantial defocus that enlarged the PSF.  An example is
shown in Fig.~\ref{fig:amalgam_focus_penalty_cgc}.  The top panel
shows the fitted ellipticity distribution in a good-seeing (blue) and
poor-seeing (red) field in GREAT3 without the penalty term, and the
bottom panel shows the same when using the penalty term on aspect
ratio.  The distribution for the poor-seeing image has a
pronounced high-ellipticity tail that is nearly removed by the penalty
term, yet the shape of the distribution  in the good-seeing image is
less altered by the addition of this term.

In some sense, the addition to the term in the $\chi^2$ is equivalent
to multiplying the likelihood, i.e., imposing a prior on the
ellipticity.  It seems that this is a way to remove or reduce noise
bias in all fields (with stronger impact on those that have poor
seeing), eliminating the need for explicit calibration factors.  For
GREAT3, the best value of shape noise $\sigma_s$ to use in the
weighting scheme and the form of the penalty term to use in the
$\chi^2$ was clearly shown to do an excellent job at shear estimation
for the particular $p(\varepsilon)$ and galaxy property distributions
used here.  However, it is unclear whether these results would
necessarily be consistently reproducible for other datasets with
different intrinsic $p(\varepsilon)$, or those with a $p(\varepsilon)$
that correlates with other galaxy properties in a way that is not
reproduced here.  For this reason, further simulations would be needed
to evaluate the generality of this procedure for real data with a
variety of properties, and confirm that the exact $\sigma_s$ and form
of the penalty term gives similar results in cancelling out noise
bias.

\subsubsection{Fair cross-branch comparison}

For the Amalgam@IAP team, it was difficult to identify a single group of
settings used for all branches. Instead, four groups of settings with
submissions in a few branches were identified:
\begin{enumerate}[1.]
\item $\chi^2$ penalty term on aspect ratio $\theta_{\rm aspect}$;
  $\sigma_s=0.5$ for weighting.
\item $\chi^2$ penalty term on $\theta_{\rm aspect}$; uniform weighting
  ($\sigma_s=\infty$).
\item $\chi^2$ penalty term on $\theta_{\rm aspect}$ and
  \sersic\ index $n_s$; $\sigma_s=0.5$.
\item No priors on model parameters; $\sigma_s=0.5$.
\end{enumerate}
The settings also differed in minor ways that have 
little impact on performance.

\begin{figure}
\begin{center}
\includegraphics[width=0.99\columnwidth,angle=0]{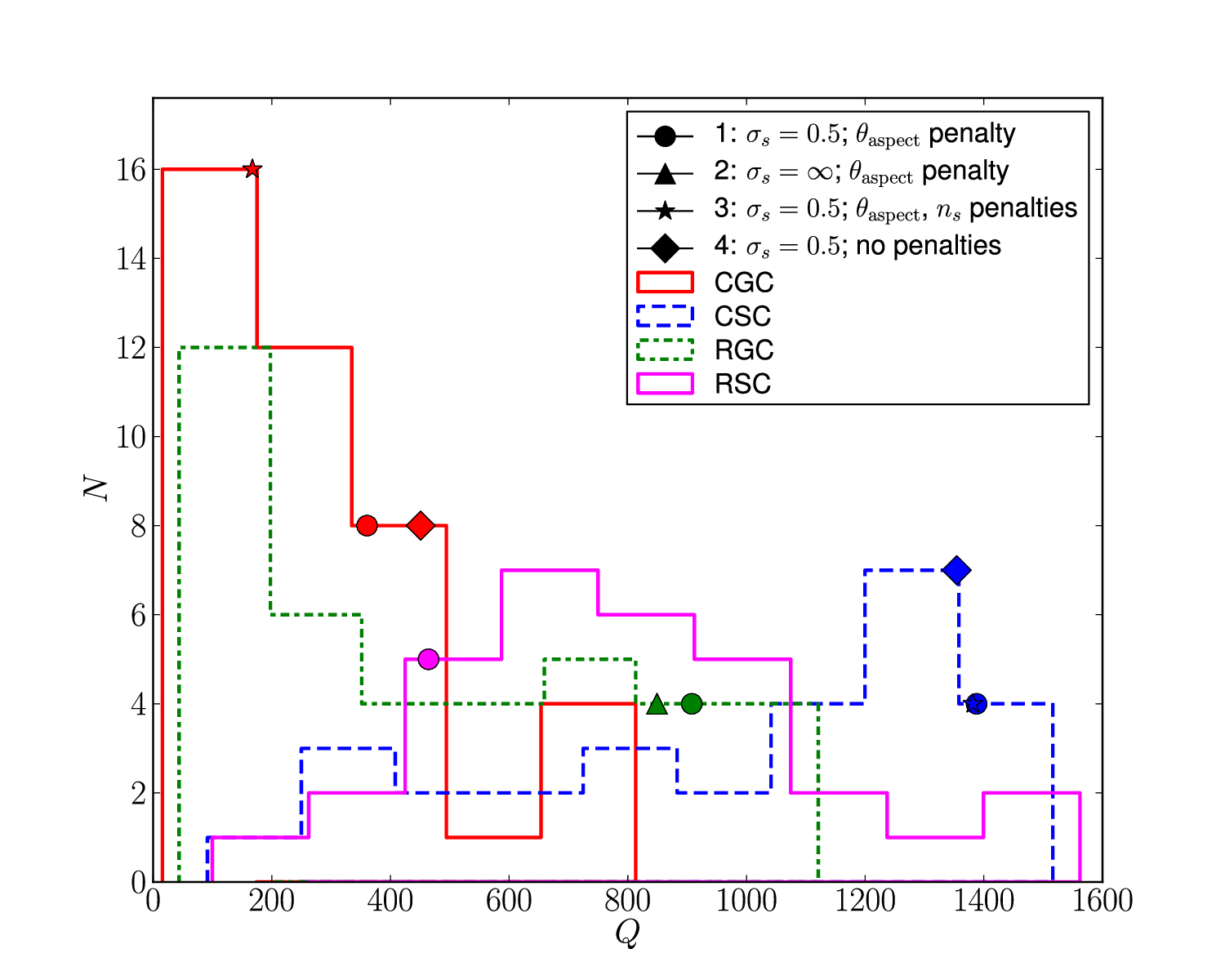}
\includegraphics[width=0.99\columnwidth,angle=0]{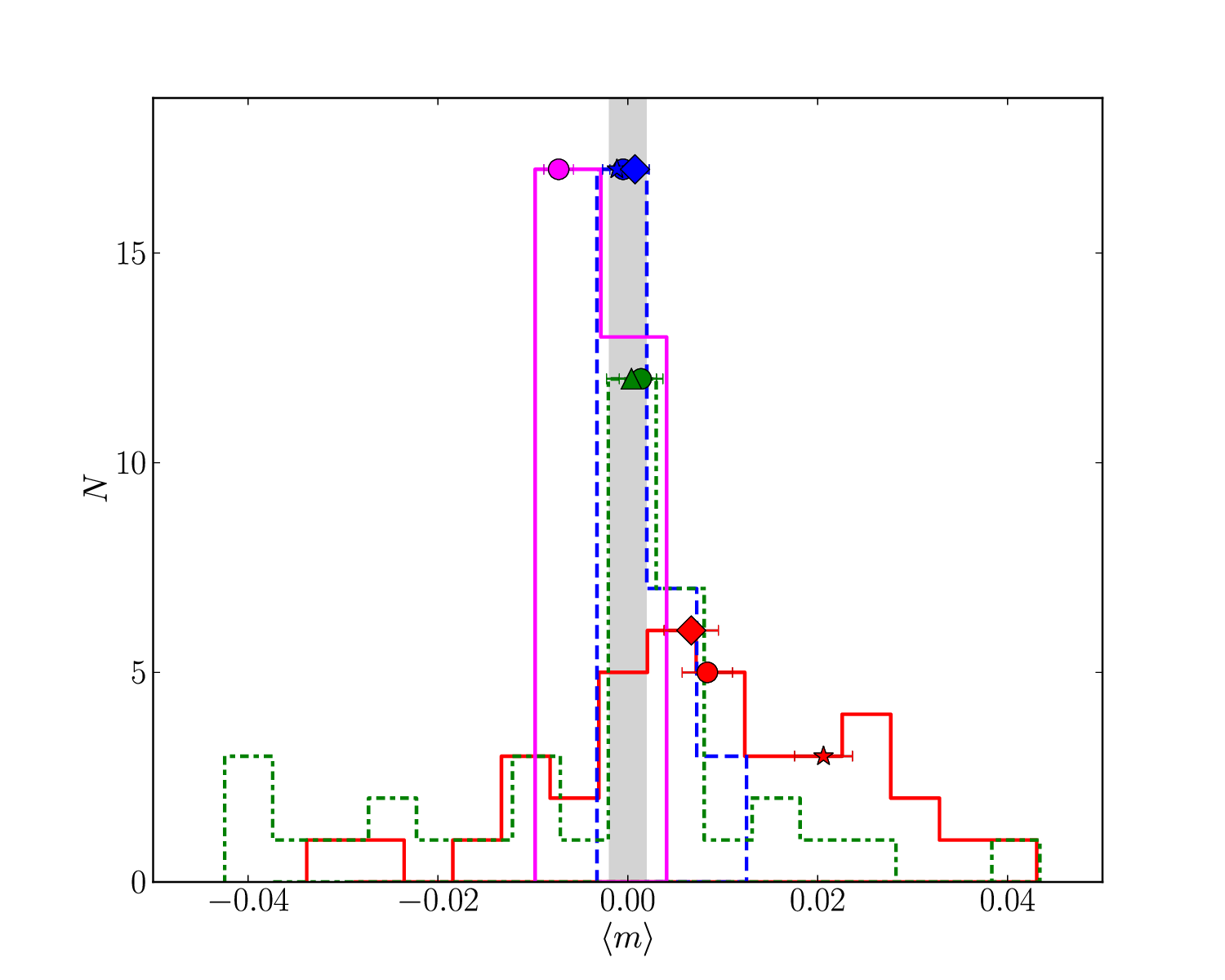}
\includegraphics[width=0.99\columnwidth,angle=0]{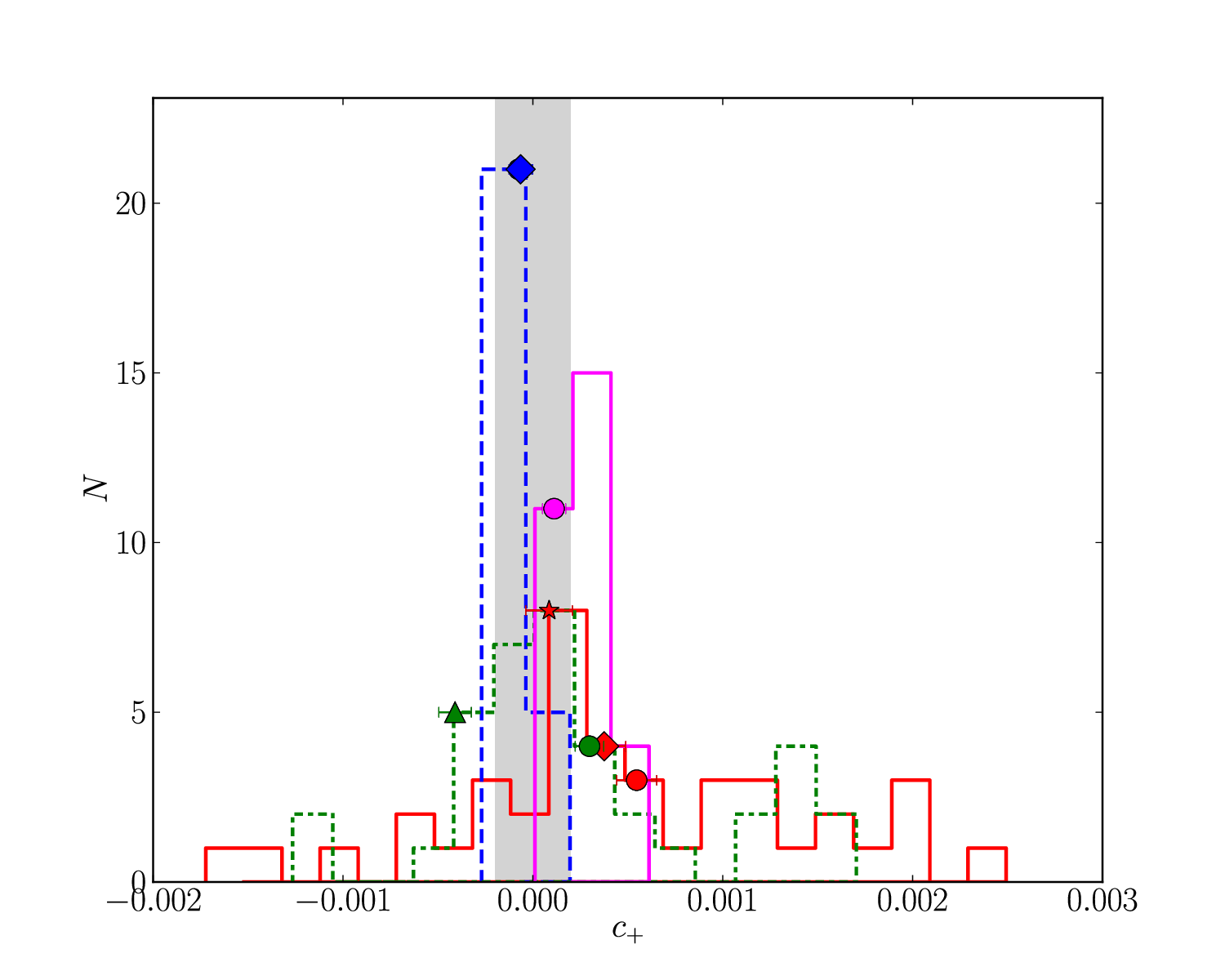}
\caption{\label{fig:amalgam_cross}\referee{{\em Top:} Histogram of $Q_{\rm c}$ values  for 
  all submissions from the Amalgam@IAP team for constant shear branches.  The 
  colored points indicate submissions that are part of the fair
  cross-branch comparisons with consistent settings, with the four settings described in the text
  indicated with different shaped points. {\em Middle,
    bottom:} The same, but for $\langle m\rangle$ and $c_+$.  The points have horizontal errorbars indicating
  their statistical uncertainty, and the shaded regions indicate the target values of $\langle
  m\rangle$ and $c_+$.  Outliers have been removed from the bottom two
  panels so that the main part of the distribution can be clearly seen.}
}
\end{center}
\end{figure}

Fig.~\ref{fig:amalgam_cross}
shows \referee{histograms of $Q_{\rm c}$}, $\langle m\rangle$, and $c_+$ values for
all Amalgam@IAP submissions in all \referee{constant shear branches, also indicating those
submissions with the aforementioned consistent settings with points}.    As
shown, for branches that include submissions with setting 1, that
submission is typically among the best in the branch\referee{, with RSC being the exception to this rule}.  This is
consistent with our previous results indicating that
$\sigma_s\sim0.5$ and the $\chi^2$ penalty term on aspect ratio
were important factors affecting the results.

Comparing the results for setting 1
and 2 \referee{in RGC, the only constant shear branch to include submissions with both settings,
  their performance seems quite consistent with each other.  However, in variable shear branches
  (not shown),} setting 1
leads to better performance, confirming the importance of the weight
including both shape and measurement noise rather than using
equal weighting.  

Comparing settings 1 and 3, we see that \referee{for CGC, setting 1 leads to
better performance due to a substantially smaller calibration bias}.
  This suggests that use of a \sersic\ $n$ penalty term is
unimportant or perhaps even harmful\referee{, though its impact is somewhat less on variable shear
  branches (not shown).  This finding may simply reflect the fact that the variable shear metric is
  less sensitive to multiplicative bias $m$}.

 Finally, settings 1 and 4 gave similar results, with comparable
$m_i$ and $c_i$.  While the use of penalty terms on $\theta_{\rm aspect}$
is helpful, that is especially true for higher $\sigma_s$
than the value used here.

In general, the results for these fairly chosen sets of submissions
are worse in CGC than in RGC. The primary reason is an average
multiplicative bias of $\langle m \rangle = 0.8\pm 0.2$ per cent in
CGC, while $\langle m \rangle$ is consistent with zero in RGC.  Since
the simulation designs in the control and realistic galaxy
experiments correspond apart from galaxy morphology, this
difference between CGC and RGC suggests a model bias due to realistic
galaxy morphology that is of that order.  This bias may be canceled
out by some other bias in RGC (perhaps noise bias, truncation bias, or
residual model bias due to mismatch between input and output
parametric models).  In contrast, the additive systematics for CGC
vs.\ RGC (setting 1) are consistent within the errors.  For space
branches, the multiplicative biases differ for RSC and CSC by $\langle
m \rangle_{\rm RSC} - \langle m \rangle_{\rm CSC} = 0.80\pm 0.15$ per
cent, suggesting that model bias due to realistic galaxy morphology
has a similar magnitude for both space and ground data.

\subsubsection{Summary}

Here we summarize the key lessons from analysis of the
Amalgam@IAP results.  First, the main factors 
that determine performance are the
magnitude of shape noise used in the weighting scheme ($\sigma_s$)
and the use of a penalty term on the aspect ratio to reduce the incidence of
spurious highly elliptical, lower $S/N$ and
resolution objects.  Using the best choices for these parameters in
all branches resulted in overall good performance, though with hints
of a model bias for ground and space data due to realistic
galaxy morphology that is slightly below a per cent.  Also, strong variation in $c_+$ with the
weighting scheme suggests that the additive systematics are a strong
function of the galaxy $S/N$.

Because of the importance of $\sigma_s$ and penalty terms in
determining performance, for the overall analysis and
comparison with other methods, we use a set of submissions 
with the same value of $\sigma_s=0.5$ and a penalty term on the
aspect ratio, with small variations in other less important
parameters\footnote{For three variable-shear branches, there were no
  submissions with $\sigma_s=0.5$.  To enable comparison in those
  branches, the Amalgam@IAP team made submissions after
  the end of the challenge using the {\em same} catalogs as 
  during the challenge, reweighted using $\sigma_s=0.5$.}.

\subsection{MegaLUT}

\subsubsection{Controlled tests of variation of parameters}

The MegaLUT team (see Appendix~\ref{app:megalut})
made many submissions with varying choices 
related to the learning sample generation, shape
measurement, input parameters for the artificial
neural network (ANN), 
architecture of the ANN, and finally the rejection of faint or
unresolved galaxies.  Here we will explore the dependence of their
results on these choices.

\begin{figure}
\begin{center}
\includegraphics[width=0.95\columnwidth,angle=0]{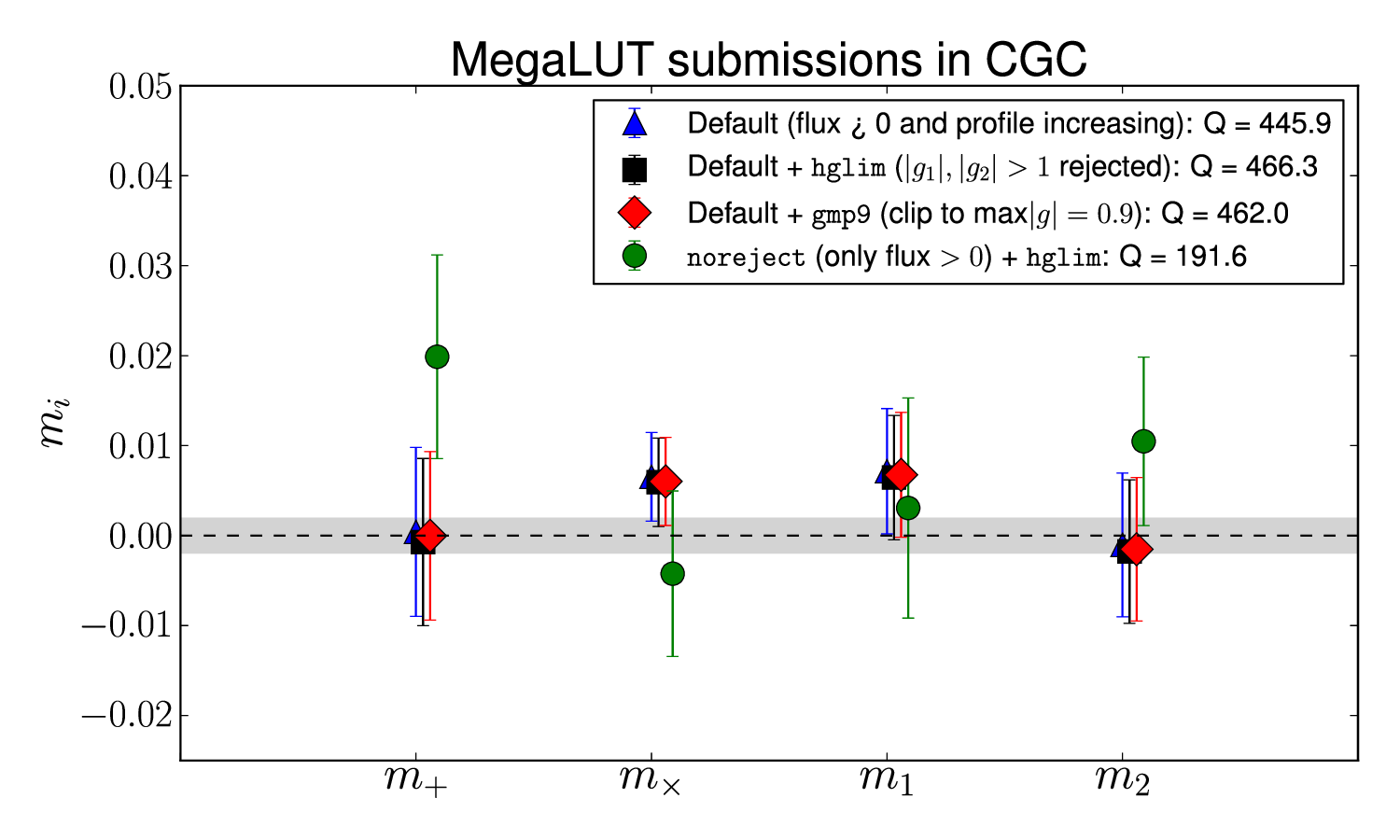}
\includegraphics[width=0.95\columnwidth,angle=0]{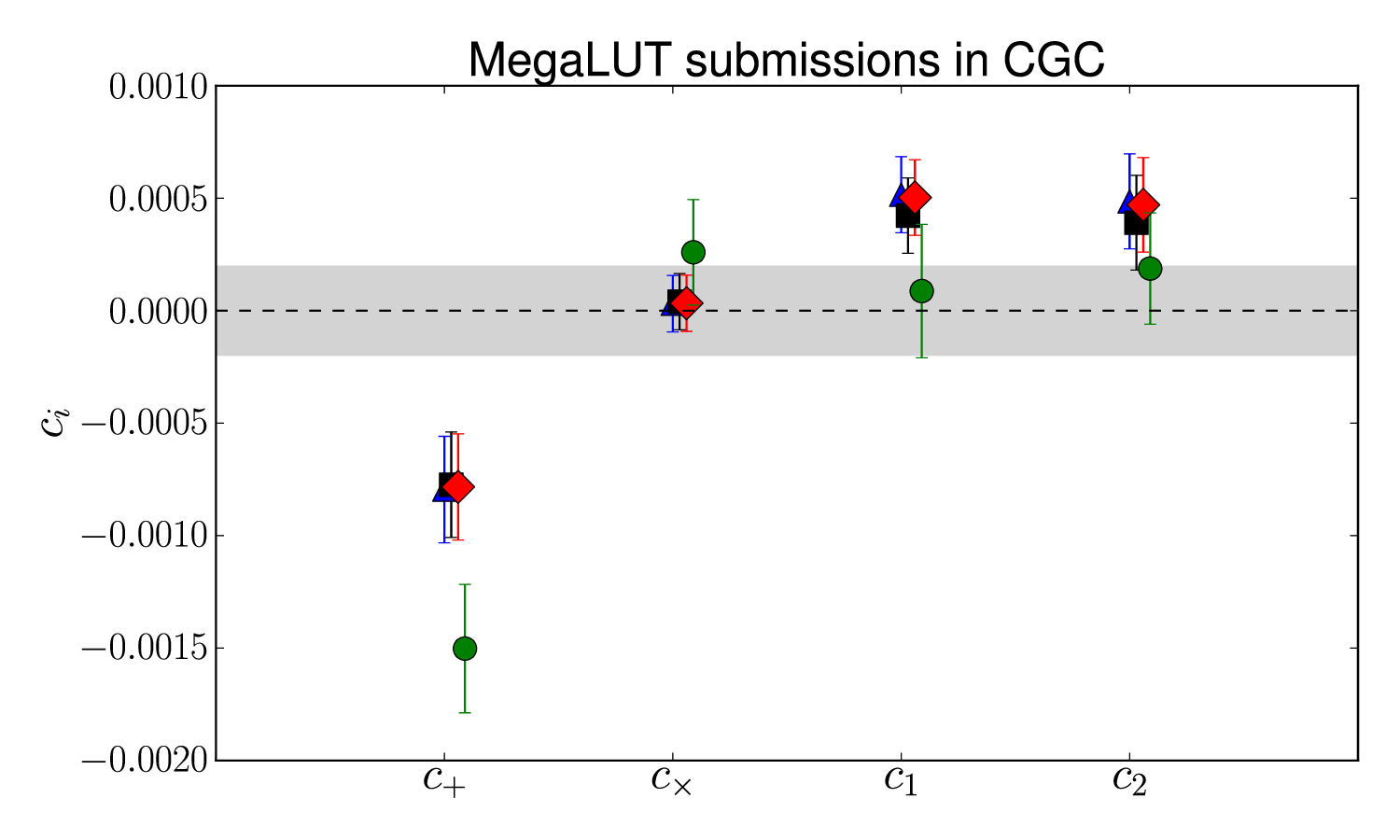}
\caption{\label{fig:focus-megalut-filtering}{\em Top:} $m_i$ values for
  four MegaLUT submissions in CGC with different
  choices for how catalogs were filtered, but otherwise the same
  settings.  {\em Bottom:} $c_i$ values.}
\end{center}
\end{figure}

First, we consider the filtering of the catalogs, comparing
four submissions to CGC that used the same
settings for all parameters except the filtering.  The $m_i$
and $c_i$ values for these four submissions are shown in
Fig.~\ref{fig:focus-megalut-filtering}, with the $Q_{\rm c}$ values
indicated in the legend.
As shown, the results for the top three options (all with default
filtering for positive flux and profile increasing) give very similar
results, regardless of other choices like rejection based on
maximum shear values, or clipping large shears (setting them to a
maximum value of $0.9$).  However, removing the default filtering and
{\em only} rejecting based on $|g_1|$ or $|g_2|>1$ gives 
significantly worse $Q_{\rm c}$.  This is 
due to both $m_i$ and $c_+$ increasing in magnitude. This submission is
only mildly correlated with the others, and the $m_i$ and $c_i$ changes are only marginally
significant ($2\sigma$). On a minor note, there is a
$2$--$3\sigma$ hint of  
non-zero $c_1$ and $c_2$, which (if real) may reflect
asymmetry in selection criteria.
Note that the default filtering option removes typically $<1$ per cent of
the galaxies.

The next test  was on CSC,
comparing two otherwise similar submissions with different choices
at the training stage.  The training sample shears 
were uniformly distributed with $|g|<1$ and with $|g|<0.7$.  The $Q_{\rm c}$ values were $289$ and
$228$ respectively, primarily because of a larger magnitude of
the (negative) calibration bias in the latter case.  This
change in $m$ is not very statistically significant ($<2\sigma$),
which is interesting because it suggests a lack of sensitivity to this
aspect of the training.

\begin{figure}
\begin{center}
\includegraphics[width=0.95\columnwidth,angle=0]{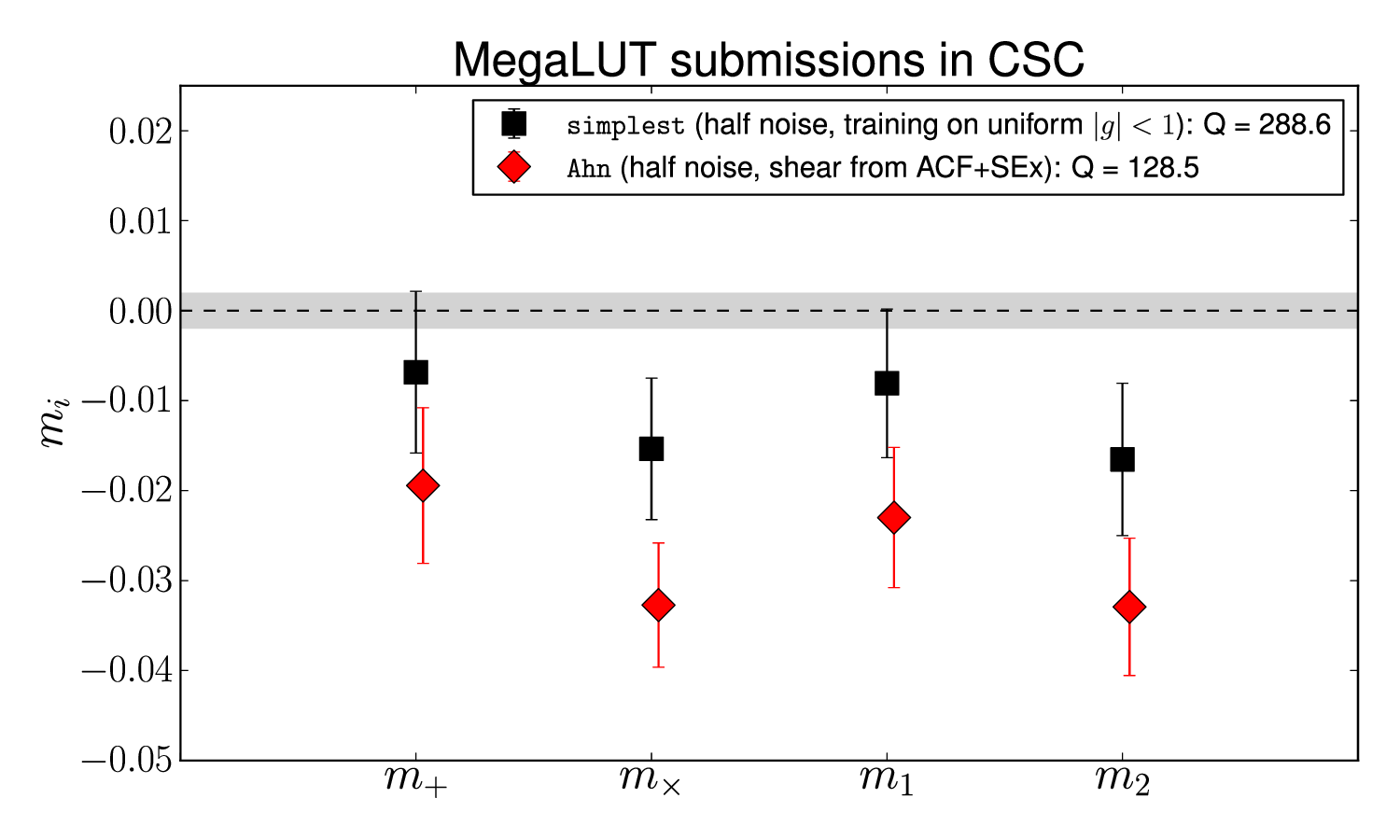}
\includegraphics[width=0.95\columnwidth,angle=0]{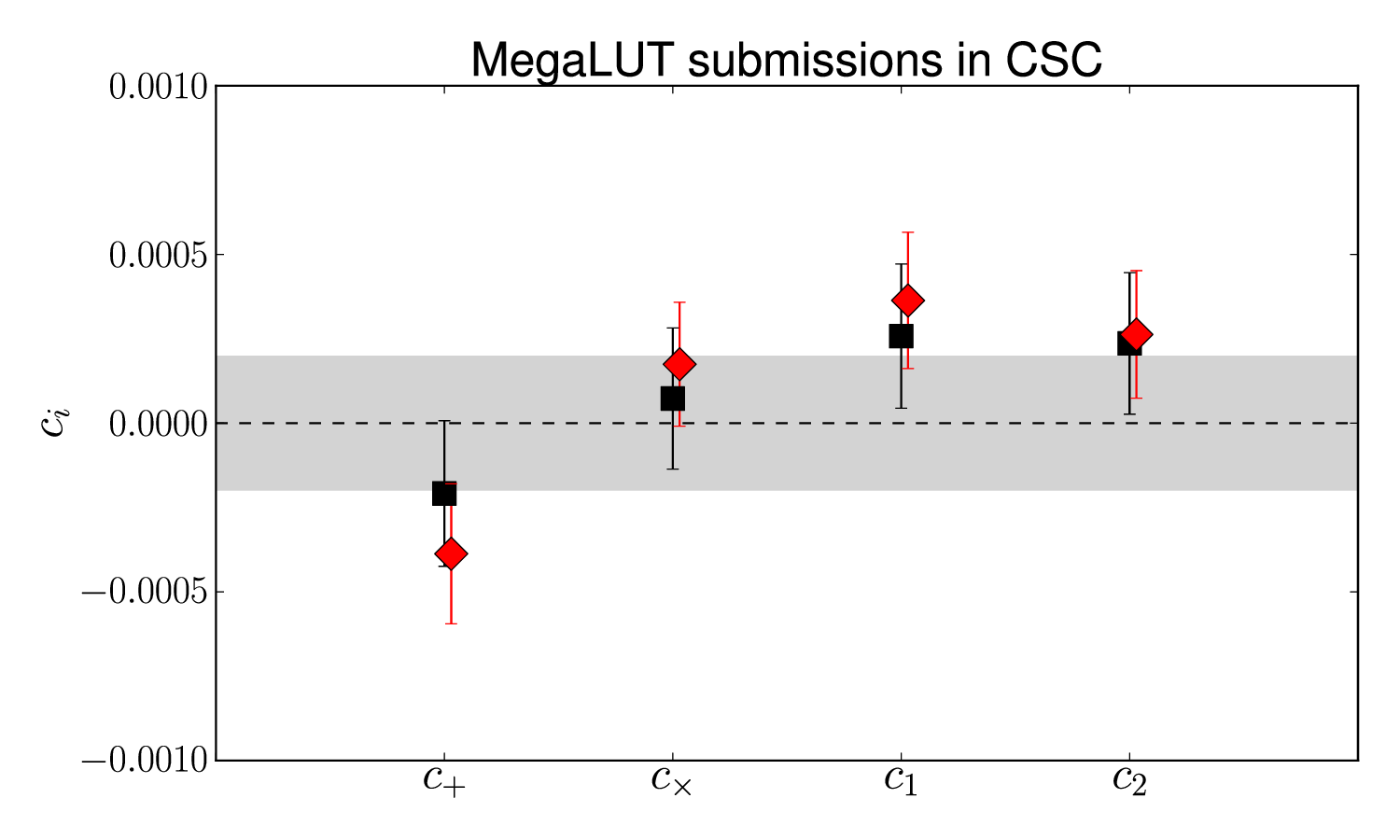}
\caption{\label{fig:focus-megalut-shape}{\em Top:} $m_i$ values for
  two MegaLUT submissions in CSC with different
  methods of measuring galaxy shapes, but otherwise the same
  settings.  {\em Bottom:} $c_i$ values.}
\end{center}
\end{figure}

Also in CSC, we compare two submission that used
different statistics of the image to describe the shape.  In one submission,
the adaptive moments routines in \galsim\ were used, effectively
fitting the image to an elliptical Gaussian; the other submission used
the moments of
the autocorrelation function \citep[ACF,][]{1997A&A...317..303V} of
the image.  The results are shown in Fig.~\ref{fig:focus-megalut-shape}.
The $Q_{\rm c}$ values are $289$ and $129$, respectively,
due to a $3\sigma$ difference in $m_i$ values
(the significance is larger than it appears on the plot due to
correlations between the submissions).
Use of the ACF gives a more negative calibration bias of
$\langle m \rangle = -2.5\pm 0.5$ per cent, compared to
$\langle m \rangle = -1.1\pm 0.5$ per cent without its use.
Apparently the ACF is not an unbiased way of compressing the
information in the image, \newtext{consistent with what was seen for two
methods using the ACF in GREAT10 \citep{2012MNRAS.423.3163K}.}

A final study performed in CSC relates to other ways of
filtering the catalogs after shear estimates have been
made,
comparing the results of the
default filtering with two other options: excluding small
objects, and using convex hull peeling \citep{eddy82}.
The exclusion of small objects changes $m_i$ and $c_i$ only slightly.
However, convex hull peeling gives substantially worse results that
are also noisier, with $Q_{\rm c}$
reduced from around $300$ to $113$, and $\langle m \rangle$ going from
$-1.1\pm 0.5$ to $3\pm 1$ per cent ($4\sigma$ significance).

\begin{figure}
\begin{center}
\includegraphics[width=0.95\columnwidth,angle=0]{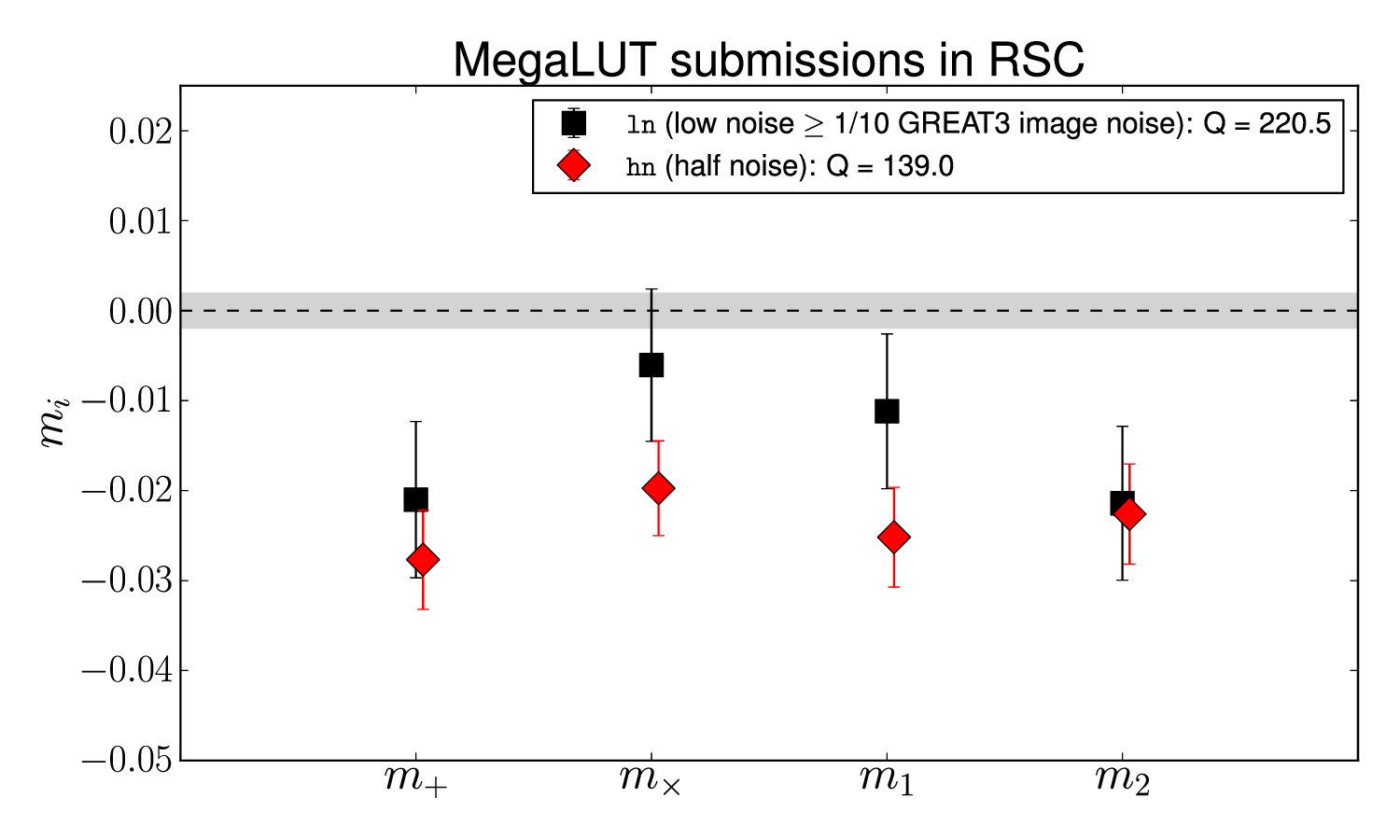}
\includegraphics[width=0.95\columnwidth,angle=0]{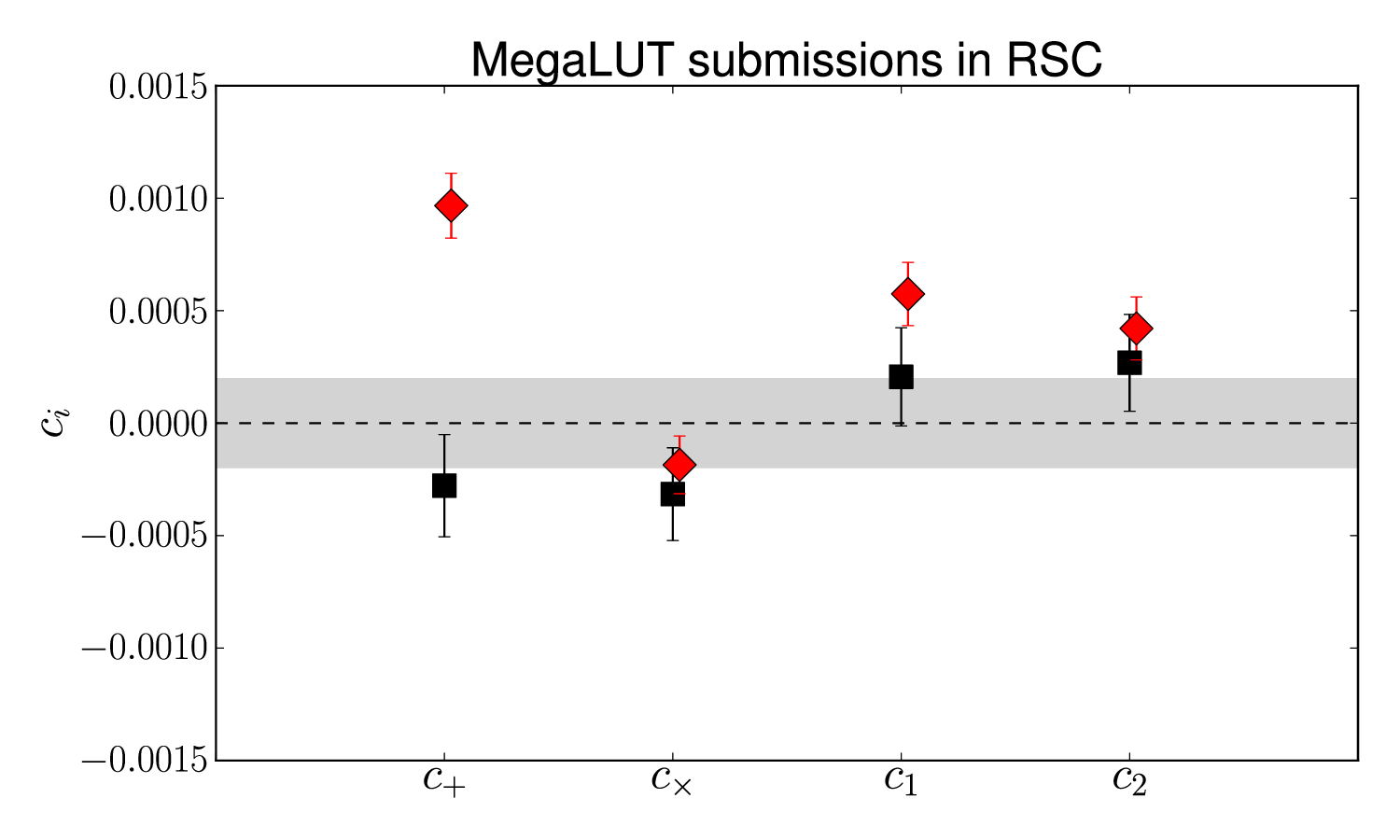}
\caption{\label{fig:focus-megalut-rsc-noise}{\em Top:} $m_i$ values for
  two MegaLUT submissions in RSC with different
 choices for how much noise to include in the training sample.  {\em
   Bottom:} $c_i$ values. 
}
\end{center}
\end{figure}

For RSC, we compared two submissions with different training options.
In one case (``half noise''), the training set images had noise that
was half the level in the GREAT3 images; in the other case, it was
``low noise'', $1/10$ the level in the GREAT3 images.
Fig.~\ref{fig:focus-megalut-rsc-noise} shows that the latter gives
significantly better performance, $Q_{\rm c}=221$ instead of $139$.
The ``half noise'' case has slightly worse $m_i$ values, and substantially
worse additive systematics of $c_+= (10 \pm 1) \times 10^{-4}$
vs.\ $c_+ = (-3\pm 2) \times 10^{-4}$ in the low noise case,
a $5\sigma$ difference given the correlations between the
submissions).
The increase in $m_i$ with increasing noise in the training sample
images could be due to the resulting noisiness in the input features
of the ANN training. This noisiness smears out any sharp structures
that the ANN regression should fit, leading to biased ANN predictions.
We speculate that the effects on $c_+$ may relate to errors in
centroiding that are larger along the PSF direction somehow being
amplified if the training sample is also noisy, but this effect
requires further study to fully understand.

The final test in RSC relates to the use of clipping the
shears, meaning setting those galaxies with estimated $|g| > g_{\rm
  clip}$ to $|g|=g_{\rm clip}$ instead of using the estimated
value.  We compare two submissions with $g_{\rm clip}=0.6$ and
$0.9$ and otherwise similar settings, and find $Q_{\rm c}$ values of
76 and 105, respectively.  While the additive systematics are
virtually identical, the submission with stronger clipping 
has worse calibration bias ($\langle m \rangle$ is more negative:
$-4.5\pm 0.5$ per cent instead of $-3.0\pm 0.5$ per cent, significant at more than
$20\sigma$ given the high correlation
between the submissions).  Not surprisingly,
aggressive clipping of shear magnitudes biases the estimated
cosmological shear low.

\subsubsection{Fair cross-branch comparison}

\begin{figure}
\begin{center}
\includegraphics[width=0.99\columnwidth,angle=0]{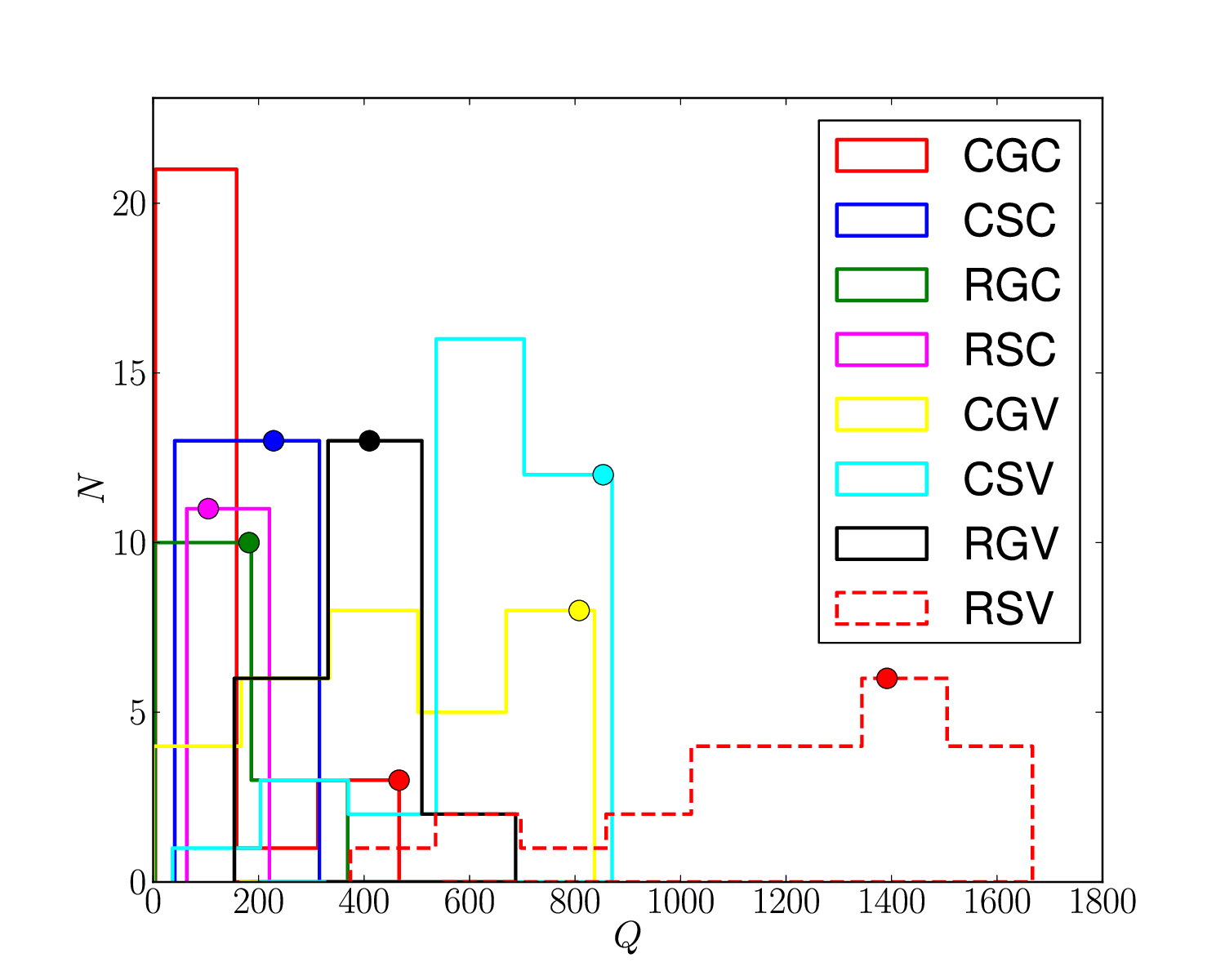}
\includegraphics[width=0.99\columnwidth,angle=0]{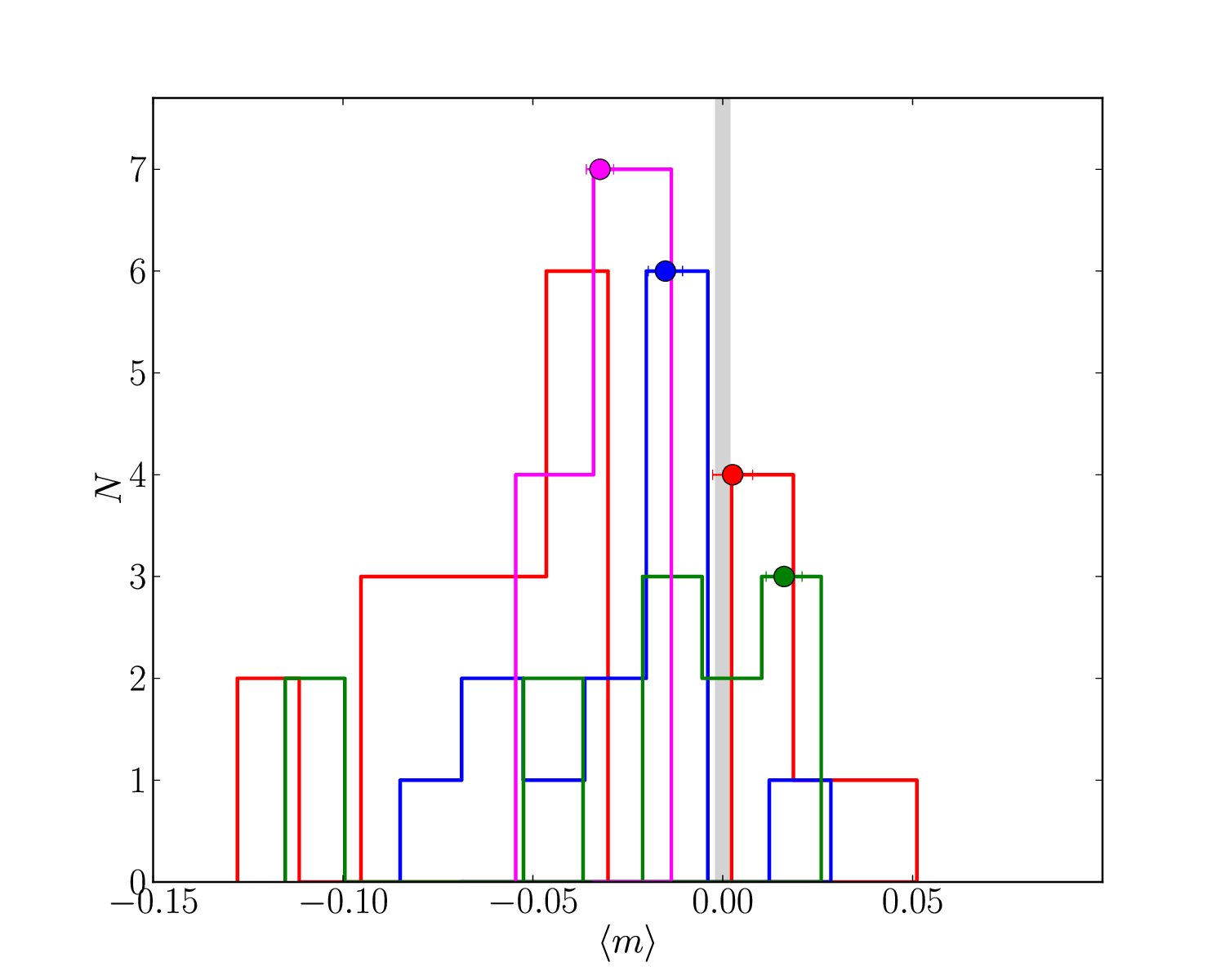}
\includegraphics[width=0.99\columnwidth,angle=0]{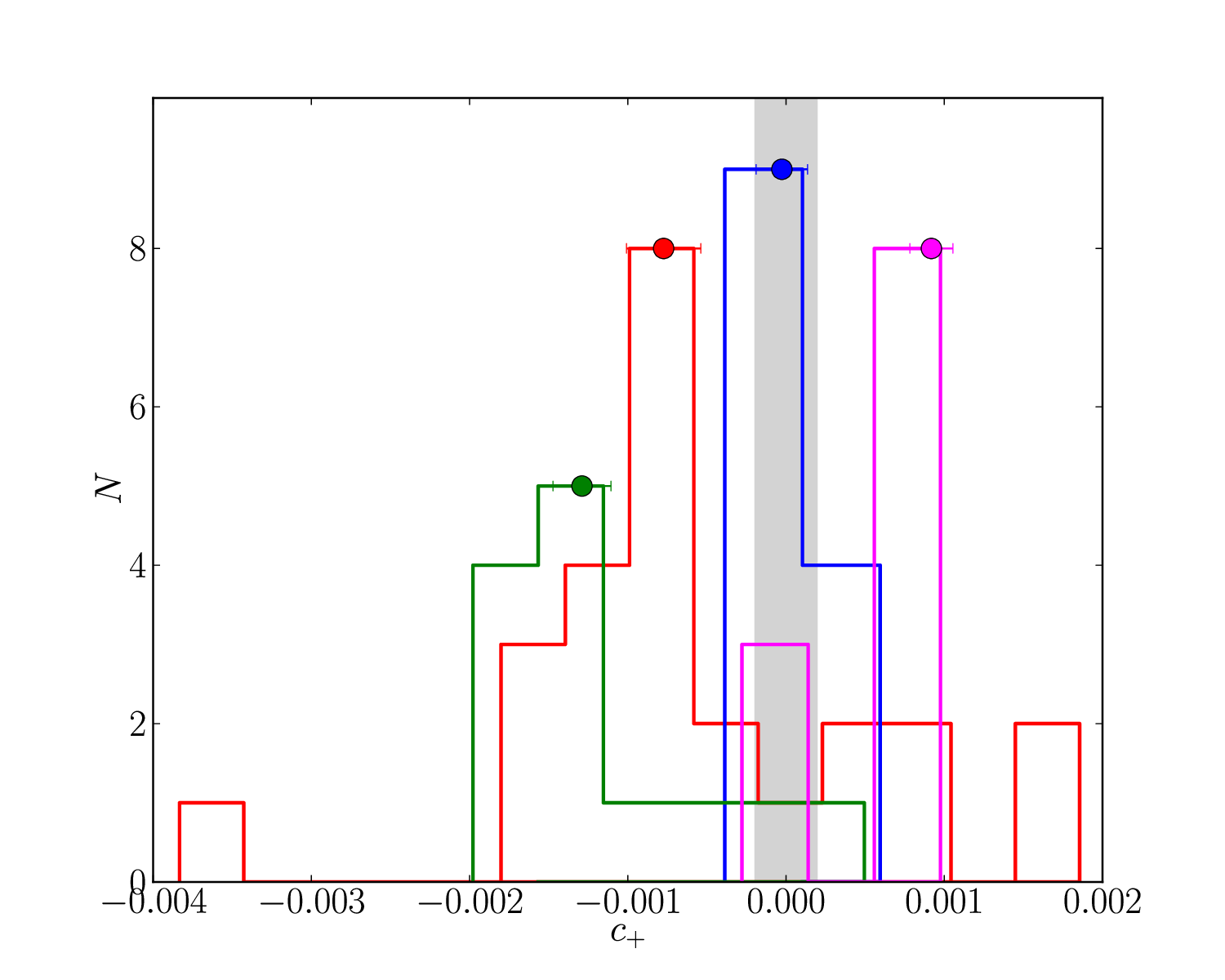}
\caption{\label{fig:megalut_cross}\referee{{\em Top:} Histograms of $Q$ values (either $Q_{\rm
    c}$ or $Q_{\rm v}$ depending on the branch) for the MegaLUT team.  The 
  colored points indicate the submissions that are part of the fair
  cross-branch comparison with consistent settings. {\em Middle,
    bottom:} The same, but for $\langle m\rangle$ and $c_+$ (respectively), which
  involves using constant-shear branches only. The points have horizontal errorbars indicating
  their statistical uncertainty, and the shaded regions indicate the target values of $\langle
  m\rangle$ and $c_+$.  Outliers have been removed from the bottom two
  panels so that the main part of the distribution can be clearly seen.}
}
\end{center}
\end{figure}

We also show results for a fair cross-branch comparison using similar
settings for the training set, filtering, and other parameters of
interest.  In this case, $|g|$ in the simulations was uniformly
distributed in a unit disk of radius $0.7$; the simulations had half
the noise of the GREAT3 simulations; galaxies with estimated $g_1$
{\rm or} $g_2$ with magnitude larger than $1$ were rejected, but no
other rejection scheme was used; and shears were clipped to a maximum
of $0.8$.  As for the other cross-branch comparisons, we show \referee{histograms 
of all submissions and points indicating  the ones in the fair comparison
set}, in Fig.~\ref{fig:megalut_cross}.

As shown in the top panel of Fig.~\ref{fig:megalut_cross}, the
MegaLUT submissions in the cross-branch comparison set are typically among their
top submissions. The top values of $Q_{\rm
  c}$ are in the range $200$--$700$, whereas the top
values of $Q_{\rm v}$ are in a higher range, $400$--$1400$.  For 
all combinations of (experiment, observation type) to which this
team submitted, the results for
variable shear were better than for constant shear.  This may reflect
the fact that the best results typically had 
$\sim 1$--$2$\% multiplicative calibration biases, to which 
$Q_{\rm c}$  is substantially more sensitive than $Q_{\rm v}$.
 For constant-shear branches, the results for
the mixed metric $Q_{\rm mix}$ were very similar to those for the
standard $Q_{\rm c}$.  Another interesting trend across branches is
that, with the exception of RSV, MegaLUT did 
better in the control experiment than in the realistic galaxy
experiment,
perhaps reflecting a preference for the parametric models used to
generate the training sample (which we explore in detail in
Sec.~\ref{subsec:overallmc}).

The middle panel of Fig.~\ref{fig:megalut_cross} shows $\langle
m\rangle$ averaged over components.  As shown, for both control and
realistic galaxy experiments, the multiplicative bias
$\langle m \rangle$ is typically positive for ground-based data
(around $1$\%) and negative for space-based data ($-1.5$\% to $-3$\%).
The magnitude of the bias is slightly larger for the realistic galaxy
experiment than for the control experiment.  The
differences are of $\langle m \rangle_{\rm RGC} - \langle m
\rangle_{\rm CGC} = 1.4\pm 0.4$ per cent for ground and $\langle m
\rangle_{\rm RSC} - \langle m \rangle_{\rm CSC} = -1.7\pm 0.3$ per
cent for space.  This may reflect differences in model bias from
realistic galaxy morphology.

Finally, the bottom panel of Fig.~\ref{fig:megalut_cross} shows the
additive bias $c_+$.  We see a statistically significant difference in
additive biases for the control and realistic galaxy
experiments, which is partly responsible for the worse performance in
realistic galaxy branches compared to control branches.  This is a
manifestation of model bias due to realistic galaxy morphology.

\subsubsection{Summary}

To summarize the MegaLUT results, we
find that good results required identification and rejection of a
small fraction of problematic galaxies.  Use of 
the image autocorrelation function led to substantially worse
performance than use of the adaptive moments (from a fit to an
elliptical Gaussian, using code in \galsim). 
An attempt to use convex hull peeling
led to substantial calibration biases and overall noisiness.  Use of
training images with $1/10$ (rather than $1/2$) the noise level of
GREAT3 reduced the 
additive systematic errors.  Finally, clipping the shears
substantially (to a maximum value of $|g|=0.6$) led to negative
calibration biases and overall worse performance.  The MegaLUT method
had overall better performance in variable-shear branches due to the
pervasive $\sim 1$ per cent calibration biases, which hurt their performance
preferentially on the constant-shear branches.  This multiplicative
calibration bias has opposite signs for ground and space data (but
similar magnitude).  We saw hints of additive and multiplicative model bias due to realistic
galaxy morphology, which we will explore in more detail in Sec.~\ref{subsec:overallmc}.

While using low noise training data led to improved performance, there was not a fair set of submissions across all branches
that used low noise.  Thus, for the overall analysis in
Sec.~\ref{sec:overall}, we will use a set of submissions with 
half noise.  However, it is important to bear in mind that this
degrades the performance of the method.

\subsection{Fourier\_Quad}

For Fourier\_Quad (see Appendix~\ref{app:fourier_quad}), the key
difference between submissions relates to the weighting scheme used
when combining per-galaxy shear estimates.  Three options were used
for GREAT3:
\begin{itemize}
\item No explicit weighting: Since the galaxy light profile 
  amplitudes scale with the flux, if this is not divided
  out, a lack of explicit weighting corresponds to {\em implicit
    weighting by $(S/N)^2$}.  In GREAT3 this
    improves performance given our use of shape noise cancellation, in a
  way that is not viable in real data where shape noise does not cancel.
\item Identifying the pairs of 90$^\circ$-degree rotated galaxies and
  dividing the $G_1$, $G_2$ and $N$ for each object
  (see Appendix~\ref{app:fourier_quad}) in the pair by the
  squared galaxy flux.  This weighting scheme is also not viable for real
  data.
\item Dividing the power spectrum of the galaxy image by the square
  of the galaxy flux, which corresponds to {\em effectively
  unweighted per-galaxy shear estimates}.
\end{itemize}

For the constant-shear branches, higher scores were achieved using the
first weighting scheme, followed closely by the second. For example,
in CGC, the top $Q_{\rm c}$ scores using the first two weighting schemes were 1202 and
1122, respectively; in RGC, 888 and 764; in CSC, 1318 and 1245; in
RSC, 1919 and 1726.  Clearly the performance was excellent with both
weighting schemes, with $m_i$ and $c_i$ values at or near the target
range.  However, since these are  
not a viable approach in real data, all comparisons with other
methods (in Sec.~\ref{sec:overall}) will use the third weighting
scheme\footnote{The submissions with that weighting scheme were
made after the end of the challenge, but in the interest of trying to
make a fair comparison with other methods, we will use them.}.

\newtext{For two reasons, Fourier\_Quad did not get high scores
in variable-shear branches.  First, unlike most of the other methods, the
shear estimators of Fourier\_Quad do not directly correspond to galaxy
ellipticities, so the method does not get the full advantage of
having zero intrinsic $E$-mode correlation 
in variable-shear branches.  Second, the way of calculating
shear correlation functions in Fourier\_Quad is still sub-optimal, as
described in App.~\ref{app:fourier_quad}. Since we wish to use results
that correspond to what would be used in real data, we do not use their
variable-shear submissions for our overall analysis.}

\subsection{sFIT}

For the sFIT team (see Appendix~\ref{app:sfit}), multiple submissions in each branch 
reflect more complete or sophisticated sets of simulations from which
to derive multiplicative and additive calibration factors to apply to
per-galaxy shear estimates. 
Thus, it is generally the case that the most
fair submission to use in each branch is the one that was submitted
last, except in a few branches with some experimental submissions
at the end.

However, comparing the results for individual submissions within a
branch provides information about the sizes of various biases. 
For example, in CGC, the $Q_{\rm c}$ value changed from
579 to 974 from the first
to the last submission.  The initial
attempt came from applying calibration based on simulations that {\em
  approximately} matched distributions in size, \sersic\ index, and
noise level, but with Gaussian PSFs rather than the real PSFs.
Despite the simplifications in the initial simulations used to derive
the calibration factors, the best-fitting $m_i$
values were $\sim 0.5\pm 0.5$ per cent in each component, and $c_+$ was
consistent with zero.  It is likely that the calibration correction in
this branch is dominated by noise bias corrections.  Later improvements involved oversampling the
\sersic\ profiles, a better PSF model (double Gaussian, which is still
not as complex as the real PSF model\footnote{Due to the 
  computational expense of
  rendering images with a full optics and atmospheric PSF, the
  simulations used to derive the calibrations by the sFIT team did not
use the full PSF model for ground branches.}), and improved \sersic\ $n$ 
distribution based on CSC, which primarily improved the score by
reducing the multiplicative bias to $0.3\pm 0.5$ per cent.
For the final submission, the average multiplicative calibration
factor over all the subfields (with a different value of calibration
depending on the PSF FWHM) was approximately $1.06$, and the magnitude
of the typical additive bias correction (which depends on the PSF FWHM
and its ellipticity) was of order $5\times 10^{-4}$.
The final results in CGV,
with $Q_{\rm v}=841.4$, resulted from directly applying the
calibration factors from the final submissions to CGC, as is
appropriate given the similarities in branch design.

In CSC, the initial basic calibration (derived in a rough way as for
CGC) led to $Q_{\rm c}=698$.  Further iterations involved narrowing
distributions of \sersic\ $n$ and $S/N$ (because the original
ones from fits to the GREAT3 data had an unphysical tail due to noise), and
ultimately achieved $Q_{\rm c}=920$.  The
processing used a $45\times 45$ postage stamp, not the full $96\times
96$, which
should result in truncation bias as in 
Sec.~\ref{subsec:focus:gfit}.  However, since the calibration
simulations also use small postage stamps, the truncation bias
should be automatically corrected.  The magnitude of the total
multiplicative bias correction for this final submission was
approximately 1.02, with an additive bias correction of order
$-2\times 10^{-4}$.

In the realistic galaxy experiment, we first consider RGC.
Interestingly, the first submission (with $Q_{\rm c}=305.4$) used
calibrations derived from simulations with real galaxy images in
\galsim.  However, the next attempt directly used the calibrations
from CGC, which do not include realistic galaxy morphology, and achieved
$Q_{\rm c}=806.9$.  This change tells us that for the sFIT
method of fitting \sersic\ profiles, the model bias due to realistic
galaxy morphology is not very
large in ground-based data, because it is in principle uncorrected in these results.  After
modifying the simulation inputs to better match the 
$p(\varepsilon)$ and size distribution in RGC, the results were as high as
$1003$, with $\langle m\rangle= 0.2\pm 0.5$ per cent, and $c_+=(1\pm
1)\times 10^{-4}$.  This suggests that residual model bias due to
realistic galaxy morphology is only
important at the $10^{-3}$ level for this method, compared to
$10^{-2}$ for the methods discussed previously. 
The best-scoring submissions in
RGV used the calibration from RGC.

In RSC, interestingly,  simulations based on real
galaxy images were necessary to improve $Q_{\rm c}$ above 
$\sim 350$.  Use of COSMOS images led to an immediate boost of
$Q_{\rm c}$ to $759$ in the first attempt, which is a
statistically significant change arising from $\langle m\rangle$
changing from $-0.9\pm 0.3$ per cent to $-0.2 \pm
0.3$ per cent, with nearly the same additive bias,
$c_+=(5.2\pm 0.8)\times 10^{-4}$.  The significance of the change
in $\langle m \rangle$ is $>20\sigma$ due to the very high correlation between
the submissions.  This suggests a statistically significant,
sub-percent model bias due to realistic galaxy morphology for this
method in 
space data.  Further attempts to improve the
simulated $p(\varepsilon)$ to match 
the GREAT3 simulations led to additional improvements to $Q_{\rm
  c}=825$, with the additive bias
remaining unchanged.  One possible cause for this residual bias is
that the calibration simulations did
not use a fully realistic PSF, which could result in slightly
incorrect additive bias corrections.

\subsection{MBI}

As described in Appendix~\ref{app:mbi}, the MBI team made submissions
using a few variations of their method.

For the Optimal Tractor and Sample Tractor, they used the
maximum-likelihood estimate of the lensed ellipticity and the average
of samples from the posterior PDFs (respectively) to derive the mean
shear for the field, typically with similar performance.  For
example, in CGC, the scores for the Optimal Tractor
submissions were 15 and 53, reflecting multiplicative biases of $24$
and $8$ per cent, and non-negligible additive systematics.  The
results for the Sample Tractor submissions were in the same range.
The results in RGC for these two cases were worse than in CGC.

However, hierarchically inferring the intrinsic ellipticity
distribution using importance
sampling from the posterior PDF for the mean 
shear, with a Gaussian $p(\varepsilon)$ in
each component, improved scores by factors of $\sim$ 3-10 in the
ground branches.  These score improvements come from a decrease in
multiplicative biases to the range $1$--$3$\%, and a reduction in 
additive systematics to within the target range.  The exception to
this trend is CSC, where the use of hierarchical inference did not 
yield significant improvement ($Q_{\rm c}$ scores were typically in the
range 90--200 regardless of method).  However, there the assumption
that the PSF can be described as a sum of three Gaussian components is
more dubious than in the ground branches, so PSF modeling may be 
the key limitation in that branch.

\newtext{The results in the ground branches for the Important Tractor
(hierarchical inference) submissions suggest that this new  method may
indeed be able to reduce some intrinsic limitations of
maximum-likelihood fitting methods (e.g., noise bias).  Noise bias
primarily arises when transforming a probability distribution for a
galaxy shape estimate into a single point estimator.  Combining the
probability distributions for all galaxies (resulting in increased
$S/N$) and applying a hierarchically inferred prior $p(\varepsilon)$
yields improved results.}

The submissions from MBI included several variants of the hierarchical
inference.  The first, called ``multi-baling'' (hierarchically
inferring the $p(\varepsilon)$ common to five subfields), led to some
improvements in scores, up to a factor of two.  In contrast, using the
deep fields to infer the $p(\varepsilon)$ did not result in an
improvement in $Q_{\rm c}$ over hierarchical inference assuming an
uninformative hyper-prior.  Finally, the MBI team made submissions
with informative prior PDFs on the lensing shear, with four different
values that seem to bracket a peak in $Q_{\rm c}$ in the CGC
branch. The highest $Q_{\rm c}$-scores obtained this way (in CGC and
MGC) were around a factor of four higher than that for an asserted
uniform prior PDF for the shear components. For example, with their
wide, default, narrow, and narrower assumed values for $\sigma_g$, the
$Q_{\rm c}$ values were $94$, $146$, $301$, and $24$, corresponding to
multiplicative biases in the range $4$\%, $3$\%, $-0.4$\%, and
$-14$\%, respectively.  In GREAT3 constant-shear simulations, the true
$p_\text{true}(|g|)\propto |g|$ (see App.~\ref{app:shear}), whereas
the MBI team used Rayleigh distributions. Their ``wide'' and
``narrower'' distributions are particularly mismatched in shape to the
true one, so the poor $Q_{\rm c}$ scores are not surprising.

After the challenge, the MBI team investigated inferring the optimal
value of $\sigma_g$ from the data directly (as opposed to from $Q_{\rm
  c}$).  This yields a factor of two improvement over an asserted
uniform prior PSF for the shear components.  It is unclear how much
better one can do in this way on GREAT3 simulations because of the
unusual $p(|g|)$, which differs from the functional form chosen by the
MBI team (and makes sense for real data).

For the overall analysis in Sec.~\ref{sec:overall}, we use the MBI
results with hierarchical inference.  While multi-baling and using
deep fields to get the $p(\varepsilon)$ may become helpful in future,
they were not fully explored in GREAT3, so we do not use them for the
overall comparison.

\subsection{COGS}\label{sec:cogs}

The COGS team made a number of submissions, using the \code{im3shape}
algorithm \citep{2013MNRAS.434.1604Z}, that are described in
Appendix~\ref{subsec:cogs}.  The submissions that used input settings
and methodology suitable for scientific analysis are labelled
\code{u7}, \code{c1}, \code{c2}, and \code{c3}.  The labels \code{c1}
-- \code{c3} denote three different schemes used to calibrate for
multiplicative biases that are expected in Maximum-Likelihood shape
estimation.  \newtext{No correction was applied for additive bias.}

In Sec.~\ref{sec:overall}, and thereafter, where we wish to draw fair
comparisons between branches and between methods, only COGS
submissions that used the \textsc{c3} calibration are used.  This
choice is made as \textsc{c3} comes closest to the approach that would
be adopted when applying \code{im3shape} to real data (see
Appendix~\ref{subsec:cogs}).

\begin{figure}
\begin{center}
\includegraphics[width=0.99\columnwidth,angle=0]{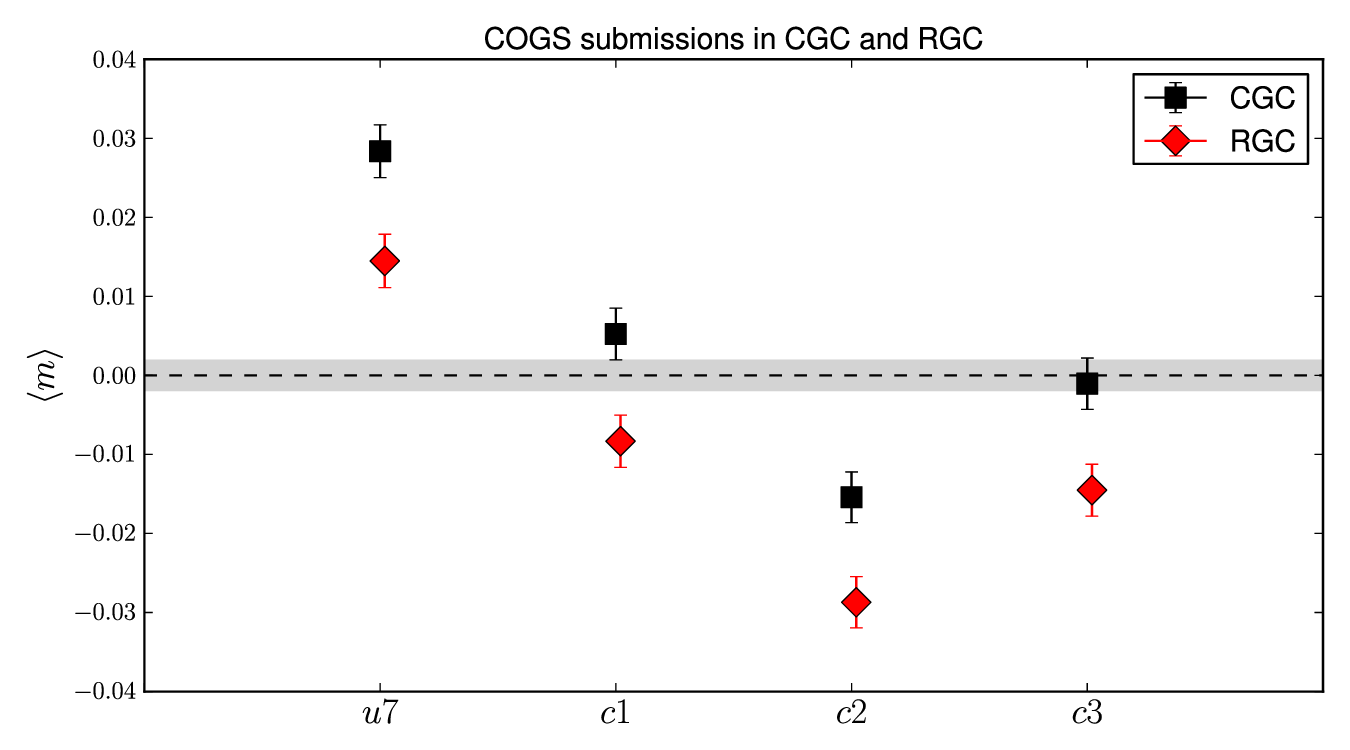}
\caption{\label{fig:cogs_focus} Averaged multiplicative
  bias $\langle m \rangle = (m_+ + m_\times )/2$ for COGS
submissions to CGC and RGC, under differing schemes for the removal of
noise bias (see Sec.~\ref{sec:cogs}).
}
\end{center}
\end{figure}

The different submissions make it possible to test for the effect of
different choices made in the noise bias calibrations, and to test for
model bias due to realistic galaxy morphology by comparing CGC and
RGC. 
Fig.~\ref{fig:cogs_focus} shows the significant impact of noise bias
calibrations on $\langle m \rangle = (m_+ + m_\times )/2$ for COGS
submissions to CGC and RGC.  The \textsc{c3} calibration, derived from
the deep data in CGV but with some outlier rejection in deep field
fits (see Appendix~\ref{subsec:cogs}), controls multiplicative biases
in CGC to within statistical uncertainties.  For \code{u7}, i.e.\
without any attempt to calibrate multiplicative bias, we find $\langle
m \rangle = 2.8 \pm 0.3$~per~cent for CGC and $\langle m \rangle = 1.4
\pm 0.3$~per~cent for RGC.  These results represent a combination of
noise, model, and other biases in the uncalibrated COGS submissions.

For each pair of submissions grouped by calibration strategy, we also
find a consistent difference in the level of multiplicative bias
between CGC and RGC results: $\langle m \rangle_{\rm RGC} - \langle m
\rangle_{\rm CGC} = -1.4 \pm 0.5$~per~cent.  This difference in
$\langle m \rangle$ can be interpreted as a difference in model bias
due to realistic galaxy morphology, for the \textsc{im3shape} galaxy
model chosen by the COGS team. It is similar in magnitude to the
effect found in other model-fitting methods.

\subsection{The role played by outliers}\label{subsec:outliers}

Several teams identified images 
with particularly challenging PSFs.  Here we consider the role played
by outliers in the challenge results, given that our metrics
$Q_{\rm c}$ and $Q_{\rm v}$ (Sec.~\ref{subsec:diagnostics}) allow
teams to weight galaxies {\em within} subfields, but not to assign
weights to the per-field 
shears before construction of the metric.  The rationale behind this
choice was that, with each subfield having fairly similar pixel noise
and the same number of galaxies, the  shear
statistics should be determined equally well for each subfield.
However, if a method has a systematic problem with a subfield, they 
cannot indicate this by giving a low (or zero)
weight, unlike in real data where they could choose to discard a 
subset of the data.

\begin{figure}
\begin{center}
\includegraphics[width=0.99\columnwidth,angle=0]{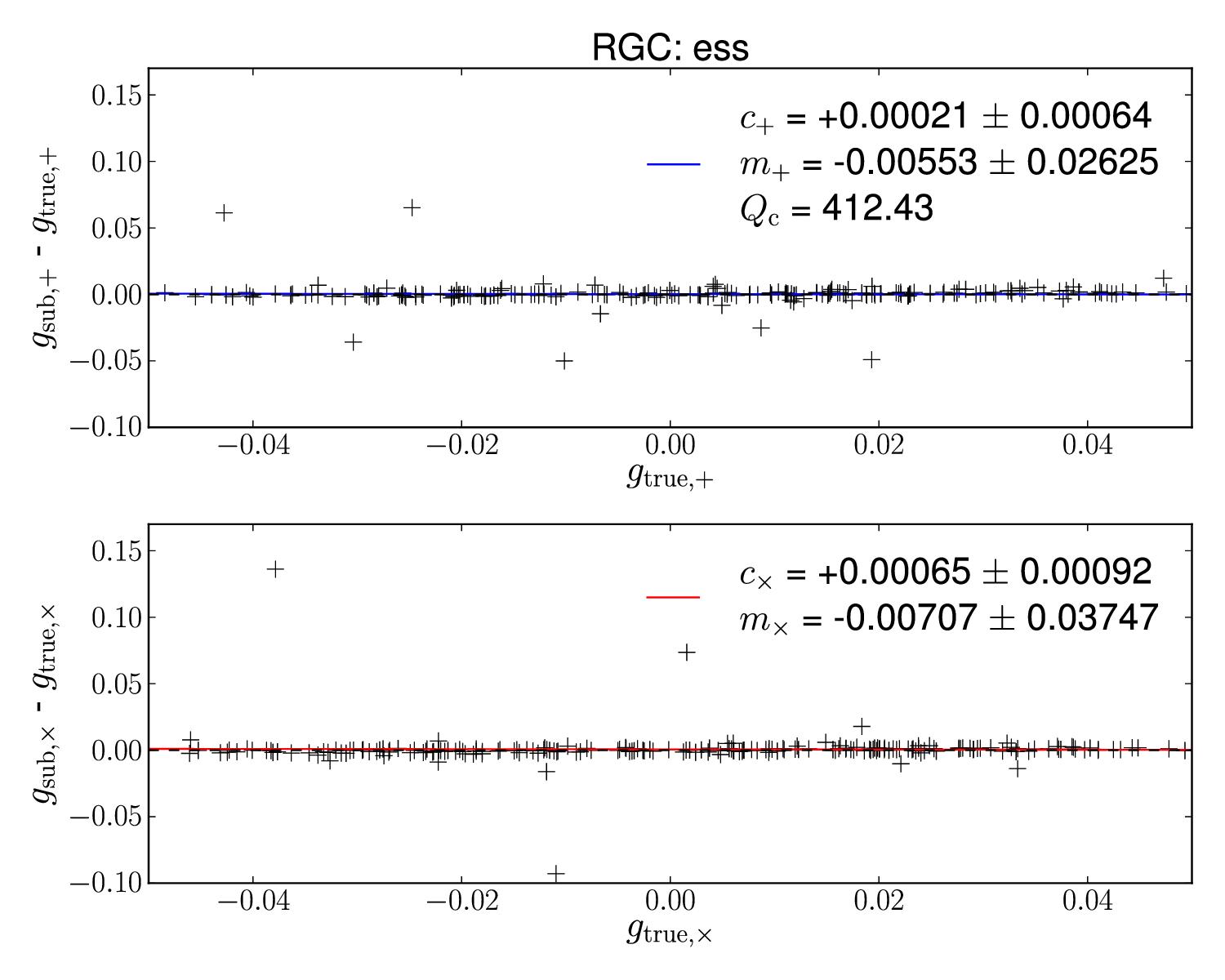}
\includegraphics[width=0.99\columnwidth,angle=0]{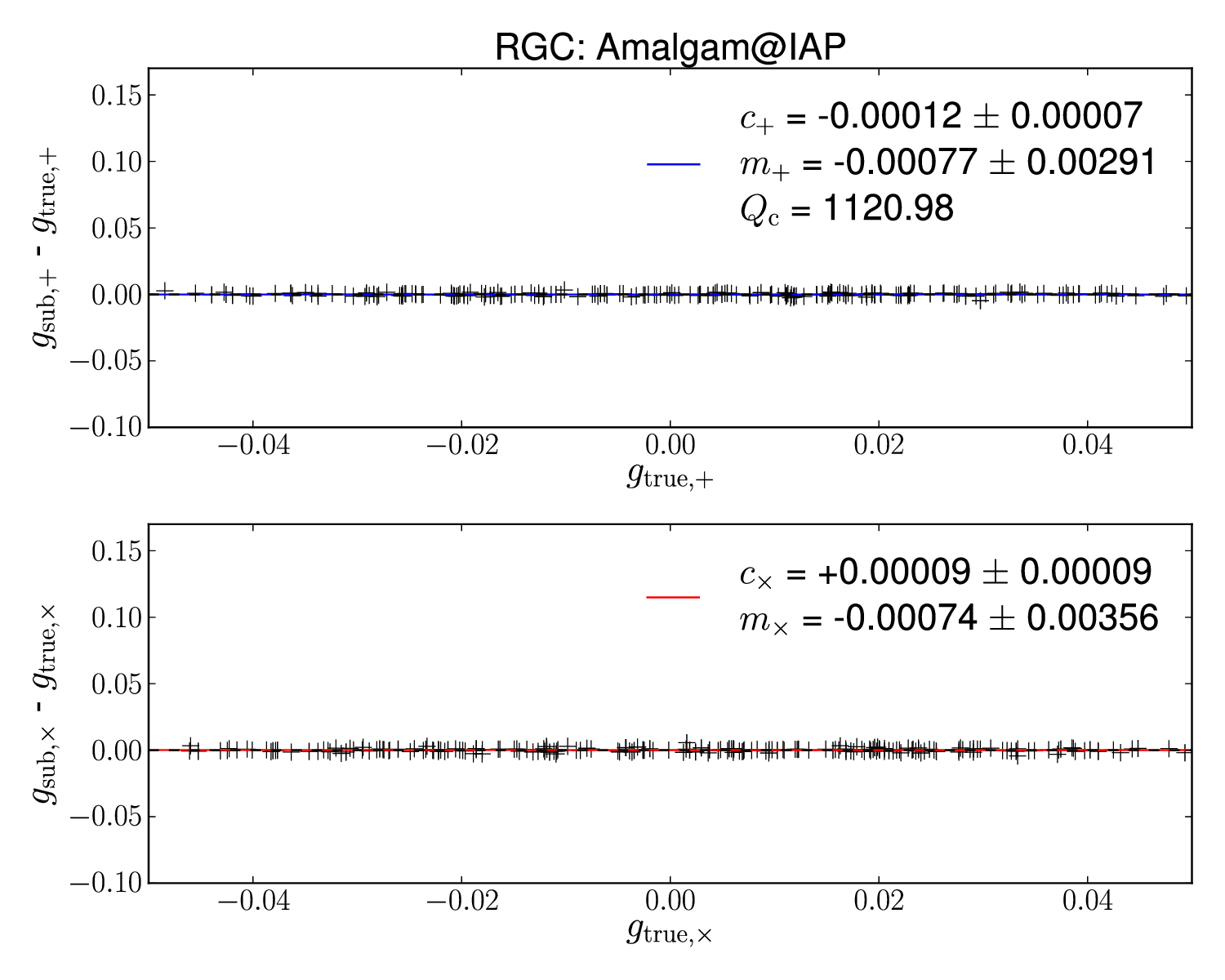}
\includegraphics[width=0.24\columnwidth,angle=0]{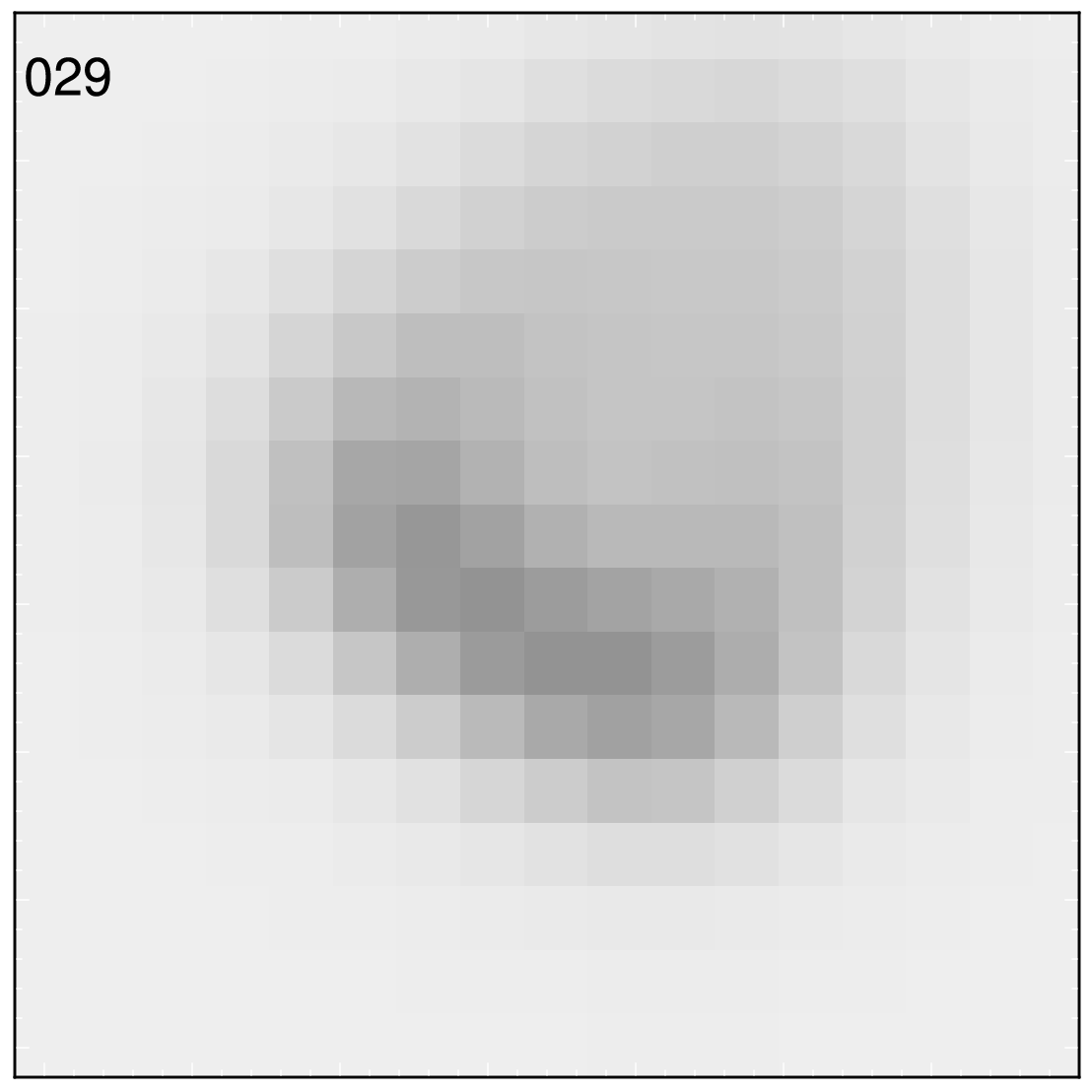}
\includegraphics[width=0.24\columnwidth,angle=0]{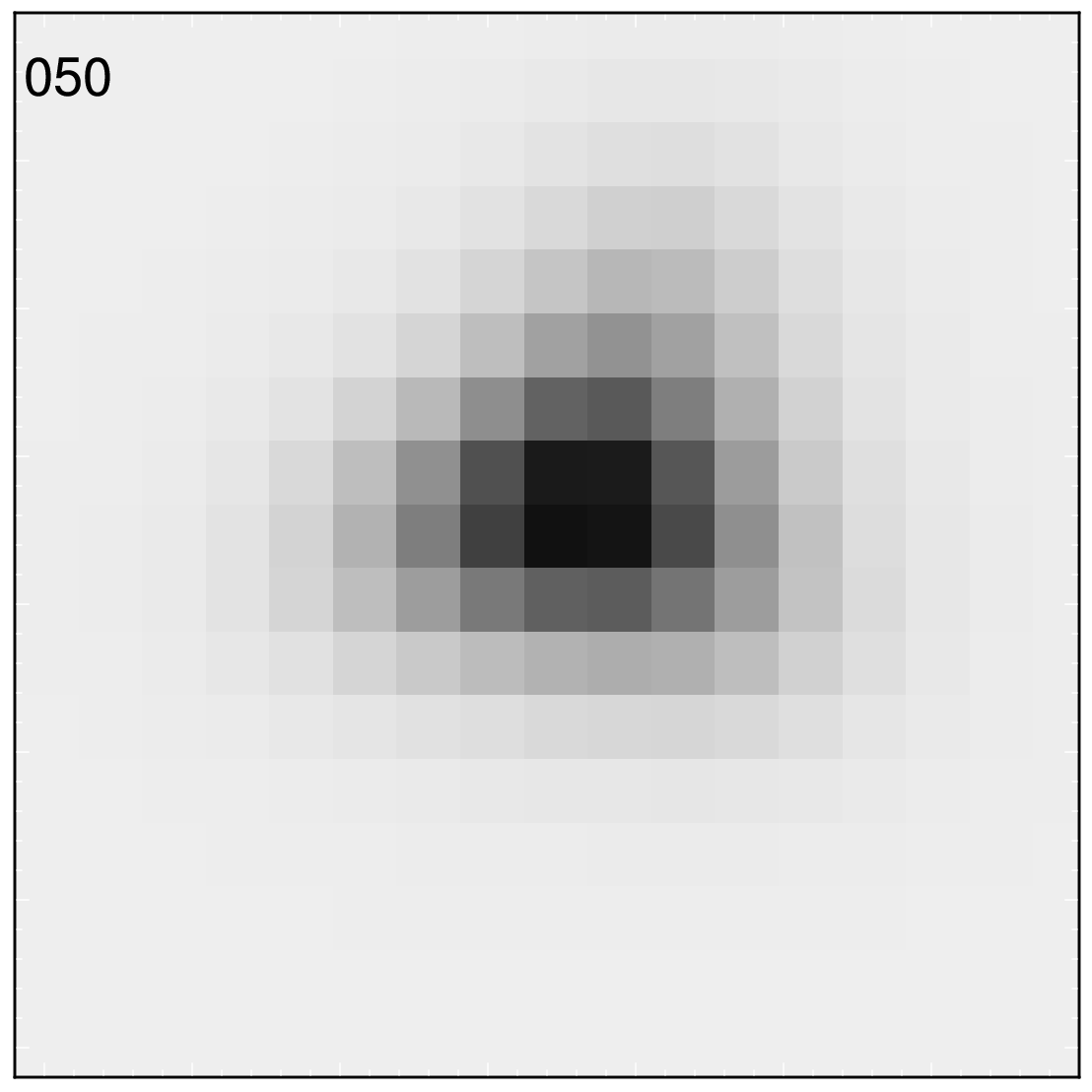}
\includegraphics[width=0.24\columnwidth,angle=0]{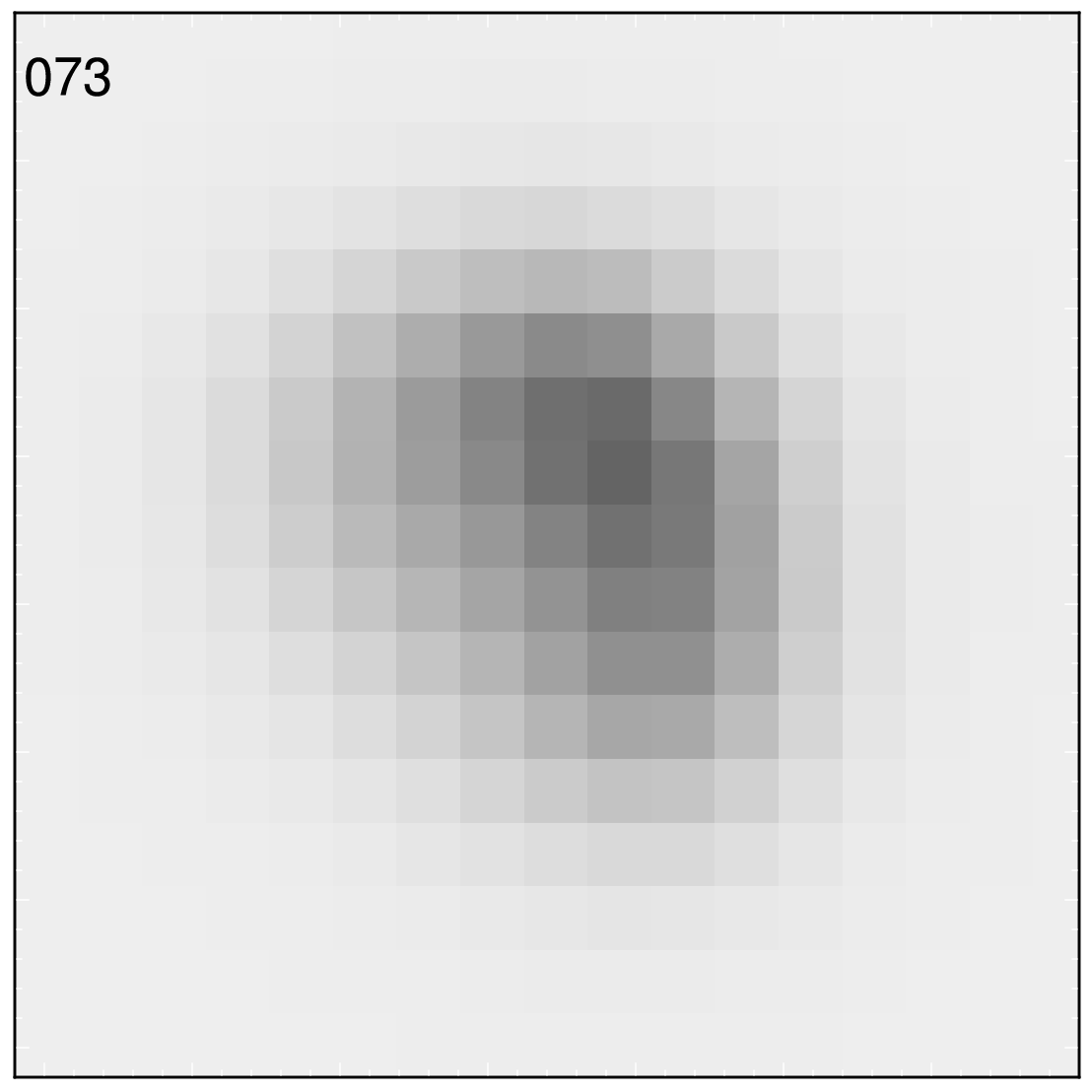}
\includegraphics[width=0.24\columnwidth,angle=0]{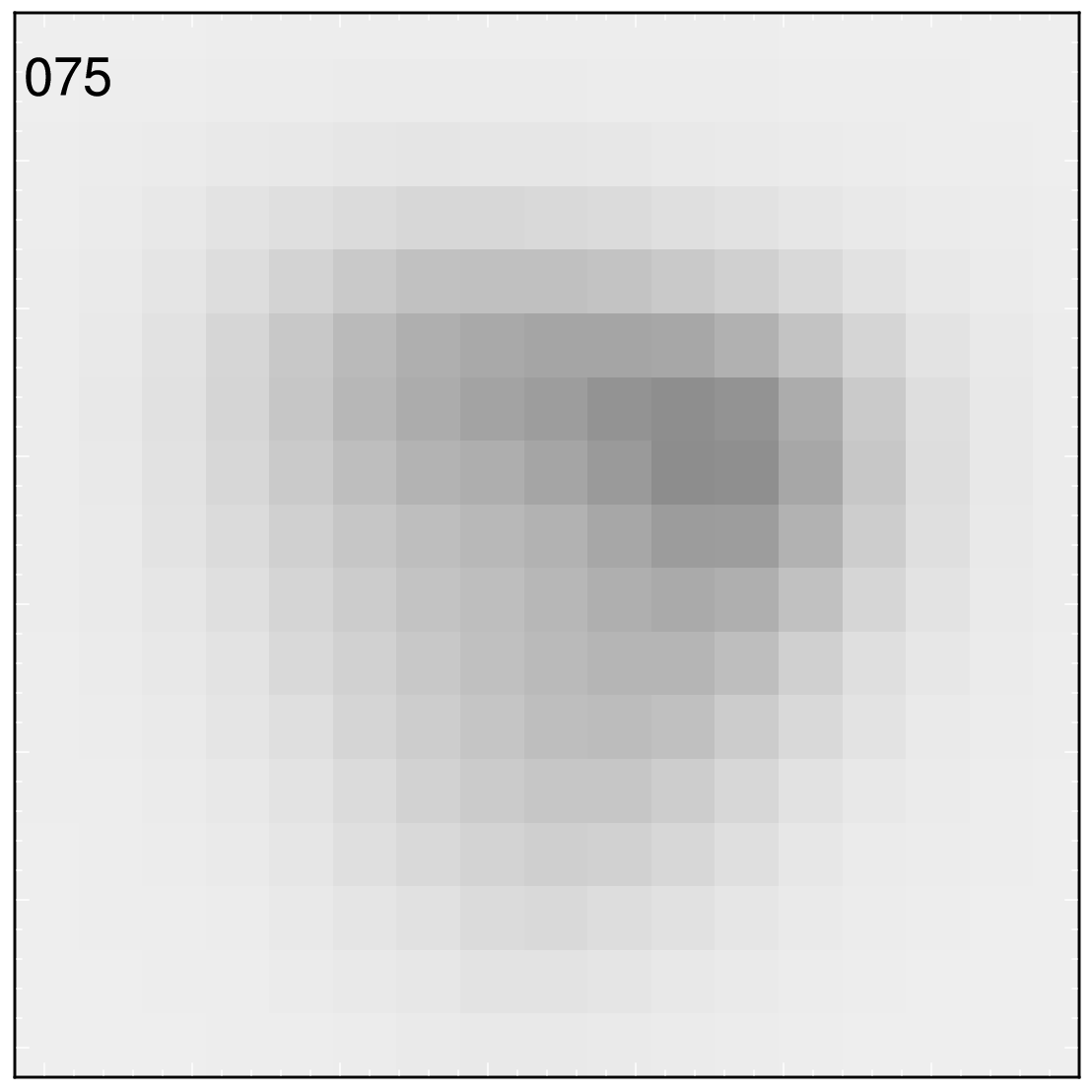}
\includegraphics[width=0.24\columnwidth,angle=0]{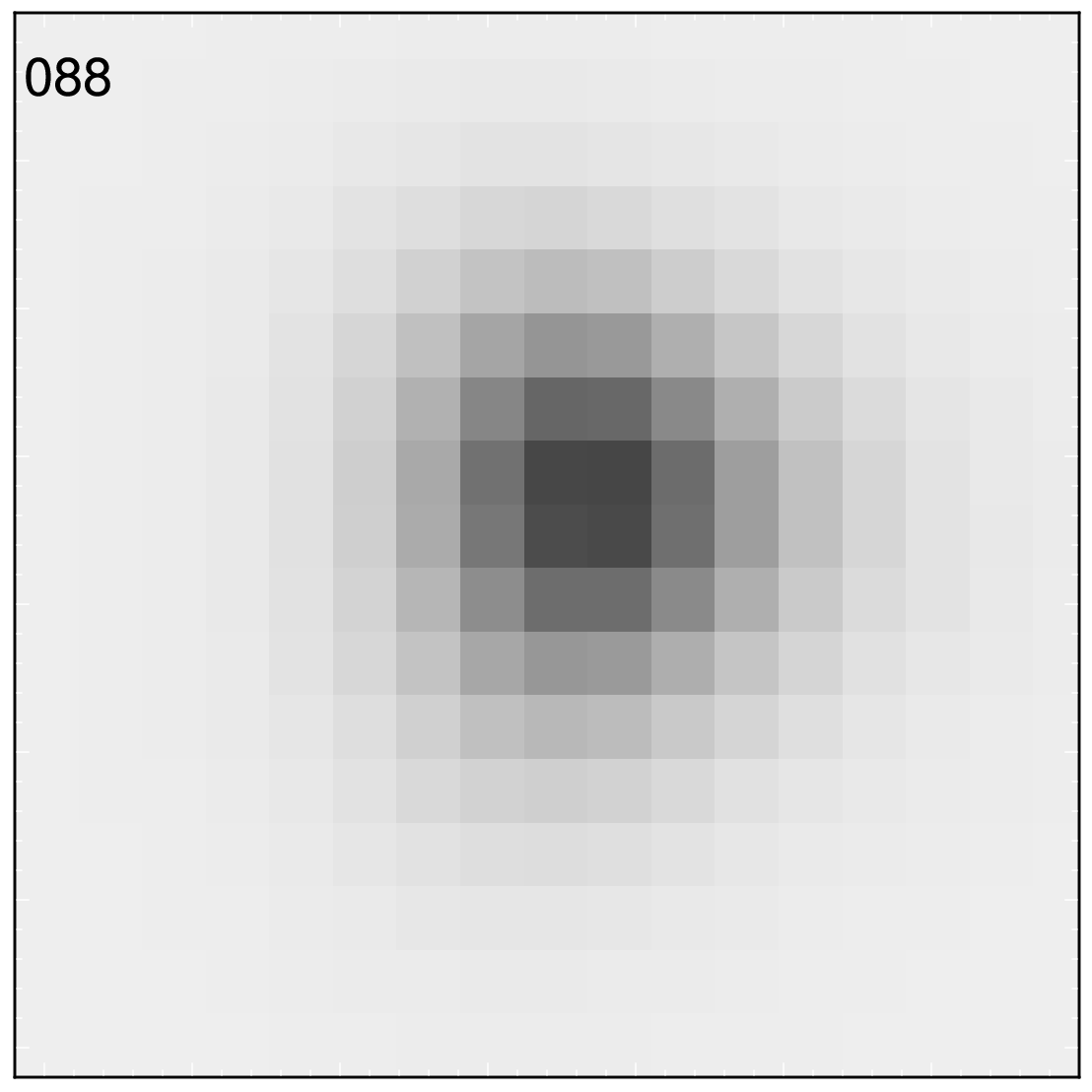}
\includegraphics[width=0.24\columnwidth,angle=0]{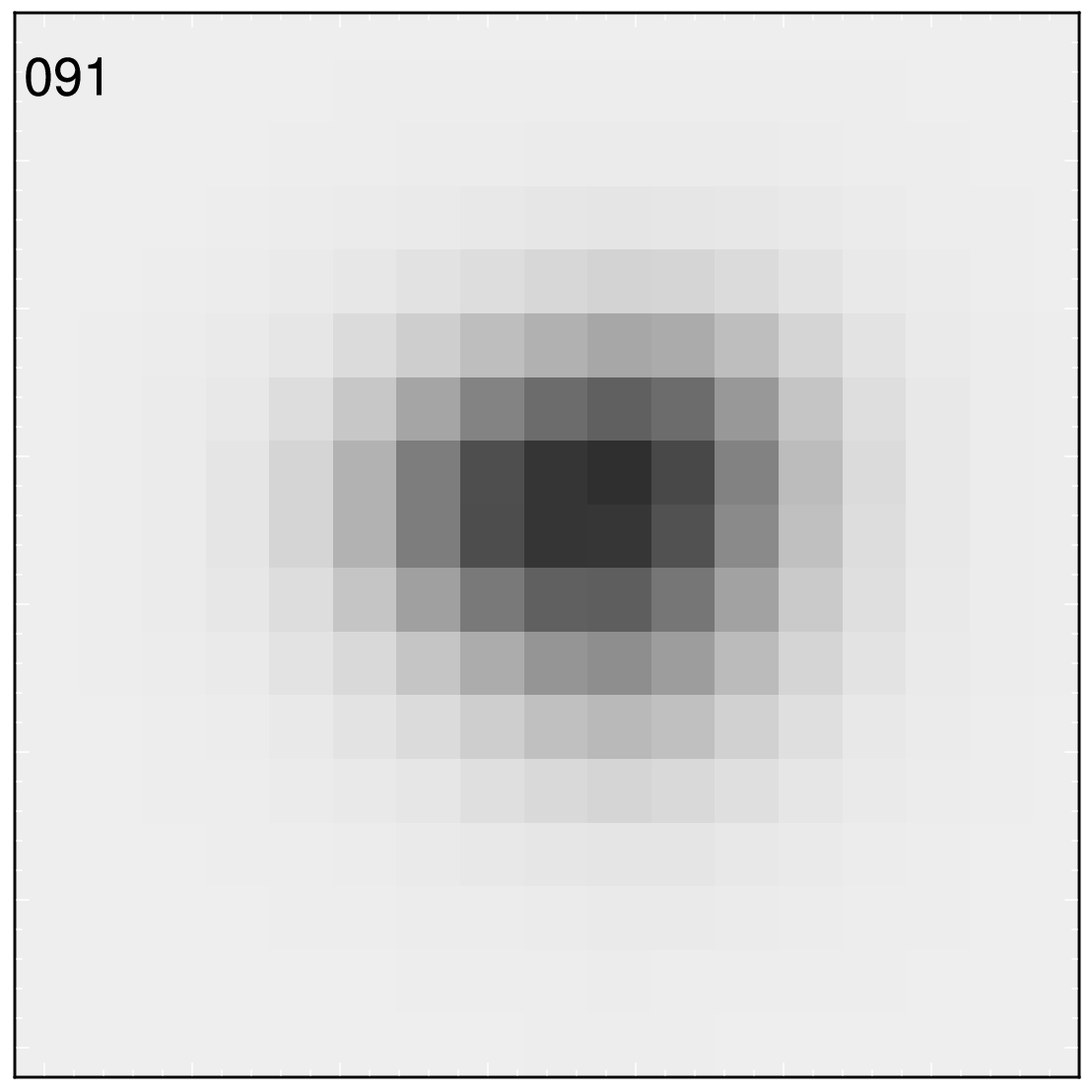}
\includegraphics[width=0.24\columnwidth,angle=0]{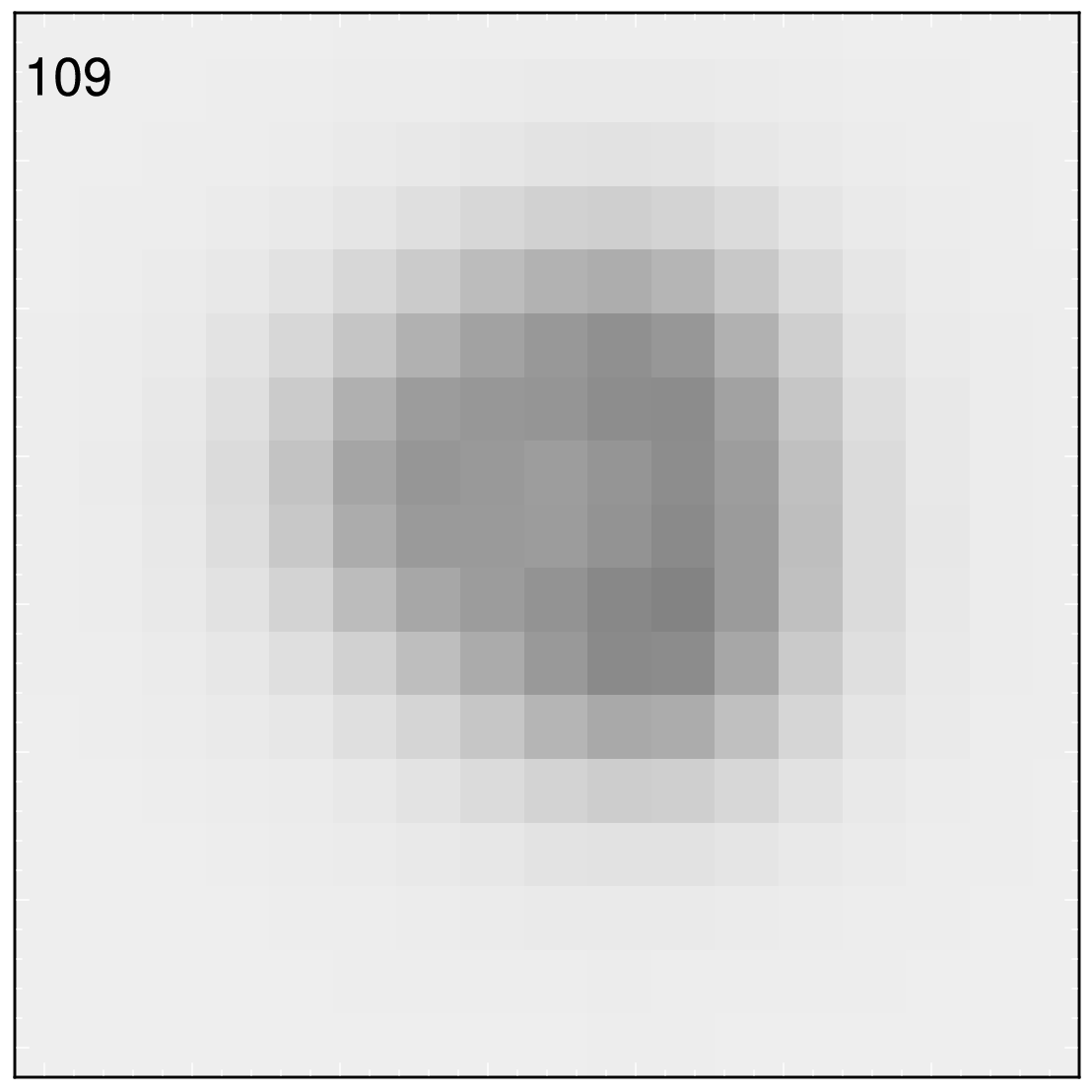}
\includegraphics[width=0.24\columnwidth,angle=0]{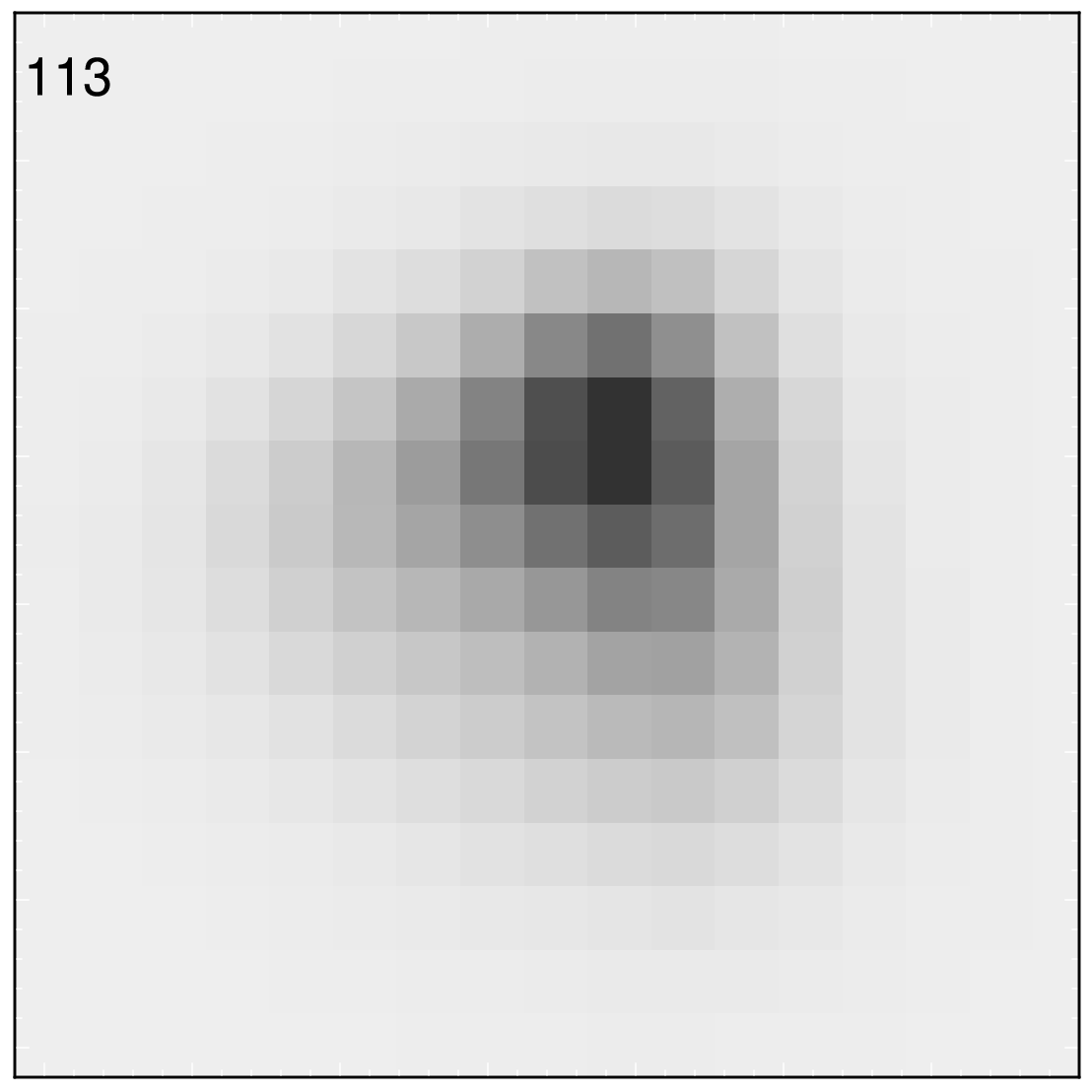}
\includegraphics[width=0.24\columnwidth,angle=0]{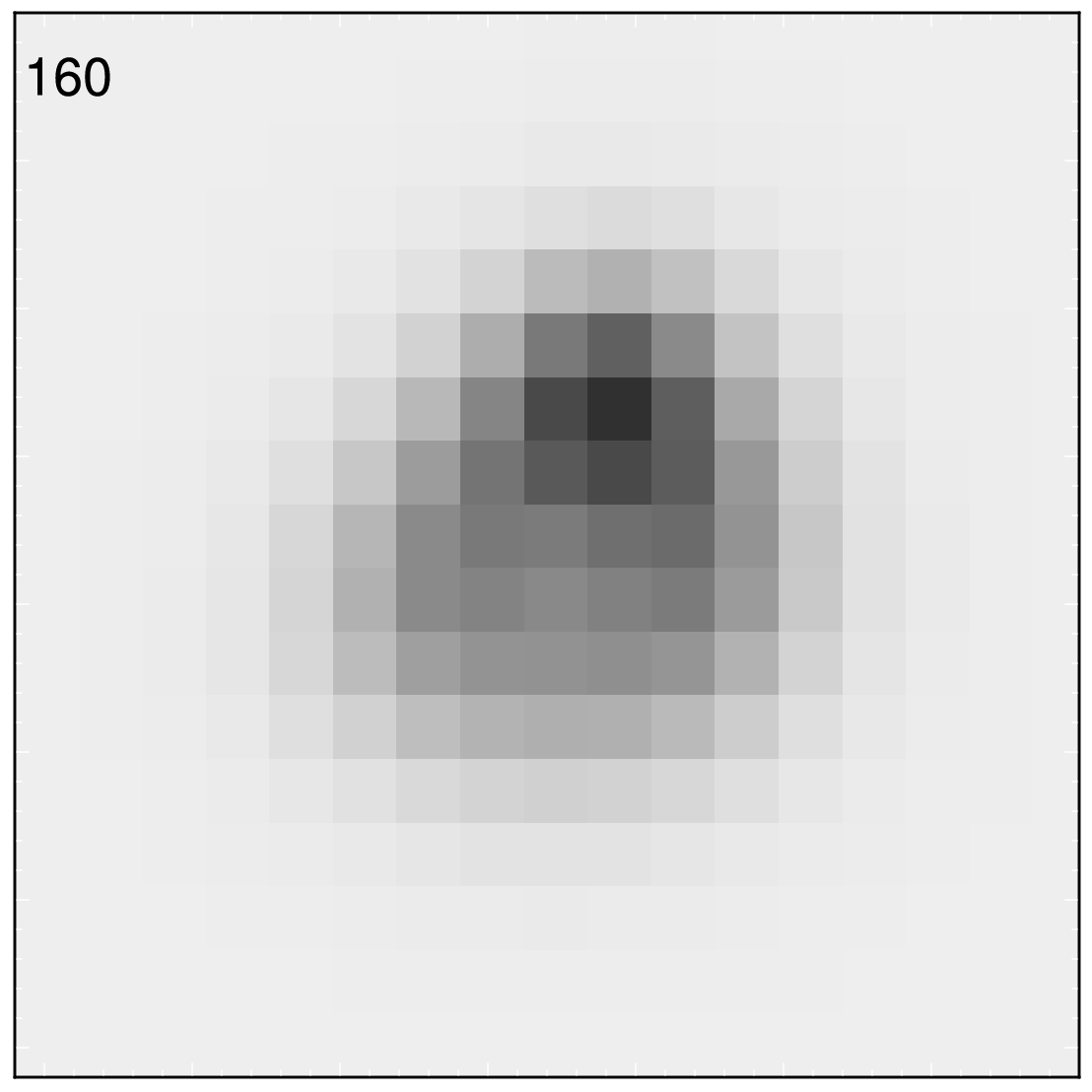}
\includegraphics[width=0.24\columnwidth,angle=0]{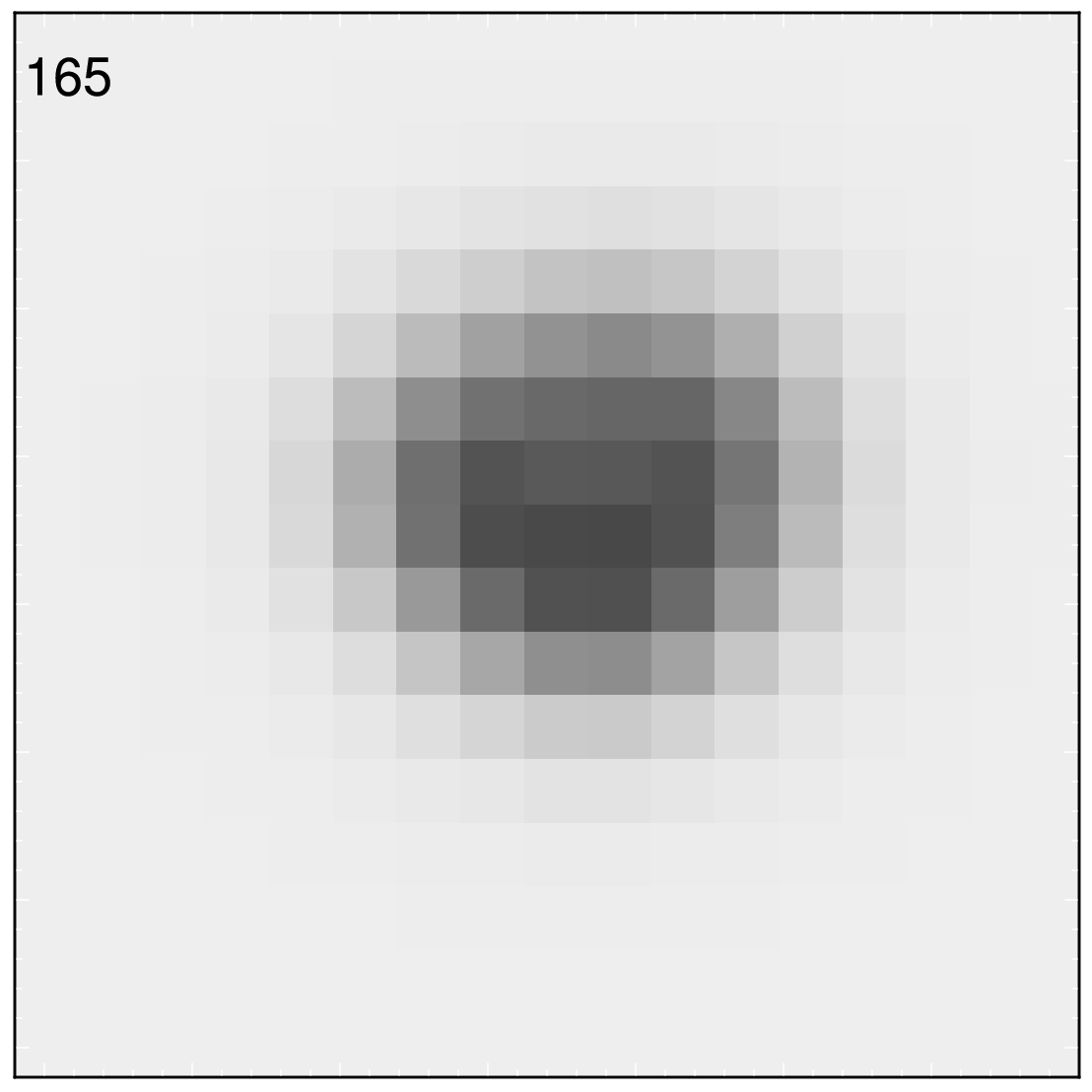}
\includegraphics[width=0.24\columnwidth,angle=0]{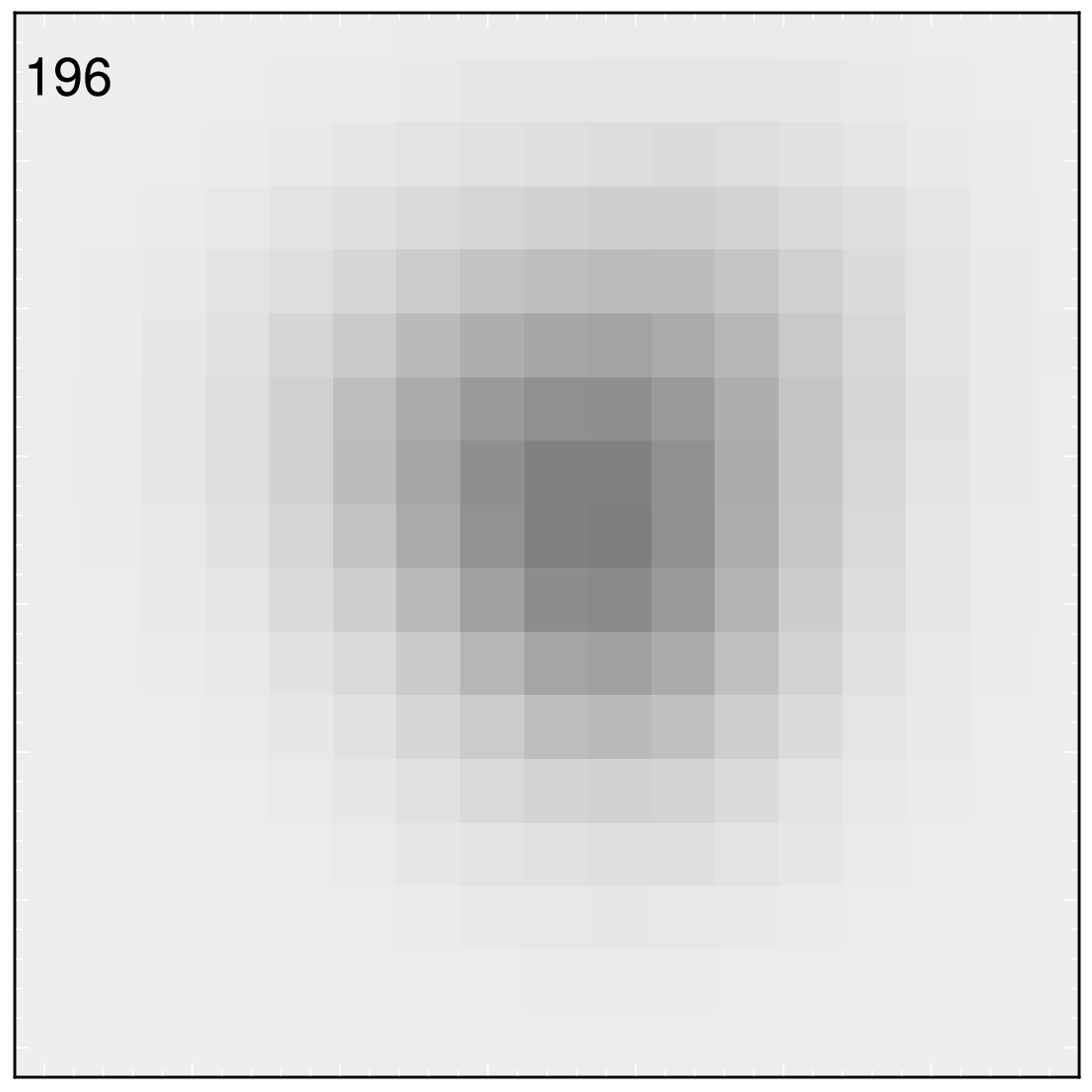}
\caption{\label{fig:outlier}{\em Top:} The difference between
  submitted and true shears vs.\ the true shears for each subfield in
  RGC, for both shear components.  The best-fit line is also shown on
  the plot, along with the $m$, $c$, and $Q_{\rm c}$ values. These
  results are for the best submission from the ``ess'' team.
  \referee{{\em Middle:}} The same, for the best submission in that
  branch from Amalgam@IAP.  \referee{{\em Bottom:} Images of the PSFs
    for the 11 subfields for which the ``ess'' results are discrepant
    at the level of $|\Delta \gamma|\ge 0.01$ in at least one shear
    components in $\ge 75$ per cent of their submissions.  Subfield
    indices are shown on the plot.  The images are shown with a
    self-consistent linear flux scaling and with the total PSF flux
    normalized to 1, so subfields with worse seeing will generally
    have a lower peak flux value.} }
\end{center}
\end{figure}
As an example,
the top panel of Fig.~\ref{fig:outlier} shows the per-subfield
submitted shear from RGC 
for the ``ess'' team.  The plotted quantities are 
used to derive $m_i$ and $c_i$ for the $Q_{\rm c}$
metric. This team has several subfields 
with highly discrepant submitted shears, well beyond the expected
standard deviation of $\lesssim 0.001$ per subfield.  This branch is
the worst case for this team, which had fewer outliers in CGC.

To explore the effect of outliers, we did a systematic test for
outliers in the submitted shears, identifying (for each branch and
team) those fields for which the submitted shears were discrepant by
more than $|\Delta g|=0.01$ in more than 75 per cent of submissions.  In
general, these subfields were consistent across methods; that is, if
two teams had a certain number of outlier fields in a given branch,
they were almost always the same set of subfields.  Those subfields
were commonly ones with higher values for the PSF defocus
\referee{(or, for the ``ess'' team, higher values of trefoil)}; we defer a
more detailed exploration of the impact of defocus on shear
systematics to Sec.~\ref{subsec:results-psf}.  For the ``ess''
team, the reason for the outliers shown in Fig.~\ref{fig:outlier} is fairly
clear: they used a sum of three Gaussian components to describe
the PSF, which makes it particularly difficult to model PSFs with
defocus \referee{or trefoil}.  In contrast, the \referee{middle} panel of that figure shows a
comparable plot for the Amalgam@IAP team, which modeled the full PSF, and
does not show significant outliers.  \referee{Finally, the bottom
  panel shows images of the PSF for the eleven subfields in RGC for
  which the ``ess'' results were seriously discrepant.  As shown, in
  about three cases the PSF has the characteristic ``donut'' shape of
  highly out-of-focus images; such data would likely be eliminated
  from a shear analysis in a real dataset.  These subfields were
  problematic for several other methods.  In other subfields, there is a triangular shape
  characteristic of trefoil, which seems to have been less problematic
  for other methods that have a more flexible representation of the PSF.}

For those teams and branches for which outlier fields
were identified, we recalculated the $m_i$, $c_i$, and $Q_{\rm c}$
values after excluding the outlier fields.
We found that while errors from the linear
regression on $m_i$ and $c_i$ decreased substantially (sometimes tens
of percent after excluding only a few percent of the subfields), the
changes in $m_i$, $c_i$, and $Q_i$ were in general not coherent.  In
many cases, results for different submissions from the same team in the same
branch would change in different directions.  There were three
combinations of branch and team with coherent changes in
results after excluding outliers (in two cases the results were almost
always worse, and in one case they were almost always better).

Several other teams had problems with outliers that were not
identified in the previous test. (Identifying them as outliers would
require a smaller threshold on $|\Delta g|$ and on the number of times
the field has a discrepancy for it to officially be called an
outlier.) These include MBI, which (like ess) used a sum of Gaussians
to describe the PSF; and MegaLUT.  We recalculated the results for these teams
after excluding the fields with the 10 per cent worst defocus in
CGC.  For MBI, the results of excluding the subfields
with the worst defocus did have a coherent effect, but 
with opposite signs in the control and realistic galaxy experiments,
increasing $Q_{\rm c}$ in the former by as much as a factor of two and
lowering it in the latter by a similar amount. 
\newtext{We speculate that the difficulty in modeling these PSFs may not lead
to some systematic overall effect because the hierarchical inference
of $p(\varepsilon)$ might be partially compensating for imperfect PSF
model fits by adjusting the galaxy model fits accordingly. For
MegaLUT, the changes in results after excluding 
the high-defocus subfields were substantially smaller than for MBI.}

Due to the generally inconclusive results of excluding outliers, with
no team showing a strong trend towards improved overall results, for
the rest of this paper we do not exclude outlier fields.  However, in
in Appendix~\ref{app:tables} we also tabulate the outlier-rejected
estimates of $\langle m \rangle$ and $c_+$ for the ess, MBI and
MegaLUT submissions, these being the teams most affected by outliers.

We note that in future challenges it may be a good idea to permit
participants to assign weights to their submitted per-subfield shears,
so they can indicate regimes in which their PSF modeling or shear
estimation does not work. Our results also suggest that PSF modeling
with a low-order decomposition into sums of Gaussians may be
inadequate to describe realistic PSFs, and can significantly affect
the shear estimates.

\section{Overall results}\label{sec:overall}

In this section, we present results for the control and
realistic galaxy experiments for all teams.

\subsection{What results are shown}\label{sec:what_results}

\begin{figure}
\begin{center}
\includegraphics[width=0.99\columnwidth,angle=0]{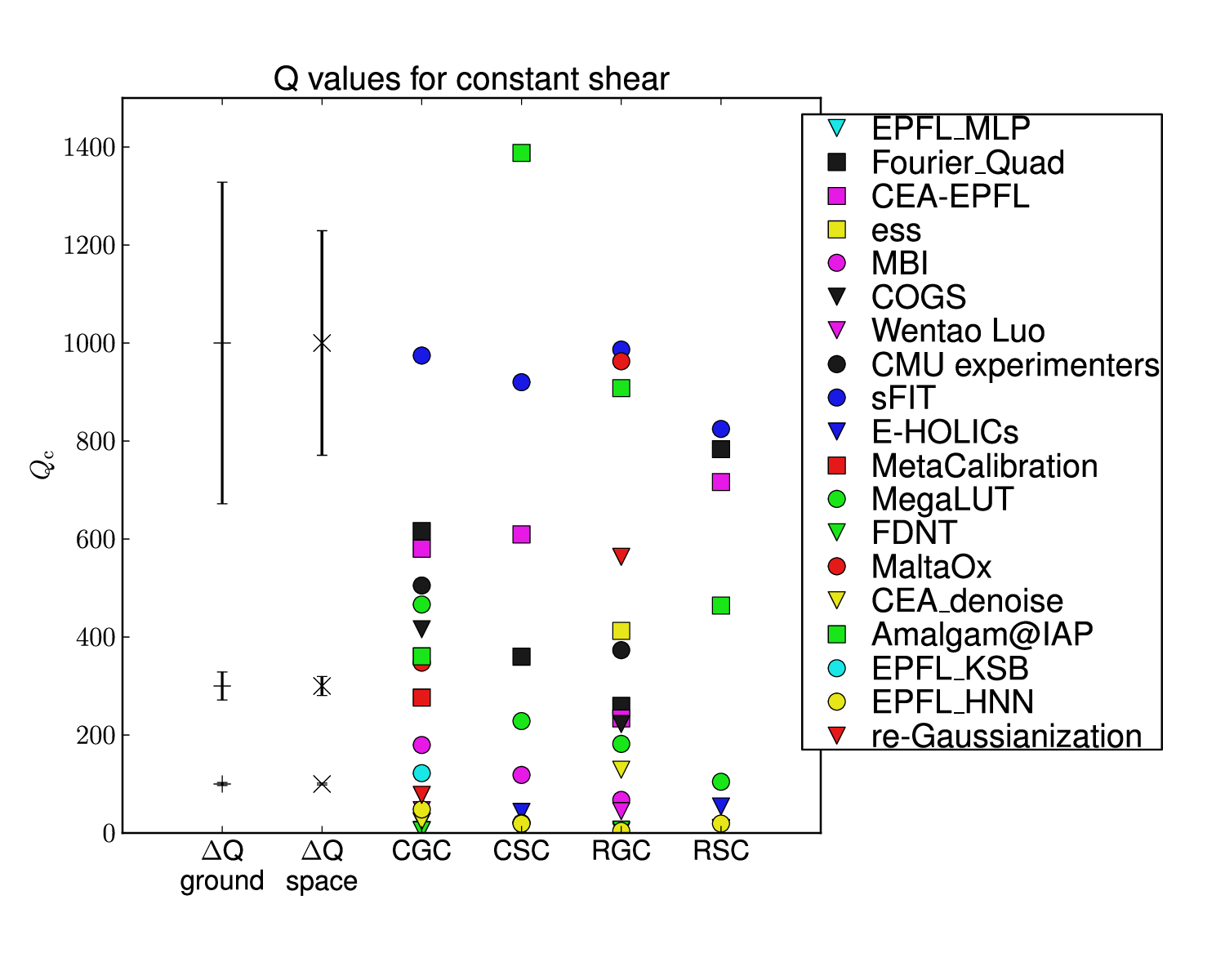}
\includegraphics[width=0.99\columnwidth,angle=0]{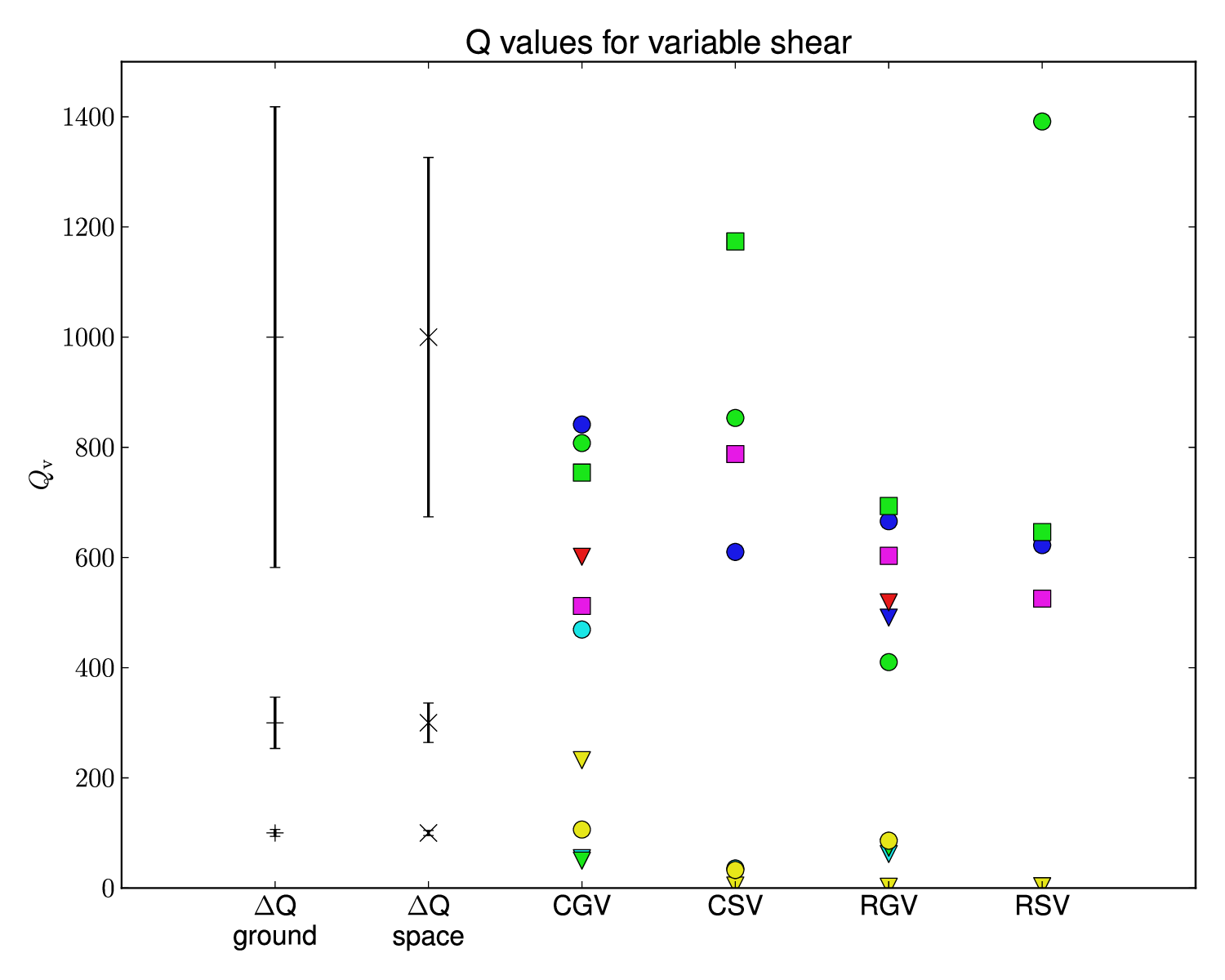}
\caption{\label{fig:basic_q_trends} $Q_{\rm c}$ (top) and $Q_{\rm v}$
  (bottom) for constant- and variable-shear branches in the
  control and realistic galaxy experiments.  The errorbars show 
  the possible range of $Q$ values for a submission with
  shear calibration biases that would nominally give a particular $Q$
  value.  As shown, the sizes of these ranges depend strongly on $Q$, 
  and are smaller for space than for ground branches.
}
\end{center}
\end{figure}
To avoid showing many submissions from each team in each branch, we
adopt a fair and consistent way to select a single submission per
branch from each team.  For the teams discussed 
in Sec.~\ref{sec:specific}, we have already stated 
what submissions will be used here.  For the remaining teams, the
selection was done as follows:

\begin{itemize}
\item FDNT: We use FDNT v1.3, with a self-consistent set of resolution
  and SNR cuts (submissions with names that include ``r12\_sn15'').
\item E-HOLICs: We use their ``snfixed200'' submission, which have a
  self-consistent set of noise bias corrections.
\item MaltaOx: We use the best results for \code{LensFit} with
  oversampled PSFs and self-calibration included.
\item ess: We only use their RGC results, with
  the priors on $p(\varepsilon)$ derived from the deep
  fields (submission name ``nfit-rgc-06-nfit-flags-02'').
\item CMU experimenters: Only one submission per branch.
\item CEA\_denoise, MetaCalibration, EPFL\_MLP /
  EPFL\_MLP\_FIT, EPFL\_KSB, EPFL\_HNN, Wentao Luo: Best submission in
  each branch.
\item GREAT3-EC (or re-Gaussianization): These results used the shear estimation example
  script described in Appendix~\ref{app:scripts}.  As noted there, for
  several reasons the results are not science-quality shear estimates
  and therefore the results have no reflection on science papers that
  use this algorithm.  However, since it is a stable algorithm in the
  public domain, and one of the few moments-based methods, we include
  it in this section to provide a basic point of comparison.
\end{itemize}

Results for the following teams are not shown in this
section: miyatake-test (for reasons
described in Sec.~\ref{sec:methods}), BAMPenn, and HSC/LSST-HSM.  The BAMPenn
results included some bugs that mean the results do not
correctly reflect the real performance of the method.  The
HSC/LSST-HSM submissions used the HSC/LSST software
pipeline with the same shear estimation method as in the GREAT3 example
scripts purely as a sanity check of the pipeline. 

\subsection{Basic $Q$ results}

\begin{figure}
\begin{center}
\includegraphics[width=0.99\columnwidth,angle=0]{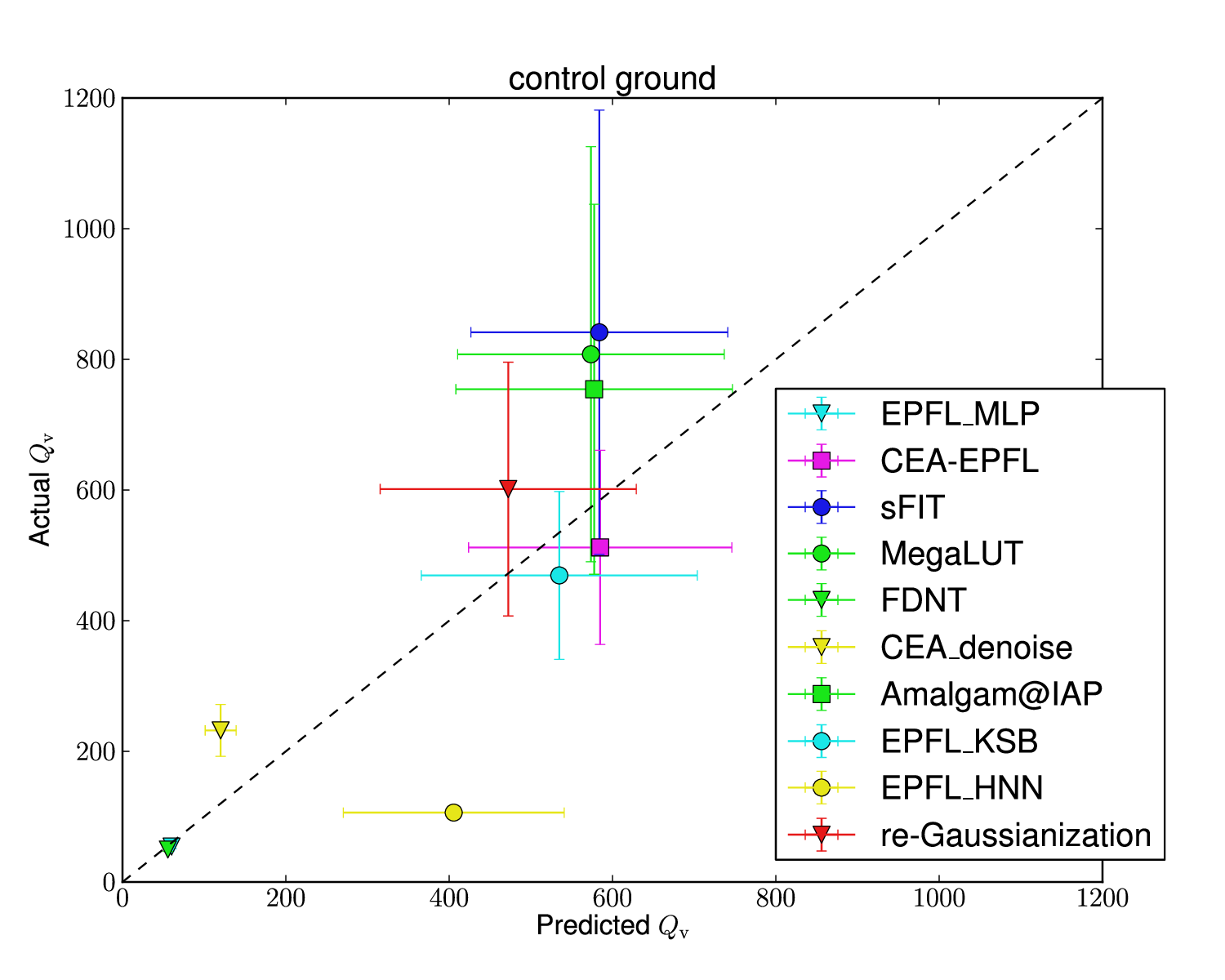}
\caption{\label{fig:qc_qv} Comparison between the $Q_{\rm v}$
  predicted from the constant-shear branch results (CGC), and the
  actual $Q_{\rm v}$ results for variable shear (CGV).}
\end{center}
\end{figure}
\begin{figure*}
\begin{center}
\includegraphics[width=0.99\columnwidth,angle=0]{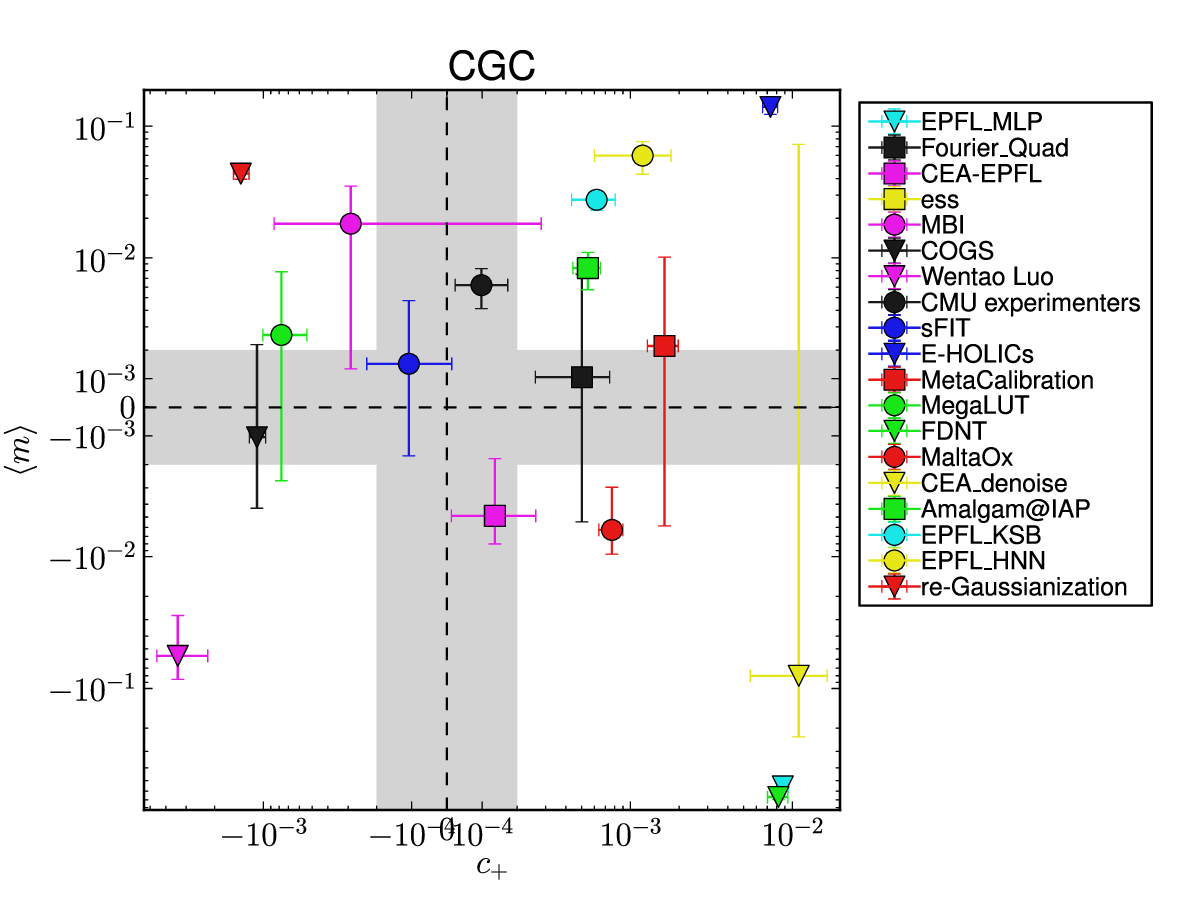}
\includegraphics[width=0.99\columnwidth,angle=0]{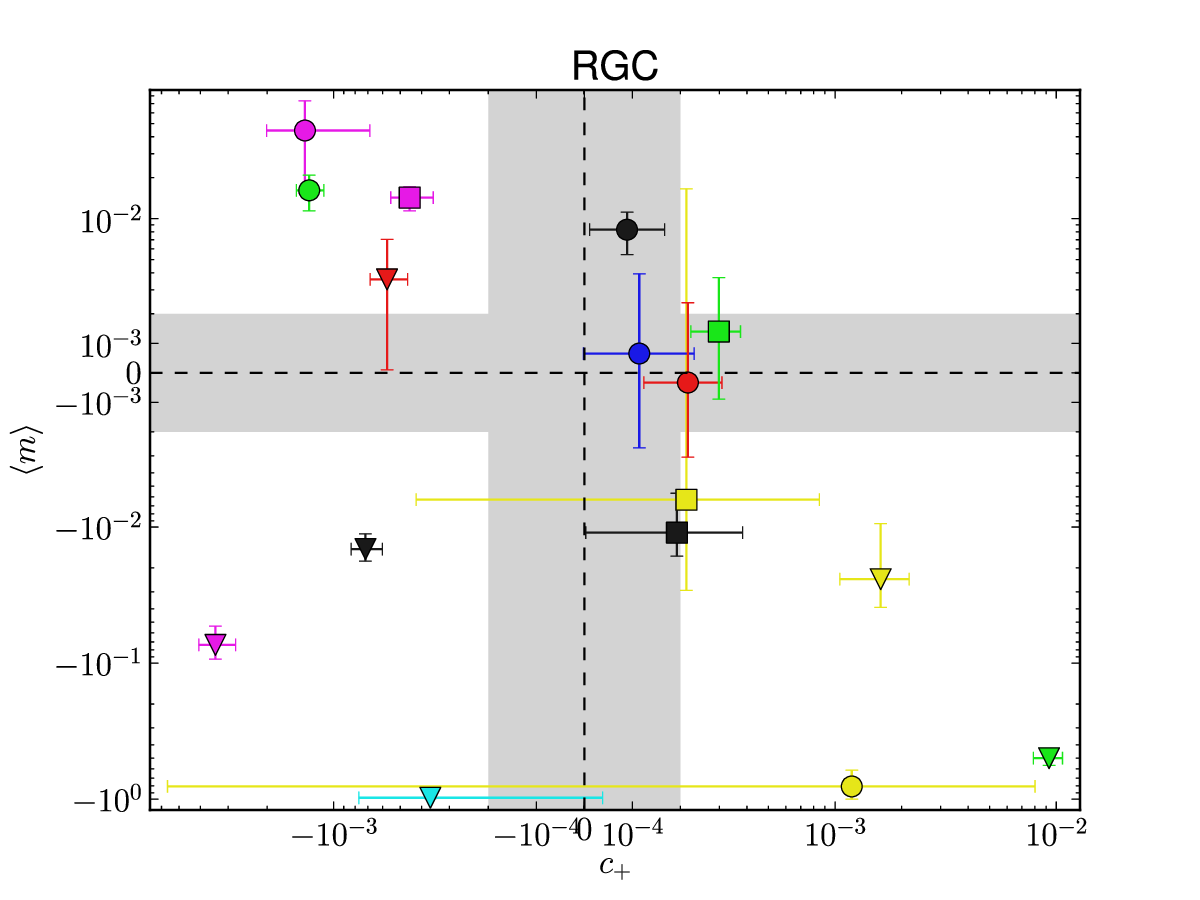}
\includegraphics[width=0.99\columnwidth,angle=0]{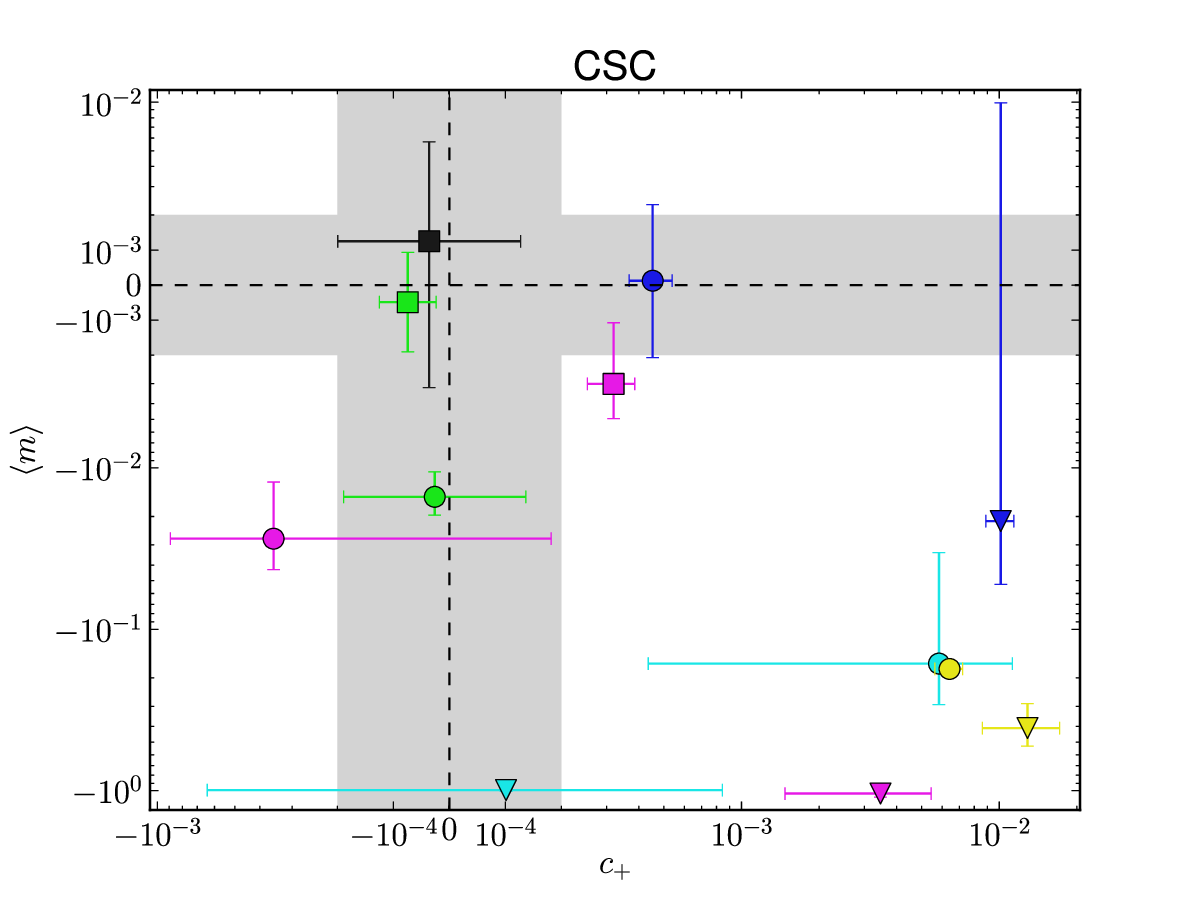}
\includegraphics[width=0.99\columnwidth,angle=0]{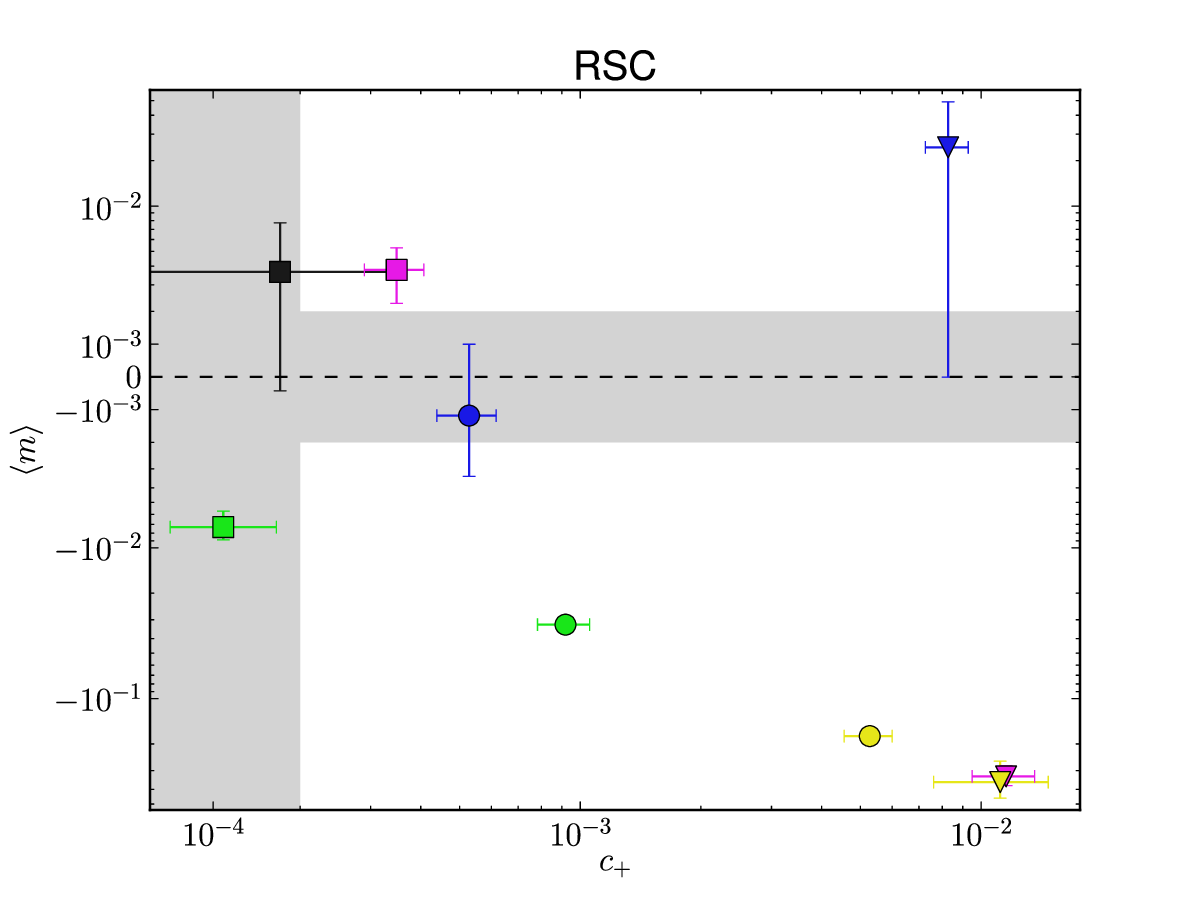}
\caption{\label{fig:basic_mc} Multiplicative and additive biases for
  constant-shear branches in the control (left) and realistic galaxy
  (right) experiments, for ground (top) and space (bottom) branches.
  For each branch, we show the averaged (over components)
  multiplicative bias $\langle m\rangle$ 
 vs.\ $c_+$, the additive bias term defined in the coordinate
  system defined by the PSF anisotropy.  The axes are linear 
  within the target region ($|m|<2\times 10^{-3}$ and $|c|<2\times
  10^{-4}$, shaded grey) and logarithmic outside that region.
}
\end{center}
\end{figure*}
\begin{figure*}
\begin{center}
\includegraphics[width=0.99\columnwidth,angle=0]{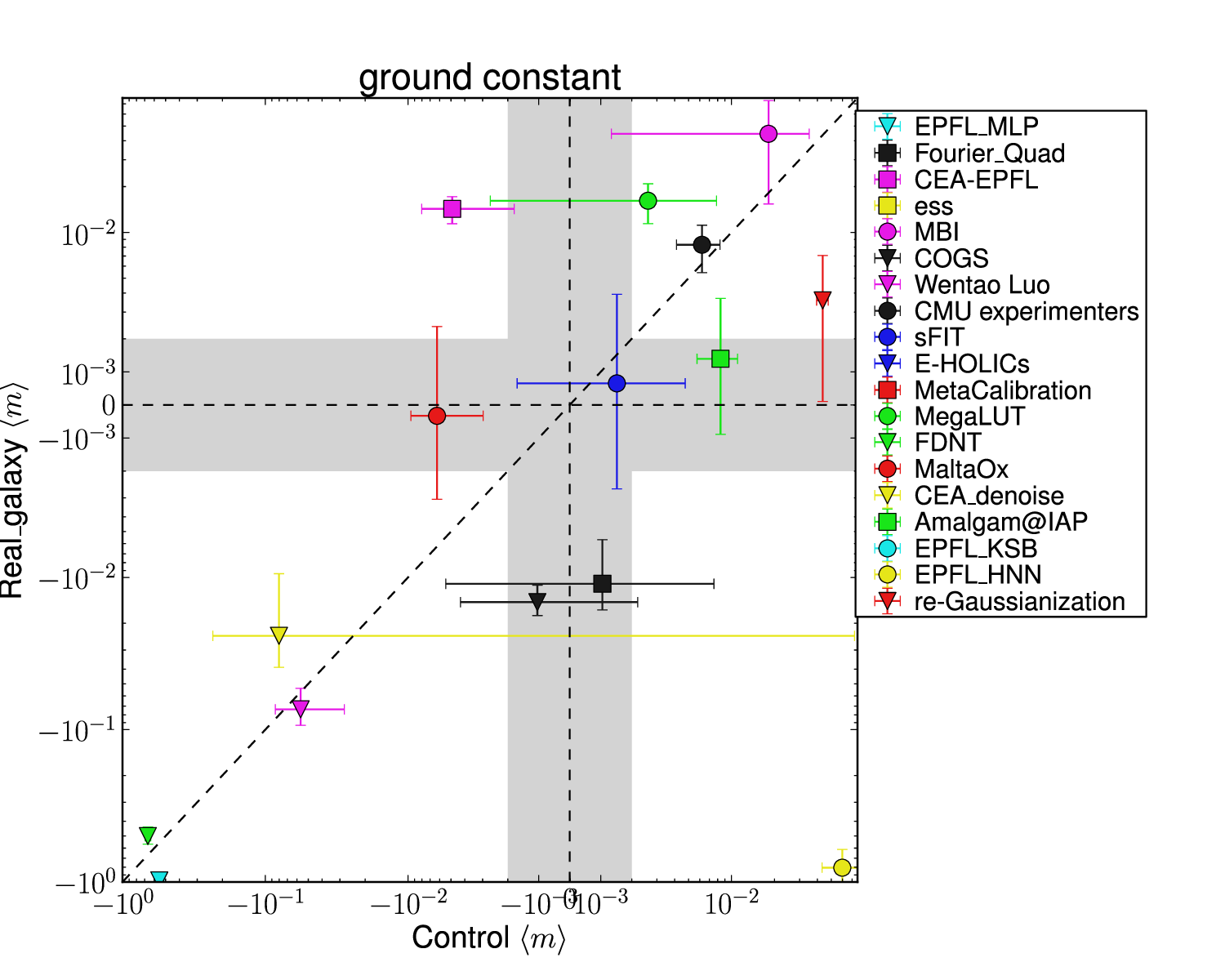}
\includegraphics[width=0.99\columnwidth,angle=0]{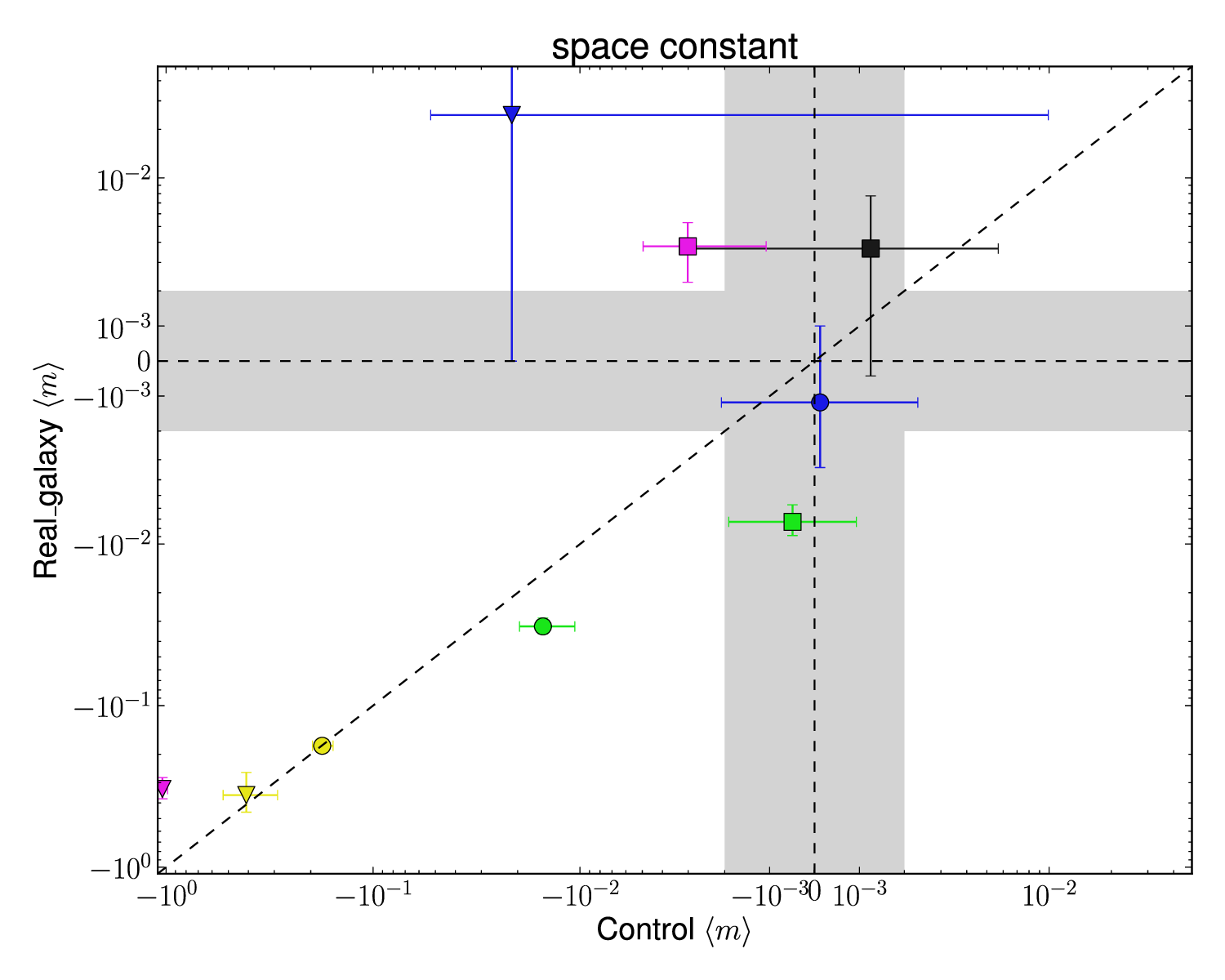}
\includegraphics[width=0.99\columnwidth,angle=0]{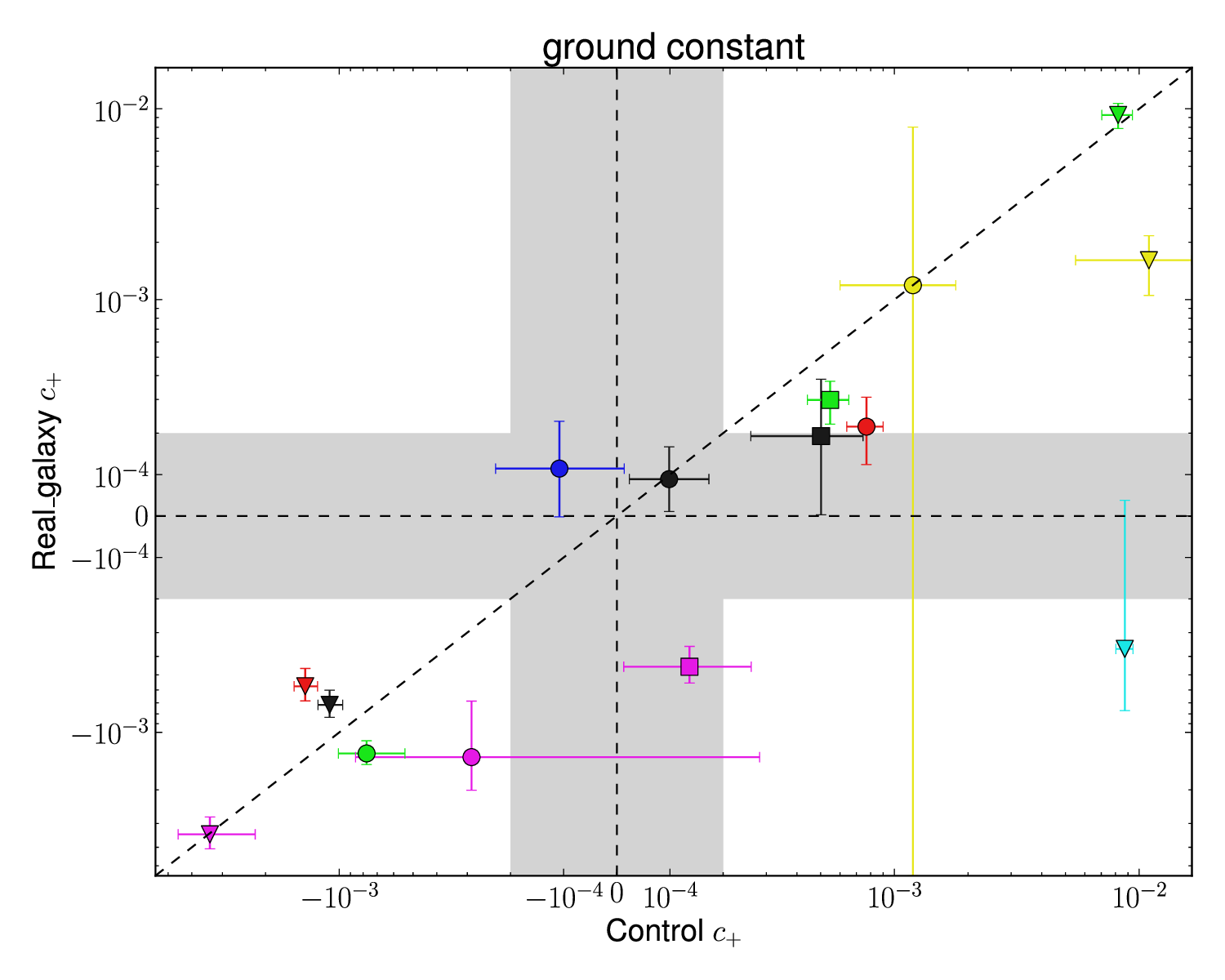}
\includegraphics[width=0.99\columnwidth,angle=0]{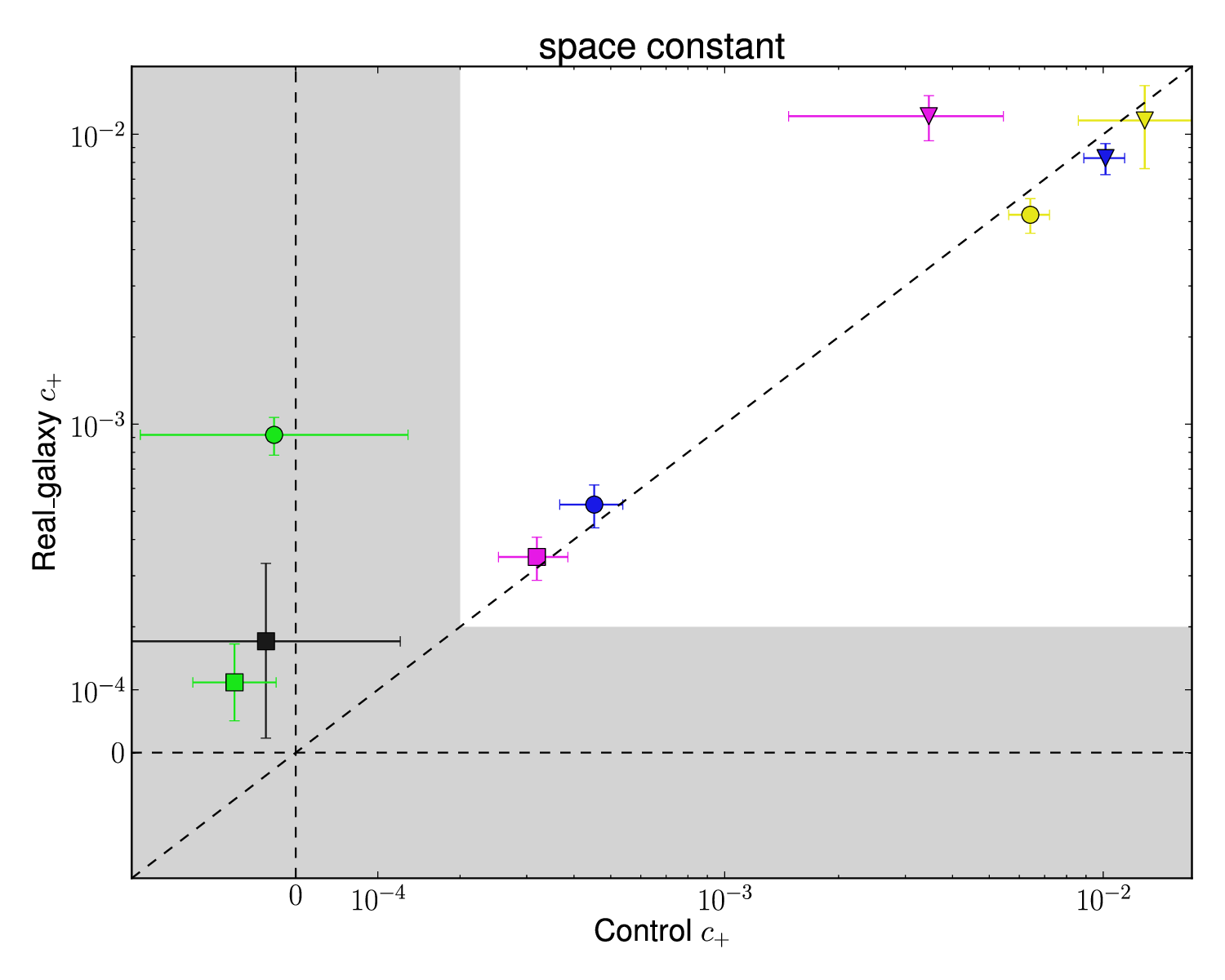}
\caption{\label{fig:mc_control_real} Left and right columns show
  results for $\langle m\rangle$ (top; averaged over components) and
  $c_+$ (bottom) for ground and space branches, respectively.  Each
  panel compares results for  control vs.\ realistic galaxy
  experiments. The axes are linear within the target region
  ($|m|<2\times 10^{-3}$ and $|c|<2\times 10^{-4}$, shaded grey) and logarithmic
  outside that region.  The black dashed line is the 1:1 line.
}
\end{center}
\end{figure*}
In this subsection, we present the $Q$ results for all teams.
Fig.~\ref{fig:basic_q_trends} 
shows $Q_{\rm c}$ and $Q_{\rm v}$ for the control and realistic galaxy
experiments.

Several trends from Sec.~\ref{sec:specific} are 
evident here.  For example, the results for sFIT are
quite consistent across all branches shown here.  The MegaLUT results
are consistently better for variable shear than for
constant shear, presumably because of a low-level $m$-type bias, to
which $Q_{\rm c}$ is more sensitive than $Q_{\rm v}$.  The results for
Amalgam@IAP and CEA-EPFL are good in many branches, but exhibit 
significant fluctuations due to partial cancellations of biases.
The results for Fourier\_Quad with a realistic weighting
scheme are quite good, but degraded compared to the results with the
unrealistic weighting schemes.

The errorbars in
Fig.~\ref{fig:basic_q_trends} show that for lower $Q$ values, the uncertainty
in $Q$ is very small.  However, near the target $Q$ values,
small uncertainties in $m$ and $c$ become large
uncertainties in $Q$.  These errorbars are quite non-Gaussian, so 
for example the difference between $Q=500$ and
$1000$ for control space branches is significantly more than the
$2\sigma$ suggested by the plot.  It is apparent that in many
branches, 2--3 teams performed well enough that the differences
between their $Q$ values (and between the target
of $\sim 1000$) are not statistically significant.

One basic question is whether the results in the constant
and variable shear branches are consistent.  We cannot
directly compare $Q_{\rm c}$ and $Q_{\rm v}$, because they respond
 to
systematic errors in different ways.  However, for a
given constant-shear submission, we can use the recovered
$m$ and $c$ values to predict $Q_{\rm v}$ \newtext{by simulating variable shear
submissions with those $m$ and $c$, and then checking their $Q_{\rm v}$.}  Comparing the predicted
$Q_{\rm v}$ with the actual one (for the same experiment and
observation type) is a valid consistency check.  We show this comparison for CGC and CGV in
Fig.~\ref{fig:qc_qv}, with a reasonable level of consistency within
the relatively large errors on the $Q_{\rm v}$, and at most
a $2\sigma$ discrepancy for one team.  The plots for the other
experiments and observation types show similar 
constant vs.\ variable 
shear consistency.

\subsection{Multiplicative and additive shear biases}\label{subsec:overallmc}

This section will focus on Fig.~\ref{fig:basic_mc}, which shows the
multiplicative and additive shear biases ($m$ and $c$) for the
constant-shear branches in the control and realistic galaxy experiment.
All $m$ and $c$ values are also tabulated in
Appendix~\ref{app:tables}. Unlike $Q_{\rm c}$, $m$ and $c$ have
well-understood errorbars.  On these plots, the errorbars are
different sizes for different methods.  In some cases, it is only an
apparent difference (due to the mixed linear and logarithmic axes),
but there is some variation in the scatter in shears that we will
explore in Sec.~\ref{subsec:noise}.
\begin{figure}
\begin{center}
\includegraphics[width=0.99\columnwidth,angle=0]{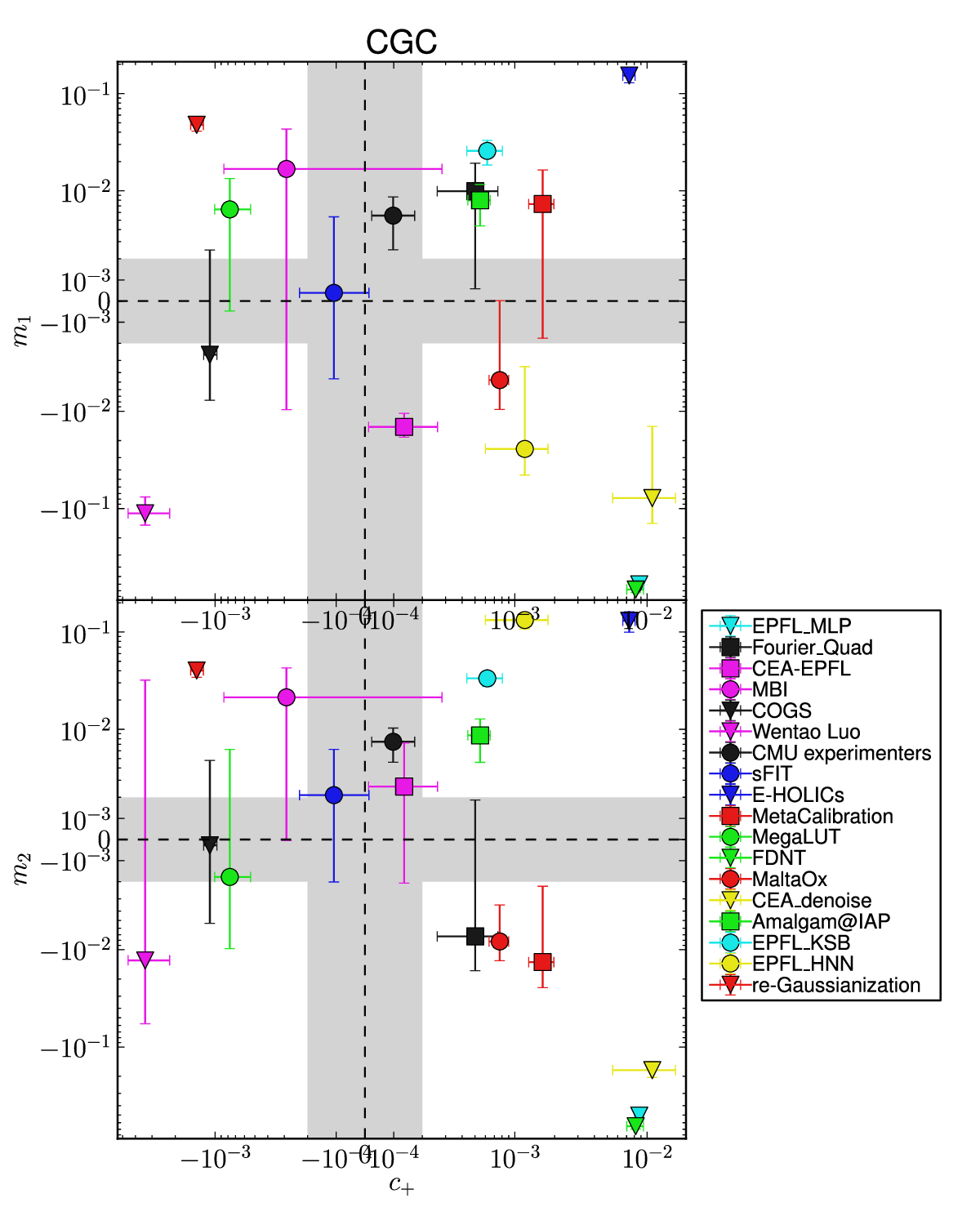}
\caption{\label{fig:basic_mc_xy} Shear biases
  for CGC, similar to Fig.~\ref{fig:basic_mc} but using $m_1$ and
  $m_2$ (defined using the pixel coordinate
  system).}
\end{center}
\end{figure}

We begin by discussing the top left panel of Fig.~\ref{fig:basic_mc},
which shows $\langle m\rangle$ (averaged over components) vs.\ $c_+$ for CGC.  Not
surprisingly, the teams that are located near 
the center of this plot (small $|m|$ and $|c|$) are the ones with 
high $Q_{\rm c}$ factors for this branch (Fig.~\ref{fig:basic_q_trends}).

A few methods (COGS, MegaLUT, MetaCalibration) are notable in
having multiplicative biases  
consistent with being in the target region, but highly significant
detections of additive bias.  Both COGS and MetaCalibration include
multiplicative bias corrections, but no additive bias corrections were
implemented by the end of the challenge period. 

\subsubsection{Impact of morphology}\label{subsubsec:mc-morphology}

Comparing the left and right sides of
Fig.~\ref{fig:basic_mc} would reveal the impact
of realistic galaxy morphology.  However, to facilitate an easier comparison,
Fig.~\ref{fig:mc_control_real} explicitly compares $\langle m\rangle$
(averaged over components) and $c_+$ values
for control vs.\ realistic galaxy experiments\newtext{, with results tabulated
in Table~\ref{tab:mc_deltas}.}  For
ground-based simulations, the $\langle m\rangle$ comparison 
 is in the top left panel.  \newtext{Many 
methods are consistent with the 1:1 line, 
 meaning that the calibration bias does not show any detectable
impact from realistic galaxy morphology. Moderate differences in
model bias due to realistic galaxy morphology can be seen 
for many teams, with typically} a $\sim$ per cent level impact of realistic
galaxy morphology on multiplicative calibration biases, although the
sign of the change in $\langle m \rangle$ depended on method.

The top right panel of Fig.~\ref{fig:mc_control_real} shows how 
$\langle m\rangle$ changes from control to realistic galaxy experiment for
space-based simulations.  \newtext{Again, some methods exhibit no significant
model bias due to realistic galaxy morphology (but note that sFIT included this effect
in their simulations, and explicitly calibrated it
out), while others have typically $\sim 1$ per
cent level calibration changes.}

The bottom left panel of Fig.~\ref{fig:mc_control_real} shows 
$c_+$ for CGC vs.\ RGC, \newtext{with everything from complete consistency to
strong differences in $c_+$ in these branches, implying that 
realistic galaxy morphology can in some cases cause additive biases.}

Finally, in the bottom right panel of Fig.~\ref{fig:mc_control_real},
the $c_+$ are consistent
between control and realistic galaxy experiments for space-based
\newtext{simulations for most methods.}  It seems
that for space simulations, removing
the PSF anisotropy is similarly difficult for both parametric and realistic
galaxy models.

\subsubsection{Impact of ground- vs.\ space-based PSF}

Comparing the top and bottom rows of Fig.~\ref{fig:basic_mc} reveals
the effects of using a space-based PSF rather than a ground-based
PSF.  \newtext{Note that the numerical values of the $c_+$ and $\langle
m\rangle$ changes are shown in Table~\ref{tab:mc_deltas}.}  
Focusing first on the control experiment (left side), the $c_+$ values
\newtext{shifted to the right (more positive) in space data for the majority of
the methods. }  
Note that if $c_+$ scales linearly with PSF
ellipticity (a model that we will validate in
Sec.~\ref{subsec:linmodel}), then $c_+$ for the
space branches should be larger than in the ground branches by a
factor of $\sim 2$.  This may explain the changes in $c_+$ for
\newtext{several teams, but not all, implying that in some cases 
the additive systematics have some additional dependence on}
the form of the PSF beyond its ellipticity.

\newtext{Comparing multiplicative biases for CGC and CSC, they are either statistically
consistent between space and ground or more negative for space
branches; 
curiously, they did not become more positive for any} 
teams.  Given the wide diversity
of methods and the apparent lack of
commonality between many that exhibit similar behavior between ground
and space data, it is difficult to draw conclusions, but the pattern
is indeed interesting.

These results were for the control experiment.  If we compare RGC
vs.\ RSC (right panels), we see that the differences in $c_+$ and
$\langle m\rangle$ between space and ground simulations in the
realistic galaxy experiment are similar to what was seen for the control
experiment for all teams except CEA\_denoise. This finding suggests
that the effect of the type of PSF (space vs.\ ground) on additive and
multiplicative biases does not typically depend on whether the
galaxies have realistic morphology or are simple parametric models.

\subsubsection{Use of pixel coordinate system}

The top left panel of Fig.~\ref{fig:basic_mc} shows $m$ vs.\ $c$ for
CGC in the coordinate system defined by the PSF anisotropy, whereas
Fig.~\ref{fig:basic_mc_xy} shows the same in the pixel coordinate
system.  In a few cases (e.g., CEA-EPFL, Fourier\_Quad, and MetaCalibration to some
degree though it is noisier), $m_1$ and $m_2$ have opposite signs, and
thus average out to something closer to zero (after rotating to the
PSF anisotropy coordinate frame) for $m_+$ and $m_\times$, resulting
in $Q_{\rm mix}<Q_{\rm c}$.

\subsection{Understanding the linear model}\label{subsec:linmodel}

In this section, we explore the linear model
for shear systematics, Eq.~\eqref{eq:linbias}, by considering some
alternative models of shear measurement bias.

It is commonly assumed that the main source of $c$-type biases 
is leakage from PSF anisotropy into galaxy shear
estimates, which should be proportional to the amplitude
of the PSF ellipticity.  (However, there are physical models that
violate this assumption, nor is this assumption completely
obvious for all methods.)
If the assumption is correct, we can write an alternative model:
\begin{equation}\label{eq:altlinbias}
g^{\rm obs}_i - g^{\rm true}_i = m_i g^{\rm true}_i + a_i g^{\rm PSF}_i
\end{equation}
\newtext{Here, the $a_i$ prefactors are average values across an entire galaxy
population that likely depend on the distribution of SNR, resolution,
morphology, and PSF type. }
In the coordinate system defined by the PSF anisotropy, $g^{\rm PSF}_+
= |g^{\rm PSF}|$ and $g^{\rm PSF}_\times = 0$.  We can therefore fit
to this new model, and if the additive errors are proportional to the
PSF anisotropy, then we should find $c_+ \propto a_+$, where the
constant of proportionality is an effective mean $|g^{\rm PSF}|$ for
that branch.

\begin{figure}
\begin{center}
\includegraphics[width=0.99\columnwidth,angle=0]{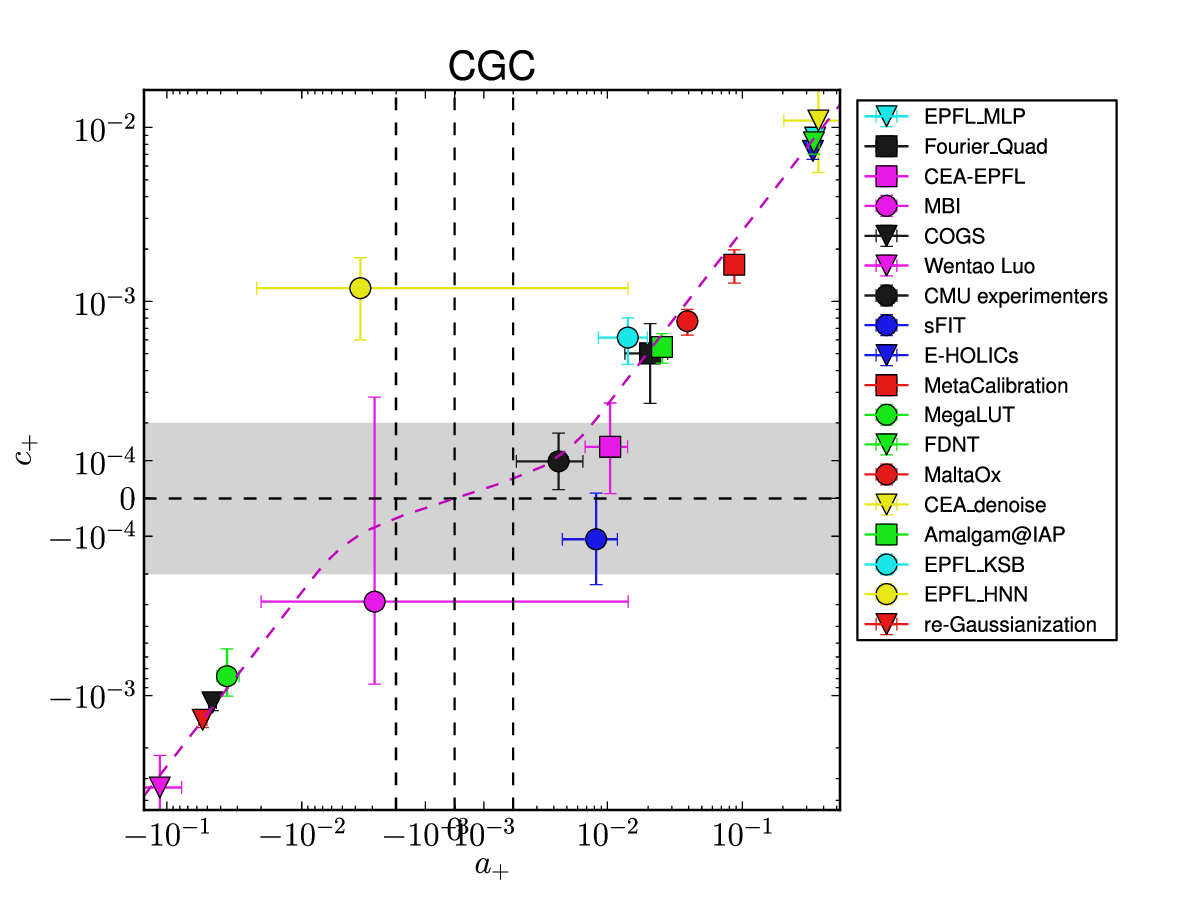}
\includegraphics[width=0.99\columnwidth,angle=0]{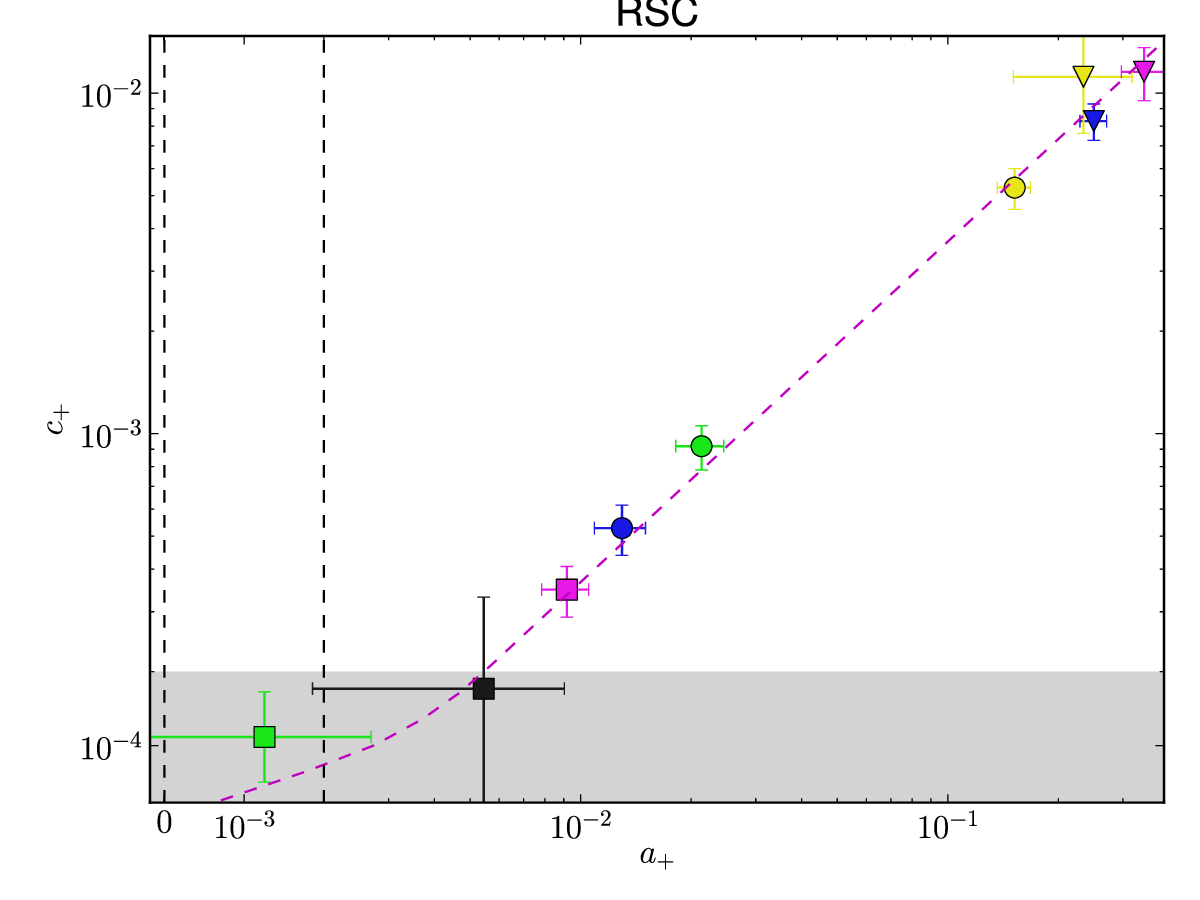}
\caption{\label{fig:ac} For CGC (top) and RSC (bottom), we compare the
  additive bias $c_+$ in the standard linear bias
  model, Eq.~\eqref{eq:linbias}, against $a_+$ for the
  alternative model in Eq.~\eqref{eq:altlinbias}.  $a_+$ is
   a constant of proportionality relating additive
  shear systematics to the PSF ellipticity.  The axes
  are linear for $|a_+|<2\times 10^{-3}$ and $|c_+|<2\times
  10^{-4}$ (where the latter is our target region for additive
  systematics, shown in grey) and logarithmic outside that region; 
  we use vertical lines
  to indicate the linear-logarithmic boundary in $a_+$.  The best-fit slope
  relating $c_+$ and $a_+$ is shown as a dashed magenta line.  It only
  appears curved because we show combined log
  and linear axes with an unequal aspect ratio.
}
\end{center}
\end{figure}

Fig.~\ref{fig:ac} compares $c_+$ and $a_+$ for CGC
(top) and RSC (bottom), though the results are quite similar for CSC
and RGC as well.  The best-fit line relating $c_+$ and $a_+$ 
goes through nearly all the points,
indicating that the linear model works well (except for 
EPFL\_HNN) for a wide variety of shear estimation methods.  The slopes of the best-fitting lines for CGC, RGC, CSC,
and RSC are 0.025, 0.016, 0.039, and 0.037, respectively,
corresponding to the effective mean per-branch $|g^{\rm PSF}|$.

$a_+$ is
essentially the fraction of PSF anisotropy that leaks into
galaxy shear estimates.  For the methods that have $c_+$ within the target region, the
$a_+$ values indicate that typically $<1$ per cent of the PSF shear
contaminates the galaxy shears.  Several methods are 
in the range of 1--10 per cent leakage, and the worst case
scenarios involve leakage of tens of per cent.  For data with a
narrower (wider) range of PSF anisotropies but otherwise similar
properties (so that $a_+$ is the same), the additive
bias $c_+$ will be better (worse) than is
shown here.  (Note that the histogram of PSF shears in GREAT3
is in Appendix~\ref{app:psf}.)

In real data, 
selection biases that correlate with PSF direction also induce 
additive systematics. While
these operate at some level in GREAT3 due to different
weights being assigned to galaxies depending on their direction with
respect to the PSF, in real data selection biases should be more
important given the need to identify galaxies.  In
that case, this simple linear model may no longer be valid.  It seems
reasonable that selection biases will cause $c_+$
to scale with $|g^{\rm PSF}|$, but it is not obvious 
that the scaling should be linear.

The success of the simple linear PSF contamination model of
Eq.~\eqref{eq:altlinbias} in describing additive bias in GREAT3,
evidenced by Fig.~\ref{fig:ac}, is striking.  However, we note that
the GREAT3 simulations were designed without many effects found in
real data that potentially cause additive bias (see
Sec.~\ref{subsec:structure} for a list) but are not directly
related to the PSF.  These may cause additive biases to show more
complex dependencies in real data.

Another question about the linear model for shear
calibration biases is whether these methods have a  nonlinear response
to shear.  This question was already addressed in
the STEP2 challenge \citep{2007MNRAS.376...13M}.  In that case, the
shears were positive in the CCD coordinate system, and the
nonlinearity test
involved a term proportional to $g_{\rm
  true}^2$.  In GREAT3, the per-component shears can be positive or
negative, so the simplest low order nonlinear terms are proportional to
$g_{\rm true}^3$ or ${\rm sign}(g_{\rm true}) g_{\rm true}^2$.  We can think of
these as being the next order beyond linear of a series expansion of
some unknown function representing the shear response.

We carried out fits with an additional term defined in either of
these two ways, and checked for nonzero prefactors for the nonlinear terms.  In general, the results for all
methods are consistent with zero.  When considering constant
shear branches in the control and realistic galaxy experiments, there are
81 submissions (across all branches and teams) that we use 
in this section, and therefore 162 fits when we use both shear
components.  Regardless of which form we use for the nonlinear
term, its prefactor differs from zero at
$>2\sigma$ for nine of the 162 fits, or 5.6 per cent, which is
consistent with what we expect if no methods have
nonlinear response.  Moreover, these $>2\sigma$ deviations are
not consistently found in any particular team, but are for a range of
teams.  We conclude that the GREAT3 results show no sign of a
nonlinear shear response for any method. However, with a maximum value of $|g_{\rm
  true}|\le 0.05$, we are not very sensitive  to nonlinear shear response, and studies that go into the cluster shear regime may need
to redo this test.

\subsection{Dependence of results on detailed PSF properties}\label{subsec:results-psf}

In this section, we check how results for each method depend on the
PSF properties within the branch.  Note that the PSF properties in the
control and realistic galaxy experiments are discussed and shown in 
Appendix~\ref{app:psf} and Fig.~\ref{fig:psf_property_dist}.

For this test, we split the subfields within a branch into those with
atmospheric PSF FWHM, defocus, or $|g^\text{PSF}|$
above and below the median values. Then we refit the submitted shears
for those subsets to estimate $m_i$ and $c_i$ values.
We can compare the  $m_i$ and $c_i$ for those with better
vs.\ worse values of seeing, defocus, and PSF shear, and compare with
the overall $m_i$ and $c_i$ for the branch.

\subsubsection{Atmospheric PSF FWHM}

\begin{figure}
\begin{center}
\includegraphics[width=0.9\columnwidth,angle=0]{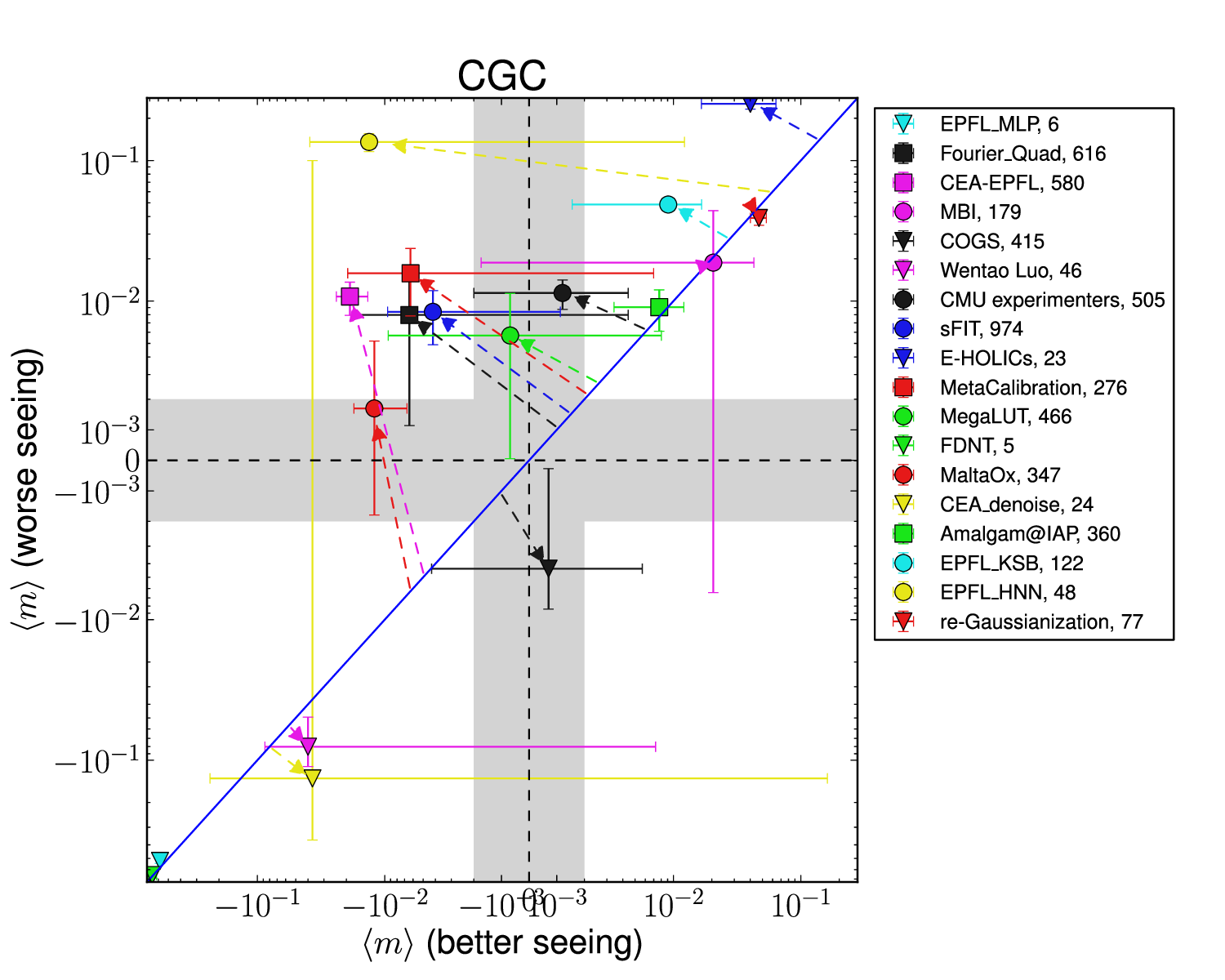}
\includegraphics[width=0.9\columnwidth,angle=0]{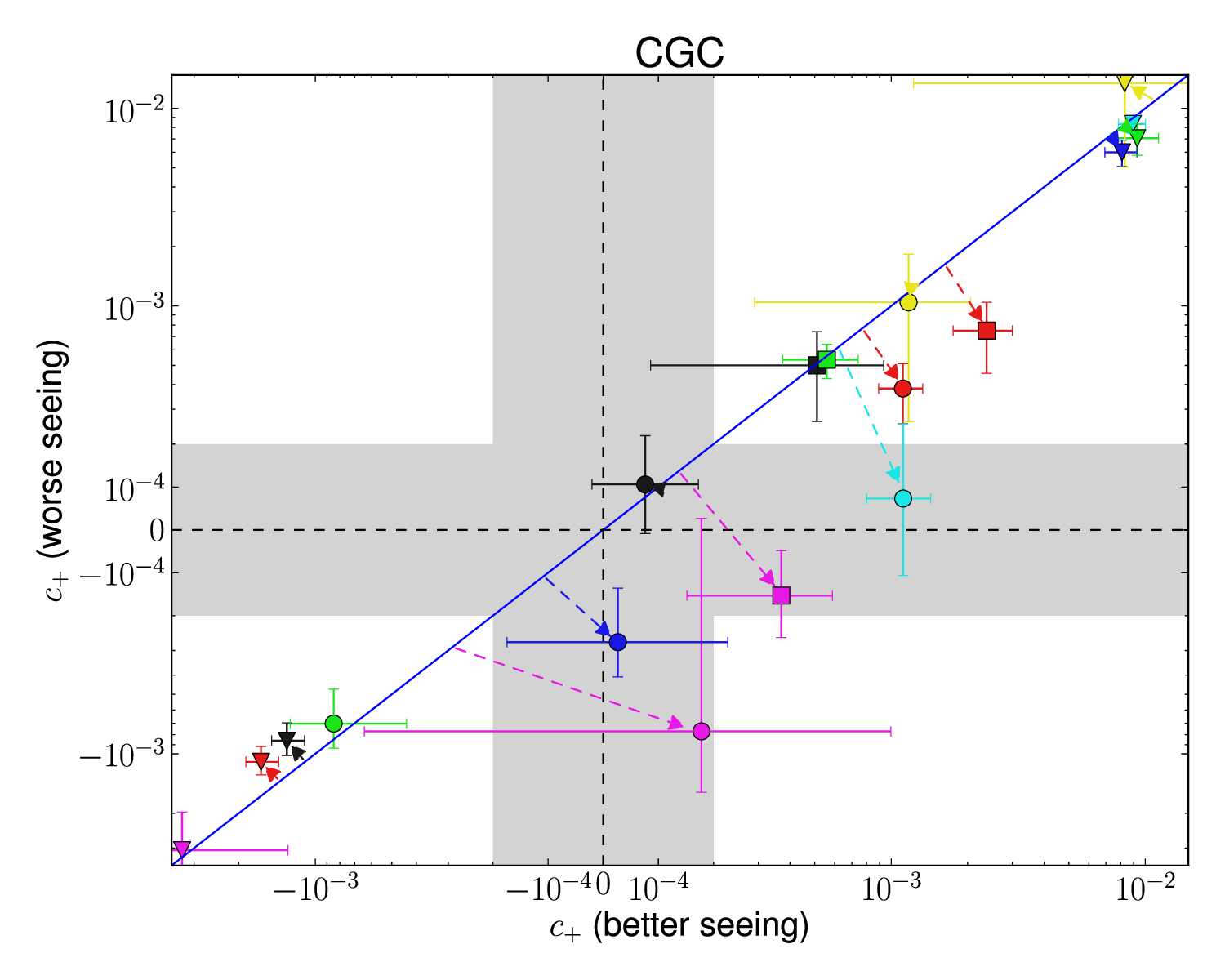}
\caption{\label{fig:cgc_seeing_split} For CGC, we show how 
  $\langle m\rangle$ (top) and $c_+$ (bottom) change when we split
  the subfields in the branch into 50 per cent above and below the
  median atmospheric PSF FWHM. The axes
  are linear within the target regions for $m$ and $c$, and
  logarithmic outside them, with the target regions shaded in grey.
  The thick blue arrows point towards the direction of reducing shear
  systematics, i.e., towards the center.  In all panels, any points
  that represent a significant change in the plotted quantity for better
  or worse seeing subfields also have an arrow showing how the results
  have changed compared to using the whole field.  \newtext{The legend gives the $Q_{\rm c}$ value for the original submission from each team.}
}
\end{center}
\end{figure}

We begin in CGC, splitting into
samples with better or worse atmospheric PSF FWHM (seeing).  The
results are in Fig.~\ref{fig:cgc_seeing_split}, in which the top panel
compares the $\langle m\rangle$ values for the 
better and worse seeing half of the subfields \newtext{(with numerical values
tabulated in Table~\ref{tab:mc_psf_deltas}).}
\newtext{The teams for which $\langle m\rangle$ differs for better vs.\ worse seeing
have a more negative (positive) calibration bias for better (worse)
seeing. }

The bottom panel of Fig.~\ref{fig:cgc_seeing_split} shows that many
teams have consistent $c_+$ for better and worse seeing, with the
\newtext{rest having a more strongly positive $c_+$ for the {\em better} seeing
subfields.}  The worse $c_+$ values for better
seeing subfields may come from the fact that the optical PSF (which is
often more elongated than the atmospheric component) dominates.  Indeed, the
correlation coefficient between PSF FWHM and $|g^{\rm PSF}|$ in
CGC (RGC) is -0.23 (-0.25), with a significance of $p=0.001$ ($3\times
10^{-4}$).  Thus, the worse seeing subfields have a
consistently rounder PSF, which can reduce additive
systematics.

The results for both the $\langle m\rangle$ and $c_+$ trends were
similar in RGC to what we have shown here for CGC, which is a point
that we will revisit in some of our later tests.

\subsubsection{Defocus}

\begin{figure}
\begin{center}
\includegraphics[width=0.9\columnwidth,angle=0]{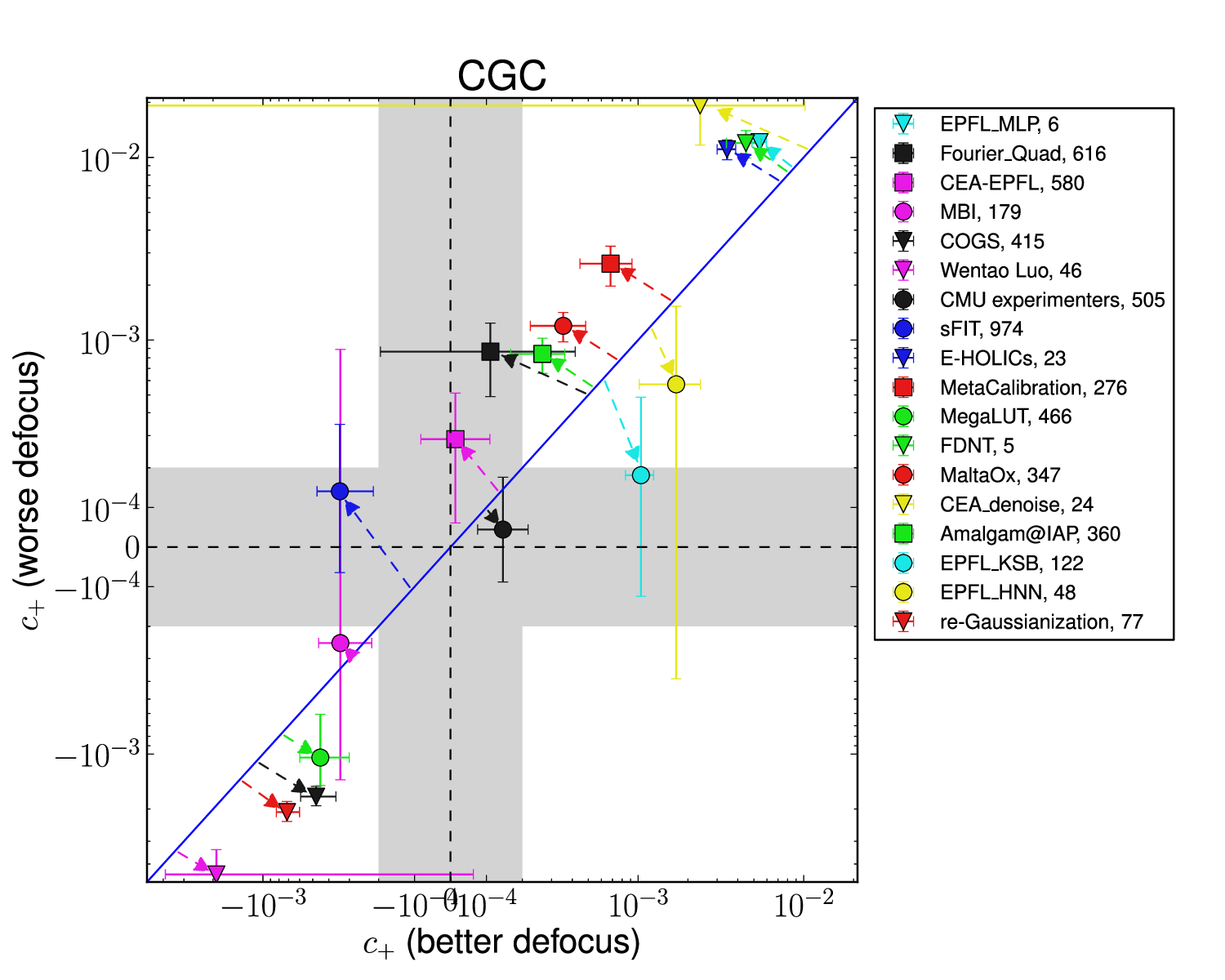}
\caption{\label{fig:cgc_defocus_g_split} For CGC, we compare additive
  biases 
  $c_+$ when splitting the subfields into those with defocus above or
  below the median. The axes
  are linear within the target region (shaded grey), and
  logarithmic outside them.
  The thick blue arrows point in the direction of reduced shear
  systematics.  Any points
  that represent a significant change compared to results for the
  entire branch have an arrow showing that change, as well.}
\end{center}
\end{figure}

In Fig.~\ref{fig:cgc_defocus_g_split}, we show how $c_+$ in CGC
\newtext{changes when we split at the median absolute value of defocus (with
results tabulated in Table~\ref{tab:mc_psf_deltas}).  The
results for many methods  exhibit}
a more strongly nonzero $c_+$ for stronger defocus.  It is not surprising that
additive systematic errors are worse when out of focus, because
defocus amplifies the effect of other aberrations like coma and
astigmatism on the PSF \citep{2011PASP..123..812S}, giving a noticably
more elliptical PSF.  
Appendix~\ref{app:psf} shows 
that we  allowed a relatively wide range of defocus values in the
ground branches, which explains why its effects are noticeable despite the fact that the atmospheric PSF is normally 
thought of as being dominant.  

The multiplicative biases $m_+$ and $m_\times$ (not shown) do not typically change when splitting by defocus, except for 
sFIT and MetaCalibration, with smaller 
changes for MBI and Amalgam@IAP.  For MBI, the
representation of the PSF as the sum of three Gaussians may be the 
limiting factor in describing out-of-focus PSFs.  For sFIT, the problem may arise from the use of simple PSFs
(rather than a range of complex PSF with varying defocus)
for the simulations used to calibrate the shears.  Explicitly deriving
calibrations for different PSFs may ameliorate this
problem.  A similar issue is likely at play for MetaCalibration, which
derived an average shear response using all subfields, 
rather than one for each PSF.
It is unclear why the calibration changes with defocus for
Amalgam@IAP, \newtext{but it may be because of difficulties in finding a
well-defined maximum likelihood for many galaxies in the more strongly
defocused cases.}

In space simulations (CSC), splitting by
defocus had qualitatively similar effects on shear
systematics as for CGC. 
However, the shifts are smaller in magnitude for space
simulations, likely because the range of defocus is much
smaller for space simulations than for ground ones (see
App.~\ref{app:psf}). 

Our findings are similar in the realistic galaxy experiment, 
suggesting that the dependence of shear systematics on defocus is
independent of realistic galaxy morphology.

\subsubsection{PSF ellipticity}

When splitting the subfields by $|g_{\rm PSF}|$, the results are
consistent with those of Sec.~\ref{subsec:linmodel}, where additive
systematics were shown to scale linearly with the PSF
ellipticity.

\subsection{Effective noise level of estimated shears}\label{subsec:noise}

\begin{figure}
\begin{center}
\includegraphics[width=0.99\columnwidth,angle=0]{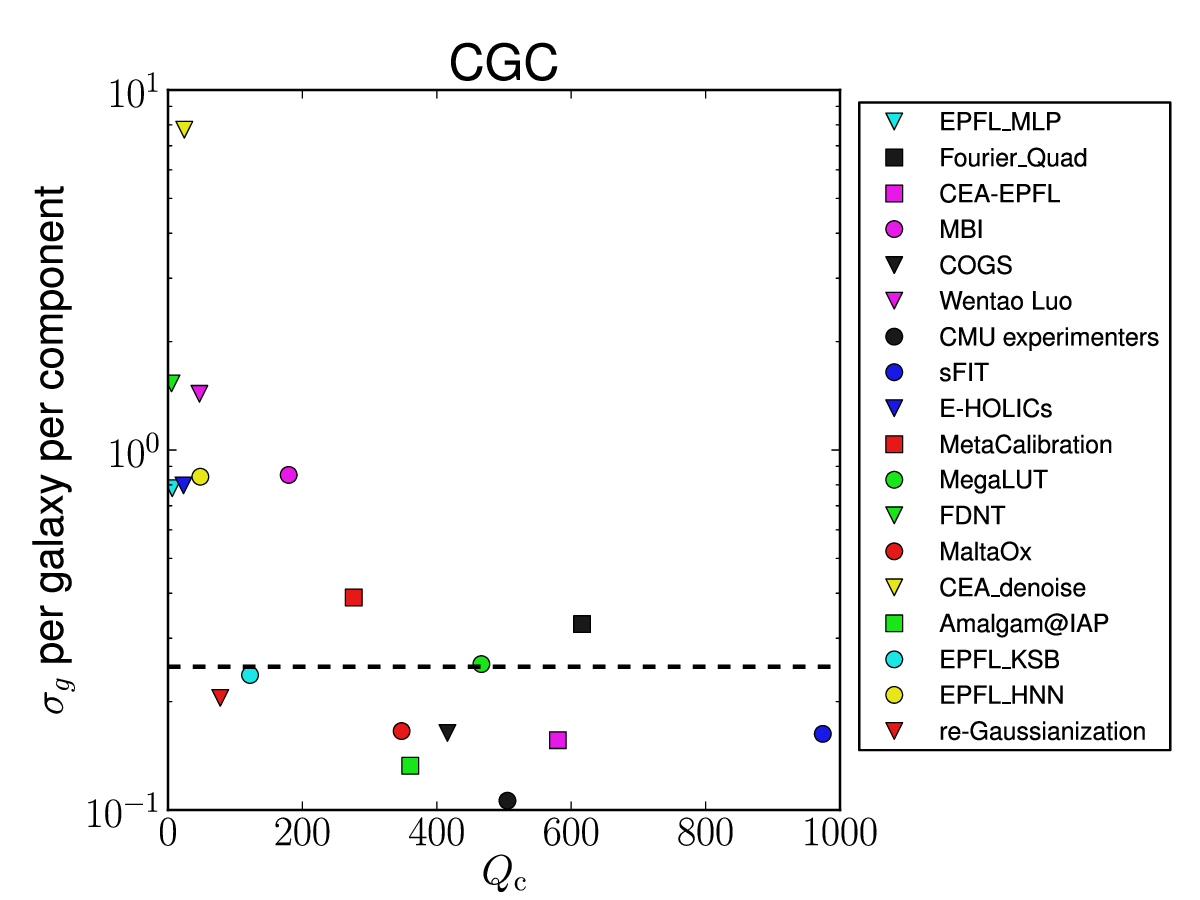}
\caption{\label{fig:sigma_g} Scatter in the estimated shears (per galaxy and per component)
  vs.\ $Q_{\rm c}$ for each method in CGC.  The horizontal line
  indicates a typical level of shape noise in realistic galaxy
  samples.}
\end{center}
\end{figure}

Here we explore the effective noise level of the estimated shears.
In principle, galaxy shapes were arranged 
in a way that cancels out shape noise, so that the dominant source of
error in the
estimated shears is measurement error due
to pixel noise.  However, the shape
noise cancellation is imperfect at low $S/N$, so that the submitted
shears include some shape noise as well. 
Fig.~\ref{fig:sigma_g} shows the per-galaxy and per-component scatter
 ($\sigma_g$) in the
estimated shears for CGC, estimated from fitting the model of
Eq.~\eqref{eq:linbias} \newtext{and finding the scatter in the shear estimates
for the subfields (then dividing by $\sqrt{10^4}$ to get a
per-galaxy value).  This scatter thus includes both the 
measurement error and any residual shape noise due to noise in the
weights, which can be seen as an additional manifestation of
susceptibility to pixel noise. 
There is a weak relationship between $\sigma_g$ and $Q_{\rm c}$,} 
with all methods that have $Q_{\rm c}\gtrsim 300$ having $0.1\lesssim
\sigma_g\lesssim 0.25$.  Methods with lower $Q_{\rm c}$
scores have higher scatter by as much as a factor of 40; the
exceptions to this rule are re-Gaussianization and EPFL\_KSB, which
notably are fairly simple moments-based methods.  In a few
cases, outliers are an issue, but even with 
$5\sigma$ clipping, the trend at low $Q_{\rm c}$ is quite evident.
This figure  for RGC looks very similar.   For the space
simulations, the effective per-galaxy $S/N$ was slightly higher, 
reducing the $\sigma_g$ values slightly, the overall trend is the
same.

The straightforward interpretation of
these results is that for methods with $Q_{\rm c} \gtrsim 300$, the
per-object measurement error is typically subdominant to 
shape noise, whereas some methods with lower $Q_{\rm c}$ 
allow significant leakage of pixel noise into the estimated shears.

\subsection{Catalog-level tests}\label{subsec:cattests}

For several teams, we carried out catalog-level tests that
involve using subsets of the galaxies.  For example, we split the
galaxies into subsamples with $S/N$ above and below the median; and
\newtext{likewise for resolution factor defined as in
\cite{2003MNRAS.343..459H} using the adaptive second moments, and
\sersic\ index\footnote{For galaxies that were represented in
  GREAT3 as a two-component model, a single-component \sersic\ 
  was used for this split.} $n$.  These splits} 
use the true (not estimated) values of these parameters, to preserve
shape noise cancellation.  The methods used for 
this test are CEA-EPFL, MegaLUT, Fourier\_Quad, re-Gaussianization, and sFIT,
which include a range of shear estimation methods.  
For Fourier\_Quad, we 
re-estimated ensemble shears for the galaxy subsets as
in Eq.~\eqref{shear_measure2}.

In general, biases such as noise bias depend on both the flux-based
$S/N$ and the resolution.  Thus, a split by a single galaxy property
may not isolate a particular bias.  Instead, these splits are a way to
estimate how much the shear systematics might change for a
particular method when dividing the galaxy sample in a way that
changes the mean $S/N$, resolution, or \sersic\ $n$.

\begin{figure*}
\begin{center}
\includegraphics[width=0.9\columnwidth,angle=0]{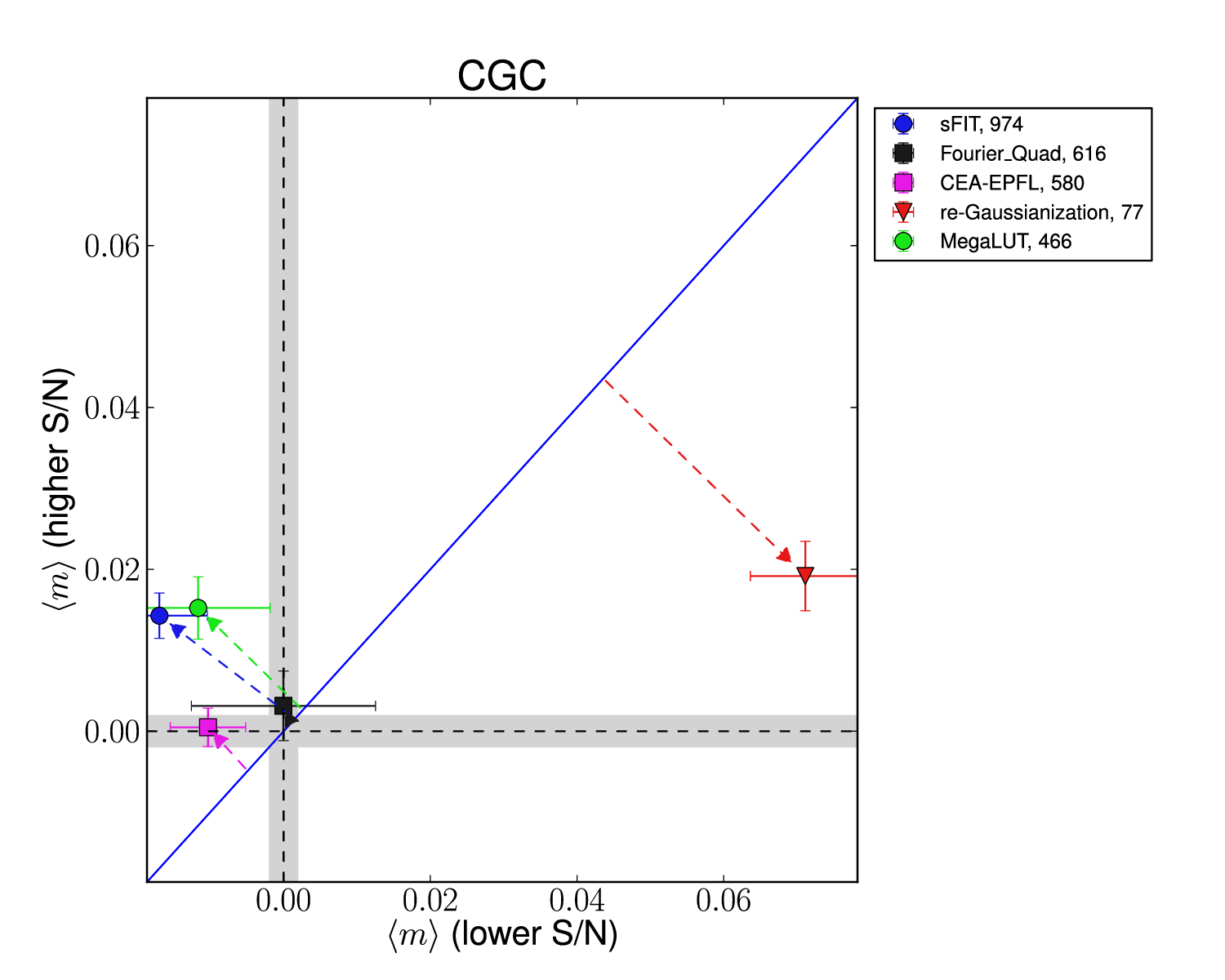}
\includegraphics[width=0.9\columnwidth,angle=0]{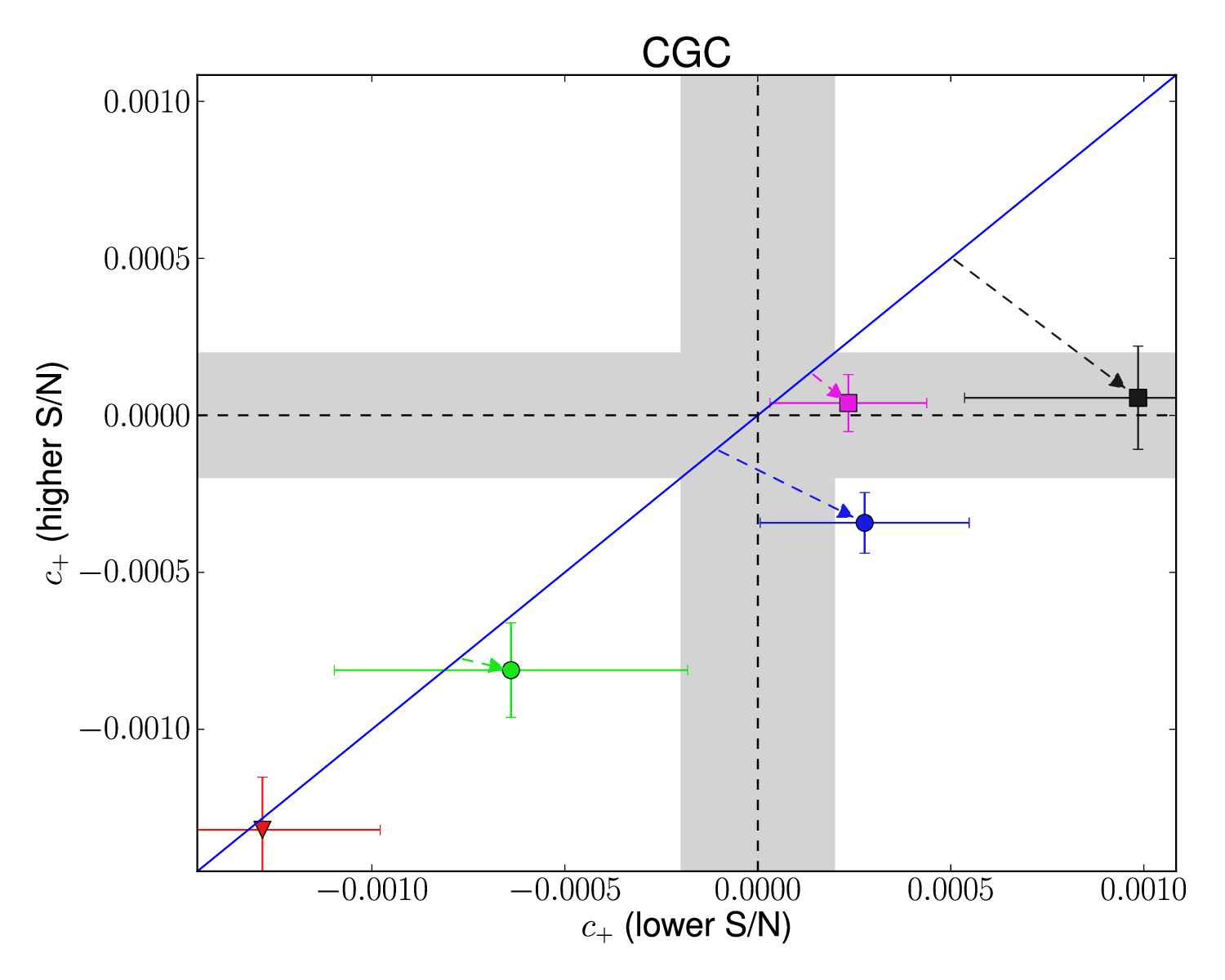}
\includegraphics[width=0.9\columnwidth,angle=0]{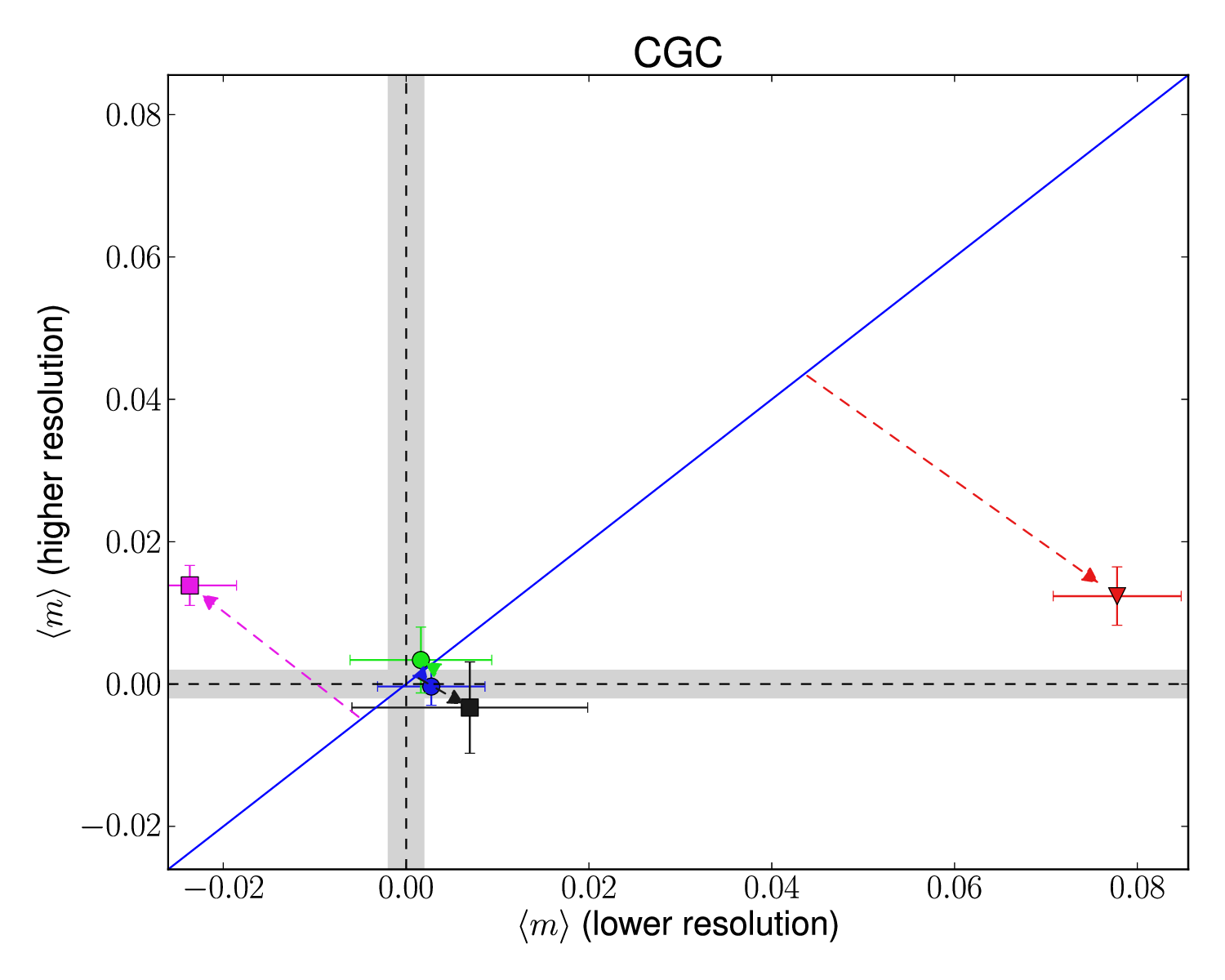}
\includegraphics[width=0.9\columnwidth,angle=0]{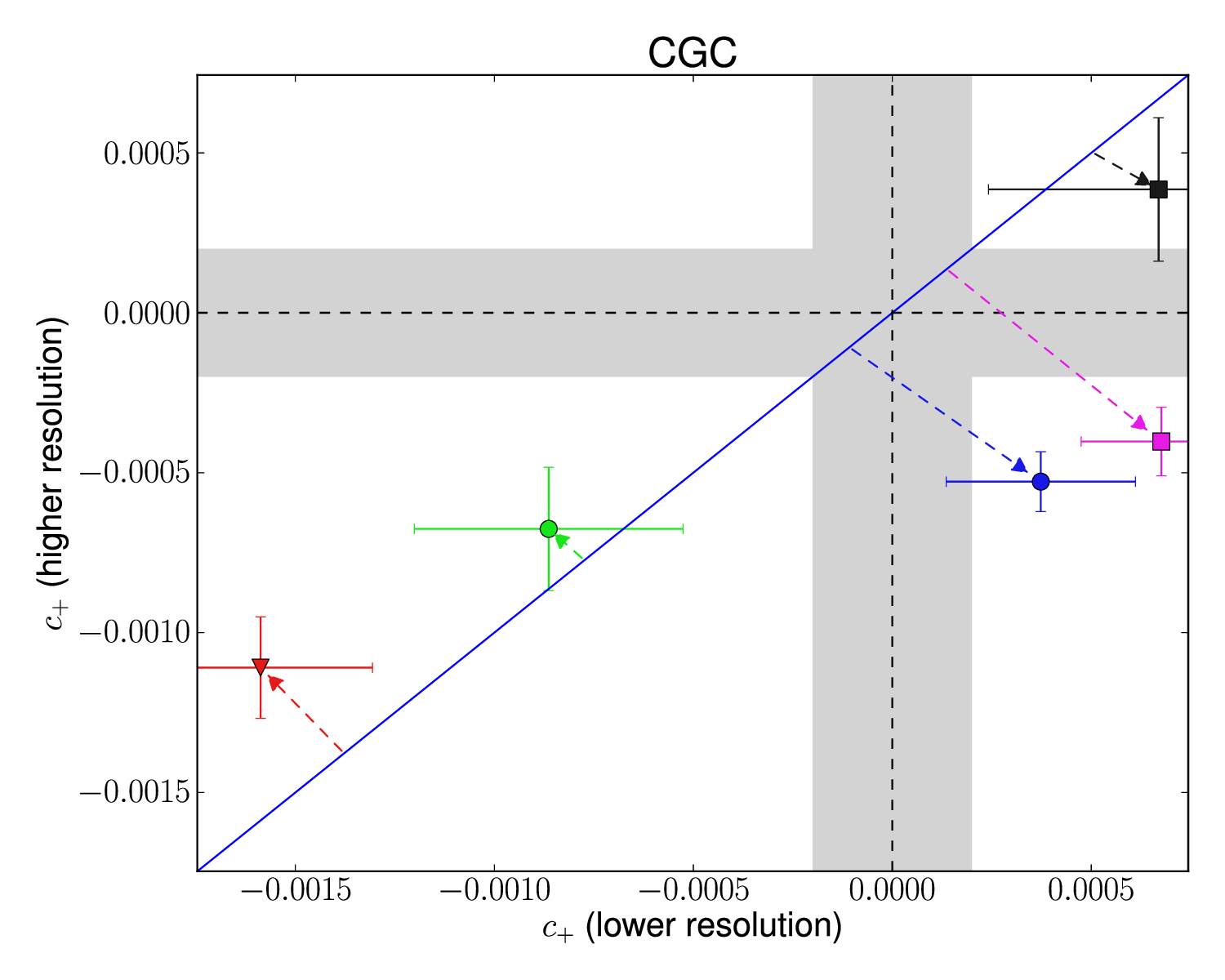}
\includegraphics[width=0.9\columnwidth,angle=0]{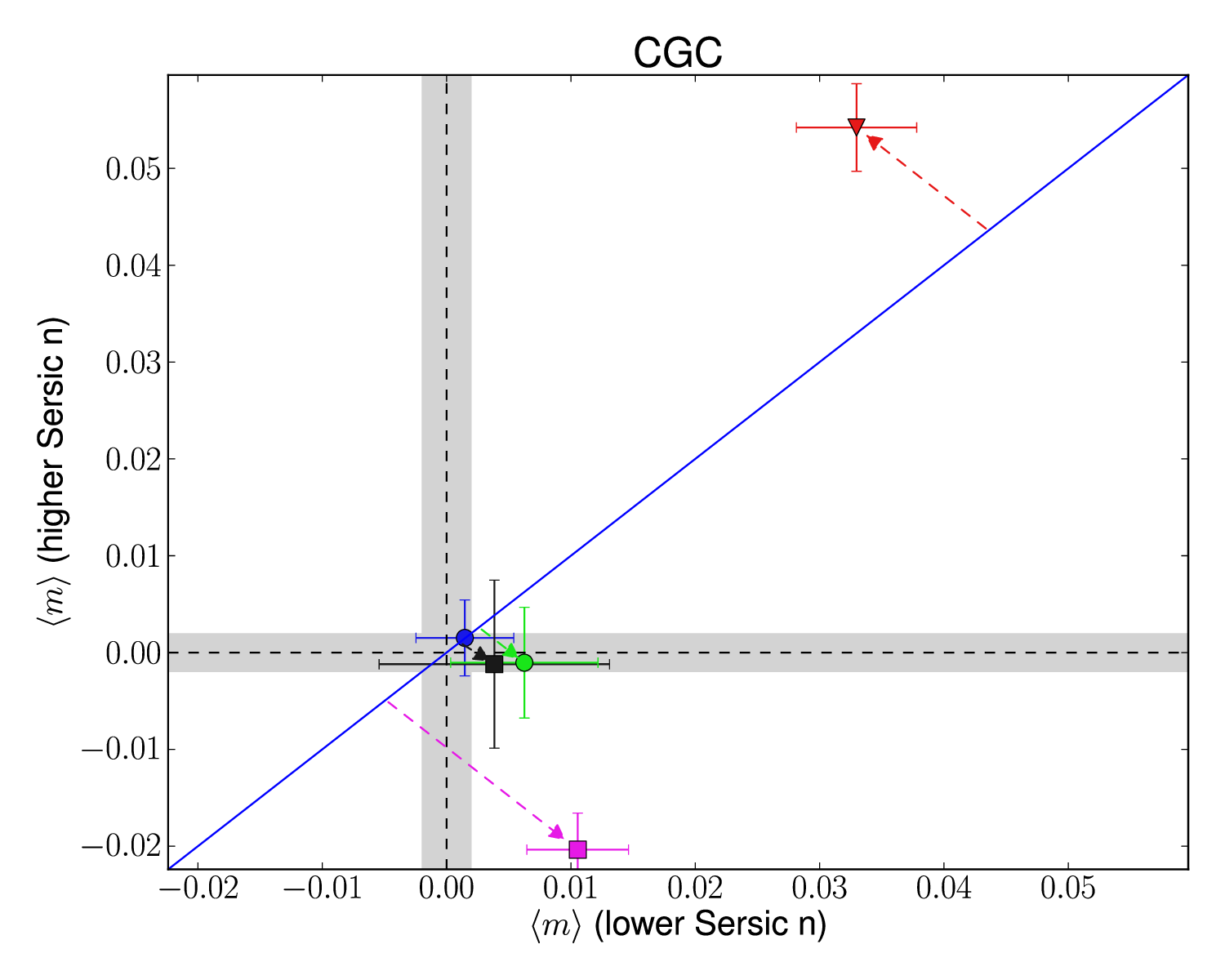}
\includegraphics[width=0.9\columnwidth,angle=0]{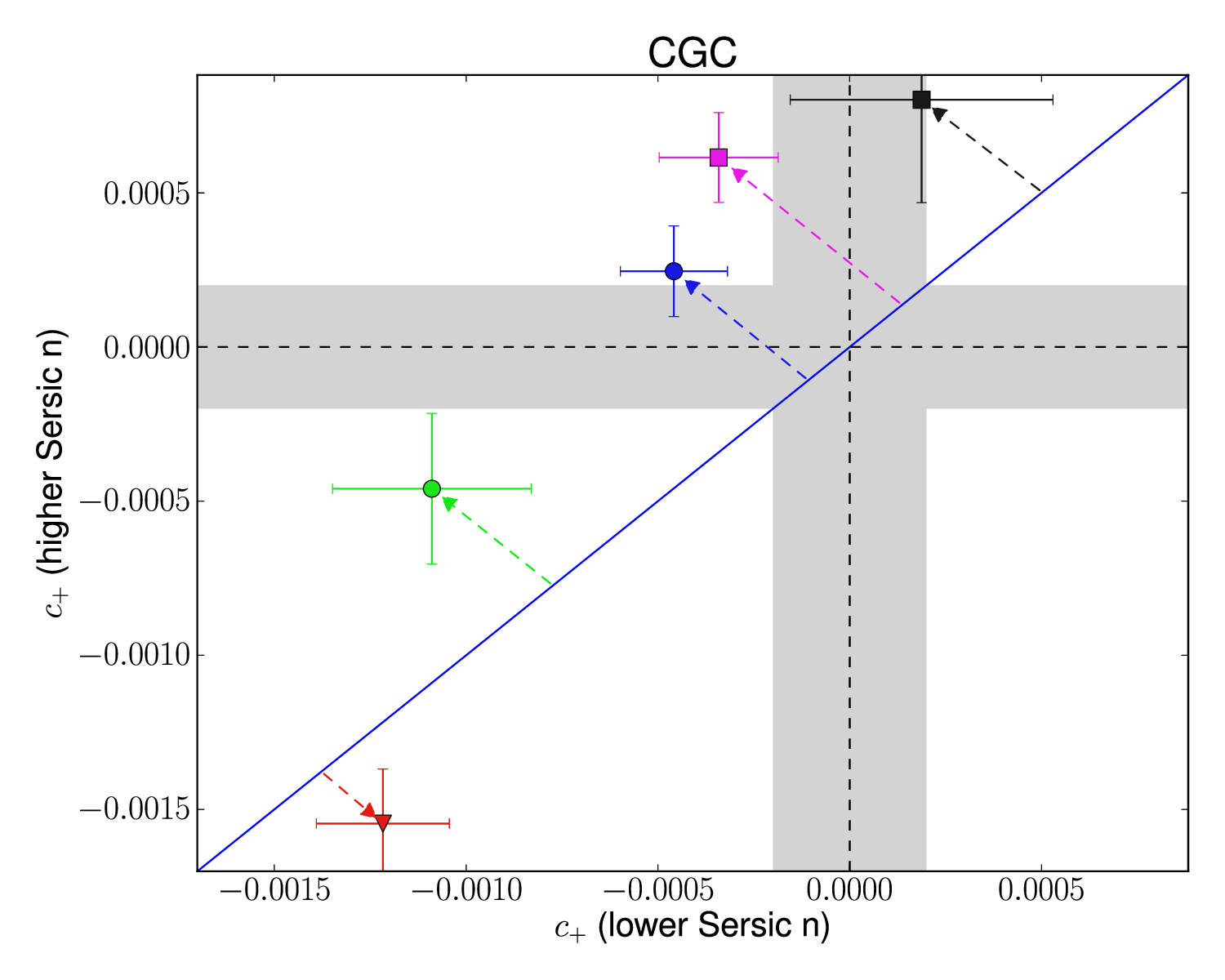}
\caption{\label{fig:cat_split} For CGC, we compare the $\langle
  m\rangle$ (left) and $c_+$ (right) values that we get by splitting
  the galaxies at the median value of $S/N$ (top), resolution
  (middle), and \sersic\ $n$.  The legend indicates the team and the
  $Q_{\rm c}$ value for the original submission using all galaxies.
  Arrows on each plot are shown for those teams for which the results
  for the subsamples differ from the overall results by more than 10
  per cent, and are drawn from the overall value to the results for
  the subsamples.}
\end{center}
\end{figure*}

Fig.~\ref{fig:cat_split} shows the results for $\langle
m\rangle$ (left) and $c_+$ (right) after dividing the galaxy
sample in CGC in these three ways.  In each case, we plot the results
for subsamples against each other, so a method that is robust to
changes in these quantities would be on the $1:1$ line. 
Methods that are not on that line must by definition move either to
the upper left or lower right.  We consider each method in turn.

The $\langle m\rangle$ and $c_+$ results for CEA-EPFL show only a mild
dependence on $S/N$, but a much greater dependence on resolution and
on \sersic\ n.  MegaLUT has less statistically significant trends,
with the most clear ones being the change $\langle m\rangle$ with
$S/N$ and the change in $c_+$ with \sersic\ $n$.  The multiplicative
bias $\langle m\rangle$ for Fourier\_Quad is quite robust to splitting
by any of the three parameters, but $c_+$ shows significant changes
for $S/N$ splits, with the change for \sersic\ $n$ being less
significant.  re-Gaussianization exhibits significant dependence on all
of $S/N$, resolution, and \sersic\ $n$, qualitatively
consistent\footnote{The magnitude of the trends is not consistent, but
  this could be because of the ways in which the example script used
  for this test differs from a science-quality measurement; see
  Appendix~\ref{app:scripts}.} with the findings in
\cite{2012MNRAS.420.1518M}, but little change in $c_+$.  Finally, for
sFIT, both $\langle m\rangle$ and $c_+$ change when splitting by all
three parameters, though the changes with $S/N$ are marginal in
significance.  Given that this team explicitly derived calibration factors to
remove additive and multiplicative biases from the entire population
(not as a function of galaxy properties), these trends are not 
surprising.  There is no reason to expect the calibration factors to be
valid for subsamples.  This exercise merely emphasizes the
necessity of rederiving them when using subsamples, or even
changing weighting schemes.

If we check in RGC whether the changes in
these methods are consistent across control and realistic galaxy
experiments, then we cannot do the comparison for sFIT due to a lack
of catalogs for that submission.  However, of the remaining methods
used, only the re-Gaussianization results when splitting the galaxy
sample are the same in CGC and RGC, which is interesting given the
significant model bias due to realistic galaxy morphology seen for
this method in Sec.~\ref{subsec:overallmc}.  The results for CEA-EPFL
when splitting by resolution and \sersic\ $n$ are the same in RGC as
in CGC, but the change in $\langle m\rangle$ when splitting by $S/N$
has the opposite sign as in CGC.  The MegaLUT method shows much
stronger trends in both $\langle m\rangle$ and $c_+$ in RGC when
splitting by all three parameters than in CGC.  Finally, for
Fourier\_Quad, the sign of the $c_+$ changes when splitting by
resolution and \sersic\ $n$ is reversed in RGC compared to CGC.

In CSC, we can check how the use of space simulations changes the
results when dividing the galaxy sample (for all but
re-Gaussianization, which only has results on ground branches).  The
CEA-EPFL and sFIT team results show different signs and/or magnitudes of
changes in shear systematics when splitting by galaxy properties in
CSC vs.\ in CGC.  For MegaLUT, the value of $c_+$ changes when
splitting by $S/N$ and resolution more significantly in CSC than in
CGC.  For Fourier\_Quad the difference in $c_+$ between subsamples in
resolution and \sersic\ $n$ changes in sign in CSC compared to CGC.
These findings suggest that essentially all teams considered here have
trends in shear systematics with galaxy properties that are different
in space vs.\ in ground data.

\subsection{\newtext{Comparison with GREAT08 and GREAT10 results}}

\newtext{In this section we compare quantitatively with the results from
GREAT08 \citep{2010MNRAS.405.2044B} and the GREAT10 galaxy challenge
\citep{2012MNRAS.423.3163K}, to the limited extent that is possible
given the different challenge designs and the lack of error analysis
in previous challenge results\footnote{Given the previous challenge data
  volumes and SNR levels, the uncertainties cannot be significantly
  smaller than the uncertainties in GREAT3.}.}

\newtext{For GREAT08, the fairest comparison is between GREAT3 CGC and GREAT08
RealNoise\_Blind bulge $+$ disk galaxy results.  We cannot compare $Q$
values since they are defined in a different way, so instead we compare
$\langle m\rangle$ and $c_+$, bearing in mind that even this
comparison is complicated by the broader, more realistic distributions
of galaxy properties in GREAT3.  In the left column, middle row of
figure C3 in \cite{2010MNRAS.405.2044B}, the number of methods that
have $|\langle m\rangle|<0.05$, $0.02$, and $0.005$ is $7$, $2$, and $0$.  In GREAT3,
these numbers are 12, 10, and 6, using only the fair
comparison sample results used throughout Sec.~\ref{sec:overall}
rather than the best submission per team for this branch.  We are also ignoring
the uncertainty on these $\langle m\rangle$ values for consistency
with how we did the calculation for GREAT08 given its lack of error estimates.  The upper
left panel of figure C4 (b$+$d) in \cite{2010MNRAS.405.2044B} suggests
that $8$ ($4$) methods have $|c_+|<1\times 10^{-3}$ ($2\times
10^{-4}$), whereas in GREAT3 CGC these numbers are 9 (3).
The latter comparison is particularly complicated by different choices for
the PSF ellipticity distribution in these challenges, since we showed
in Sec.~\ref{subsec:linmodel} that for essentially all methods, $c_+$ is linearly
proportional to PSF ellipticity.}

\newtext{For GREAT10, the simplest comparison is with the inferred $m$ and $c$
values in table 3 of \cite{2012MNRAS.423.3163K}, again ignoring noise
due to the fact that no uncertainties are quoted.  However, two of the
better-performing submissions in that table have no $m$ or $c$
estimates, since they used a power spectrum analysis.  In the
absence of more information we will include them in the best category
that we consider, $|\langle m\rangle|<0.005$ and $c_+<2\times
10^{-4}$.  Given this choice, the number of methods in GREAT10 with
$|\langle m\rangle|<0.05$, $0.02$, and $0.005$ is $7$, $5$, and $2$,
which should again be compared with 12, 10, and 6 in GREAT3.  All 12
methods in table 3 of \cite{2012MNRAS.423.3163K} had $c$ values within
$2\times 10^{-4}$, with the range of PSF ellipticities being different
from that in GREAT3, but not to a very large extent.}

\newtext{The GREAT3 results show that significant progress has been made in
controlling multiplicative biases since GREAT08 and GREAT10, with the
situation for additive biases being less clear.  However, additive
biases are easier to identify in real data (for example, using 
star-galaxy cross-correlations), so this situation fairly reflects the
community's focus on the more pernicious multiplicative biases. Given that, as
discussed in Appendix~\ref{app:sn}, the GREAT3 simulations have a
realistic $S/N$ distribution with an effective cutoff of $12$, this
improvement in control of multiplicative biases is a significant
achievement reflecting tremendous progress in the weak lensing
community as a whole.}

\section{Lessons learned about shear estimation}\label{sec:lessons}

  In this section, we discuss lessons learned about shear
  estimation based on the analyses in Sections~\ref{sec:specific}
  and~\ref{sec:overall}.  Our focus is on results that
  are more general than just a single method; conclusions for
  individual methods can be drawn from earlier plots and
  discussion.

\subsection{What do we learn about shear estimation in general?}

Many teams that participated in GREAT3 used
model-fitting methods, which must make choices about
which pixels to use for the fitting.  The results in
Sec.~\ref{subsec:focus:gfit} highlight the importance of 
truncation bias due to use of overly-small modeling
windows.  Truncation bias can potentially be several per cent
(multiplicative bias), and also is a source of additive
\newtext{bias; its magnitude makes it relevant for present-day surveys, and
could potentially be worse in the case of blends (which might lead to
the choice of a more restricted modeling window). }  
These model-fitting methods 
make choices about which models to use, with two popular options being a
single \sersic\ model (Amalgam@IAP, sFIT, MBI) and a sum of a bulge and
disk \sersic\ models with fixed $n$ (COGS, \code{gfit}). The
good performance of these methods suggests that use of
\sersic\ profiles can reduce model bias that is observed with, e.g.,
shapelets or other models that do not describe galaxy light profiles
as well as \sersic\ profiles.  

Several methods of calibration were successful for 
model-fitting methods: 
external simulations for which the inputs were iteratively updated until
the output galaxy properties match those in the GREAT3 data (sFIT),
derivation of calibration corrections from a deep subset of the same
data (COGS), and addition of a penalty term to the $\chi^2$ to reduce
noise bias (Amalgam@IAP).  External simulations are always limited by their realism,
though use of iterative methods seems to be helpful.
Calibration corrections from deep data do not, in principle, 
require external validation.  Addition of a
penalty term to the $\chi^2$ does require external simulations
to check that the penalty term really 
removes the noise bias.

Our results in Sec.~\ref{subsec:linmodel} confirm the applicability of the
linear model for shear calibration biases in the $|g|\lesssim 0.05$
regime for all methods that participated in
GREAT3.  Several methods showed
tendencies for multiplicative biases defined in the pixel coordinate
system to differ between the component along the pixel axes and along
their diagonals, similar to what was seen in
e.g. \cite{2007MNRAS.376...13M}. In all cases, the additive biases
$c_+$ were linearly proportional to the 
amplitude of the PSF ellipticity (of order 0.1 per cent of the PSF
ellipticity for the best methods, and more typically 1--5 per cent).
It is possible that some biases in real surveys but
not GREAT3 would violate this pattern (e.g., selection biases
that depend on the PSF anisotropy).

The results for many methods show a dependence on PSF properties like
the FWHM, defocus, and ellipticity.  In some cases, the results seem
to have been calibrated to work on average, so that they are worse for
better or worse quality data than for the
challenge overall.
Defocus tends to result primarily in additive (not multiplicative)
systematics.  Some methods are particularly
sensitive to outliers in defocus, which results in more 
complicated-looking PSFs; it is difficult to assess to what extent
that sensitivity is intrinsic to the PSF correction method (because
those PSFs violate one of its assumptions) vs.\ arising from 
how the PSFs are modeled (because of limitations of 
the PSF modeling software).  \newtext{Some future surveys will have additional
diagnostic data regarding PSFs; these results suggest that it may be
helpful to incorporate this information in the PSF modeling and shear
estimation process.}

When splitting galaxy samples
by $S/N$, resolution, or \sersic\ $n$, we observe statistically
significant trends for the five methods that were considered; these
trends are sensitive to real galaxy morphology (control
vs.\ realistic galaxy experiment) and the type of data (space vs.\
ground).  In contrast, the variation in shear systematic errors due to data
properties like atmospheric PSF FWHM or defocus was fairly robust to
realistic galaxy morphology.

Comparing ground vs.\ space data, additive systematics seem
to be more important for the latter.  In space branches, several 
teams saw their $c_+$ become significantly more positive, which 
contributed towards there being almost entirely positive $c_+$
submissions in space branches. However, not all the teams with
negative $c_+$ in the ground branches submitted to the space branches.

Finally, the effective noise level of the shear estimates (measurement
error due to pixel noise) showed a weak inverse relationship with $Q$.
For the majority of the methods (especially those with $Q_{\rm c}\gtrsim 200$), the values of
$\sigma_g$ per component were fairly consistent across
methods.  This confirms the general tendency to select shear
estimation methods based on their multiplicative and
additive biases, rather than separately considering their measurement
errors.

\subsection{The impact of realistic galaxy morphology}

Many methods, including some that performed extremely well, show a
small but statistically significant change in model bias due to
realistic galaxy morphology, with order of magnitude 
1 per cent.  Realistic galaxy morphology can also result in additive
systematics.  Our findings for the order of magnitude of this effect
for multiple methods is consistent with the finding for the
\code{im3shape} software \citep{2014MNRAS.441.2528K}.
For some methods, realistic galaxy morphology was more important for
space branches than for ground (e.g., the sFIT team had to 
explicitly calibrate out the bias due to realistic galaxy morphology
only for space). 

One key limitation in lessons learned about realistic
galaxy morphology in GREAT3 is
that, since its impact is relatively small (typically detected at $\sim
3\sigma$), it is hard to distinguish between space and ground results or
clearly identify trends with other data properties.
However, this in itself is good news for future surveys, since it
provides an indication that model bias due to realistic galaxy
morphology may rank behind other effects, such as
noise bias, in terms of its direct impact on shear measurements.

\newtext{In real data with a substantially deeper source population than is
represented in the sample of galaxies from COSMOS used as the basis
for the GREAT3 simulations, these results will have to be revisited 
due to the larger fraction of irregular galaxies at higher redshift
\citep[e.g.,][]{2005ApJ...625..621B}.}

\section{Conclusions}\label{sec:conclusions}

We have presented results for the control and realistic galaxy
experiments of the GREAT3 challenge, the goal of which was to test
ensemble shear estimation given a galaxy population with a realistic
distribution of size, $S/N$, ellipticity, and morphology, and with a
(known) fairly complicated PSF.
A key result is that, within the ability of the simulations to determine
systematics at this level and bearing in mind that some
effects are not included in them,
a range of methods can now carry out shear estimation with
systematics errors around the level required by Stage IV dark energy
surveys.

We have explored how the
results for each team depend on the galaxy and PSF 
properties; and explored the impact of realistic galaxy
morphology by comparing the control and realistic galaxy branches.  Our conclusions on these points are summarized in
Sec.~\ref{sec:lessons}, with the main one being that shear systematic
errors due to realistic galaxy morphology are, for those methods for
which we have a clear detection, typically of order $\sim 1$ per cent.
 While significant enough that future surveys must take these effects
 into account, this source of model bias error is subdominant when
 compared to the level of noise bias expected for similar galaxy
 populations to those in GREAT3
 \citep[e.g.,][]{2012MNRAS.427.2711K,2012MNRAS.424.2757M,2012MNRAS.425.1951R}. 
In Paper II, we will use the other
branches of the challenge to explore whether these overall 
results from Sec.~\ref{sec:lessons} carry over to the case where the PSF is
not known.

Treating the participants as a fair subset of the community,
it seems that model-fitting methods now dominate the field in
both popularity and (broadly) performance.  Some differences between
methods may relate to implementation details rather
than true issues with a method.  Unlike a decade ago, moments
methods are now a minority. However there are some highly interesting
alternative methods, for which we have seen the introduction and/or
evident maturity in GREAT3 (some based on \cite{2014MNRAS.438.1880B};
MetaCalibration; self-calibration for \code{LensFit} as carried out by
the MaltaOx team; hierarchical inference as done by the MBI team;
machine learning based methods like MegaLUT; and Fourier\_Quad),
adding variety and quality to the field.  This includes the
introduction of some teams that just infer ensemble shears (MBI, BAMPenn, ess,
Fourier\_Quad) rather than per-object shears; however, a demonstration of these methods on
variable shear data will be crucial for their more general acceptance.

Choices related to calibration of shears were quite varied, 
with some teams that aim for an unbiased
measurement (e.g. BAMPenn, ess, MBI) and others that apply 
calibrations in a variety of ways.  Aside from external
simulation-based calibrations, which are subject to the 
limitation that the calibrations are only as good as the simulations,
a few more sophisticated options were tried.
These include iterative external simulations that get updated until
the outputs match those in the dataset that is being analyzed (sFIT),
analysis of a deep subset of the same data (COGS), and
self-calibration using manipulations of the images themselves
(MaltaOx, MetaCalibration).  These alternatives appear promising,
and avoid some of the objections to the most basic brute-force
calibration.  The utility of the deeper data to
several teams, either for calibrations or deriving galaxy property
distributions, suggests that future surveys may find it useful to have
a deeper subsurvey, as indeed many already intend to do.
Several teams used self-calibration methods
(MetaCalibration and MaltaOx) and hierarchical-inference (MBI) methods 
that in principle could be used to remove the biases in many other
shear estimation methods.  These newer methods
were not among the very top performers, but did impressively well
for new implementations, so it will be interesting to
follow their future development. 

We also have a number of conclusions about
GREAT-type challenges based on the GREAT3 challenge
process.  Unfortunately, the variable shear simulations were 
less powerful than originally intended at detecting systematic
biases in the shear fields.  Despite our best efforts in attempting
to define a metric with a reasonably small variance, $Q_{\rm v}$ was
noisier than $Q_{\rm c}$, the constant-shear metric.
However, for the methods that submitted results to
constant and variable shear branches, the results were consistent with the
estimated shears having the same underlying biases (within the
errors), as we would expect.  Future challenges that want to determine
biases with variable shear fields may require 
substantially larger data volumes than in GREAT3.
Future challenges may also want to allow participants
to assign weights to downweight data that they do not want to use,
rather than requiring shear estimates for all fields.

After the end of the challenge, we found that use of a metric
based on systematics in the coordinate system defined by the PSF
anisotropy resulted in accidental preference for methods with
calibration biases in the coordinate system defined by the pixel frame
that were related as $m_1\approx -m_2$.  While this had little effect
on the challenge itself, it highlights the fact that a 
challenge with a public leaderboard including $Q$ values 
(even without any 
multiplicative and additive biases) cannot be considered truly blind.
Participants sometimes made choices based on
feedback from the leaderboard, which at times was useful in helping
them avoid completely futile pathways, but at times may have involved tuning
to low levels of noise rather than making real conclusions. Thus, if the goal
is a truly blind challenge (which helps 
evaluate existing methods rather than assisting the development of
new ones), then we recommend that future challenges consider some
change in the public leaderboard.  For example, the public
leaderboard could use a subset of the data, with the
real leaderboard that uses all the data being released
only after the end of the challenge.  An alternative would be to tell participants
a range in which their $Q$ values fall (e.g.,
$0<Q<200$, $200<Q<400$, and so on).  Both options would give participants a
basic idea of their results (allowing them to check, e.g., shear conventions and avoid
submitting junk by accident) while not encouraging them to potentially
tune to the noise.  

A final point for future challenges and even 
planning for future surveys relates to the importance of the
$S/N$ definition.  It is quite common to use galaxies above some $S/N$ limit,
but in GREAT3, we found that depending on the $S/N$ definition,
the effective $S/N$ can vary by nearly a factor of two.  For
example, as stated in the handbook, we initially set a $S/N>20$ limit
to ensure that most teams would
be able to compute shears for all galaxies, with shape noise
effectively canceled.  The disadvantage of this limit was that we
would not 
dig too deeply into the noise bias-dominated regime.
However, we found in practice (see Appendix~\ref{app:sn}) that our
$S/N$ estimator \newtext{was so optimal as to be completely unachievable in
practice, given that it assumes perfect knowledge of the light
profile. Our tests showed that the lower $S/N$ limit using  more}
practical estimators is around $12$.
On the positive side, this meant that the results have a more
realistic level of noise bias, but on the negative side, it meant that the simulations
were less powerful in constraining shear systematics.
This finding highlights the importance of how $S/N$ is defined both for future
challenges and for parameter forecasts and mission specifications
for future lensing surveys.

In conclusion, GREAT3 has led to substantial progress
in quantifying shear systematics for a wide range of methods,
including traditionally recognized effects like noise and model bias
due to mismatch between assumed and real galaxy light profiles 
in the control branch, but also newer effects like truncation bias and
model bias due to realistic morphology, the latter of which was enabled by the use of {\em
  HST} data for the simulations.  The results show that the field has
made significant advances in the years since the end of the GREAT10
challenge, \newtext{particularly in controlling multiplicative biases, and that community challenges can be beneficial by inspiring the creation or development of
new shear estimation methods.}  Within this field, there are both new and established
methods that are now capable of handling weak lensing data from upcoming Stage
III surveys, provided adequate care is taken over identified sources of
bias.   Although development will be needed in many areas, 
\newtext{the GREAT3 results provide new reasons to be optimistic about delivering reliably accurate shear estimates at Stage IV survey accuracy.}

\section*{Acknowledgments}

We thank Gary Bernstein and Mike Jarvis for providing helpful feedback
on this paper, Peter Freeman for providing guidance on the statistical interpretation of
results\referee{, and the anonymous referee for making suggestions that improved the presentation of
  results in the paper}.  
  We thank the PASCAL-2 network for its sponsorship of the challenge.
  This project was supported in part by NASA via the Strategic
  University Research Partnership (SURP) Program of the Jet Propulsion
  Laboratory, California Institute of Technology; and by the IST
  Programme of the European Community, under the PASCAL2 Network of
  Excellence, IST-2007-216886.  This article only reflects the
  authors' views. This work was supported in part by the National
  Science Foundation under Grant No. PHYS-1066293 and the hospitality
  of the Aspen Center for Physics.

RM was supported \referee{during the development of the GREAT3 challenge} 
in part by program HST-AR-12857.01-A, provided by
NASA through a grant from the Space Telescope Science Institute, which
is operated by the Association of Universities for Research in
Astronomy, Incorporated, under NASA contract NAS5-26555, and in part
through an Alfred P. Sloan Fellowship from the Sloan Foundation; \referee{her work on the final
  analysis of results was supported by the Department of Energy Early Career Award Program}.
BR, JZuntz, and TKacprzak acknowledge support from the European Research Council in the form of
a Starting Grant with number 240672. HM acknowledges support from Japan Society for the Promotion of Science (JSPS) Postdoctoral Fellowships for Research Abroad and JSPS Research Fellowships for Young Scientists. 
The Amalgam@IAP Team (AD, EB, RG) acknowledges the Agence Nationale de la
Recherche (ANR Grant ``AMALGAM'') and Centre National des Etudes
Spatiales (CNES) for financial support.  MT acknowledges support from
the Deutsche Forschungsgemeinschaft (DFG) grant Hi 1495/2-1. TKuntzer), MGentile, HYS, and FC
acknowledge support from the Swiss National Science Foundation (SNSF)
under grants CRSII2\_147678, 200020\_146813 and 200021\_146770.  Part
of the work carried out by the MBI team was performed under the
auspices of the U.S. Department of Energy at Lawrence Livermore
National Laboratory under contract number DE-AC52-07NA27344 and SLAC
National Accelerator Laboratory under contract number
DE-AC02-76SF00515.  HYS acknowledges the
support by a Marie Curie International Incoming Fellowship within the
$7^\text{th}$ European Community Framework Programme, and NSFC of
China under grants 11103011. JEM was supported by National Science
Foundation grant PHY-0969487. JZhang is supported by the national
science foundation of China (Grant No. 11273018, 11433001), and the
national basic research program of China (Grant No. 2013CB834900,
2015CB857001).  J-LS, MK, FS, and FMNM were supported by the
European Research Council grant SparseAstro (ERC-228261). EMH is
grateful to Christopher Hirata for insightful discussion and feedback
on the MetaCalibration idea.

Contributions: RM and BR were co-leaders of the challenge itself, and
coordinated the analysis presented in this paper.  In the rest of this
listing, people whose names are given as initials  or first
initial-last name (when initials are ambiguous) are co-authors on
the paper, and those who are not have their names listed in full.  JB,
Chihway Chang, FC, MGill, Mike Jarvis, HM, RN, JR, MS,
and JZuntz were members
of the GREAT3 Executive Committee, which helped to design the
simulations and run the challenge.  
The other co-authors were members
of teams that participated in the challenge:
\begin{itemize}
\item Amalgam@IAP: EB, AD, RG
\item BAMPenn: RA, Gary Bernstein, MM
\item EPFL\_gFIT: MGentile, FC
\item CEA-EPFL: MGentile, FS, MK, J-LS, FNMN, SP-H, FC
\item CEA\_denoise: MK
\item CMU\_experimenters: RM
\item COGS: BR, JZuntz, TKacprzak, Sarah Bridle
\item E-HOLICs: YO
\item EPFL\_HNN: GN, FC
\item EPFL\_KSB: HYS
\item EPFL\_MLP: GN
\item FDNT: RN
\item Fourier\_Quad: JZhang
\item HSC-LSST-HSM: JB, RM
\item MBI: DBard, DBoutigny, WAD, DWH, DL, PJM
  JEM, MDS
\item MaltaOx: LM, IFC, KZA
\item MegaLUT: TKuntzer, MT, FC
\item MetaCalibration: EMH, RM
\item Wentao\_Luo: WL
\item ess: ESS
\item sFIT: MJJ
\end{itemize}

\bibliographystyle{mn2e2}
\bibliography{../papers}

\appendix

\section{GREAT3 challenge details}\label{app:challenge}

In this appendix, we summarize some details of the GREAT3 challenge
that were not included in the handbook.

\subsection{Galaxy intrinsic ellipticity distribution} \label{app:intrinsic-pe}

The galaxy intrinsic ellipticity distribution, or $p(\varepsilon)$, is
important, since many methods make assumptions about or try to infer
it. \newtext{ We measure this distribution for the GREAT3 galaxy samples using parametric fits to COSMOS galaxies.}

The galaxy selection in each
subfield has three goals: first, it should 
roughly preserve the joint size, $S/N$,
morphology, and ellipticity distributions of real galaxy samples;
second, each subfield should have a similar $S/N$ cutoff (which
depends on the PSF as well as the pixel noise); and finally, the galaxies
should be sufficiently resolved that essentially all
methods can measure them.  In 
ground branches, where the PSF size varies substantially from
subfield to subfield, it is not obvious that the galaxy
population will have the same $p(\varepsilon)$ in each subfield after these cuts.

\begin{figure}
\begin{center}
\includegraphics[width=0.99\columnwidth,angle=0]{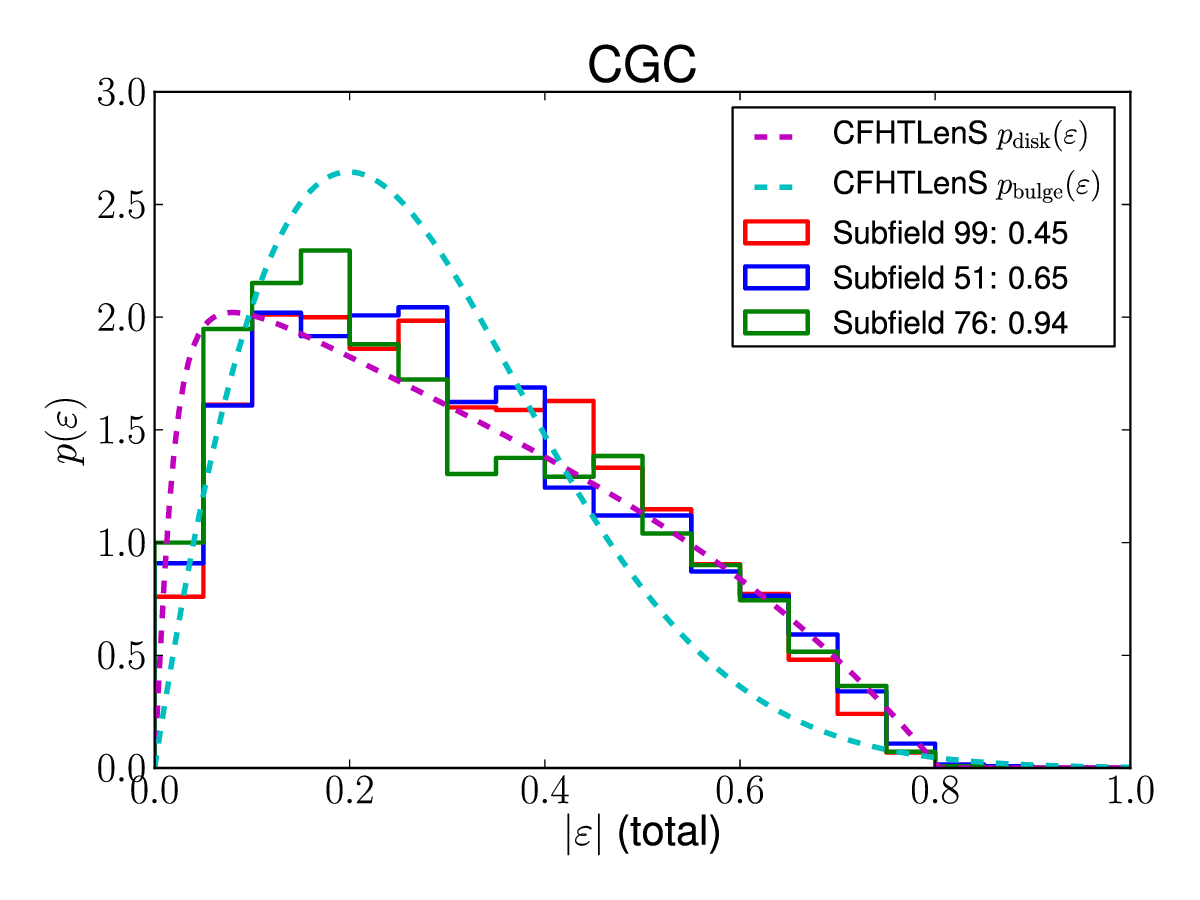}
\includegraphics[width=0.99\columnwidth,angle=0]{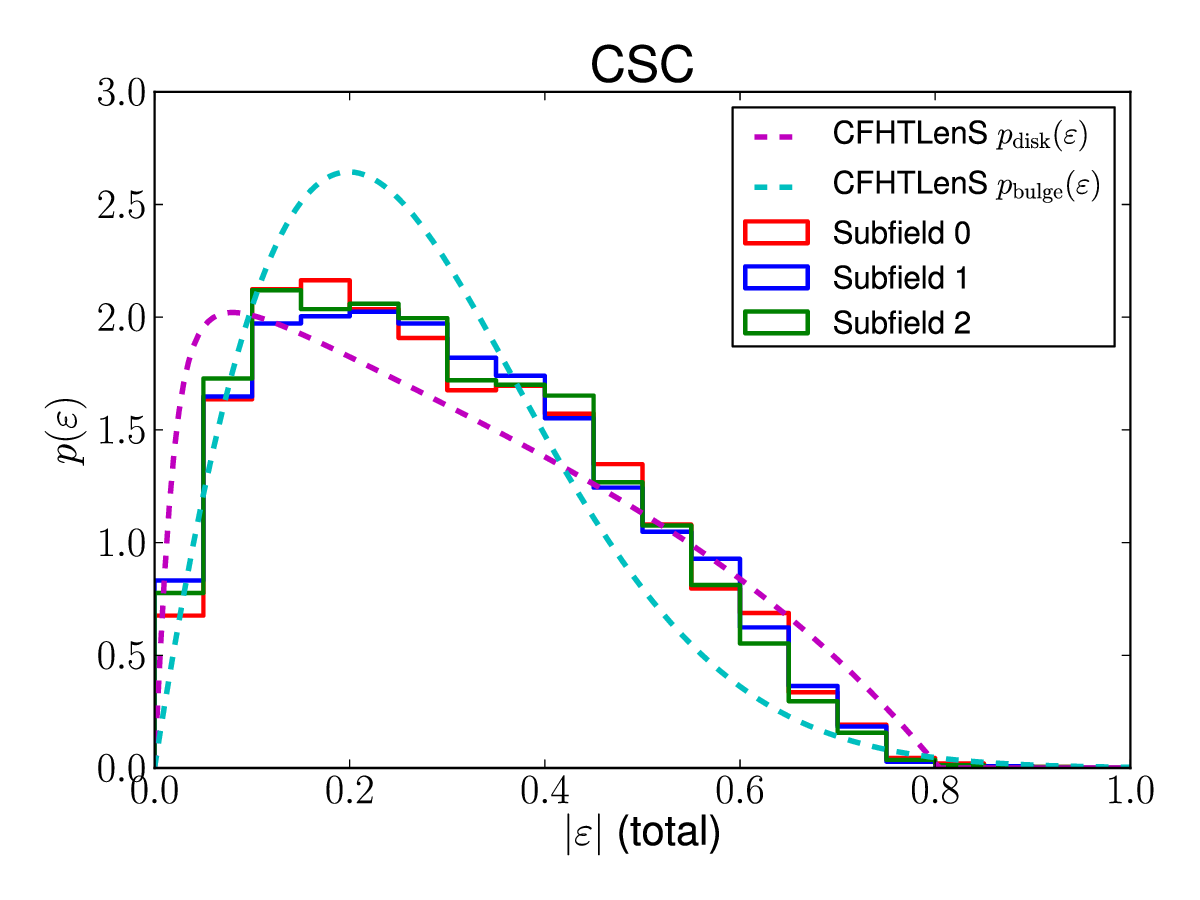}
\caption{\label{fig:pe} The intrinsic ellipticity distribution
  $p(\varepsilon)$ for CGC (top) and CSC (bottom), for three
  subfields. For CGC, the legend shows the subfield index and
  atmospheric PSF FWHM.  \newtext{Both panels show intrinsic ellipticity
  distributions for disk and bulge galaxies from
  \protect\cite{2013MNRAS.429.2858M}.}
}
\end{center}
\end{figure}

In Fig.~\ref{fig:pe}, we show the $p(\varepsilon)$ for
several subfields in CGC and CSC, with several apparent trends. 
First, the $p(\varepsilon)$ are similar for space and 
ground branches.  Second, within different subfields in CSC, there are
small fluctuations in the $p(\varepsilon)$, but
these appear consistent with noise.  For ground branches, the PSF FWHM results in quite different
populations being represented in each subfield.  For this figure, we
deliberately show one subfield with atmospheric PSF FWHM around
the median, along with the subfields with the minimum and maximum
values of PSF FWHM.  Thus, we have maximized population differences
due to our FWHM-dependent galaxy selection process.  However, 
$\langle\varepsilon\rangle$ is only slightly smaller in the worst
seeing subfield than for the more
typical and best subfields, and part of the difference here is 
due to statistical fluctuations.  The
results are similar for the realistic galaxy experiment, and for 
variable shear branches.  Thus, the $p(\varepsilon)$ are
largely stable within and across branches.  \newtext{Moreover, they are reasonably
consistent with a linear combination of observationally-motivated
distributions for bulges and disks derived in a completely different
way and used in \cite{2013MNRAS.429.2858M}, as
shown on the plot.}

\begin{figure}
\begin{center}
\includegraphics[width=0.99\columnwidth,angle=0]{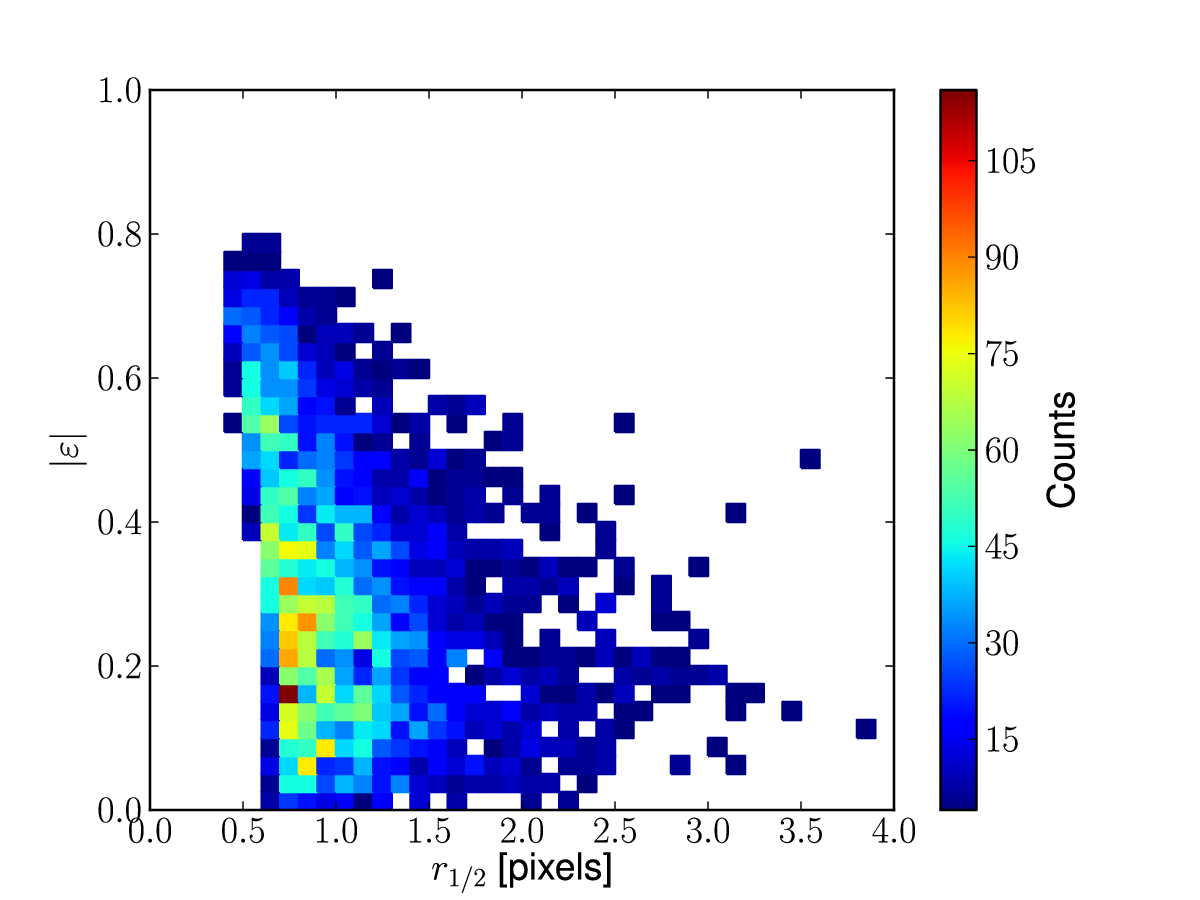}
\caption{\label{fig:pe-size} The 2D histogram of galaxy half-light
  radius $r_{1/2}$ and ellipticity magnitude $|\varepsilon|$ for
  subfield 51 in CGC, which has atmospheric PSF FWHM around the median
  value.}
\end{center}
\end{figure}

The resolution cut is slightly ellipticity-dependent
for the smallest galaxies, as shown in 
Fig.~\ref{fig:pe-size} (the 2D distribution of half-light radius and
ellipticity).  In general, $\lesssim 5$ per
cent of the galaxies are small enough to be affected by this problem. 
Also, this effect is irrelevant
in space branches, where the cuts remove very few 
galaxies.

\subsection{Lensing shears}\label{app:shear}

Here we describe the distributions from which the lensing shears were
drawn.

In constant-shear branches, the lensing shears had random
orientations, with magnitudes between $0.01\le |g|\le 0.05$.  The
distribution of magnitudes within this range is $p(|g|)\propto |g|$,
which emphasizes higher shear values and thus increases our
sensitivity to systematic errors in the shear.

In variable shear branches, each galaxy had an applied shear and
magnification according to a shear power spectrum.  The 
shear power spectrum came from interpolation between tabulated ones
for a particular cosmological model with three median redshifts
$z_\text{med}=0.75$, $1.0$, and $1.25$.  However, the power spectrum was 
altered in two ways.  First, the amplitude was doubled, 
to increase our sensitivity to multiplicative biases.  Second, 
\newtext{to make the power spectrum one that cannot be guessed by participants, we}
added a term corresponding to a sum of shapelets 
with randomly chosen amplitudes (of order 10 per cent of the
original power spectrum amplitude).  For more details, see the
publicly available simulation scripts on the GREAT3 GitHub
page.

\subsection{Atmospheric and optical PSF properties}\label{app:psf}

While the handbook contained
details on many inputs to the PSF models, here we show the 
outputs that are relevant for tests carried out in
this paper, especially in
Sec.~\ref{subsec:results-psf}.

\begin{figure*}
\begin{center}
\includegraphics[width=0.99\columnwidth,angle=0]{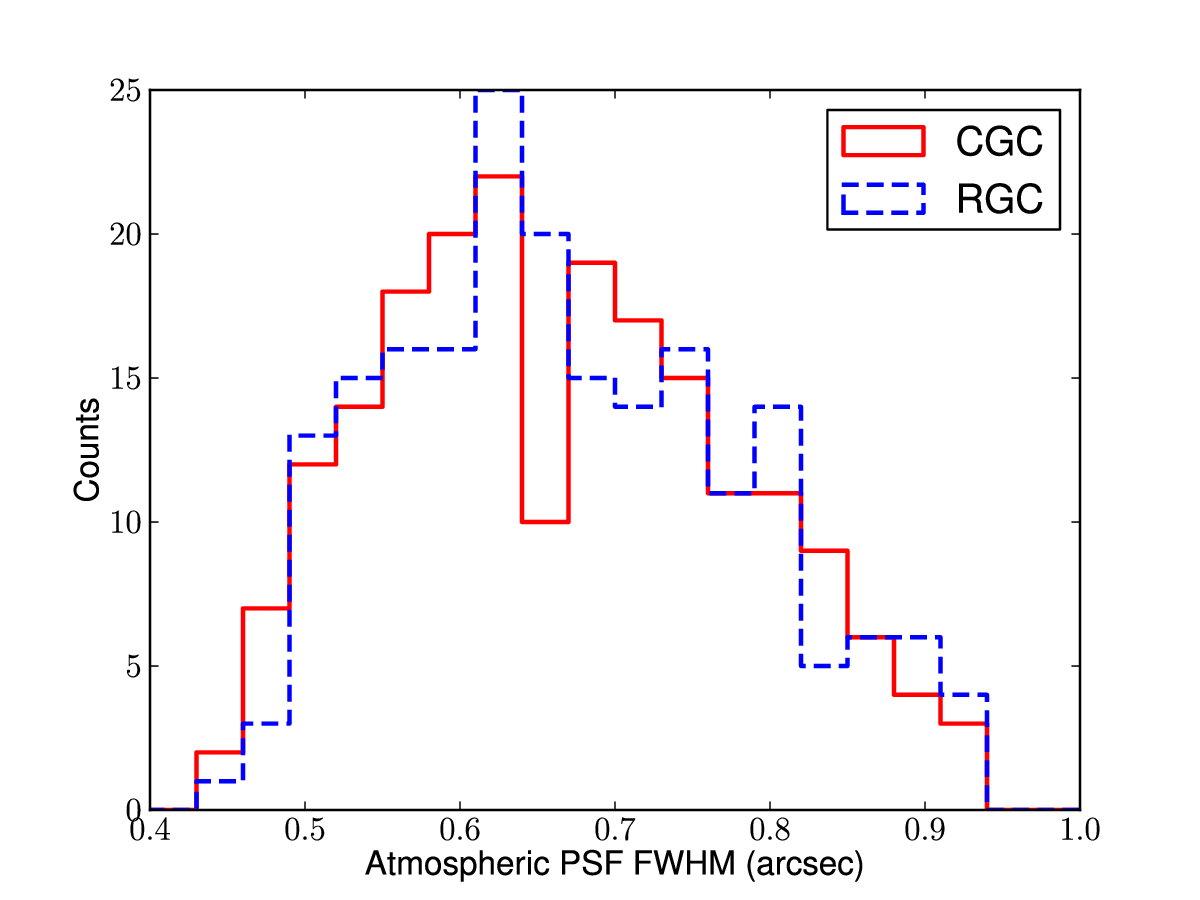}
\includegraphics[width=0.99\columnwidth,angle=0]{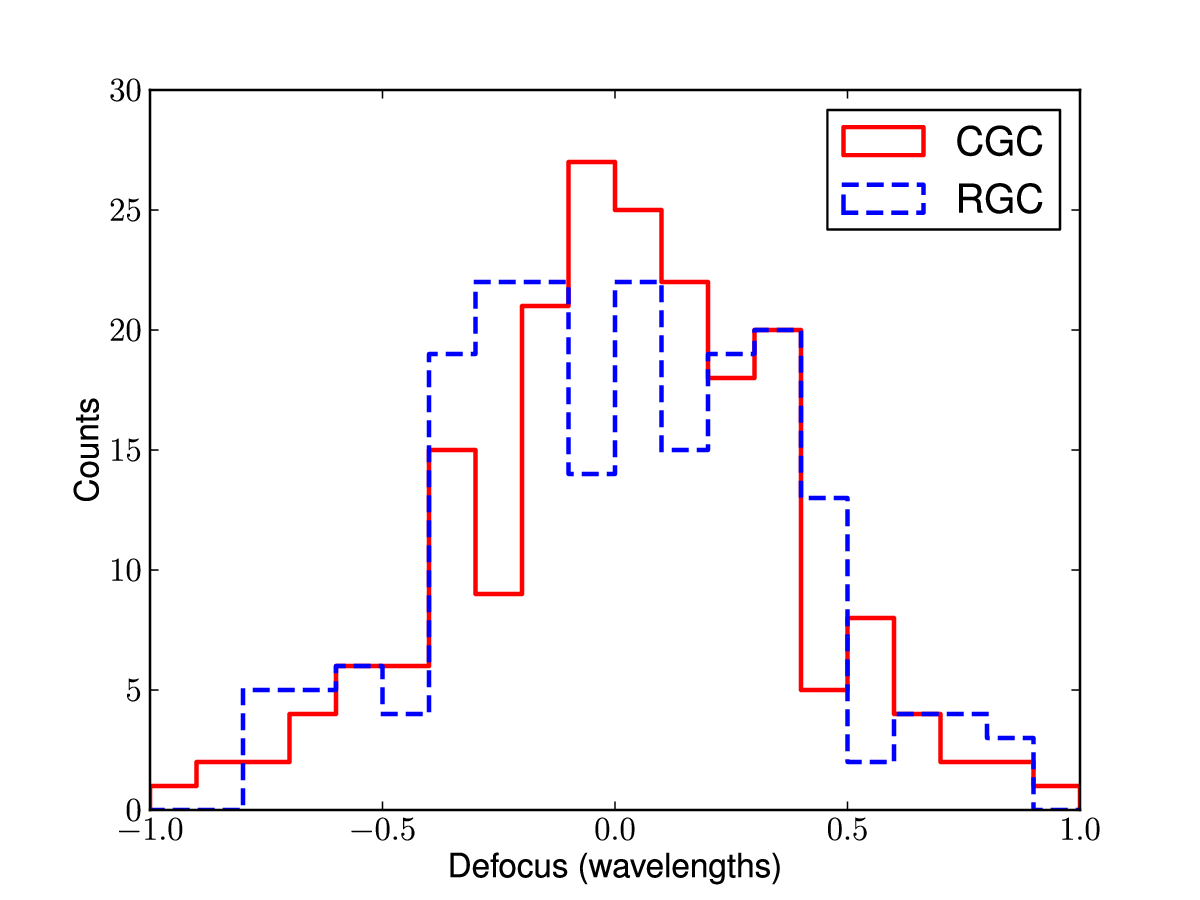}
\includegraphics[width=0.99\columnwidth,angle=0]{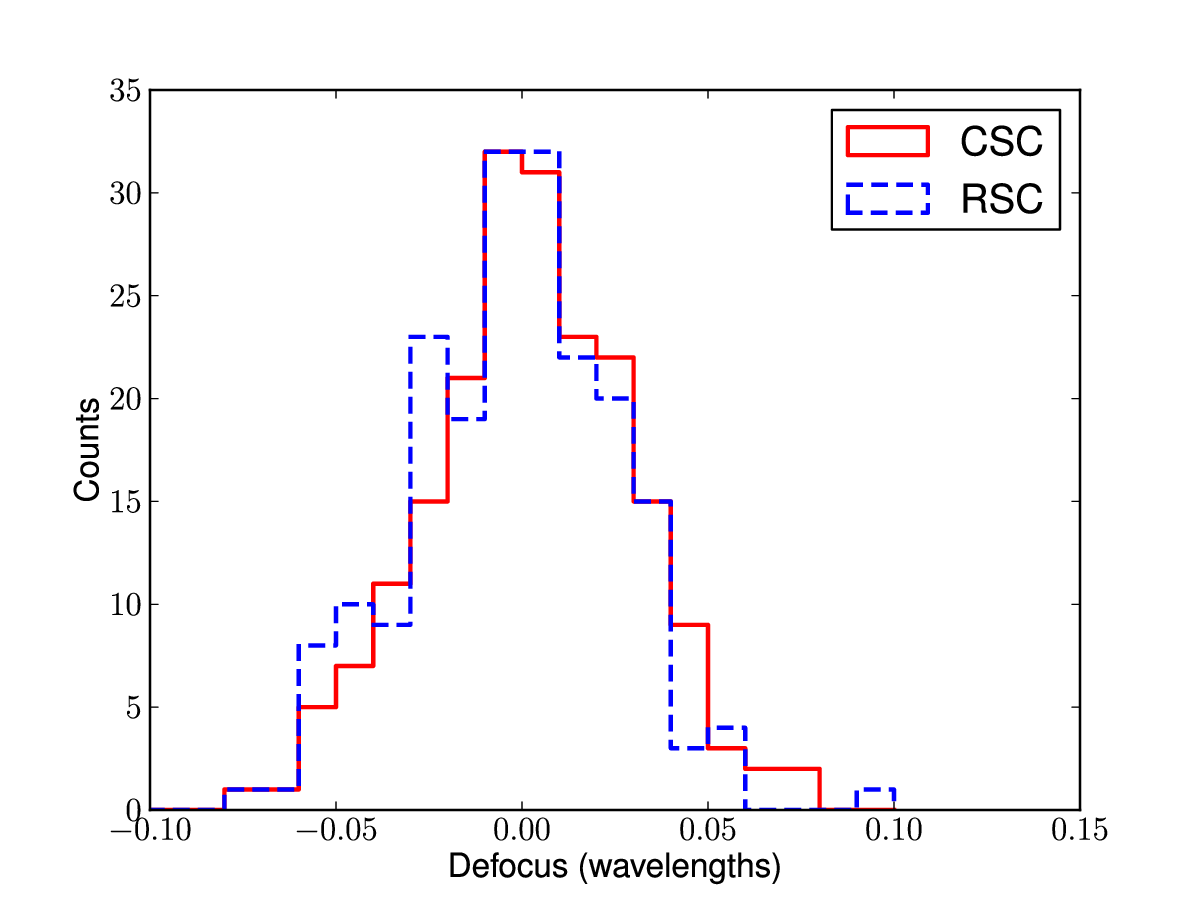}
\includegraphics[width=0.99\columnwidth,angle=0]{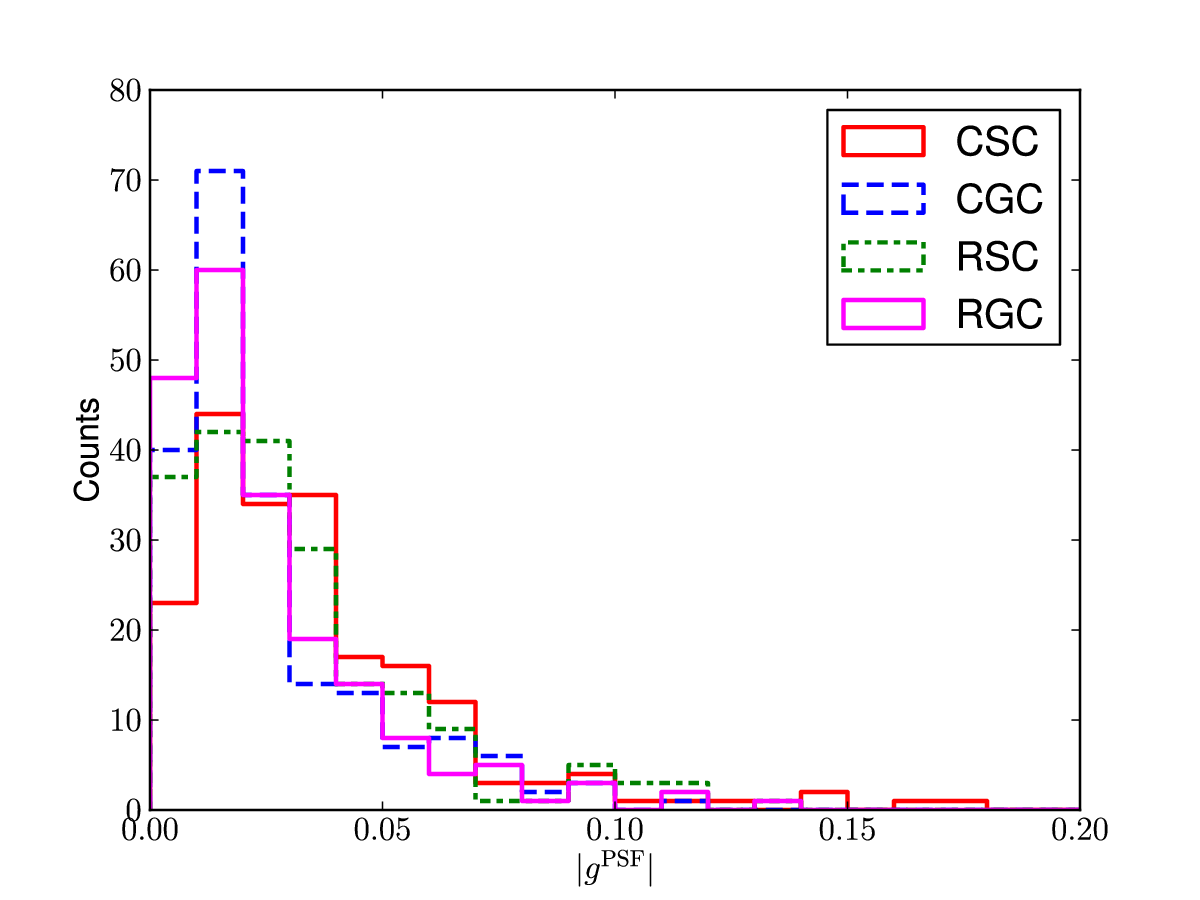}
\caption{\label{fig:psf_property_dist} 
  Distributions of PSF properties across all subfields in various
  branches.  {\em Top:} Seeing (left) and defocus
  distributions (right) for CGC and RGC.  {\em Bottom left:}
  Defocus distributions for CSC and RSC; note the smaller dynamic
  range compared to the ground branches.  {\em Bottom right:}
  Distribution of PSF shear in the four constant-shear branches in the
control and realistic galaxy experiments.
}
\end{center}
\end{figure*}

Fig.~\ref{fig:psf_property_dist} shows the distributions of the seeing
(atmospheric PSF FWHM) in two branches; the
defocus for ground and space-based simulations; and
finally the effective PSF ellipticities including all components.
As shown (top left), the seeing distributions in CGC and RGC are
consistent, modulo small noise fluctuations.
This consistency is important for the comparison between
control and realistic galaxy experiments, since consistency in PSF
properties leads to consistency in the simulated galaxy populations.

The top right panel of Fig.~\ref{fig:psf_property_dist} shows the
distribution of defocus values for the optical PSF in the ground-based
simulations.  CGC and RGC are again consistent, with most
subfields have a maximum defocus of $1/2$ wave, but with a tail to
higher values.  The subfields that
seemed most problematic in Sec.~\ref{subsec:outliers}
are those with higher defocus values, which suggests that identifying
and removing such data could be advantageous.
The bottom left panel shows the defocus distribution for
simulated space-based data, and as expected, the simulated
distribution is roughly a factor of ten narrower than for ground data.  Moreover, CSC and RSC are consistent, which facilitates 
comparison between control and realistic galaxy experiments.

Finally, Fig.~\ref{fig:psf_property_dist} (bottom right) 
shows the distributions of effective PSF shear for four branches.
Typically this quantity is $\lesssim 0.05$, consistent with real data;
\newtext{two-sided KS tests show that the PSF shears are consistent between
pairs of branches that are meant to represent the same data type
(e.g., CSC and RSC, CGC and RGC).}
In both ground and
space simulations, there is a positive correlation between the
absolute value of defocus and $g^{\rm PSF}$, $\sim 0.33$ in both
cases (with a $p$-value  of
order $10^{-7}$).

\subsection{Galaxy $S/N$ distributions}\label{app:sn}

The galaxy $S/N$ distribution in the GREAT3 simulations is important
because it determines the level of noise bias, an
important systematic error for shear estimation. The handbook 
states that the galaxies have $S/N\ge 20$, which is 
higher than the cutoff that is used by many methods in real data.
However, the $S/N$ estimator used
to impose that cutoff is an optimal one that assumes perfect
knowledge of the galaxy profile (which is unachievable in real data).
Thus, to relate the 
quoted $S/N$ cutoff to what is used in real
data, we must use a more realistic $S/N$ estimator.

\begin{figure}
\begin{center}
\includegraphics[width=0.99\columnwidth,angle=0]{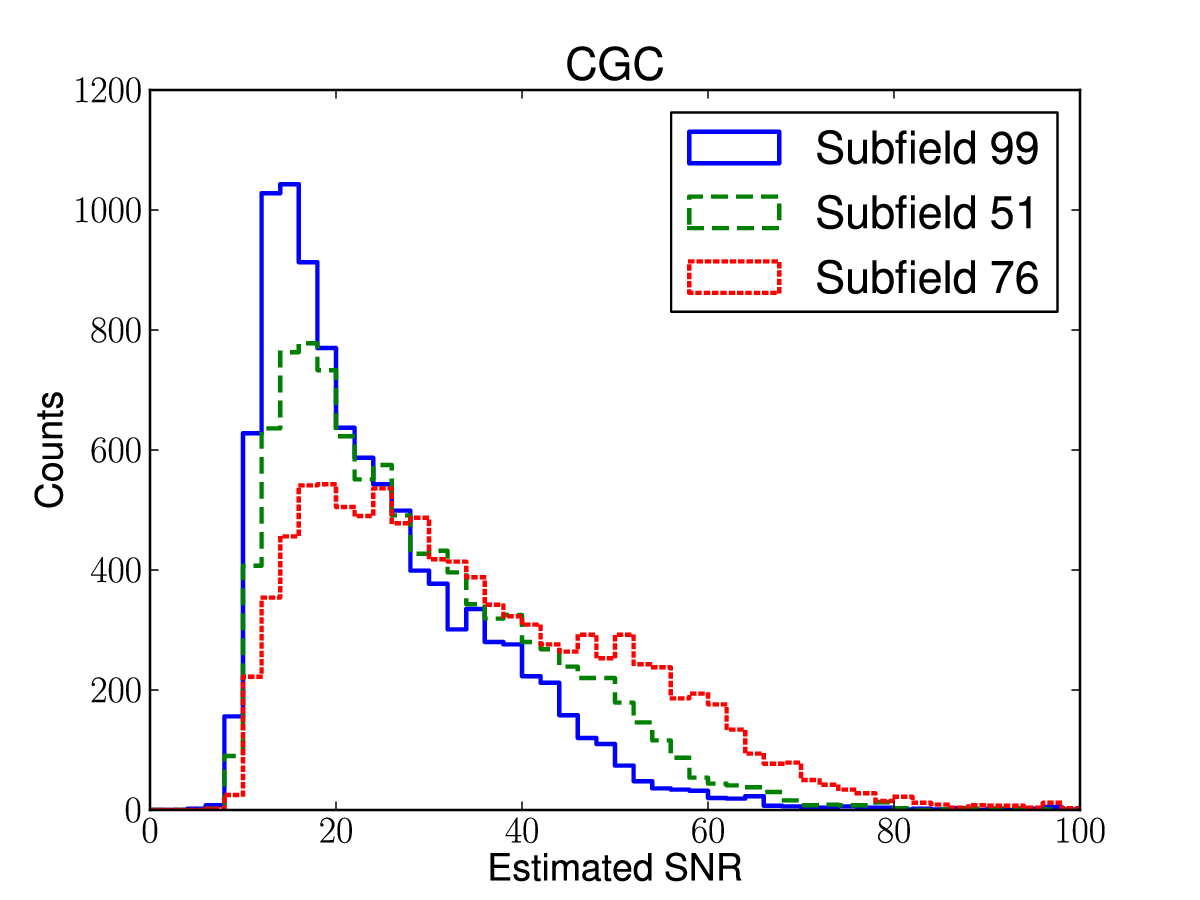}
\includegraphics[width=0.99\columnwidth,angle=0]{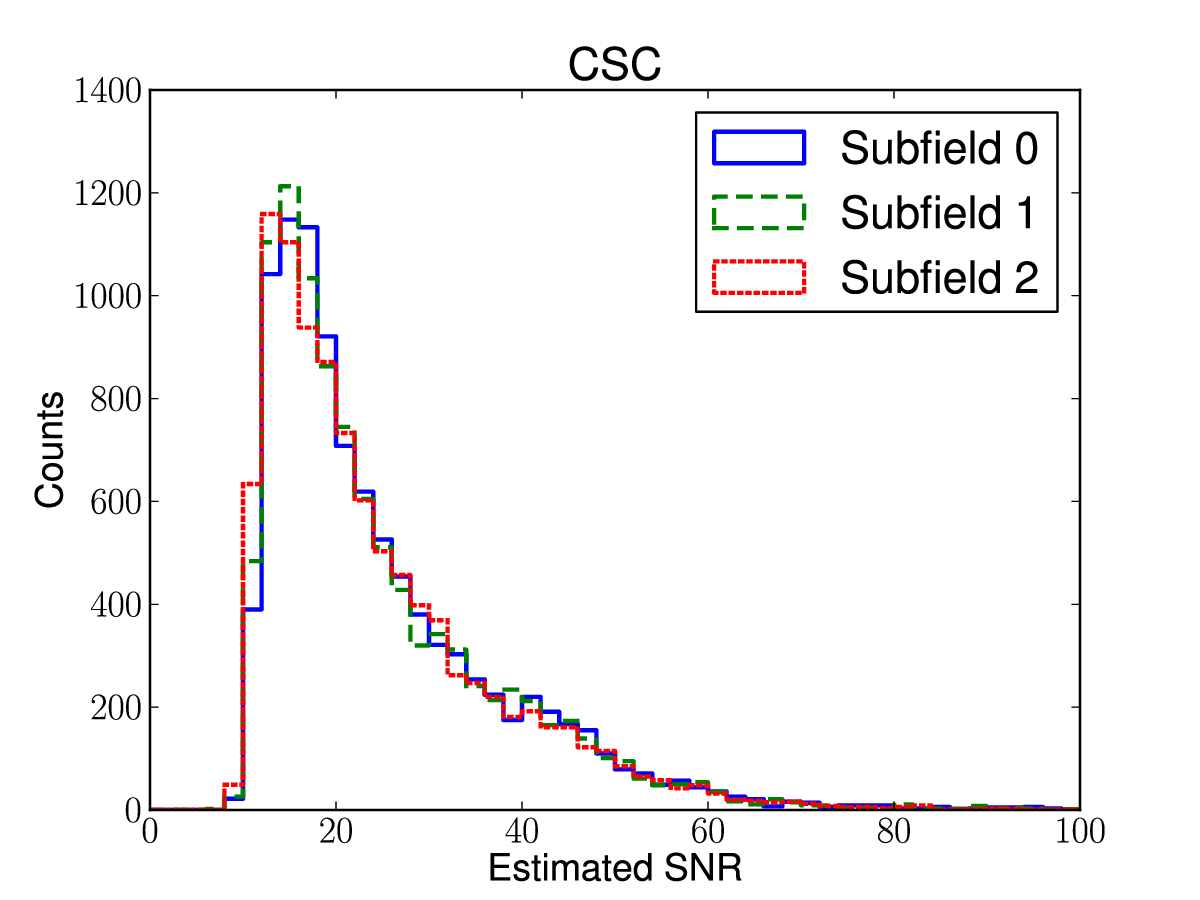}
\caption{\label{fig:snr_dist} The distribution of galaxy $S/N$ 
   using the second $S/N$ estimator described in the text, for
  three subfields in CGC (top) and CSC (bottom).}
\end{center}
\end{figure}

For this purpose, we considered two $S/N$ estimators.  One is the
$S/N$ within an elliptical Gaussian aperture determined using the
best-fitting elliptical Gaussian model for the PSF-convolved galaxy.
Another is the ratio of \code{sextractor} outputs
\texttt{FLUX\_AUTO}$/$\texttt{FLUXERR\_AUTO}. 
Fig.~\ref{fig:snr_dist} shows $S/N$
distributions using the second definition for several
subfields in ground (top) and space (bottom) branches.

As shown, the $S/N$ distribution is quite uniform across subfields in
space branches.  The 5th percentile for $S/N$  is $\sim 12$.  In contrast, the $S/N$ distribution
for ground branches varies with the subfield; the ones shown 
here are the same as in Fig.~\ref{fig:pe}, with maximal 
variation in the atmospheric PSF FWHM. 
Subfields with worse seeing typically have higher
average galaxy $S/N$.  The 5th percentile $S/N$ value is 11.3, 12.0,
and 13.5 for subfields with the best, median, and worst atmospheric
PSF FWHM.  If we use the elliptical Gaussian-based $S/N$
estimate, then the plots shift slightly to the right (higher
$S/N$), with a lower limit of $\sim 14$ instead of
12 for space branches.  This is still a far cry from the nominal $S/N>
20$ limit 
using the optimal estimator, which highlights the need for care in
comparing predictions with different estimators.

\section{Example scripts}\label{app:scripts}

  The GREAT3
  Executive Committee distributed a shear estimation example script
  (called simple\_shear.py) on the GREAT3 GitHub page.  This example
  script estimates per-galaxy shears for all
  galaxies and outputs them as catalogs in the
  format expected by  the publicly available
  presubmission scripts.  Teams could take this code to
  do the bookkeeping while substituting their per-galaxy shear
  estimation routine in place of the one in the example script.

  The example script uses the \galsim\ \citep{2014arXiv1407.7676R}
  implementation of the re-Gaussianization \citep{2003MNRAS.343..459H}
  PSF correction method; see those papers for more details of the algorithm and
  implementation.  Because the script is
  a simple and fast example (not meant to 
  get a science-quality shear estimate), it applies
  only a simply-derived calibration correction that does not include
  all known systematics.    \newtext{For the ``shear responsivity''
  \citep{2002AJ....123..583B} describing how galaxies with a particular distortion
  respond to a lensing shear, the script uses an
  overly-simplistic expression rather than a more accurate one (both available in the above reference).  It also uses a simple but inaccurate way of
  estimating the RMS distortion of the galaxy population, rather than more accurated but more complicated methods that are available in the literature \citep[e.g.,][]{2012MNRAS.425.2610R} as an
  input to the responsivity calculation.  Finally, the default
  settings for initial guess of object size lead to convergence to a
  local minimum for the space branches that cuts out the outer parts
  of the PSF, resulting in very wrong shear estimates (but accurate
  centroid estimates).  Fine-tuning the initial guesses is necessary
  for this script to give reasonable results on space simulations.}

\section{Shear estimation methods}\label{app:methods}

\subsection{Amalgam@IAP}\label{app:amalgam}

\subsubsection{PSF modelling}
\label{sec:psf}

The PSF modeling was performed using the \psfex\
package\footnote{\url{http://astromatic.net/software/psfex}}
\citep{2011ASPC..442..435B} to compute the PSF model for
the star postage stamps.  The PSF
modeling procedure starts by normalising and re-centering point-source
images to a common ``PSF grid'' using a regular image
resampling technique. The coefficients of a set of basis functions of
point-source coordinates $X_c(\boldsymbol{\theta})$ (simple
polynomials) are adjusted in the $\chi^2$ sense to every PSF
``pixel'' to compute a coarse PSF model and its spatial
variation, in the form of a set of tabulated PSF components 
$\boldsymbol{\phi}_c$. 

The model is further refined by adding corrections
$\Delta\boldsymbol{\phi}_c$ by
minimising the following cost function over all pixels $i\in {\cal
  D}_s$ from all point sources $s$:

\begin{equation}
\begin{split}
E(\Delta\boldsymbol{\phi}_{1,2,...}) &= \\
&\!\!\!\!\!\!\!\!\sum_s \sum_{i \in {\cal D}_s}
        \frac{\left(p_{i} - f_s \sum_c X_c(\boldsymbol{\theta})
        \left[\phi_{c\,i}'(\boldsymbol{x}_s)+\Delta\phi_{c\,i}'(\boldsymbol{x}_s)\right]\right)^2}{\sigma_{i}^2} \\
    &+ \sum_c
    \frac{\|\Delta\boldsymbol{\phi}_{c}\|^2}{\sigma_{\phi}^2},
\end{split}
\end{equation}
where $p_i$ is the value of pixel $i$, with uncertainty $\sigma_i$,
and $f_s$ the flux of point source $s$. $\sigma_\phi$ sets the
amplitude of the regularisation term. In practice,
$\sigma_\phi\approx 10^{-2}$ represents a good compromise between
fidelity and robustness of the solution.

The prime indicates a resampled version of the PSF components; e.g.,
the value of pixel $i$ with coordinates $\boldsymbol{x}_i$ in the
image of PSF $\boldsymbol{\phi}$ resampled at the point-source
position $\boldsymbol{x}_s$ with PSF sampling step $\eta$:

$$\phi_{i}'(\boldsymbol{x}_s) = \sum_j h_s\left(\boldsymbol{x}_j
        - \eta(\boldsymbol{x}_i - \boldsymbol{x}_s)\right)\phi_j,$$
where $h_s(\boldsymbol{x})$ is the interpolation function.

The version of \psfex\ used for the GREAT3 challenge is identical
to v3.17.1 except for the interpolation
function, which is either a Lanczos-4 or Lanczos-5 kernel instead of
the default Lanczos-3. Support for
measurement vectors as PSF dependency parameters ({\tt PSFVAR\_KEYS})
was added early in the challenge to allow \psfex\ to map PSF variations
as a function of any set of columns in an ASCII list, through
\sextractor's {\tt ASSOC}  mechanism.

The \psfex\ configuration used for GREAT3 differs from the default 
one in a few minor ways.  The first difference is in the use 
of super-resolution, adopting  a
constant sampling step $\eta$ of 0.6 image pixels for all
branches. This sampling step offers the best compromise
between robustness and accuracy given the limited number of PSF images
for branches with a constant PSF.  Also, the full star postage stamp
size is used 
for each branch.  PSF variations are modelled using 0th and 5th degree
polynomials of star coordinates for constant and variable PSF
branches, respectively.  Finally, the noise on point
source images is assumed to be purely additive, setting {\tt
  PSF\_ACCURACY} to 0.

\subsubsection{Galaxy shape measurement}
\label{sec:gal_meas}

Galaxy shapes are measured using 
\sextractor\footnote{\url{http://astromatic.net/software/sextractor}}
v2.19.15 
\citep{1996A&AS..117..393B,2011ASPC..442..435B}.  The measurement
process involves independently fitting each galaxy image with a 
\sersic\ model convolved with the local PSF model
from \psfex. To avoid galaxy detection problems, 
the Amalgam@IAP team used a detection image to 
explicitly tell \sextractor\ about the gridded galaxy positions.

The vector of \sersic\ model parameters $\boldsymbol{\theta}$
includes the $(x,y)$ centroid position, amplitude, effective radius, aspect
ratio, position angle and \sersic\ index.
Physically meaningful constraints (e.g., amplitude $> 0$)
are imposed on all parameters except position angle through a change
of variables $\boldsymbol{\theta} \rightarrow
\boldsymbol{\theta}'$. For instance, for the aspect ratio (parameter
$\theta_{\rm aspect}$) the Amalgam@IAP team instead constrain the transformed parameter
$\theta_{\rm aspect}'$ defined as
\begin{equation}
\theta_{\rm aspect}' = \ln \frac{\ln \theta_{\rm aspect} - \ln
  0.01}{\ln 100 - \ln \theta_{\rm aspect}}.
\end{equation}
Individual ellipticities from
the aspect ratio and position angle of the best-fitting galaxy
model 
are used directly. \sextractor\ also extracts the associated uncertainties
and their correlation coefficient from the covariance matrix of the
fitted parameters.

The fit itself is achieved by minimising a quadratic cost function
with the Levenberg-Marquardt algorithm using the 
\code{LevMar}\footnote{\url{http://users.ics.forth.gr/~lourakis/levmar/}} library. 
The cost function is the weighted sum of squared residuals plus a
quadratic penalty term 
\begin{equation}\label{eq:amalgam-cost-function}
E(\boldsymbol{\theta}) = \chi^2(\boldsymbol{\theta}') + \sum_i
\frac{(\theta_i'-\mu_{\theta_i})^2}{\sigma^2_{\theta_i}}.
\end{equation}
where the sum is over galaxy model parameters $i$.  
The version of \sextractor\ used by default has $\sigma_{\theta_i} \equiv
\infty$ for all parameters (no penalty).

The fitting process typically converges in 50-100 iterations.
Compared to the latest publicly available version of the package, the
following changes were made to the \sextractor\ code for  GREAT3: 
\begin{itemize}
\item Fitting area (normally set automatically) is limited to
  the size of the GREAT3 galaxy images to avoid overlapping with
  neighbouring galaxies.
\item  Sampling of the model is forced to 0.3 image pixel, instead of
  the default which depends on the input PSF model. 
\item  The step used in difference approximation to the Jacobian in
  \code{LevMar} is set to $10^{-4}$. 
\item  Penalty parameters for the aspect ratio are set to
  $\mu_{\theta_{\rm aspect}} = 0$ and $\sigma_{\theta_{\rm aspect}} =
  1$ to disfavour very large ellipticities for the most poorly
  resolved objects, without significantly affecting the results for
  more resolved galaxies. 
\item  The default, modified $\chi^2$ (which is more 
  robust for partially overlapping objects) is replaced with a
  regular $\chi^2$. 
\end{itemize}

Finally, the \sextractor\ configuration used by the Amalgam@IAP team
reflects the details of the
GREAT3 simulations: the background is set to 0 ADU; the {\tt GAIN} is
set to 0 (equivalent to infinite); and the {\tt MASK\_TYPE} detection
masking parameter is usually NONE.

\subsubsection{Galaxy weighting}\label{app:amalgam-weight}

The Amalgam@IAP team used a modified inverse-variance weighting scheme
based on the full covariance matrix from \sextractor\ (approximated by the Hessian calculated by the
\code{LevMar} minimization engine) to account for possible
covariance between parameters and for differences in the recovery of
$e_1$ and $e_2$ components. This covariance matrix forms the 
basis for the per-galaxy shear covariance matrix. To avoid
giving too much weight to high $S/N$ objects, the Amalgam@IAP team added a constant
$\sigma_s^2$ to the diagonal entries. For constant-shear branches,
they used the full per-object covariance
$\bmath{C}_i$ to estimate the shear as 
$$\hat\gamma =  \left( \sum_k C^{-1}_k \right)^{-1}  \sum_j  C^{-1}_j \boldsymbol{e_j},$$
using the 2-vector $\boldsymbol{e_j}$ and $2\times 2$ matrix
$\bmath{C}^{-1}$. In practice, the 
difference between using the full covariance matrix and its
isotropic approximation was small.

For variable shear branches, the Amalgam@IAP team used the provided \code{corr2}
code with isotropized scalar weights defined as  
$$w_i=\frac{2}{\sigma_{i,1}^2+\sigma_{i,2}^2+2 \sigma_s^2},$$ where the
denominator represents the quadrature sum of measurement error 
and shape noise.

\subsection{BAMPenn}\label{app:bampenn}

This team used the Bayesian Fourier Domain (BFD) method
from \cite{2014MNRAS.438.1880B}, which relies on 
weighted moments calculated in Fourier space and a
prior for the noiseless distribution of galaxy moments (e.g.,
from deep data).  Weighting is implicit rather than explicit in this
Bayesian calculation.  The ensemble shear from the mean of the 
Bayesian posterior should be unbiased in the limit that many 
galaxies are used for shear estimation, potentially 
avoiding noise biases that can plague maximum-likelihood
methods.  It does not result in a per-object shear estimate.

The submissions made during the challenge period came from an immature software
pipeline and had errors that were identified
after the fact.  Currently, the machinery is in place only for a
constant-shear analysis, not variable shear.

\subsection{EPFL\_gfit}\label{subsec:gfit}

All submissions by the EPFL\_gfit team used the \gfit\
method.  A few submissions also used a wavelet-based \code{DWTWiener}
denoising code from \cite{2011A&A...531A.144N}, integrated into \gfit. 
The \gfit\ method uses a maximum-likelihood, forward model-fitting algorithm to measure
galaxy shapes. 
An earlier version of \gfit, used in the GREAT10 galaxy
challenge \citep{2010arXiv1009.0779K,2012MNRAS.423.3163K}, was 
described in \cite{2012arXiv1211.4847G}. The version used 
 in GREAT3 is completely new, written in Python and  relies on the
 \code{NumPy}, 
\code{SciPy} and \code{PyFits} libraries.  The software has a modular 
design, so that additional galaxy models and minimizers can be
plugged in fairly easily. The behavior of \gfit\ is controlled though
configuration files.

\gfit\ requires catalogs generated via an automated process from input galaxy and PSF mosaic
images by \sextractor\ 
\citep{1996A&AS..117..393B}.  
The following galaxy models, for which images are generated using \galsim, are currently supported:
\begin{enumerate}[(a)]
 \item A pure disk \sersic\ model.
 \item A sum of an exponential \sersic\ profile (\sersic\
   $n=1$) to model the disk and a de Vaucouleurs \sersic\
   profile ($n=4$) to model the bulge.
    The disk and
   bulge share the same centroid and ellipticity. 
\item A model similar to the previous but with
                a varying disk \sersic\ index.
\end{enumerate}

Almost all GREAT3 submissions used the second galaxy
model, with the following eight parameters: galaxy centroid, total flux, flux
fraction of the disk, bulge and disk 
radii, and ellipticity.

Fitting can be performed with two minimizers, using input
\sextractor\ catalogs to get initial guesses for galaxy
centroids, fluxes and sizes.  The first minimizer is
the \code{SciPy} Levenberg-Marquardt non-linear least-squares implementation.
The second is a simple coordinate descent minimizer (SCDM), a loose
implementation of the Coordinate Descent
algorithm.  In the SCDM, the model parameters 
are sequentially varied in a cycle, to explore 
all directions in parameter space. After each cycle, the
change in the
objective function is measured and the sense of variation maintained or reversed. The step size for
each parameter is dynamically adjusted based on previous iterations. The
algorithm is, by nature, slow but quite robust, with a failure rate
below 1/1000 on GREAT3 images. 
Several stopping conditions are available and can be combined. 

\newtext{The EPFL\_gfit submissions used a simple weighting scheme that was one
of the options used by CEA-EPFL (below), involving constant weighting
for all galaxies except those that have unusually large fit residuals,
which are rejected entirely (typically $<1$ per cent of the galaxies).}

\subsection{CEA-EPFL}\label{subsec:cea-epfl}

The CEA-EPFL team used 
an object-oriented framework written in Python and
usable in other contexts than GREAT3 with minimal changes, including: 

\begin{itemize}
\item Galaxy shape measurement (\gfit\ from Sec.~\ref{subsec:gfit}).
\item Weight calculation (\code{sfilter}).
\item PSF estimation (star shape measurement, PSF interpolation, PCA decomposition and reconstruction).
\item Image coaddition routines.
\item Wavelet-based tools for deconvolution, denoising, 
  coaddition, and super-resolution.
\end{itemize}
\gfit\ was described in Sec.~\ref{subsec:gfit}, but the remaining 
pipeline elements used in GREAT3 are described
below.

\subsubsection{Weighting scheme}

The \code{sfilter} tool uses catalogs produced by \gfit\ to assign a
weight to each galaxy.
 
In GREAT3, two weighting schemes were used.  The simpler scheme
involved eliminating entries with large fit residuals by giving them
weights of zero.  The more complex scheme involved assigning weights
based on PCA
analysis of the RMS between ellipticities
fitted by \gfit\ on GREAT3 data and those obtained
after running \gfit\ on GREAT3-like simulated data.  The galaxy
simulations were created using \galsim\ with GREAT3-like PSF, noise
and $S/N$; the galaxy parameters were motivated by the outputs
from a \gfit\ analysis of the RSV 
branch. A PCA decomposition was then performed on a vector with first
component $|\Delta e| = \sqrt{ (e_{1,\text{out}} - e_{1,\text{in}})^2 +
  (e_{2,\text{out}} - e_{2,\text{in}})^2 }$. The other PCA components
were either (a) flux, disk and bulge radii, disk fraction, \code{gfit}
output parameters or (b) \sextractor\ FWHM, size, $S/N$, flux, and \gfit\ disk and bulge radii.
   
The first component, $|\Delta e|$, was plotted against various
PCA components to select a cut-off value $v_0$ that separated regions of low and
high $|\Delta e|$. A weight $w_\text{low}$ was assigned to all galaxies with $v <
v_0$, with $w_\text{low} = 0.6$ for choice (b), and $w_\text{low} =
0.2$ for choice (a).

\subsubsection{PSF estimation}

\newtext{For the three experiments with constant PSFs that were provided for
the participants, the CEA-EPFL team used the provided PSFs directly.}
The \code{spredict} tool was used to estimate the PSF at the
positions of galaxies in the variable PSF and full experiments. The version of \code{spredict} used in GREAT3
supports two PSF models:
\begin{itemize}
\item An elliptical Moffat profile, based on maximum-likelihood
  fitting using \galsim\ to generate images.  \newtext{This was used
  in a few submissions to the ground branches.}
\item A data-driven model based on PCA decomposition of selected PSF
  images (with  sufficiently
     high $S/N$, either $>20$ or $>30$) into either 10, 15, or 20  PCA
     components. 
\end{itemize}
More details of these algorithms will appear in Paper II.

\subsubsection{Differences between GREAT3 submissions}

The differences between submissions in a given branch arose mainly
from the size of the postage stamps used for the fits;
constraints placed on galaxy model parameters; minimizer options;
weight functions; choice of galaxy models (though most used the
second one in Sec.~\ref{subsec:gfit}); and occasionally
attempts to include wavelet-based denoising.

\subsection{CEA\_denoise}

The CEA\_denoise team denoised the GREAT3 galaxy images using a publicly available, multi-scale
wavelet-based code \code{mr\_filter}, based on \cite{2006A&A...451.1139S}. They then measured unweighted second
moments of the denoised galaxy images and noiseless PSF images
using \sextractor. Finally, they 
 corrected for PSF convolution by 
subtracting the PSF moments from the galaxy moments, as proposed by 
\cite{2000ApJ...536...79R} and \cite{2011MNRAS.412.1552M}.

The CEA\_denoise team varied the denoising options (such as using 2 vs.\ 3 wavelet
scales), and selected the denoising methods by comparing the original
and filtered galaxies by eye. Strong denoising often resulted in
blurry galaxies with correlated noise features around the galaxies.

No weighting was applied to the measured shears.

\subsection{CMU\_experimenters}

The stacking method used by CMU\_experimenters was a simple modification of
the example script described in App.~\ref{app:scripts}.  The basic steps were galaxy registration, stacking, and PSF correction of the stacked
image.

First, CMU\_experimenters measured the weighted first moments
(centroids) for all galaxy images. They used the default \galsim\
interpolation routines to shift each galaxy so the centroid would be
at the exact center of the postage stamp.
Next, they stacked all $10^4$ galaxies in a
single GREAT3 image using a simple
unweighted average.  Finally they used \galsim\ routines for PSF
correction (re-Gaussianization) to estimate
the PSF-corrected distortion $\hat{e}$.  The shear estimate for the field is 
simply $\hat{g}=\hat{e}/2$, since the stacked
object is effectively round in the absence of a shear.  There is a
calibration factor of 1.02 
for the intrinsic
limitations of re-Gaussianization 
\citep{2012MNRAS.420.1518M}.

\subsection{COGS}\label{subsec:cogs}

All submissions from the COGS team used the 
\code{im3shape} galaxy model fitting code described in
\cite{2013MNRAS.434.1604Z}.

\subsubsection{Galaxy model fitting}

The COGS team used a two component galaxy model, with a de
Vaucouleurs bulge (\sersic\ $n=4$) and an exponential disk ($n=1$). 
The two components were constrained to have the same
half-light radius, centroid, and ellipticity.  The seven free
parameters in the fit were therefore total flux, bulge-to-total flux
ratio, radius, centroid ($x,y$), and ellipticity.

The best-fitting model was identified by minimizing the squared
residual between data and model image, using the \code{LevMar}
implementation of the Levenberg-Marquardt algorithm (see
\citealt{2013MNRAS.434.1604Z} for details).  The parameter settings
and optimizer termination criteria are given in the \code{im3shape}
initialization
file\footnote{\url{https://github.com/barnabytprowe/great3-public/wiki/COGS-.ini-file}}
used for GREAT3. The full galaxy postage stamps were used for all
fits.

One important parameter for \code{im3shape} is the \code{upsampling},
the internal super-resolution at which profiles are drawn and FFT
convolutions performed.  For speed, early submissions used the native
resolution, which causes artifacts in the modeling and increases
biases. Later COGS submissions set \code{upsampling}~=~7.  These
submissions required similarly upsampled PSF images, which 
were generated via bicubic interpolation across the noise free
PSF images provided with the GREAT3 data.  These entries with
\code{upsampling}~=~7 can be considered to be the baseline set of COGS
submissions with high precision input settings.  These submissions are
referenced by their label \code{u7} in this paper.

%
%

\subsubsection{Noise bias calibration}\label{sss:cogs-noise}

Some \code{im3shape} submissions include a multiplicative calibration
factor to correct for expected noise biases in Maximum-Likelihood shape
estimation.  These can be grouped under the following three labels:
\begin{itemize}
\item \code{c1}: A correction for an isotropic multiplicative bias
  $\langle m \rangle = 0.0230$ is applied.  This expected noise bias
  was estimated in simulations performed by \citet[table
  2]{2014MNRAS.441.2528K} using a galaxy population that differs
  somewhat from that in GREAT3.
\item \textsc{c2}: A correction for an isotropic multiplicative bias
  $\langle m \rangle = 0.0330$ is applied.  This bias was estimated
  using the CGV deep data. The ellipticity of galaxies in the CGV deep
  fields was measured using \code{im3shape} (with
  \code{upsampling}~=~7).  These images were then degraded by adding
  noise to match the regular (non-deep) GREAT3 images, and
  re-measuring the ellipticities.  By fitting a polynomial including a
  constant, linear, and cubic term to $\varepsilon_{1,\text{deep}}$
  vs.\ $\varepsilon_{1,\text{degraded}}$, the COGS team estimated a
  calibration factor $m(\varepsilon)$ and then calculated an expected
  calibration bias of $\langle m \rangle = 0.033$ based on
  $p(\varepsilon_\text{deep})$.  The CGV deep data was used since it
  exhibited less variation in the image properties than 
   the deep CGC data.  This possibly relates to the
  relatively strong seeing variation identified in the deep CGC data,
  discussed in Appendix~\ref{app:ess}.
\item \code{c3}: A correction for an isotropic multiplicative bias
  $\langle m \rangle = 0.02943$ is applied.  This factor was estimated
  using the CGV deep data in a similar manner to \code{c2}, but using
  only galaxy models with best-fitting $|\varepsilon_{\rm deep}| <
  0.9$ in the deep data to estimate $m(\varepsilon)$.  This removal of
  outliers was found to provide a better fit to the (most numerous)
  galaxies with lower ellipticity values.
\end{itemize}
No calibration was made for additive biases due to noise, although
these are expected where PSFs are anisotropic
\citep{2012MNRAS.427.2711K}.

\subsubsection{Differences between GREAT3 submissions}
The main differences between submissions were the correction of bugs
in the interface between \code{im3shape} and the GREAT3 data format,
the upsampling, and noise bias calibrations applied.
Early submissions used low accuracy settings for rapid basic
validation of the GREAT3 data, and are unsuitable as a basis for careful
scientific analysis.

However the later set of submissions (with labels \code{u7},
\code{c1}, \code{c2}, and \code{c3} as described above) can be
used for fair scientific comparison, and to explore systematic errors
in the \code{im3shape} approach more generally.
All galaxies were given uniform statistical weights when generating submissions.

\subsection{E-HOLICS}

The E-HOLICs method
\citep{2011ApJ...730....9O,2012ApJ...748..112O,2013ApJ...771...37O}
is a moment-based method based on the KSB method 
\citep{1995ApJ...449..460K}.   One important improvement
of the E-HOLICs method compared to KSB 
is its use of an elliptical (not circular) weight function.

In the E-HOLICs analysis of GREAT3 data, all galaxies that were used
for the analysis were uniformly weighted.  However, galaxies with estimated ellipticities $>1$ were
rejected (i.e., given zero weight).  The E-HOLICs
 team applied a
correction for systematic error due to pixel noise as derived in the above references, with different
submissions having different corrections.  

\subsection{EPFL\_HNN}

The EPFL\_HNN method deconvolves the data by the given PSF,
represented by linear algebra formalism as a Toeplitz matrix. This
allows for solution of the convolution equation by applying the
Hopfield Neural Network (HNN) forward recurrent algorithm. At each
iteration, the selected neurons of the network (image pixels) are
updated to minimize the energy function. To measure the
ellipticity of galaxies in deconvolved images, the second order moments
of the image autocorrelation function are used \newtext{ \citep{Nurbaeva2014arXiv}.}

HNN is an unsupervised neural network, so input galaxy stamps could be
initialized to zero. To reduce CPU time, the
observed data was used as input. 
The output consists of reconstructed images of the
deconvolved galaxies, their autocorrelation functions,
and an ellipticity catalog.  All galaxies received equal
weighting when calculating the average shears, and no calibration
correction was applied.

Differences between submissions in each branch include:
\begin{itemize}
\item the size of the effective galaxy postage stamp size;
\item the pixel updating value (a smaller number gives finer reconstruction, while increasing the iteration number and CPU time); and 
\item filtering (removing  the galaxies for
  which the HNN
  algorithm failed to converge).
\end{itemize}

\subsection{EPFL\_KSB}

The EPFL\_KSB team used an implementation of the KSB method
\citep{1995ApJ...449..460K,1997ApJ...475...20L,1998ApJ...504..636H} 
based 
on the KSBf90 pipeline \citep{2006MNRAS.368.1323H}.
The KSB method parametrises galaxies and stars according to their
weighted quadrupole moments. In the standard KSB method, a Gaussian
filter of scale length $r_g$ is used, where $r_g$ is galaxy
size. The EPFL\_KSB team 
also tried other weighting functions.

The main assumption of the KSB method is that the PSF 
can be described as a small but highly anisotropic distortion
convolved with a large circularly symmetric function. 
With that assumption, the shear can be recovered to
first-order from the observed ellipticity of each galaxy via
\begin{equation} \label{eqn:weight}
\gamma=P_{\gamma}^{-1}\left(e^{\rm obs}-\frac{P^{\rm sm}}{P^{\rm sm*}}e^{*}\right),
\end{equation}
where asterisks indicate quantities that should be measured from the
PSF model at that galaxy position, $P^{\rm sm}$ is the
smear polarisability (see \citealt{2006MNRAS.368.1323H} for definitions) 
and $P_\gamma$ is the correction to the shear
polarisability that includes the smearing with the isotropic component
of the PSF. The ellipticities are constructed from 
weighted quadrupole moments, and the other quantities involve
higher order moments. All definitions are taken from
\cite{1997ApJ...475...20L}. 
The shear contribution from each
galaxy is weighted according to the quadrature sum of shape noise and
measurement error, calculated as in appendix A of
\cite{2000ApJ...532...88H}.

Submissions from this team fall into two categories: those using the
standard KSB Gaussian filter, and those using a combination of KSBf90 and a
multiresolution wiener filter with bspline wavelet
transform  \citep[\code{mr\_filter}, ][]{2006A&A...451.1139S}.  The latter submissions tended to perform
better.  Among the first type of submissions, the better-performing
ones use a polynomial fitting formula for $P_\gamma$ based on the
galaxy size and $S/N$, and rejection of galaxies with extremely large values of
$P_\gamma$.

\subsection{EPFL\_MLP}

The EPFL\_MLP team's method involved training a Multilayer Perceptron
(MLP) Neural Network to measure galaxy shapes. The MLP is a
feedforward neural network with one hidden layer \newtext{ \citep{haykin2009neural,Rojas:1996:NNS:235222}.}  The arctangent
function is used as an activation function.  The input data are the
set of neurons, represented by the galaxy image pixels.  The output is
the ellipticity catalog.  The MLP is trained on simulated
data with the standard back-propagation algorithm.  

MLP works in two passes.  During the forward pass, the weight matrix
is applied to the training set, the output is compared to the desired
result to obtain the error gradient and to average them over the batch
set.  During the backward pass, the weight updates $\Delta w$ are
calculated from the gradient descent method using the learning rate.

This method uses a batch learning scheme, where the input data is a batch
of galaxy stamps and the weights are updated based on the error rate averaged
over the batch.

For each submission, the EPFL\_MLP team varied the following parameters: the number
of neurons in the hidden level, the learning rate, the epoch number
(an epoch corresponds to one forward pass and a backward pass), 
the batch number (batch learning improves stability by averaging), the
momentum rate ($\mu$ indicates the relative importance of the
  previous weight change on the new weight increment).

The training set consists of galaxy images, simulated using \galsim\ with the
following parameters:
disk and bulge half-light radii,
ellipticity modulus $|e|$,
orientation angle, 
galaxy total flux, 
bulge ratio, and 
signal-to-noise ratio.

Both bulge and disk have the same centroid and ellipticity.  No
weighting was applied to shear estimators, and no calibration factors
were applied.

\subsection{FDNT}

This team used an implementation of the Fourier-domain nulling
technique (FDNT, \citealt{2010MNRAS.406.2793B}).  This method
estimates a per-galaxy shear in the 
Fourier domain after PSF effects have been removed by Fourier
division (equivalent to deconvolution in real space).  This team's
approach was to then apply bias corrections based on image
simulations.  The bias is a function of (1) $S/N$, (2) resolution, (3)
PSF shape, (4) radial flux distribution of the galaxy, and (5) radial
flux distribution of the PSF. Additive bias was found to be directly
proportional to the PSF shape.

In some cases, galaxies were weighted according to the combination of
shape noise (determined from the deep data) and shape measurement
uncertainty.  All FDNT submissions (v0.1 through v1.3) have the
wrong bias corrections applied, and hence all results submitted during the
challenge period are not indicative of the real performance of this
method once this error is corrected.

\subsection{Fourier\_Quad}\label{app:fourier_quad}

This team used Fourier-space methods described in a sequence of papers 
\citep{2008MNRAS.383..113Z,2010MNRAS.403..673Z,2011MNRAS.414.1047Z,2011JCAP...11..041Z,2013arXiv1312.5514Z}.
The shear estimators for the two components of the reduced shear $g_1$
and $g_2$ are defined based on the Fourier transform of the galaxy
image. There are three quantities: $G_1$, $G_2$, and $N$, based on 
multipole moments of the spectral density distribution of the galaxy
image in Fourier space:
\begin{eqnarray}
\label{shear_estimator}
G_1&=&-\frac{1}{2}\int \mathrm{d}^2\mathbf{k}\,(k_x^2-k_y^2)T(\mathbf{k})M(\mathbf{k})\\ \nonumber
G_2&=&-\int \mathrm{d}^2\mathbf{k}\,k_xk_yT(\mathbf{k})M(\mathbf{k})\\ \nonumber
N&=&\int \mathrm{d}^2\mathbf{k}\,\left[k^2-\frac{\beta^2}{2}k^4\right]T(\mathbf{k})M(\mathbf{k})
\end{eqnarray}
where
\begin{eqnarray}
T(\mathbf{k})&=&\left\vert\tilde{W}_{\beta}(\mathbf{k})\right\vert^2/\left\vert\tilde{W}_\text{PSF}(\mathbf{k})\right\vert^2\\ \nonumber
M(\mathbf{k})&=&\left\vert\tilde{f}^S(\mathbf{k})\right\vert^2-\left\vert\tilde{f}^B(\mathbf{k})\right\vert^2 \label{shear_estimator_dis_para}
\end{eqnarray}
and $\tilde{f}^S(\mathbf{k})$, $\tilde{f}^B(\mathbf{k})$, 
$\tilde{W}_\text{PSF}(\mathbf{k})$, and  $\tilde{W}_{\beta}(\mathbf{k})$  are the Fourier
transforms of the galaxy image, an image  of background noise,  the
PSF image, and an isotropic Gaussian function of scale radius $\beta$, 
respectively. 
The latter is defined
as
\begin{equation}
\label{beta}
W_{\beta}(\mathbf{x})=\frac{1}{2\pi\beta^2}\exp\left(-\frac{\left\vert\mathbf{x}\right\vert^2}{2\beta^2}\right).
\end{equation}
The factor $T(\mathbf{k})$ is used to convert the form of the PSF to
an isotropic Gaussian function. The value of 
$\beta$ should be at least slightly larger than
the original PSF $W_\text{PSF}$ to avoid singularities
in the conversion. If the intrinsic galaxy images are statistically
isotropic, the ensemble averages of the shear estimators
defined above recover the shear to second order in
accuracy, i.e., 
\begin{equation}
\label{shear_measure}
\frac{\left\langle G_j\right\rangle}{\left\langle N\right\rangle}=g_j+O(g_{1,2}^3)
\end{equation}
for $j=1, 2$. 
Note that ensemble averages are taken for $G_1$, $G_2$, and $N$
separately; these should be weighted averages, as we will
discuss in Sec.~\ref{subsubsec:fq-great3}. In practice, $G_1$, $G_2$, and $N$ are calculated using
discrete Fourier transforms.

In the presence of source Poisson noise, the method is
modified/extended by adding more terms into the shear estimators to
keep them unbiased. Statistically, the Poisson noise has a
scale-independent spectral density in Fourier space. Its amplitude can
be estimated at the large wave-number limit, at which the source
spectrum is subdominant due to filtering by the PSF. The estimated
Poisson noise spectrum can then be subtracted from the spectral
density of the image on all scales. This operation is particularly
suitable for these shear estimators, as the ensemble averages are taken
directly on the spectral density. Finally, the same procedure should
be repeated in the neighboring image of background noise, as the
Poisson noise in the source image is partly due to the background
photons. Removing the source Poisson noise
effect requires modification of the definition of $M(\mathbf{k})$ in
Eq.~\eqref{shear_estimator_dis_para} to
\begin{equation}
\label{shear_estimator_dis_para3}
M(\mathbf{k})=\left\vert\tilde{f}^S(\mathbf{k})\right\vert^2-F^S-\left\vert\tilde{f}^B(\mathbf{k})\right\vert^2+F^B
\end{equation}
with
\begin{equation}
\label{shear_estimator_dis_para4}
F^{S,B}=\frac{\sum_{\vert\mathbf{k}_j\vert>k_c}\left\vert\tilde{f}^{S,B}(\mathbf{k}_j)\right\vert^2}{\sum_{\vert\mathbf{k}_j\vert>k_c}
1} 
\end{equation}
where $k_c$ is a value at which the Poisson noise amplitude
dominates over the source signal, typically $\sim
3/4$ of the Nyquist wavenumber. 

\subsubsection{GREAT3 Experience}\label{subsubsec:fq-great3}

In GREAT3, the PSF for constant-shear branches was determined by stacking the spectral
densities of the nine provided PSF images. Several different
weighting schemes were used, for each of which the weight is 
a function of the total source flux $F$ 
(rather than the shape parameters) to avoid introducing
systematic biases. Shear estimation for the $j$th component was
carried out via
\begin{equation}
\label{shear_measure2}
\frac{\sum_i G_{j,i}W_i}{\sum_i N_i W_i}=g_j.
\end{equation}
Since the background noise in GREAT3 images is uncorrelated, its power spectrum in Fourier space is
scale-independent. Thus, its
contamination can be directly removed using the source image
itself, without using a neighboring background image, 
rewriting Eq.~\eqref{shear_estimator_dis_para3} as
\begin{equation}
\label{shear_estimator_dis_para5}
M(\mathbf{k})=\left\vert\tilde{f}^S(\mathbf{k})\right\vert^2-F^S.
\end{equation}

\newtext{Three weighting options were tried:}
\begin{enumerate}
\item \newtext{$W = 1$, for which the contribution to the shear signal scales
  as $(S/N )^2$, guaranteeing equal weights for the galaxies in each
  90$^\circ$ rotated pair and maximizing shape noise
  cancellation. However, in terms of contribution to the ensemble
  shear signal, the bright galaxy pairs are much more important than
  the faint ones.}
\item \newtext{$W=(S/N )^{-2}$ for galaxies that can be easily identified as
  90$^\circ$ rotated pairs by sorting the galaxy luminosity
  distribution. For two galaxies in a pair, their average flux is used
  for calculating $W$. For galaxy pairs that are too faint to be
  identified, $W=(S_{min}/N)^{-2}$, where $S_{min}$ is the minimum
  galaxy flux from the identified pairs.}
\item \newtext{$W=(S/N )^{-2}$ for all galaxies without identifying pairs.}
\end{enumerate}

\newtext{The first two weighting options are effective in GREAT3 due to its shape noise
cancellation, which is not relevant for real data. The
last weighting scheme is applicable to real data, though it is not yet
optimal.}

To calculate the shear-shear correlation function using the shear
estimator defined in Eq.~\eqref{shear_estimator}, the Fourier\_Quad
team would ideally use \citep{2011MNRAS.414.1047Z}
\begin{equation}
\label{shear_corr}
\langle\gamma_j(\mathbf{x})\gamma_j(\mathbf{x}+\Delta\mathbf{x})\rangle=\frac{\sum_i G_j(\mathbf{x}_i)G_j(\mathbf{x}_i+\Delta\mathbf{x})}{\sum_i N(\mathbf{x}_i) N(\mathbf{x}_i+\Delta\mathbf{x})}
\end{equation}
The above formula is similar (but not equivalent) to the usual shear-shear correlation calculation using 
ellipticities $\varepsilon_{1,2}$ and weights $W$:
\begin{multline}
\label{shear_corr_old}
\langle\gamma_j(\mathbf{x})\gamma_j(\mathbf{x}+\Delta\mathbf{x})\rangle=\\
\frac{\sum_i \varepsilon_1(\mathbf{x}_i)\varepsilon_1(\mathbf{x}_i+\Delta\mathbf{x})W(\mathbf{x}_i) W(\mathbf{x}_i+\Delta\mathbf{x})}{\sum_i W(\mathbf{x}_i) W(\mathbf{x}_i+\Delta\mathbf{x})}
\end{multline}

To use the GREAT3 presubmission script, the Fourier\_Quad team
converted $G_1, G_2, N$ to per-galaxy $\varepsilon_1, \varepsilon_2,
W$ via $\varepsilon_j = G_j/N$ and $W=N$.  This choice had several 
drawbacks, the main one of which is that for lower $S/N$ sources, $G_1$, $G_2$, $N$ can take both
positive and negative values, due to the subtraction of the background
noise contribution in Eq.~\eqref{shear_estimator_dis_para5}. As a
result, the $\varepsilon_{1,2}$ can be extremely noisy
($|\varepsilon_{1,2}|\gg 1$), which is not a problem 
if the shear correlation is calculated using
Eq.~\eqref{shear_corr}.    The proof of concept for 
variable shear estimation using this method is the subject of ongoing work.

\subsection{HSC-LSST-HSM}

The HSC/LSST-HSM team  attempted
to reproduce the results of the publicly released shear estimation
example script, but using the
HSC/LSST pipeline for the bookkeeping and a slightly older version of
the re-Gaussianization method \citep{2003MNRAS.343..459H}.  From a scientific perspective the
results should be the same, so this is primarily a sanity
check that the HSC pipeline has no bugs that would 
cause re-Gaussianization to perform differently.

The HSC/LSST pipeline was used for the preliminary parts of the data
processing, which in this case was mostly just bookkeeping.  
Only the first PSF image in the constant PSF branches was used, after
shifting it 
by $(-0.5, -0.5)$ pixels using
5th-order Lanczos interpolation to match the conventions of the
HSC/LSST pipeline.   Objects were selected by cutting out postage
stamps according to the provided galaxy catalog; the HSC/LSST
pipeline object detection routines  were not used.  
Then, an early implementation of re-Gaussianization that is part of
the HSC pipeline was run.   Shear responsivity,
weighting, and an additional calibration factor of 0.98 were all done
in a way identical to the publicly released example script that uses
the \galsim\ implementation of re-Gaussianization.

\subsection{MBI}\label{app:mbi}

The MBI team carried out a hierarchical (multi-level) Bayesian joint
inference (MBI) of the shear and the intrinsic ellipticity
distribution given the image pixel data, assuming simply-parametrised
galaxy models, simply-parametrised PSF models, and a simply
parametrised $p(\varepsilon)$. The team's goal was 
to begin the exploration of this new approach to shear measurement in a
realistic setting, without expecting to be competitive given the
simplicity of its PSF and galaxy models, but hoping to learn something
by comparing various hierarchical inferences with the
standard maximum likelihood estimates.  \newtext{A paper describing
this method \citep{2014arXiv1411.2608S} gives an overall picture of
the MBI framework and several ideas for improvements beyond the
implementation used in the GREAT3 challenge.}

The MBI team modeled the PSF with a mixture of three Gaussians using the star
image data.  Galaxies are modeled as elliptical \sersic\ profiles
(using constrained Gaussians mixtures;  
\citealt{2013PASP..125..719H}) with six parameters: position, effective radius, \sersic\ index,
and two (lensed) ellipticity components. The \code{Tractor} software
developed by Lang and Hogg (Lang et al. {\em in prep.}) was used for these low-level individual galaxy
inferences: the posterior PDF for each galaxy's model parameters is
sampled using the ensemble MCMC sampler \code{emcee} \citep{2013PASP..125..306F} starting near the mode of the posterior found by a simple non-linear
least squares optimizer. These individual galaxy model
inferences are carried out in embarrassingly parallel fashion.

The intrinsic (pre-lensing) galaxy $p(\varepsilon)$ is
modeled as a Gaussian in both components,
centred on zero and with width $\sigma_{\varepsilon}$. This parameter is
inferred jointly for both shear components for each field by
importance sampling\footnote{Generally, importance sampling is
  a process for estimating the properties of some distribution despite
only having samples generated from a different distribution.  Because
of this difference, the samples that are drawn must be reweighted.}
the \code{emcee} outputs with a flat hyperprior on
$\log{\sigma_{\varepsilon}}$ (assuming an
uninformative prior on lensed ellipticity), using the standard relation
between shear, intrinsic and observed ellipticity.  The 
best results use this simple Gaussian prior; 
a double Gaussian did not improve accuracy.  For GREAT3
submissions, the MBI team reported the posterior mean estimates of the shear
components. They only entered the constant-shear and constant-PSF branches of the
challenge, where their simple assumptions are 
valid and no PSF interpolation is required. Of the six branches
fitting this description, they did not submit to two (RSC and
MSC) due to lack of time.

MBI team submissions are labeled as follows:
\begin{itemize}
\item Optimal Tractor: The shear estimator is the mean of the
maximum likelihood galaxy lensed ellipticity estimates for all 
galaxies in the field.
\item Sample Tractor: The shear estimator is the mean of all
  samples from all galaxies' lensed ellipticity posterior
  PDFs.
\item Important Tractor: Submissions derived from the importance
  sampling analysis, assuming an independent Gaussian $p(\varepsilon)$ in each field.
\end{itemize}

Some 
submissions  experimented with other aspects of the method.  For
example, those labeled ``multi-baling'' involved
inferring a $p(\varepsilon)$ common to five fields
at a time.  Submissions labeled ``deep'' 
used the deep fields to obtain a hyper-prior on the
$p(\varepsilon)$ width parameter
$\sigma_{\varepsilon}$, which was then asserted during the importance
sampling of the wide fields.  The MBI team additionally experimented with
informative prior PDFs for the lensing shear, asserting the
shear components to have been drawn from a Gaussian distribution 
centred on zero with width $\sigma_g$.

The MBI team attempted no explicit calibration of any kind.  Finally, we note that their approach is general, and can easily be
attached to other shape measurement algorithms.

\subsection{MaltaOx}

The Malta-Oxford team based their measurements on the \code{lensfit}
algorithm \citep{2013MNRAS.429.2858M}.
This method measures the likelihood of PSF-convolved galaxy
models fitted to the pixel data for individual galaxies, adopting a
Bayesian marginalisation over nuisance parameters but using a
frequentist likelihood estimate of ellipticity for each 
galaxy.  Shear for the constant-shear branches was estimated from the
weighted mean of galaxy ellipticity values. 

 The galaxy models 
were two-component exponential disk plus de Vaucouleurs bulge, with
fixed relative ellipticity and scale length.  The 
galaxy position, scale length, total flux and 
bulge fraction were nuisance parameters.  For GREAT3, the priors for the
marginalisation over galaxy scale length  were obtained
by running \code{lensfit} on the GREAT3 deep data,
and fitting a lognormal distribution to the measured scale
lengths, accounting for the ellipticity-dependent size cut
(App.~\ref{app:intrinsic-pe}) in 
the fitting process. The ellipticity prior was similarly derived from
\code{lensfit} fits to the GREAT3 deep data, although it 
only enters the final shear estimate as part of the weight function. The individual galaxy weight is
an inverse variance weight, defining the variance as the quadrature sum
of ellipticity measurement variance and 
shape noise (see \citealp{2013MNRAS.429.2858M} for 
details).

For CGC and RGC, where noise-free
PSFs were provided, the MaltaOx team used a modified version of the \code{lensfit} PSF
modeling code to convert the nine images for each subfield into
a single oversampled PSF model in a pixel basis set.  In the one
variable PSF branch that they entered, they used the most
recent \code{lensfit} PSF modellng code without modification.  However, the 
data format required many modifications to work with
this code, so they lacked time to optimise the assumed scale length of
variation of the PSF.

When used for CFHTLenS \citep{2013MNRAS.432.2433H}, noise
bias was calibrated using simulations that matched the 
observations.  For GREAT3, the MaltaOx team wanted to test a new
self-calibration method (to be described in a future paper), integral to the likelihood measurement
process, that does not rely on external data or simulations.  The
final MaltaOx submissions 
used this self-calibration method.  A final post-measurement step to
isotropise the weights, to remove $S/N$-dependent
orientation bias, was also applied.

\subsection{MegaLUT}\label{app:megalut}

MegaLUT uses a supervised machine learning technique to estimate galaxy shape
parameters by measuring the PSF-convolved, noisy galaxy 
images. The method can be seen as a detailed empirical
calibration of {\em a priori} inaccurate shape measurement algorithms, such
as raw moments of the observed galaxy image. The 
distinctive feature of MegaLUT is to completely leave it to the
machine learning algorithm to ``deconvolve'' and correct crude shape
measurements for the effects of the PSF and for noise bias, instead of calibrating
only the residual biases of {\em a priori} more accurate techniques. In this
way, the input to the machine learning algorithm is close to the recorded
information of each galaxy, avoiding potential information loss
from deconvolutions. A further advantage of this approach is
its very low computational cost, due to the use of simple shape
measurements.

\subsubsection{MegaLUT implementation for GREAT3}

To build the learning samples on which MegaLUT is trained for GREAT3,
the MegaLUT team used simple \sersic\ profiles to represent the
galaxies. They can therefore train the algorithm to directly predict
the \sersic\ profile parameters, in  
particular the ellipticity. For
branches with constant known PSFs, this training was performed 
separately for each PSF. The measurements are based on
\sextractor\ \citep{1996A&AS..117..393B}, the adaptive moments
implemented in \galsim\ \citep{2014arXiv1407.7676R, 2003MNRAS.343..459H}, and, for
some submissions, on moments of the discrete autocorrelation function
\citep[ACF,][]{1997A&A...317..303V}. The most fundamental change 
in MegaLUT with respect to its implementation for GREAT10
\citep[described in][]{2012A&A...544A...8T} is the machine learning
itself. MegaLUT now uses feed-forward artificial neural networks
(ANNs), which are trained interchangeably via the SkyNet
\citep{2014MNRAS.441.1741G} or FANN \citep{nissen03} 
implementations.  The method works in effectively the same way for
control and realistic galaxy branches, and for ground- and space-based
branches. 

For multiepoch branches, the MegaLUT team
coadded the images with the stacking algorithm
provided by the GREAT3 EC. For the pre-deadline
submissions, the coaddition process was not simulated in the learning
sample, and MegaLUT could therefore not learn about related
biases, which will be the subject of further work. Regarding the
variable PSF branches, the MegaLUT team developed an approach that
incorporates PSF interpolation into the machine learning. In
essence, the galaxy position is included as an input
to the ANN, which is trained using PSFs at various locations. Prior to
the deadline, this treatment of variable PSF 
branches was not sufficiently mature to be used as a proof of concept
of this novel approach.

\newtext{The MegaLUT team submissions do not use the deep datasets, and do not
weight the per-galaxy shear estimators, aside from rejections
following simple criteria. The time per galaxy listed in
Table~\ref{T:methods} for this method, 20~ms, includes the overhead
involved in generating a typical-sized training dataset as well as the
training of the ANN.  However, once the ANN has been trained, 
the shear estimation per galaxy takes roughly 3~ms.}

\subsubsection{Differences between submissions}

Multiple submissions within a branch differ in 
the learning sample generation, the shape measurement, the
selection of ANN input parameters, the ANN architecture,
and the rejection of faint or unresolved galaxies. 
The distribution of shape parameters of the learning sample does
not have to closely mimic the ``observations'', as it does not act as
prior. For those parameters that do affect the shape measurement
output, the distributions used to generate the learning sample merely
define the region in parameter space over which the machine learning
can perform an accurate regression.

\subsection{MetaCalibration}

The philosophy behind the MetaCalibration method is
that since shear systematics depend on the galaxy
population and PSF model, all shear
systematics corrections should be determined directly 
from the images themselves (rather than from
independently-generated simulations).   In practical terms, the 
method involves constructing a model 
of the image with the shear as a
parameter. Varying the shear parameter allows a direct 
measurement of the shear response from
the difference between pipeline outputs with and without the
additional shear.  For GREAT3, this team using re-Gaussianation as the
shear estimation method, but in principle MetaCalibration could be
used for any method.

In detail, inspired by \cite{2000ApJ...537..555K}, the MetaCalibration
team constructs the
model sheared image by deconvolving the 
original image by the PSF model, applying a small shear to the
deconvolved image, and then convolving the result with a slightly
enlarged version of the original PSF. The final, lossy step is
required because the applied shear moves noisy modes inside the PSF
kernel window; reconstructing a sheared version of the
original image would require access to information on scales
hidden by the original PSF. The measured sensitivity is correct for
the version of the image with the enlarged PSF, so the final shear
measurements are performed on the reconvolved image, with an enlarged
PSF but no applied shear. This procedure should allow us to measure
shear calibration biases for any shear measurement pipeline; for
GREAT3, the MetaCalibration team used the \galsim\ implementation 
of re-Gaussianization, but the approach could be 
applied to self-calibrate any other shear estimation method.

Since the per-object response is quite noisy, using a per-object
response or even a per-image mean over 10000 
galaxies proved unstable. The entire set of images for a given
branch was used to  model the shape of the likelihood
curve and derive the shear response.

This approach was used to directly
calibrate out multiplicative systematics from the data.  An extension
of the method to remove additive bias 
was not implemented before the end of the challenge. 
Also, the anisotropic correlated noise in the images with added shear
was not whitened or made four-fold symmetric; there are plans to test the
effects of this limitation as well, with an updated version of
\galsim\ that can impose symmetry on the final noise
field.

\subsection{Wentao\_Luo}

This team used an independent implementation of the re-Gaussianization
method \citep{2003MNRAS.343..459H}.  Given the choice of applying the
PSF dilution correction to the re-Gaussianized
image or the version after application of a rounding kernel, they used
the latter as it was found during tests on STEP2 \citep{2007MNRAS.376...13M} images to
give better performance.  For the rounding kernel, a $5\times 5$
kernel was constructed following \cite{2002AJ....123..583B}.

Due to convergence issues, only $\sim 60$ per cent of the galaxies had
estimated shapes, and a further size cut reduced the number to $\sim
30$ per cent, resulting in quite noisy submitted results.

Submissions were made using two weighting schemes.  The first, from
\cite{2005MNRAS.361.1287M}, is inverse variance weighting using the
quadrature sum of shape noise and measurement error due to pixel
noise.  The second is an ellipticity-dependent weight from
\cite{2002AJ....123..583B} ($w=1/\sqrt{e^2+2.25\sigma_e^2}$, using
the measurement error due to pixel noise).  The former led to
better results than the latter, by roughly a factor of $\sim 2$ in $Q$
score. 

\newtext{The shear responsivity (to convert from distortion to shear) was
calculated as in \cite{2002AJ....123..583B}, and no additional
calibration factors were applied.}

\subsection{ess}\label{app:ess}

The ess team implemented the Bayesian model-fitting (BMF) shear measurement algorithm introduced by
\cite{2014MNRAS.438.1880B}.  For general details about the
implementation\footnote{\url{https://github.com/esheldon/ngmix}}, see \cite{2014MNRAS.444L..25S}.  The only details of importance that are not
in \cite{2014MNRAS.444L..25S} are about PSF fitting, prior determination and choice of
models.

For constant PSF branches, the ess team fit three unconstrained 
Gaussians to one of the provided PSF images using an Expectation Maximization (EM)
algorithm chosen for its high level of stability.
\newtext{For subfields without strong defocus, the residual of the model with
the PSF was typically consistent either with random noise or had a triangular
shape perhaps due to trefoil in the PSF (which cannot easily be represented by
the adopted PSF model).  In fields with strong defocus, the residuals
were quite bad; see Sec.~\ref{subsec:outliers} for a further discussion
of this point.}

A number of different galaxy models were used, including full \sersic\ profiles, but the
best performing on the realistic galaxy branches was a simple exponential disk.
  \newtext{The fits were carried out using the full $48\times 48$ postage stamps. } 
Fits to the deep field images were used to estimate priors on the size and flux.
The joint size-flux distribution averaged over all deep fields in the
branch was then parametrized by sums of
Gaussians, again fit using an EM
algorithm.

For ellipticity, the ess team tried fitting the deep fields and using the galaxy
model fits provided by the GREAT3 team based on fitting the
COSMOS HST data at full resolution to a \sersic\ model
\citep{2012MNRAS.421.2277L}.  The latter approach led to better
results than the former.

Because the \cite{2014MNRAS.438.1880B} algorithm breaks down at high
shear, the ess team iterated the solution
on the constant-shear fields, expanding the Taylor series about the result
from the previous iteration. In the absence of additive errors, this iteration
converges in three iterations even for $\sim 10$ per cent shears, but since
the results did have some 
additive bias, full convergence was not possible.

The ess team worked primarily with the realistic galaxy branch because 
performance on the control branch was rather poor.  Their estimates of
galaxy properties on the deep fields for the
CGC branch suggested a strong variation in their statistical
properties both within the branch and compared to RGC.   Priors are crucial for
\cite{2014MNRAS.438.1880B}, 
and this variation may
have resulted in poor performance.  The RGC deep fields
seemed more
uniform in their properties according to this team's analysis.
An analysis after the fact using the truth tables showed that the
atmospheric PSF FWHM for the deep fields in both branches had the same
mean value, but a dispersion of 0.12\arcsec\ vs.\ 0.08\arcsec\ for CGC
and RGC, supporting the claim that the deep fields in CGC exhibited
more variation than in RGC.

\subsection{sFIT}\label{app:sfit}

The sFIT (shapes from iterative training) method is a set of principles to use simulations to characterize systematic errors in shear estimation. The principles of the
method are:
\begin{enumerate}[(a)]
\item Shear estimation consists of two steps: initial ellipticity
  estimation (which must be highly repeatable) and application of calibration.
\item Shear calibration is derived via image simulation.
\item Simulated galaxies must have properties matching those
  in real data (in this case, the GREAT3 data).
\item Each step in image processing affects the calibration
  factor. This includes image coaddition, PSF estimation and
  interpolation, handling of under-sampling, etc.
\end{enumerate}

\newtext{A more detailed description will be presented in Jee \& Tyson (in prep.).}

\subsubsection{Implementation of the sFIT Method}
 
For shear calibration using image simulations, the three important
questions are: 1) How well 
does the simulation match reality? 2) How far can the galaxy model be
simplified? (i.e., minimization of the number of calibration 
parameters), and 3) What is the requirement for the initial
ellipticity measurement method?

\textbf{Initial ellipticity measurement:} The sFIT team uses forward-modeling to
obtain the initial ellipticity estimate for each galaxy, by 
convolving the galaxy model with the PSF and minimizing the difference
between the simulated and actual galaxy image. The choice of
galaxy model is important. The sFIT team experimented with a wide range of galaxy
models, examining their stability (convergence rate), speed, bias, and
measurement noise. Perhaps the simplest parametrization is an
elliptical Gaussian as used in the Deep Lens Survey
\citep[DLS,][]{2013ApJ...765...74J}. The strength of this model
includes the high convergence rate, speed, and small measurement
error. The drawback is that it requires rather a large
calibration factor, of order 10 per cent.  Although in principle 
 a calibration factor can be derived for this choice, it is preferable
if the corrections that are being applied are small. 
Another option is the bulge$+$disk
model, which may be regarded as the opposite extreme to the elliptical
Gaussian approach. This sophisticated representation
of galaxy profiles reduces the bias, but with an unacceptably poor
convergence rate (fails for $\sim 20$ per cent of the GREAT3
galaxies) and slow speed ($\sim$10 sec per object). The
increase in the number of parameters 
also increases noise bias for faint galaxies. The compromise
that was adopted for GREAT3 is a single \sersic\ representation, which
is a 
one-parameter extension to the elliptical Gaussian model used for DLS. Without
any external calibration, the model introduces a
reasonably small multiplicative bias ($\sim ~2$ per cent).  The model
converges 
$\sim 98$ per cent of the time, and takes $\sim 1$~second per galaxy.

\textbf{Image simulation method:} 
The sFIT team used \galsim\ to perform its image simulations. Although the team
already has a high-fidelity image simulator used for DLS, there are
merits in using \galsim\ for the GREAT3 challenge. First,
the GREAT3 data are generated with \galsim.  Were \galsim\ to
make some unknown systematic error when representing galaxies under
shear, the potential impact on competitive performance is best minimized by
using the same simulator to make images (while the scientific value in
identifying a discrepancy is, unfortunately, sacrificed).

Second, for the real galaxy branches, it is important to
match the galaxy properties. This team's DLS image simulator uses galaxy
images in the Ultra Deep Field (UDF), which detects faint galaxies
down to 30th mag at the $10\sigma$ level. Clearly, these galaxies are
different from those in GREAT3. 

The sFIT team used \sersic\ fits to the GREAT3 data to estimate distributions of
galaxy sizes, ellipticity, \sersic\ indices, PSF properties, and noise
level. Then, they ran \galsim\ with input parameters based on these
measurements \newtext{ by drawing values from parametrized distributions. }
 It is not trivial to guess the input parameters that will 
generate images that closely match the GREAT3 data, since the noise in
the GREAT3 data means that  the observed distributions deviate
\newtext{from the true inputs (they are wider than the inputs, with shifted means). Several iterations
were required before the mean, width, and tail shape of the distribution agreed well with the observed one.}

\textbf{Calibration:} 
Many details such as properties of galaxies and PSFs,
method of image reduction, implementation details of ellipticity
measurement, noise level, etc. all affect shear
calibration. However, for practical purposes, the number of
parameters in the calibration process must be limited.
The sFIT team avoided calibration against implementation details by
keeping the size of the
postage stamp images, the over-sampling ratio, the centroid constraint
method, etc. fixed throughout the challenge.

The galaxy properties are important parameters. However, individual
measurements are noisy. Thus, instead of a per-galaxy correction based
on each galaxy's properties, shear calibrations were derived based on
aggregate statistics and applied to the entire population (an exception is made
for variable shears; see below).

The most important parameters are the PSF properties such as
ellipticity, size, kurtosis, etc. Even with perfect
knowledge of PSF, galaxy ellipticities still have both 
additive and multiplicative bias, which increases with the size of
the PSF. In their GREAT3 analysis, the sFIT team ignored kurtosis and characterize
the PSF in terms of its ellipticity and FWHM. They modeled the
variation of both additive and multiplicative errors as a function of
PSF FWHM using second-order polynomials. Variable shear branches do
require a 
per-galaxy correction using the PSF
properties at the galaxy location to estimate the correction
factors (but not using the individual galaxy-fitting results).

\textbf{Weighting Scheme:} The ellipticity measurement code used by
the sFIT team outputs
ellipticity uncertainties by evaluating the Hessian matrix. Unfortunately,
these ellipticity uncertainties are somewhat correlated
with galaxy shapes, so if the ellipticity
uncertainties are used directly to evaluate individual weights, the
shapes would be correlated with the weights. To avoid this problem,
the sFIT team derived average $S/N$
vs.\ ellipticity uncertainty relations, and converted per-galaxy $S/N$ values into
ellipticity uncertainties. Then, the weights are evaluated from the
equation $w=1/(\sigma_e^2 + \sigma_\text{SN}^2)$, where $\sigma_e$ is the ellipticity uncertainty
derived from the $S/N$ value, and $\sigma_\text{SN}$ is the intrinsic ellipticity dispersion per
component.

\subsubsection{GREAT3 submission policy}

To avoid tuning to the GREAT3 simulations in too much detail, the
sFIT team 
tried to minimize the number of submissions.  Submissions were
made in the following cases:

\begin{itemize}
\item When obvious mistakes were found, such as applying calibration
  factors to the wrong branch.
\item When better calibrations become available. Since 
  shear calibration requires significant computing time, occasionally the
  sFIT team took 
  shortcuts to reduce computing
  time. However, if this shortcut resulted in poor performance, 
  they revisited the problem and performed brute-force
  simulations to obtain calibration parameters directly.
\item For many variable shear branches, results 
  improved when galaxies are unweighted. Thus, the sFIT team experimented
  with their weighting scheme (by turning on/off) for almost every
  variable shear branch (except for VSV, where they
  achieved the highest score with just one submission).
\end{itemize}

\section{Cross-branch comparison of submissions}\label{app:tables}
Tables~\ref{tab:mc_general_ground}~and~\ref{tab:mc_general_space}
provide 
estimates of $c_+$ and the component-averaged $\langle m\rangle$ for all submissions described in
Sec.~\ref{sec:what_results} in branches CGC, RGC, CSC, and RSC.
\newtext{Tables~\ref{tab:mc_deltas} and~\ref{tab:mc_psf_deltas} show the
changes in $c_+$ and $\langle m\rangle$ when comparing across branches
and within branches while splitting by PSF properties, respectively.}

\begin{table*}
  \begin{center}
    \caption{\label{tab:mc_general_ground} Additive bias $c_+$ and
      component-averaged multiplicative bias $\langle m \rangle$ for the 
      submissions selected for the fair cross-branch comparison
      (see Sec.~\ref{sec:what_results}) in ground branches CGC and RGC.}
    \input{mc_table_general_ground}

  \end{center} {\small $^1$Outlying values in the submitted shears
    were removed from the submission and scores recalculated, as described
    in Sec.~\ref{subsec:outliers}.} \\
  {\small $^2$The worst 10 per cent of fields by PSF defocus value were
    removed and scores recalculated, as described in Sec.~\ref{subsec:outliers}.}
\end{table*}

\begin{table*}
  \begin{center}
    \caption{\label{tab:mc_general_space} Additive bias $c_+$ and
      component-averaged multiplicative bias $\langle m \rangle$ for the submissions
      selected for the fair cross-branch comparison
      (see Sec.~\ref{sec:what_results}) in space branches CSC and RSC.}
    \input{mc_table_general_space}

  \end{center}
\end{table*}

\begin{table*}
  \begin{center}
    \caption{\label{tab:mc_deltas}\newtext{ Change in additive bias $\Delta c_+$ and
      component-averaged multiplicative bias $\Delta \langle m
      \rangle$ across branches, for the submissions
      selected for the fair cross-branch comparison.  The ordering of branch
      labels indicates the order in which the bias results are subtracted.}}
    \input{mc_table_deltas}

  \end{center}
\end{table*}

\begin{table*}
  \begin{center}
    \caption{\label{tab:mc_psf_deltas} \newtext{Change in additive bias $\Delta c_+$ and
      component-averaged multiplicative bias $\Delta \langle m
      \rangle$ within CGC, when splitting by atmospheric PSF FWHM and optical PSF defocus, for the submissions
      selected for the fair cross-branch comparison.}}
    \input{mc_table_psf_deltas}

  \end{center}
\end{table*}

\end{document}

%% file: mc_table_general_ground.tex
\begin{tabular}{lcccc}
\hline\hline
~ & CGC & CGC & RGC & RGC \\
Team & $10^4 \, c_{+}$ & $10^3 \, \langle m \rangle$& $10^4 \, c_{+}$ & $10^3 \, \langle m \rangle$\\
\hline
Amalgam@IAP & $5.5 \pm 1.1$ & $8.4 \pm 2.6$ & $3.0 \pm 0.8$ & $1.4 \pm 2.3$ \\
CEA\_denoise & $109.6 \pm 54.6$ & $-80.2 \pm 153.0$ & $16.1 \pm 5.6$ & $-24.2 \pm 14.7$ \\
CEA-EPFL & $1.4 \pm 1.2$ & $-4.9 \pm 3.1$ & $-4.5 \pm 1.0$ & $14.3 \pm 2.9$ \\
CMU experimenters & $1.0 \pm 0.7$ & $6.2 \pm 2.1$ & $0.9 \pm 0.8$ & $8.3 \pm 2.9$ \\
COGS & $-11.0 \pm 1.3$ & $-1.0 \pm 3.2$ & $-7.2 \pm 1.2$ & $-14.5 \pm 3.3$ \\
E-HOLICs & $73.3 \pm 7.7$ & $139.8 \pm 16.8$ & --- & ---  \\
EPFL\_HNN & $11.9 \pm 5.9$ & $59.8 \pm 16.6$ & $11.9 \pm 68.3$ & $-807.0 \pm 195.0$ \\
EPFL\_KSB & $6.2 \pm 1.8$ & $27.6 \pm 4.7$ & --- & ---  \\
EPFL\_MLP & $87.3 \pm 7.0$ & $-553.3 \pm 16.0$ & $-3.7 \pm 4.0$ & $-977.9 \pm 11.9$ \\
ess & --- & ---  & $2.1 \pm 6.4$ & $-6.3 \pm 22.9$ \\
ess (outlier clipped$^1$) & --- & ---  & $4.2 \pm 1.4$ & $24.3 \pm 3.7$ \\
Fourier\_Quad & $5.0 \pm 2.4$ & $1.1 \pm 6.5$ & $1.9 \pm 1.9$ & $-11.0 \pm 5.4$ \\
FDNT & $82.0 \pm 11.8$ & $-665.5 \pm 30.3$ & $92.8 \pm 14.0$ & $-500.3 \pm 64.8$ \\
MaltaOx & $7.7 \pm 1.3$ & $-6.3 \pm 3.3$ & $2.2 \pm 0.9$ & $-0.3 \pm 2.7$ \\
MBI & $-2.9 \pm 5.7$ & $18.2 \pm 16.8$ & $-13.5 \pm 6.6$ & $44.5 \pm 29.1$ \\
MBI (outlier clipped$^2$) & $1.9 \pm 2.1$ & $5.6 \pm 6.6$ & $-16.0 \pm 3.5$ & $85.3 \pm 10.8$ \\
MegaLUT & $-7.7 \pm 2.3$ & $2.6 \pm 5.3$ & $-12.9 \pm 1.8$ & $16.2 \pm 4.7$ \\
MegaLUT (outlier clipped$^2$) & $-9.7 \pm 1.5$ & $9.6 \pm 3.9$ & $-11.0 \pm 1.6$ & $19.6 \pm 4.3$ \\
MetaCalibration & $16.2 \pm 3.5$ & $2.1 \pm 8.0$ & --- & ---  \\
re-Gaussianization & $-13.8 \pm 1.5$ & $43.6 \pm 4.0$ & $-5.7 \pm 1.1$ & $3.6 \pm 3.5$ \\
sFIT & $-1.1 \pm 1.2$ & $1.5 \pm 3.2$ & $1.1 \pm 1.2$ & $0.7 \pm 3.3$ \\
Wentao Luo & $-33.8 \pm 11.7$ & $-56.6 \pm 28.6$ & $-34.2 \pm 6.5$ & $-73.6 \pm 20.0$ \\
\hline
\end{tabular}

%% file: mc_table_general_space.tex
\begin{tabular}{lcccc}
\hline\hline
~ & CSC & CSC & RSC & RSC \\
Team & $10^4 \, c_{+}$ & $10^3 \, \langle m \rangle$& $10^4 \, c_{+}$ & $10^3 \, \langle m \rangle$\\
\hline
Amalgam@IAP & $-0.7 \pm 0.5$ & $-0.5 \pm 1.4$ & $1.1 \pm 0.6$ & $-7.3 \pm 1.6$ \\
CEA\_denoise & $128.7 \pm 42.8$ & $-409.6 \pm 120.4$ & $111.6 \pm 35.5$ & $-358.0 \pm 98.6$ \\
CEA-EPFL & $3.2 \pm 0.7$ & $-3.0 \pm 1.9$ & $3.5 \pm 0.6$ & $3.8 \pm 1.5$ \\
E-HOLICs & $101.4 \pm 12.6$ & $-21.3 \pm 31.3$ & $82.7 \pm 10.2$ & $24.5 \pm 24.5$ \\
EPFL\_HNN & $64.2 \pm 7.9$ & $-176.1 \pm 20.1$ & $52.8 \pm 7.2$ & $-177.3 \pm 17.6$ \\
EPFL\_KSB & $58.4 \pm 54.0$ & $-163.1 \pm 129.6$ & --- & ---  \\
EPFL\_MLP & $1.0 \pm 7.4$ & $-992.4 \pm 19.2$ & --- & ---  \\
Fourier\_Quad & $-0.4 \pm 1.6$ & $1.3 \pm 4.4$ & $1.8 \pm 1.5$ & $3.7 \pm 4.1$ \\
MBI & $-3.5 \pm 5.4$ & $-27.4 \pm 15.2$ & --- & ---  \\
MegaLUT & $-0.3 \pm 1.6$ & $-15.1 \pm 4.5$ & $9.2 \pm 1.4$ & $-32.3 \pm 3.6$ \\
sFIT & $4.5 \pm 0.9$ & $0.1 \pm 2.2$ & $5.3 \pm 0.9$ & $-1.2 \pm 2.2$ \\
Wentao Luo & $34.6 \pm 19.9$ & $-1041.6 \pm 56.5$ & $115.5 \pm 20.5$ & $-328.2 \pm 49.6$ \\
\hline
\end{tabular}

%% file: mc_table_deltas.tex
\begin{tabular}{lcccccc}
\hline\hline
~ & \multicolumn{2}{c}{RGC $-$ CGC} & \multicolumn{2}{c}{RSC $-$ CSC} & \multicolumn{2}{c}{CSC $-$ CGC} \\
Team & $10^4 \, \Delta c_{+}$ & $10^3 \, \Delta \langle m \rangle$& $10^4 \, \Delta c_{+}$ & $10^3 \, \Delta \langle m \rangle$& $10^4 \, \Delta c_{+}$ & $10^3 \, \Delta \langle m \rangle$\\
\hline
Amalgam@IAP & $-2.5 \pm 1.3$ & $-7.0 \pm 3.5$ & $1.9 \pm 0.8$ & $-6.8 \pm 2.1$ & $-6.2 \pm 1.2$ & $-8.9 \pm 3.0$ \\
CEA\_denoise & $-93.5 \pm 54.9$ & $56.0 \pm 153.7$ & $-17.1 \pm 55.6$ & $51.5 \pm 155.6$ & $19.2 \pm 69.3$ & $-329.4 \pm 194.7$ \\
CEA-EPFL & $-5.9 \pm 1.6$ & $19.2 \pm 4.2$ & $0.3 \pm 0.9$ & $6.8 \pm 2.5$ & $1.8 \pm 1.4$ & $1.9 \pm 3.7$ \\
CMU experimenters & $-0.1 \pm 1.1$ & $2.1 \pm 3.5$ & --- & ---  & --- & ---  \\
COGS & $3.8 \pm 1.7$ & $-13.5 \pm 4.6$ & --- & ---  & --- & ---  \\
E-HOLICs & --- & ---  & $-18.6 \pm 16.1$ & $45.9 \pm 39.7$ & $28.1 \pm 14.7$ & $-161.2 \pm 35.5$ \\
EPFL\_HNN & $-0.0 \pm 68.5$ & $-866.8 \pm 195.7$ & $-11.4 \pm 10.7$ & $-1.3 \pm 26.7$ & $52.3 \pm 9.9$ & $-235.8 \pm 26.1$ \\
EPFL\_KSB & --- & ---  & --- & ---  & $52.2 \pm 54.1$ & $-190.7 \pm 129.7$ \\
EPFL\_MLP & $-91.0 \pm 8.1$ & $-424.7 \pm 19.9$ & --- & ---  & $-86.3 \pm 10.2$ & $-439.1 \pm 25.0$ \\
Fourier\_Quad & $-3.1 \pm 3.1$ & $-12.1 \pm 8.4$ & $2.1 \pm 2.2$ & $2.4 \pm 6.0$ & $-5.4 \pm 2.9$ & $0.2 \pm 7.9$ \\
FDNT & $10.7 \pm 18.3$ & $165.2 \pm 71.5$ & --- & ---  & --- & ---  \\
MaltaOx & $-5.5 \pm 1.6$ & $5.9 \pm 4.3$ & --- & ---  & --- & ---  \\
MBI & $-10.6 \pm 8.7$ & $26.3 \pm 33.6$ & --- & ---  & $-0.7 \pm 7.8$ & $-45.6 \pm 22.7$ \\
MBI (outlier clipped$^2$) & $-17.9 \pm 4.1$ & $79.7 \pm 12.7$ & --- & ---  & $-11.2 \pm 5.9$ & $-36.5 \pm 17.6$ \\
MegaLUT & $-5.2 \pm 3.0$ & $13.5 \pm 7.1$ & $9.4 \pm 2.1$ & $-17.2 \pm 5.8$ & $7.5 \pm 2.9$ & $-17.7 \pm 6.9$ \\
MegaLUT (outlier clipped$^2$) & $-1.3 \pm 2.2$ & $10.0 \pm 5.8$ & $9.2 \pm 2.3$ & $-14.2 \pm 6.2$ & $9.5 \pm 2.3$ & $-26.6 \pm 6.3$ \\
re-Gaussianization & $8.0 \pm 1.9$ & $-40.0 \pm 5.3$ & --- & ---  & --- & ---  \\
sFIT & $2.2 \pm 1.7$ & $-0.9 \pm 4.6$ & $0.8 \pm 1.2$ & $-1.3 \pm 3.1$ & $5.6 \pm 1.5$ & $-1.4 \pm 3.9$ \\
Wentao Luo & $-0.5 \pm 13.4$ & $-17.0 \pm 34.9$ & $80.9 \pm 28.5$ & $713.4 \pm 75.2$ & $68.4 \pm 23.1$ & $-985.0 \pm 63.3$ \\
\hline
\end{tabular}

%% file: mc_table_psf_deltas.tex
\begin{tabular}{lcccc}
\hline\hline
~ & \multicolumn{2}{c}{better $-$ worse atmospheric PSF FWHM} & \multicolumn{2}{c}{better $-$ worse optical PSF defocus} \\
Team & $10^4 \, \Delta c_{+}$ & $10^3 \, \Delta \langle m \rangle$& $10^4 \, \Delta c_{+}$ & $10^3 \, \Delta \langle m \rangle$\\
\hline
Amalgam@IAP & $0.2 \pm 2.1$ & $-1.3 \pm 5.2$ & $-5.8 \pm 2.1$ & $9.6 \pm 5.2$ \\
CEA\_denoise & $-51.2 \pm 109.6$ & $97.9 \pm 307.4$ & $-168.5 \pm 108.7$ & $234.7 \pm 304.2$ \\
CEA-EPFL & $5.2 \pm 2.4$ & $-29.5 \pm 5.9$ & $-2.7 \pm 2.5$ & $17.8 \pm 6.2$ \\
CMU experimenters & $-0.3 \pm 1.5$ & $-10.2 \pm 4.2$ & $1.0 \pm 1.5$ & $-0.3 \pm 4.2$ \\
COGS & $-4.4 \pm 2.5$ & $5.0 \pm 6.4$ & $12.3 \pm 2.4$ & $4.1 \pm 6.3$ \\
E-HOLICs & $21.0 \pm 14.8$ & $-213.1 \pm 31.7$ & $-76.4 \pm 14.4$ & $-100.6 \pm 31.6$ \\
EPFL\_HNN & $1.3 \pm 11.8$ & $-148.8 \pm 32.3$ & $11.3 \pm 11.8$ & $-35.3 \pm 33.2$ \\
EPFL\_KSB & $10.4 \pm 3.6$ & $-39.6 \pm 9.0$ & $8.6 \pm 3.7$ & $-11.3 \pm 9.4$ \\
EPFL\_MLP & $5.9 \pm 14.0$ & $-60.5 \pm 31.8$ & $-67.0 \pm 13.2$ & $69.3 \pm 30.4$ \\
Fourier\_Quad & $0.1 \pm 4.9$ & $-14.4 \pm 12.8$ & $-7.5 \pm 4.8$ & $-6.4 \pm 13.0$ \\
FDNT & $22.2 \pm 23.6$ & $-11.9 \pm 60.2$ & $-75.0 \pm 23.0$ & $-73.8 \pm 59.5$ \\
MaltaOx & $7.3 \pm 2.5$ & $-13.8 \pm 6.4$ & $-8.4 \pm 2.5$ & $-1.3 \pm 6.5$ \\
MBI & $9.5 \pm 11.4$ & $1.8 \pm 33.6$ & $-0.9 \pm 11.4$ & $4.2 \pm 34.1$ \\
MegaLUT & $-1.4 \pm 4.7$ & $-6.4 \pm 10.4$ & $6.0 \pm 4.6$ & $18.7 \pm 10.5$ \\
MetaCalibration & $16.3 \pm 6.9$ & $-22.0 \pm 15.4$ & $-19.4 \pm 6.9$ & $30.2 \pm 15.7$ \\
re-Gaussianization & $-5.4 \pm 3.0$ & $7.9 \pm 8.0$ & $13.6 \pm 2.8$ & $-26.7 \pm 7.8$ \\
sFIT & $3.0 \pm 2.4$ & $-12.6 \pm 6.4$ & $-4.8 \pm 2.4$ & $13.7 \pm 6.4$ \\
Wentao Luo & $-2.7 \pm 23.4$ & $40.1 \pm 56.4$ & $26.5 \pm 23.2$ & $135.7 \pm 56.9$ \\
\hline
\end{tabular}